\documentclass[10pt]{iopart}

\usepackage{color}
\usepackage[dvipsnames,svgnames,table]{xcolor}
\usepackage[colorlinks=true,linkcolor=blue,urlcolor=blue,citecolor=blue]{hyperref}
\usepackage{diagbox}
\usepackage{siunitx}
\usepackage{graphicx}
\usepackage{array}
\usepackage{lipsum}
\usepackage{bm}
\usepackage{subfigure}
\usepackage{tabularx}
\usepackage{braket}
\usepackage{ulem}
\usepackage{verbatim}
\usepackage{orcidlink}
\usepackage{doi}
\usepackage{amssymb}
\usepackage{tablefootnote}
\usepackage{booktabs}
\usepackage{multirow}

\usepackage{geometry}

\usepackage{lipsum}
\usepackage{mathrsfs}

\usepackage[numbers,square,sort&compress]{natbib}

\usepackage{titlesec}
  \titlespacing*{\section}{0pc}{1em}{0em}

\newenvironment{noverticalspace}
 {%
  \par 
  \offinterlineskip 
 }
 {\par}

\begin{document}

\title[Roadmap RE-induced wall damage]{Runaway electron-induced plasma facing component damage in tokamaks}


\author{S.~Ratynskaia\,\orcidlink{0000-0002-6712-3625}$^1$, 
M.~Hoelzl\,\orcidlink{0000-0001-7921-9176}$^2$, 
E.~Nardon\,\orcidlink{0000-0003-0427-2292}$^3$,
P.~Aleynikov\,\orcidlink{0009-0002-3037-3679}$^2$,
F.J.~Artola\,\orcidlink{0000-0001-7962-1093}$^4$,
V.~Bandaru\,\orcidlink{0000-0003-4096-1407}$^5$,
M.~Beidler\,\orcidlink{0000-0002-7385-3886}$^6$,
B.~Breizman\,\orcidlink{0000-0002-7908-6497}$^7$,
D.~del-Castillo-Negrete\,\orcidlink{0000-0001-7183-801X}$^7$,
M.~De Angeli\,\orcidlink{0000-0002-7779-7842}$^8$,
V.~Dimitriou\,\orcidlink{0000-0003-4823-0350}$^{9}$,
R.~Ding\,\orcidlink{0000-0003-2880-9736}$^{10}$,
J.~Eriksson\,\orcidlink{0000-0002-0892-3358}$^{11}$,
O.~Ficker\,\orcidlink{0000-0001-6418-9517}$^{12}$,
R.S.~Granetz\,\orcidlink{0000-0002-6560-1881}$^{13}$,
E.~Hollmann\,\orcidlink{0000-0002-6267-6589}$^{14}$,
M.~Hoppe\,\orcidlink{0000-0003-3994-8977}$^1$,
M.~Houry\,\orcidlink{0009-0002-5349-6743}$^{3}$,
I.~Jepu\,\orcidlink{0000-0001-8567-3228}$^{15}$,
H.R.~Koslowski\,\orcidlink{0000-0002-1571-6269}$^{16}$,
C.~Liu\,\orcidlink{0000-0002-1571-6269}$^{17}$,
J.R.~Martin-Solis\,\orcidlink{0000-0003-0458-4405}$^{18}$,
G.~Pautasso$^2$,
Y.~Peneliau,\orcidlink{0000-0002-4282-9621}$^{3}$,
R.A.~Pitts\,\orcidlink{0000-0001-9455-2698}$^{4}$,
G.I.~Pokol\,\orcidlink{0000-0003-1473-0736}$^{19}$,
C.~Reux\,\orcidlink{0000-0002-5327-4326}$^{3}$,
U.~Sheikh\,\orcidlink{0000-0001-6207-2489}$^{20}$,
S.A.~Silburn\,\orcidlink{0000-0002-3111-5113}$^{15}$,
T.~Tang\,\orcidlink{0000-0002-2898-7658}$^{10}$,
R.A.~Tinguely\,\orcidlink{0000-0002-3711-1834}$^{13}$,
P.~Tolias\,\orcidlink{0000-0001-9632-8104}$^1$,
E.~Tomesova\,\orcidlink{0000-0002-0381-9244}$^{12}$,
R.~Villari\orcidlink{0000-0001-7972-1676}$^{21}$
}

\address{
$^1$KTH Royal Institute of Technology, Stockholm, SE-100 44, Sweden
$^2$Max Planck Institute for Plasma Physics, Garching b. M. and Greifswald, Germany
$^3$CEA, IRFM, F-13108 Saint-Paul-lez-Durance, France
$^4$ITER Organization, Saint Paul Lez Durance Cedex, France
$^5$Department of Mechanical Engineering, Indian Institute of Technology Guwahati, India
$^6$Fusion Energy Division, Oak Ridge National Laboratory, Oak Ridge, TN 37831, USA
$^7$Institute for Fusion Studies, The University of Texas at Austin, USA
$^8$Institute for Plasma Science and Technology - CNR, Milan, Italy
$^{9}$Institute of Plasma Physics and Lasers - IPPL, University Research and Innovation Centre, Hellenic Mediterranean University, Rethymno, GR-74150 Greece
$^{10}$Institute of Plasma Physics, HFIPS, Chinese Academy of Sciences, Hefei 230031, China
$^{11}$Department of Physics and Astronomy, Uppsala University, SE-75237 Uppsala, Sweden
$^{12}$Institute of Plasma Physics of the CAS, CZ-18200 Praha 8, Czech Republic
$^{13}$Plasma Science and Fusion Center, Massachusetts Institute of Technology, Cambridge, MA, USA
$^{14}$University of California - San Diego. San Diego, CA, 92093 USA
$^{15}$UKAEA, Culham Campus, Abingdon, OX14 3DB, UK
$^{16}$Forschungszentrum J\"ulich GmbH, Institute of Fusion Energy and Nuclear Waste Management -- Plasma Physics, 52425 J\"ulich, Germany
$^{17}$State Key Laboratory of Nuclear Physics and Technology, School of Physics, Peking University, Beijing 100871, China
$^{18}$Universidad Carlos III de Madrid, Avenida de la Universidad 30, 28911-Madrid, Spain
$^{19}$Department of Nuclear Techniques, Faculty of Natural Sciences, Budapest University of Technology and Economics, 1111-Budapest, Hungary
$^{20}$Swiss Plasma Center (SPC), Ecole Polytechnique Federale de Lausanne (EPFL), CH-1015 Lausanne, Switzerland
$^{21}$ENEA, Department of Fusion and Technology for Nuclear Safety and Security, Frascati (Rome), Italy}

\ead{srat@kth.se, 
matthias.hoelzl@ipp.mpg.de,
eric.nardon@cea.fr}

\vspace{10pt}

\clearpage
\begin{abstract}
This Roadmap article addresses the critical and multifaceted challenge of plasma-facing component (PFC) damage caused by runaway electrons (REs) in tokamaks, a phenomenon that poses a significant threat to the viability and longevity of future fusion reactors such as ITER and DEMO. The dramatically increased RE production expected in future high-current tokamaks makes it very difficult to avoid or mitigate REs in such devices when a plasma discharge terminates abnormally. Preventing damage from the intense localised heat loads they can cause requires a holistic approach that considers plasma, REs and PFC damage. Despite decades of progress in understanding the physics of REs and the thermomechanical response of PFCs separately, their complex interplay remains poorly understood. This document aims to initiate a coordinated, interdisciplinary approach to bridge this gap by reviewing experimental evidence, advancing diagnostic capabilities, and improving modelling tools across different scales, dimensionalities, and fidelities. Key topics include RE beam formation and transport,  damage mechanisms in both brittle and metallic PFCs, and observed effects in major facilities such as JET, DIII-D, WEST and EAST. The Roadmap emphasises the urgency of predictive, high-fidelity modelling validated against well-diagnosed controlled experiments, particularly in the light of recent changes in ITER's wall material strategy and the growing importance of private sector fusion initiatives. Each section of the Roadmap article is written to provide a concise overview of one area of this multidisciplinary subject, with an assessment of the status, a look at current and future challenges, and a brief summary. The ultimate goal of this initiative is to guide future mitigation strategies and design resilient components that can withstand the intense localised loads imposed by REs, thus ensuring the safe and sustainable operation of the next generation of fusion power plants.

\end{abstract}

\ioptwocol
%
%
%
%
%


\section*{Introduction}
\author{S. Ratynskaia$^1$, M. Hoelzl$^2$, E. Nardon$^3$ }
\address{
$^1$KTH Royal Institute of Technology, Stockholm, SE-100 44, Sweden\\
$^2$Max Planck Institute for Plasma Physics, Boltzmannstr. 2, 85748 Garching b. M, Germany\\
$^3$IRFM, CEA Cadarache, F-13108 Saint-Paul-lez-Durance, France}

Back in 2001, a comprehensive review on plasma-wall interactions by Gianfranco Federici and coworkers pointed out that\,\cite{Federici_2001}: ``Although in recent years significant progress has been made in characterizing runaway electrons generation, there are still uncertainties in the quantification of material damage". More than twenty years later, the latter remains valid: runaway electron (RE) induced damage on plasma facing components (PFCs) is still insufficiently understood. Meanwhile, the topics of RE generation \& transport and thermomechanical PFC response have independently matured to a large degree. The ambition of the present roadmap is to initiate a coordination between the practitioners in these two distinct topics that will lead to a deeper understanding of the complex physics and multi-faceted consequences of RE impacts on PFCs. 

\begin{figure}
    \centering
    \begin{noverticalspace}
    \includegraphics[width=\linewidth]{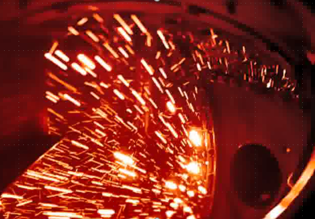} \\
    \includegraphics[width=\linewidth]{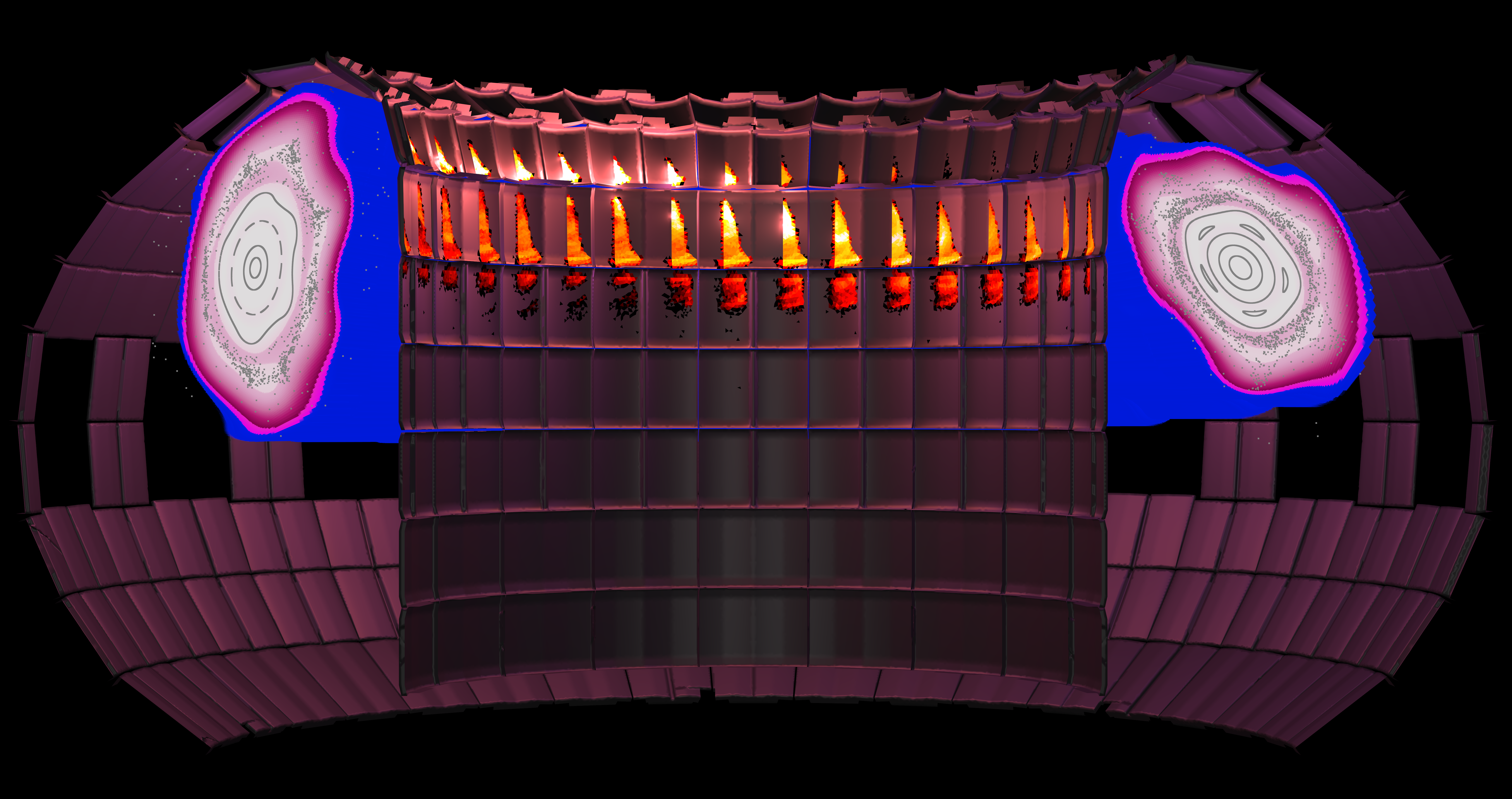} \\
    \includegraphics[width=\linewidth]{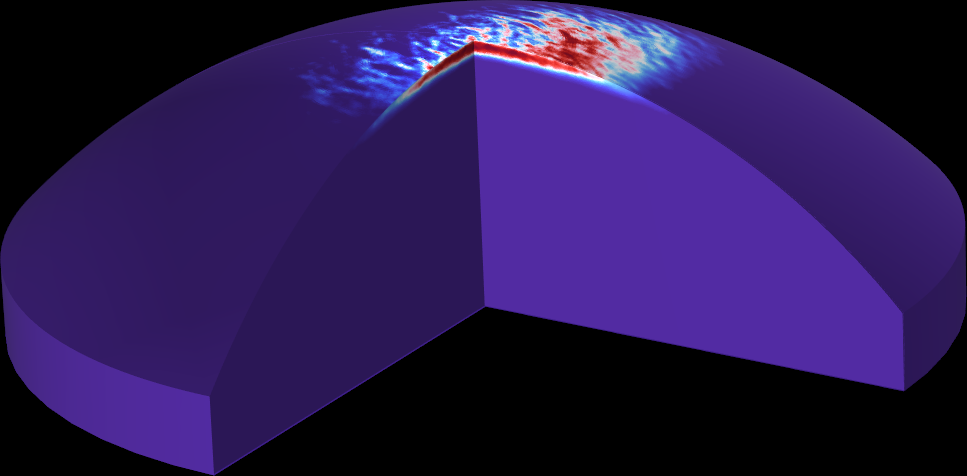}
    \end{noverticalspace}
    \caption{\textbf{Top:} Visible camera image of the material ejected from a metal wall in Alcator C-Mod as a result of a RE beam impact on the plasma facing components. Courtesy of R.~Granetz. 
    \textbf{Middle:} JOREK simulation of a RE beam termination and the resulting load to plasma facing components during a RE beam termination in ITER~\cite{Bandaru_2024,Bergstrom_2024}. Courtesy of H.~Bergstr{\"o}m.
    \textbf{Bottom:} Brittle failure calculations in a graphite PFC exposed to a strong RE beam~\cite{Ratynskaia_2025}. Courtesy of T.Rizzi.
    }
    \label{fig:intro}
\end{figure}

\subsection*{Importance and timeliness}

Recent years have seen dramatic growth in privately funded fusion companies. Long-awaited breakthroughs concerning ignition (NIF-USA)\,\cite{ICF_2024} and fusion energy produced (JET-EUROfusion)\,\cite{Maslov_2023} made headlines, attracting even more capital to an already multi-billion dollar industry \cite{Dinklage_2025}. Transition from experimental facilities to commercial power plants involves the upgrade from short pulses to long plasma operation, which is only viable if the vessel power handling capability is not compromised. In the most technologically mature magnetic confinement concept of tokamaks, the provision of PFCs with sufficient lifetime is one of the major technological obstacles to be overcome\,\cite{Pitts_2017,Pitts_2019}. This issue is central to the construction of the ITER reactor and the engineering design of the DEMO power plant. Besides enhanced energy efficiency and favorable plasma operations, extended PFC lifetime implies direct economic advantages (infrequent tile replacements)\,\cite{Pitts_2025, Krieger_2018} and fewer safety concerns (limited dust production) \,\cite{Ratynskaia_2022_b, Ratynskaia_2022_c, Beckers_2023}. Consequently, frenetic R{\&}D efforts have focused on the reliable modelling of PFC damage. 

The PFC longevity and power handling capabilities are mainly threatened during unplanned transients such as edge-localized modes or major disruptions\,\cite{Pitts_2025,Krieger_2025}. Given that incident plasma particles (electrons, ions) have typical energies in the low keV range and thus a submicron depth range\,\cite{Berger_1992}, such scenarios essentially imply surface heating. Dedicated EUROfusion and ITPA coordinated experimental activities\,\cite{Coenen_2013,Coenen_2017,Krieger_2017,Krieger_2018,Jepu_2019,Corre_2021,Corre_2023} have provided a wealth of empirical data on metallic PFC melting induced by surface heat loads, which have guided the development of self-consistent physics models based on thermoelectric magnetohydrodynamics within the shallow water approximation that enable high-fidelity simulations
\cite{Ratynskaia_2020,Ratynskaia_2021,Thoren_2018,Thoren_2021,Ratynskaia_2022_a,Ratynskaia_2024,Paschalidis_2023,Paschalidis_2024}. Successful validation against multiple tokamak experiments\,\cite{Krieger_2025} has lent confidence in their predictive power\,\cite{Coburn_2020,Coburn_2021,Coburn_2022,Paschalidis_2024b}.

Conversely, the understanding of the potentially most disastrous source of damage -- the impact of REs on PFCs -- remains in its infancy~\cite{Pitts_2025, Krieger_2025}.  The volumetric nature of the energy deposition of REs into matter\,\cite{Berger_1992} together with the fast timescales and high energy densities associated with RE impacts\,\cite{Lehnen_2015} result in complex PFC responses beyond incompressible hydrodynamics. Thus, the potentially catastrophic consequences of RE-induced PFC damage for tokamak operation include deep volumetric melting, in-vessel loss-of-coolant accidents\,\cite{Lehnen_2015}, debris expulsion, as well as non-local cratering\,\cite{Krieger_2025}. This has been evidenced in current tokamaks worldwide, which have reported explosive PFC events (see Fig.~\ref{fig:intro} (top)). REs will pose a more severe threat to future fusion devices such as ITER, where more stored magnetic energy will be available for conversion to RE energy, in particular \textit{via} the powerful avalanche mechanism~\cite{Boozer_2015,Boozer_2017,Breizman_2019}. An important experimental discovery from recent years giving some hope in this respect is the possibility to make RE beam terminations ``benign'' (i.e., harmless to PFCs) by injecting pure deuterium into the beam~\cite{Reux_2021b,Paz-Soldan_2021}. Unfortunately, the physics of RE-PFC interaction and its practical consequences are currently poorly explored, largely due to the absence of controlled experiments capable of providing sufficient input and empirical constraints for theoretical modelling\,\cite{Krieger_2025} and due to the large extrapolation steps from contemporary devices to ITER and future fusion reactors\,\cite{Boozer_2015,Boozer_2017,Breizman_2019}.

The very recent ITER re-baselining, that foresees tungsten (W) as the plasma-facing material in both the divertor and the first wall, has brought the topic of RE-induced PFC damage to the forefront\,\cite{Pitts_2025}. According to the modified research plan, the Start of Research Operation (SRO) campaign will use inertially cooled temporary W panels\,\cite{Pitts_2025,Loarter_2024}. This will allow experience to be gained concerning the emergence, mitigation and avoidance of major disruptions and REs, without risking damage to the actively cooled costly W panels that will be installed for the DT campaigns. The ITER re-baselining further motivated ITPA and EUROfusion coordinated experimental activities. The first controlled experiments on RE-induced damage in instrumented graphite domes were recently performed in DIII-D\,\cite{Hollmann_2025} and the first controlled experiments on RE-induced damage in instrumented W tiles have been scheduled in the ASDEX Upgrade (AUG) and WEST tokamaks within the EUROfusion Tokamak Exploitation Work Package. 

This is in resonance with recent achievements in the predictive and interpretative modelling of formation and transport of REs towards the vessel wall. In view of such simulation capabilities, a whole hierarchy of models has emerged with entirely different complexities, computational costs, and predictive capabilities. These range from fast lower dimensional models (e.g., Refs.~\cite{Hoppe_2021b,Martin-Solis_2017}) to the computationally much more expensive but also more self-consistent 3D global non-linear codes that can capture the complex mutual interaction between REs and large-scale plasma instabilities (e.g., Refs.~\cite{Liu_(Chang)_2018,Bandaru_2019}). Results for an ITER RE beam termination are shown in Fig.~\ref{fig:intro} (middle). RE Besides the spatial dimensionality, these models differ also in terms of their ability to resolve the phase-space distributions and RE wall deposition. Since RE induced material damage requires not only the deposition patterns, but also the energy and incidence angles of the incoming particles, predictive capabilities are boosted if all this information is available from high fidelity plasma and RE simulations for the calculation of wall heating and damage. Such integrated workflows are a key for a full understanding of the circumstances under which highly localized RE induced damage can occur, on one hand, or the much more desirable benign termination with broad deposition and an absence of pronounced hot-spots, on the other hand. 

The dramatic increase in potential RE avalanche gain during disruptions between present devices and their future successors such as ITER or DEMO power plants leads to qualitative changes of the expected RE dynamics~\cite{Breizman_2017}. Indeed, while avoiding large RE beams is relatively easy in present machines (where one instead typically needs a dedicated ``recipe'' to generate such beams), models suggest that in ITER and DEMO, multi-MA RE beams may be very difficult or even impossible to avoid in case of disruption at full plasma current~\cite{Vallhagen_2024}. Furthermore, a possibly strong re-avalanching may take place after a partial RE termination event and, consequently, repetitive cycles of re-formation and termination may arise in future high-current tokamaks. Predicting the consequences and optimizing the mitigation strategies in this context cannot rely on experiments alone, but requires a wholistic predictive view onto the involved processes originating from theory and simulation. Such input is needed to constrain plasma operation in view of risk mitigation when ITER moves towards full-current operation, and it is essential for designing well targeted and dimensioned engineering solutions like sacrificial limiters, which are considered in order to protect the first wall from RE impacts in various future devices. The landscape of different simulation tools with very complementary capabilities is overall developing towards the ability of providing such input in the coming years, although there are still clear gaps in the present capabilities that will need to be addressed.

Thus, the conditions are mature and the timing is critical for major advances in the genuinely multi-physics problem of the PFC response to RE impacts by synergetic theoretical and experimental efforts. 

\begin{table*}[h!]
\small
\centering
\begin{tabularx}{\textwidth}{@{}rXXX@{}}
\hline
\textbf{Device} & \textbf{Material/PFC type} & \textbf{RE-PFC interaction} & \textbf{Ejection of debris} \\
\hline
JET & CFC, e.g. UDP & Only wetted areas known,  hot spots $\sim$10 cm & Unclear \\
TCV & Carbon walls & No damage identified & Observations not yet linked to RE impact \\
COMPASS & Carbon and graphite, e.g.\ limiters, probes & Strong erosion, cratering, surface and bulk cracks & Yes \\
& BN support structure & Strong erosion, BN fragments found & Yes \\
TEXTOR & Graphite, e.g.\ limiter, probes, probe housing & Strong erosion, cracks & Unclear \\
Tora Supra & CFC walls & Black markings, no signs of major damage & Yes \\
WEST & BN inner wall and limiter tiles & Strong erosion & Yes \\
\hline
\end{tabularx}
\caption{Accidental RE-induced damage; brittle PFCs}
\end{table*}


\begin{table*}[h!]
\small
\centering
\begin{tabularx}{\textwidth}{@{}rXXX@{}}
\hline
\textbf{Device} & \textbf{Material/PFC type} & \textbf{RE-PFC interaction} & \textbf{Ejection of debris} \\
\hline
JET & Be UDP & Splashed melt pools over tile's apexes & Yes, possibly droplets \\
Alcator C-Mod & Mo wall tiles & Melt damage & Yes \\
FTU & Mo and TZM limiters & Severe melt damage, cratering on surrounding PFCs by high-speed debris impacts &  Yes, $\sim1$ km/s fast  solid debris\\
WEST & W divertor tiles & Trailing edges melting, splashes & Yes \\
& W outer limiter tiles & Splashed molten pools & Yes \\
EAST & W tiles of the main limiter& Splashed molten pools & Yes \\
\hline
\end{tabularx}
\caption{Accidental RE-induced damage; metal PFCs}
\end{table*}


\subsection*{Historical background}

One of the first evidence of intense RE-PFC interaction was reported 50 years ago in connection to neutron measurements of the ion temperature in the TFR CEA tokamak\,\cite{TFR_1975}. In particular, large neutronic fluxes of non-thermonuclear origin were found to be correlated with the presence of REs in plasmas. Already existing measurements of the neutron yields from $10-30\,$MeV electron bombardment of high-Z solids\,\cite{Barber_1959}, led to the interpretation of these fluxes as being photoneutrons. In fact, RE transport leads to the generation of high energy photons through Bremsstrahlung which can be absorbed by nuclei that are subsequently de-excited by emitting neutrons. Two years later, direct observations of photoneutrons that emanated from a W limiter after RE incidence were reported in the Princeton PLT tokamak\,\cite{Strachan_1977}. Ten years later, the phenomenon was confirmed by activation and metallographic analyses of the Frascati Tokamak (FT) stainless steel limiter after intense interaction with REs\,\cite{Maddaluno_1987}.

\begin{figure}
    \centering
    \includegraphics[width=8.1cm]{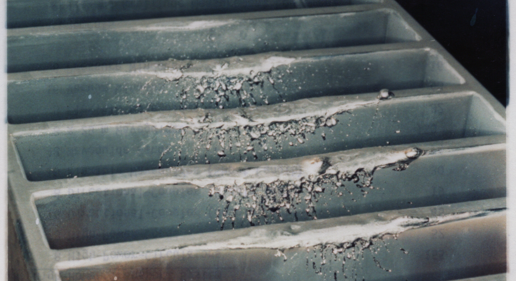}
    \caption{RE-induced damage of a stainless steel antenna from 1983 ASDEX experiments on Lower Hybrid Heating~\cite{ASDEX_1983,Leuterer_1985} due to $\ge$5 MeV electrons that also caused nuclear material activation. Courtesy of Dr.\ F.\ Leuterer.}
    \label{fig:ASDEX1983}
\end{figure}

Since then, the empirical database of RE-induced PFC damage in tokamaks with graphite (Fig.~\ref{fig:intro} shows modeling of a brittle failure resulting from an RE beam impact) and metallic composition has significantly expanded. Among early documentation of damage of metallic PFCs is the evidence from ASDEX experiments on Lower Hybrid Heating. Upon extraction of the antenna, made of stainless steel with 8 mm walls, significant melting and melt splashing was found at the front-ends of the central waveguides, shown in Fig.~\ref{fig:ASDEX1983}. The interpretation that the observed damage is caused by high-energy runaway electrons impacts was proposed back in 1983 \cite{ASDEX_1983,Leuterer_1985}. 
In addition to instances of in-vessel component degradation of various degrees, there have also been reports of costly damage events, such as severe material loss from signal cables of soft x-ray tomography arrays in Alcator C-Mod (see Sec. 3), water leakage in Tore Supra\,\cite{Nygren_1997}  and the toroidal field coil quench in WEST\,\cite{Torre_2019}. The risk of forcing a device out of operation with possibly costly repairs is the prime reason for the fact that by far the majority of RE-induced PFC damage observations are a result of \textit{accidental events}. The lack of \textit{controlled} RE-driven damage experiments has crucially impeded the development of RE-induced damage models.

\begin{table*}[h!]
\centering
\begin{tabularx}{\textwidth}{l X X }
\hline
\textbf{Device/Year} & \textbf{Experimental details} & \textbf{Results}  \\
\hline
DIII-D 2023-2024 & Graphite dome-shaped sacrificial limiters,  with thermal couple and radiation detectors & Top of the limiter blown away, explosive release of debris \\
WEST  spring 2025 & Inner limiter W tile with thermal couple & Conducted April 2025 \\
AUG  summer 2025 & W sample, with thermal couple and  radiation detectors, mounted on MEM & Upcoming \\
\hline
\end{tabularx}
\caption{First controlled RE-induced damage experiments}
\end{table*}


\subsection*{Classification of possible impact scenarios}

RE beam impacts are typically associated to their motion towards PFCs, which can be either intentional or due to a loss of position control. There then seems to exist two types of impact scenarios. In \textit{scrape-off} impacts, magnetic surfaces are well preserved. This type can be subdivided into two variants depending on whether the beam remains axisymmetric or not. The impacted PFC area for \textit{scrape-off} impacts is small and presumably determined by a combination of the beam trajectory, RE Larmor radius, orbit shifts, beam distortion and PFC geometry, but a good understanding remains to be developed. The timescale of \textit{scrape-off} impacts is set by the speed at which the beam moves `into' PFCs. On the other hand, in \textit{stochastic} impacts, magnetic stochasticity develops as a consequence of plasma instabilities (often denoted as a burst of MHD activity). In this case, the amplitude and growth rate of the stochasticity may determine both the impact time scale and deposition area. The former may be much smaller and the latter much larger than for \textit{scrape-off} impacts. \textit{Stochastic} impacts with strong and fast growing stochasticity are connected to benign terminations. Note that a given impact might combine the \textit{scrape-off} and \textit{stochastic} types. For instance, it may start as \textit{scrape-off} and later turn into \textit{stochastic}. Observations and modelling of some impacts e.g. on TEXTOR\,\cite{Lehnen_2015} or JET\,\cite{Chen_2021} suggest that they are of the \textit{scrape-off} type. \textit{Scrape-off} impacts are also often the baseline assumption in predictions for future devices (see e.g. Section~\ref{:sec10}). However, it is presently not fully clear what type of impacts take place in existing devices or are to be expected in future ones. This topic should be a focus of future work.

\subsection*{Structure of the Roadmap}
RE-induced PFC damage is a genuine multiphysics problem. The complete workflow for the modelling of material damage driven by RE incidence needs to address (i) the properties of the RE beam (current density profile, energy and pitch angle distribution) and its companion plasma (composition, density, temperature) before the impact, (ii) the transport of REs to the wall and the determination of their impact characteristics (location and momentum), (iii) the transport of the REs and the particle shower products inside the PFC until their complete thermalization, (iv) the thermomechanical response of the PFC to the combined RE volumetric load. This constitutes a highly complex modelling chain, with each block characterized by its own uncertainties and simplifying assumptions, that requires controlled well-diagnosed experiments for its validation. The synergy between modelling and experiments is imperative and is reflected in the structure of the present roadmap.

Experimental evidence is an essential element in the understanding of the physical mechanisms behind PFC damage. Sections~\ref{:sec1}--\ref{:sec3} are surveying RE-induced PFC damage evidence in all major tokamaks, including devices which are no longer in operation, see Tables 1-2. These sections also highlight challenges in designing and conducting controlled experiments to provide sufficient empirical constraints to the physics models, see Table 3. 

This is followed by Section~\ref{:sec4}, which  describes present and future diagnostic techniques that are capable of characterizing the incident REs (energies, pitch or impact angles) and the PFC response (e.g. temperature response, total energy received). 

Generation and transport of REs before and during their impacts on the wall is a critical input to the modelling of PFC thermo-mechanical response. Sections~\ref{:sec5}--\ref{:sec7} describe the tools available for this and their further developments. More precisely, Section~\ref{:sec5} concentrates on models that aim to characterize the RE beam properties prior to its termination, Section~\ref{:sec6} addresses reduced models for disruptions and beam terminations, and Section ~\ref{:sec7} describes computationally expensive high fidelity models in which the 3D plasma dynamics is self-consistently coupled to the RE dynamics.

The modelling of PFC damage is described in Section~\ref{:sec8} focusing on the Monte Carlo simulation of the electromagnetic shower that is induced by relativistic electron passage into condensed matter and on the finite element modelling of the thermomechanical PFC response to the generated volumetric loads.

Section~\ref{:sec9} is dedicated to PFC activation due to interaction with REs with incident energies above the kinematic threshold of photo-nuclear reactions. 

The roadmap is concluded with discussions on future reactors, Section~\ref{:sec10} providing the perspective of ITER and Section~\ref{:sec11} the perspective of DEMO and other tokamaks that are being built or designed world-wide. 

Note that all acronyms and abbreviations used in this article are  listed in a Table in the Appendix.

\clearpage 
\section{RE-induced PFC damage in JET}\label{:sec1}
\author{I. Jepu$^1$, C. Reux$^2$}
\address{
$^1$ UKAEA, Culham Campus, Abingdon, OX14 3DB, UK\\
$^2$ IRFM, CEA Cadarache, F-13108 Saint-Paul-lez-Durance, France }

\subsection*{Status}

In 40 years of operation, from 1983 until 2023, the Joint European Torus (JET)~\cite{Keilhacker_1999, Litaudon_2017} tokamak has been pivotal in the advancement of fusion research, being the perfect testbed for different operation scenarios directly applicable to future fusion devices such as ITER and DEMO~\cite{Loarte_2007, Federici_2016}.
During its lifetime, JET has operated with two main wall configurations as plasma-facing components (PFC). Between 1985 and 2009, JET’s PFCs were predominantly carbon fiber composite (\textit{JET-C})~\cite{Coad_1997}, while from 2010 to 2023 JET's PFCs were upgraded to the metallic wall configuration, comprising beryllium (Be) in the main chamber and pure tungsten (W) and tungsten coated carbon fiber composites (W-CFC) in the divertor, also known as \textit{JET-ILW} (JET with a W divertor and main chamber Be)~\cite{Matthews_2007, Matthews_2011, Matthews_2013}.

\paragraph{JET-C era.} 
During \textit{JET-C} the PFCs were predominantly carbon fiber composites (CFC), and runaway electron (RE) generation was frequently observed during major disruptions. This wall material had several implications for runaway electron physics. The low atomic number of carbon, decreased impurity radiation, which enabled high performance operations, but also altered the dynamics of the RE beam interaction with the wall. Under these conditions, experimental studies revealed that unmitigated plasma disruptions typically produced consistent RE avalanches~\cite{Gill_2002}. The rapid current quench phases were induced by the significant radiation of carbon accelerated electrons via the Dreicer mechanism~\cite{Helander_2002} and subsequent avalanche multiplication. In this regime, although REs were frequently produced, the high resistance of carbon to sublimation and erosion often led to damage that was more spread out, especially when compared to the severe damage observed on metallic surfaces.
In several dedicated experiments~\cite{Riccardo_2010}, a series of major disruptions were triggered using either Massive Gas Injection (MGI) or by a slow constant puff of impurity gas~\cite{Plyusnin_2012}. These experiments allowed the comparison of “fast” versus “slow” disruptions. Fast disruptions produced by rapid MGI were characterized by a steep current quench (typically on the order of 3–7 ms), leading to induced electric fields several hundred times larger than the Dreicer field. Such large fields could accelerate electrons rapidly. Up to 60–70\% of the pre-disruption plasma current could be carried by a beam of runaway electrons with energies reaching 10 – 15 MeV and currents on the order of 1 MA~\cite{Plyusnin_2012}. These quantitative measurements have been verified against diagnostic signals such as hard X-ray (HXR) emission, neutron bursts, and spectroscopic data. Complementary data from neutron diagnostics and fast visible cameras further constrained the runaway electron density and current. Ultimately, the RE beam could reach the first wall and lead to severe damage.

\begin{figure}
    \centering
    \includegraphics[width=6.1cm]{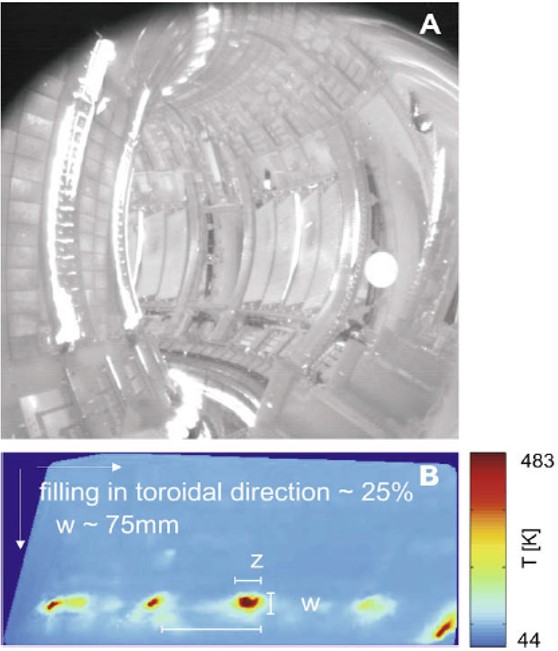}
    \caption{Heat load distribution during the disruption of JET pulse no. 68782. Over-exposed overview frame for visualization (a) and  temperature rise at the upper dump plate due to runaway impact (b). Reproduced with permission from Ref.~\cite{Lehnen_2009}.
    }
    \label{fig:JET-C}
\end{figure}

\paragraph{Evidence of PFC damage in JET-C era.}
The most noticeable result of runaway electron generation in the \textit{JET-C} configuration was the formation of high-power levels of hard radiation, gamma rays, and some neutrons when the REs interact with the vessel walls. The SXR images in Ref.~\cite{Gill_2000} show that REs interaction with the wall is confined to small areas with poloidal widths below 10 cm on the upper or lower vessel parts, depending on the direction of the vertical movement of the beam. Observation of discrete pulses (bursts) of HXR emission during the impact duration suggests that the runaway current channel itself is filamented~\cite{Hender_2007}. In a comprehensive study described in Ref.~\cite{Lehnen_2009}, analysing close to 20000 JET pulses, RE detection was done by using the increasing neutron rate during current quenches. This way, it was determined that for 8\% of these pulses runaway electrons were detected in limiter and divertor configuration. However, observations of the heat load during RE production, and consequently damage to the PFCs were rarely documented. Figure~\ref{fig:JET-C} shows the first RE impact measured with a wide Infrared (IR) camera reported in Ref.~\cite{Lehnen_2009}. For this particular case, a runaway electron current of 0.48 MA was produced. The entire recording consisted of two frames only.

During the thermal quench heating of the divertor and inner limiter is observed in the first frame of this recording, while in the second frame, taken 20 ms later, a temperature increase is noticed on the upper dump plates (UDPs), which consequently was attributed to the RE beam impact with the PFCs. Figures~\ref{fig:JET-C} (a) and (b), show the two frames. In Figure~\ref{fig:JET-C}(a) the thermal radiation is presented on the inner limiter and lower inner divertor during the disruption that led to the RE generation, while Figure~\ref{fig:JET-C}(b) shows the IR image of the heated upper dump plates as a direct consequence of the interaction with the RE beam. Based on the IR imaging, it was assessed that the temperature increase between the two frames was of 530 K, while the wetted surface area heated by the RE impact was found to be $\sim 0.3$ m$^2$. It was noted that the inhomogeneous heat load might have been caused by a small tile misalignment, which can have a considerable effect at low angles of incidence. 

Similar type of dump plate damage induced by REs was also discussed in Ref.~\cite{Arnoux_2011}. In this work it was noted that for 17 JET pulses studied, the red localized spots, similar to those shown in Figure~\ref{fig:JET-C}(b), are footprints of the distinct impacts measured on UDPs, implying that the dump plate geometry defines the heat load distribution. In addition, it was reported that the temperature increase measured on the CFC tiles scales with the square of the runaway electron current.

As a concluding remark, RE experiments in \textit{JET-C} highlighted the potential of high RE currents and associated PFC damage to the PFC including high temperature spots, localized erosion as well as dust generation. These effects, although far less obvious in a carbon wall environment as compared with a metallic wall configuration, were considered of high importance for the long-term integrity of the PFCs and plasma purity. The results in the \textit{JET-C} served as a benchmark for further runaway electron studies~\cite{Plyusnin_2006, Bazylev_2011}.  

\paragraph{JET-ILW era.}
With the introduction of \textit{JET-ILW} configuration, the RE production changed significantly. It was observed that with this new configuration, the RE production was drastically reduced as compared with the \textit{JET-C}. As reported in Ref.~\cite{Reux_2015}, only two out of 7000 \textit{JET-ILW} pulses up to 2014 JET operations (ILW1 and ILW2) showed low-energy REs during a low density current ramp-up due to a failure to the gas introduction system. The total absence of the REs was attributed to slower current quenches. In fact, RE generation was only experimentally triggered using a massive gas injection (MGI) with argon. It should be noted that in this wall configuration, REs were not produced as a consequence of an unmitigated plasma disruption but rather induced for this purpose only. The main difference in comparison to disruptions with carbon wall consists of a lower radiated energy fraction during the disruption process, resulting in a larger plasma temperature after the thermal quench, which in turn is affecting the runaway electron beam formation. The study in Ref.~\cite{Papp_2013} found that the ILW slows down the current quench during disruptions. This slower current quench reduces the likelihood of RE avalanche formation, particularly during spontaneous disruptions triggered by slow argon injection. Although the RE domain is similar to the \textit{JET-C} conditions, the metallic wall influences the plasma cooling and radiation characteristics~\cite{Lehnen_2013}. However, when disruption mitigation techniques such as massive gas injection (MGI) were deliberately applied, RE beams with currents up to several hundred kiloamperes were observed~\cite{ReuxJNM_2015}. JET experiments in the ILW era have employed advanced diagnostics, such as fast visible and infrared (IR) cameras, hard X-ray detectors, and spectroscopic techniques to capture the evolution of RE beams and their interaction with PFCs.

\paragraph{Evidence of PFC damage in JET-ILW era.}
With the installation of ILW in 2010, due to the lower melting point of beryllium covering the main chamber, a significant consideration was given to the operation scenarios to avoid a significant PFC damage due to unmitigated plasma disruptions and RE impacts. 
The values of new tolerable heat loads were more restrictive, with Be and W melting occurring at $\simeq$20 MJ/(m$^{2}$s$^{1/2})$ and  $\simeq$50 MJ/(m$^2$s$^{1/2}$, respectively ~\cite{Riccardo_2010, deVries_2012}. 

Between 2010 when the ILW replaced the carbon configuration, and 2016, three main campaigns, also known as ILW1-ILW3, were run. Each of the campaigns had different operational purposes: ILW1 (2010-2012) focused mainly on the influence of the new wall to the plasma operation, material migration and fuel retention, ILW2 (2012-2014) focused on high power scenarios and disruption mitigations by massive gas injection, while the ILW3 (2014-2016) aimed at an increase of fusion performance (see Ref.~\cite{Jepu_2019} and reference therein). 

During ILW2, a comprehensive study on the RE generation and mitigation was performed and fully described in Refs.~\cite{Reux_2015, ReuxJNM_2015}. For this purpose, REs were induced using one disruption mitigation valve (DMV1) by delivering a massive Argon gas injection. First evidence of PFCs damage due to low RE currents was reported in Ref.~\cite{ReuxJNM_2015} showing that the Be tile melting was unlikely below an RE current of 150 kA. It was found that most of the plasmas with observed REs were in divertor elongated configuration, which consequently meant that RE beams affected UDPs. In Ref.~\cite{Reux_2015} it was reported that electron impacts on Be UDPs resulted in energy deposition over a depth of approximately 2 ± 1 mm, with only a portion of the beam’s total energy being absorbed by an individual tile. Instead, a considerable fraction of the beam current continued and struck the next tile in the toroidal direction. Additionally, the impacts from these lower currents were found to be quite symmetrical along the toroidal axis. As in the \textit{JET-C} case described in Ref.~\cite{Lehnen_2009}, these observations were made using a wide-angle infrared camera. A similar pattern emerged, with localized hot spots along the toroidal direction on some of the UDPs, directly caused by the RE beam impact, and covering a similar area of interaction as reported in Ref.~\cite{Arnoux_2011}. It was found in Ref.~\cite{Reux_2015} that the impact from long-lasting and less vertically unstable RE beams in a limiter configuration present a different scenario. The beam strikes the upper part of the JET Inner Wall Guard Limiter (IWGL) over an area of roughly 10 cm² per tile. 

    \label{fig:Fig3_JET-ILW_post-RE}

\begin{figure}
    \centering
    \begin{noverticalspace}
    \includegraphics[width=4.75cm]{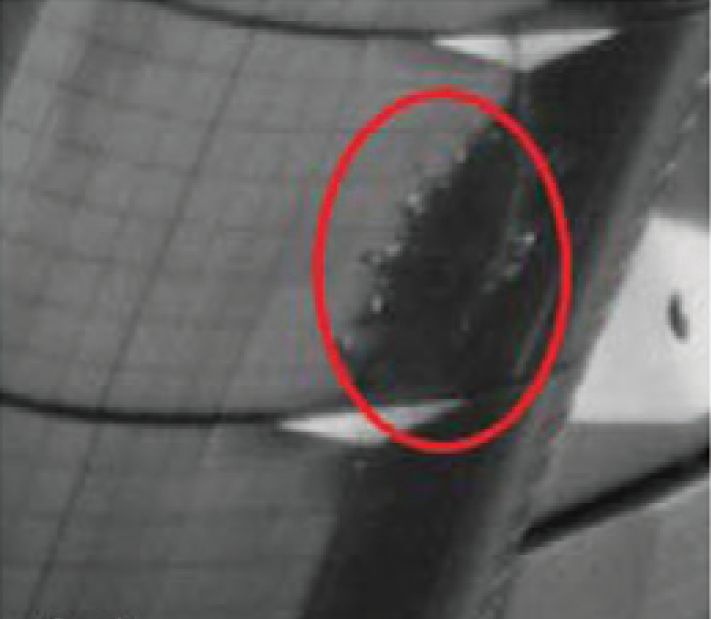} \\
    \includegraphics[width=4.75cm]{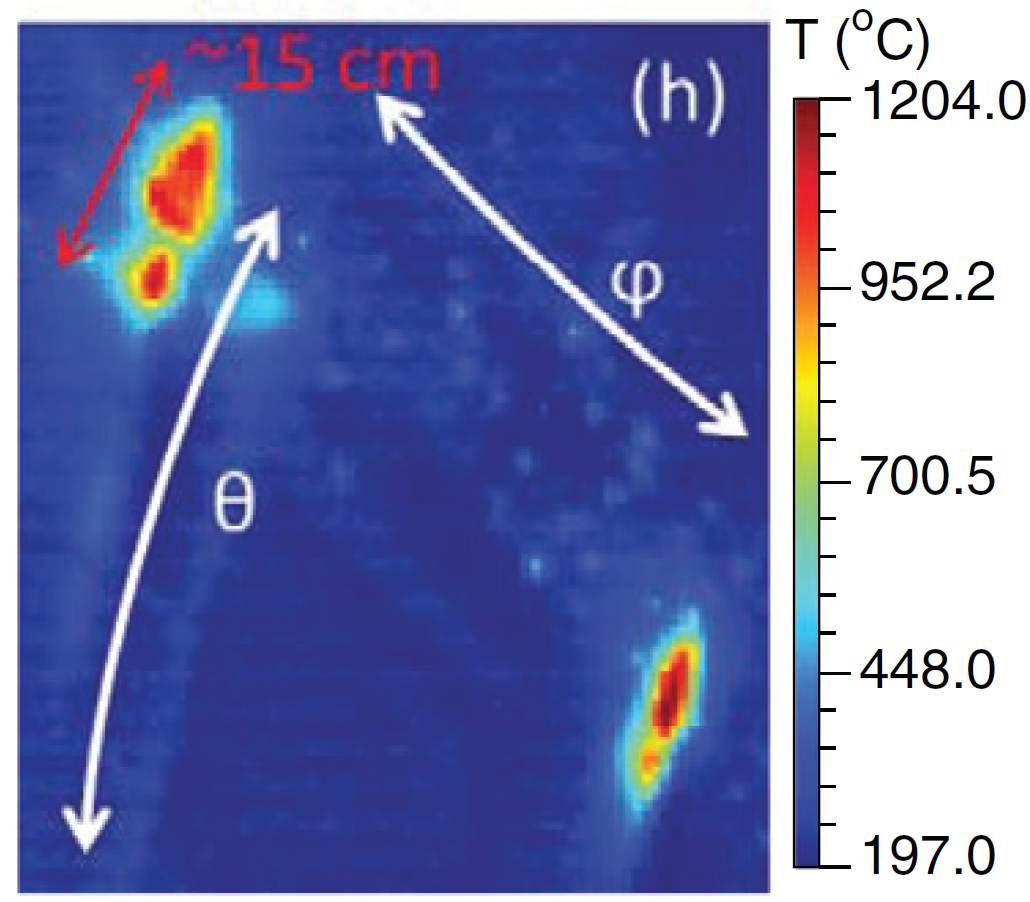} \\
    \end{noverticalspace}
    \caption{RE beam impacts on Be tiles; visual inspection of the damage and an IR image of the impact. Reproduced with permission from Ref.~\cite{Reux_2015}. 
    }
    \label{fig:Fig3}
\end{figure}
A fast infrared camera recorded a peak temperature of 1400°C, which is above the melting point of Be (1287°C). Following the RE impact, a significant material ejection was observed from the inner limiter, see  Fig.~\ref{fig:Fig3}.
Be droplets observed from both inner limiters showed in Fig.~\ref{fig:Fig3} most likely have ended in the divertor area. Similar type of Be droplets, observed via IR cameras, were seen post unmitigated plasma disruptions events affecting the UDPs~\cite{Jepu_2019,Jepu_2024}. Given the limited camera resolution and that these observed droplets are close to the pixel resolution limit, accurate estimations of droplet size and overall Be mass ejected are not feasible. Meanwhile, the metallic splashes and solidified droplets found on various wall components and in the divertor area were in the range from a few nm to 500 µm ~\cite{Jepu_2019,Jepu_2024,Moon_2019,Rubel_2018, Fortuna_2017} . 

\begin{figure}
    \centering
    \includegraphics[width=\linewidth]{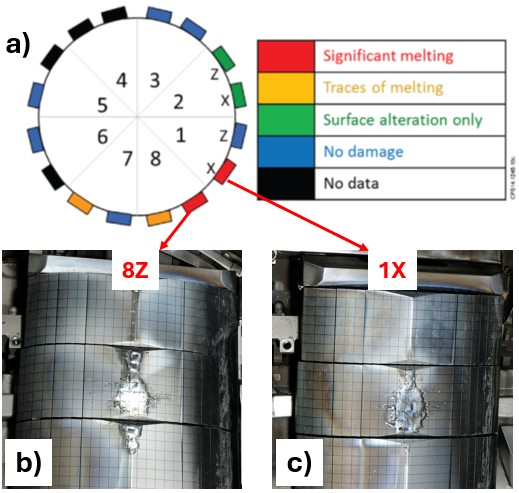}
    \caption{Map of RE-induced tile damage on the IWGL (a) and images of the two most affected tiles, (b) and (c). Reproduced with permission from Ref.~\cite{Reux_2015}.
    }
    \label{fig:Fig2_JET-ILW}
\end{figure}

Localized melting was also confirmed by in-vessel visual inspections (IVIS). Furthermore, a toroidal damage map presented in Refs.~\cite{Reux_2015, Jepu_2024} revealed the complexity of the observed RE-induced PFC modifications. The damage extended across several adjacent limiters along the toroidal direction before gradually fading, see Figure~\ref{fig:Fig2_JET-ILW}(a).  During the 2017 JET intervention, several components were exchanged. Among those removed was the central component of the 1XR18 IWGL tile, depicted in Figure~\ref{fig:Fig2_JET-ILW}(c). The first damage assessment was carried out during the remote post-experimental inspection while results of the morphological and structural analysis of this tile were reported in Ref.~\cite{Jepu_2024}. 

\begin{figure}
    \centering
    \includegraphics[width=6.1cm]{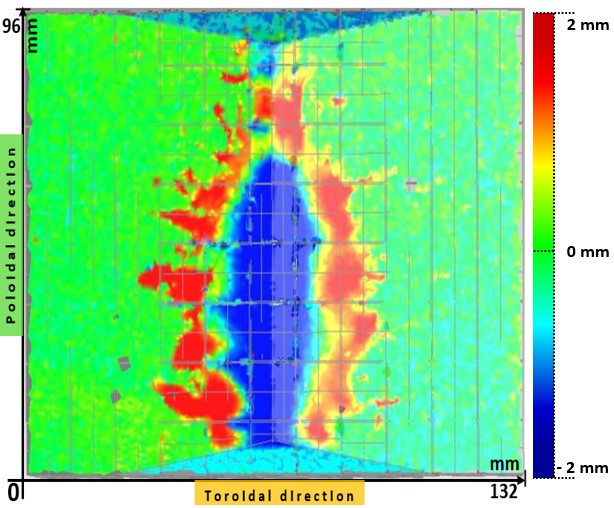}
    \caption{The 3D profiling reconstructed damage of Be tile (shown in Fig.~\ref{fig:Fig2_JET-ILW}(c)) affected by RE impacts. Reproduced with permission from Ref.~\cite{Jepu_2024}.)
    }
    \label{fig:Fig_JET-ILW_3D_RE-damage}
\end{figure}

The result of 3D profiling measurement performed on the removed tile is shown in Figure~\ref{fig:Fig_JET-ILW_3D_RE-damage} where 'zero' on the scale refers to the undamaged or 'reference' surface. Displaced and re-solidified Be melt can be seen as 1-2 mm high elevations, along with  1-2 mm deep "valley". The estimated area below the reference was  $\sim 11$ cm$^{2}$ while the area above the reference was close to 14 cm$^{2}$. The overall damage area was assessed to be $\approx$21\% of the total surface area. The analysis confirmed that the REs caused significant but very localized damage featuring melt splashing and material ejection from the impact areas yet without major effects on the surrounding PFCs. Evidence of molten material bridging gaps between adjacent castellations  but with no accumulation of the molten material into the gaps was also reported in Ref.~\cite{Jepu_2024}. The analysis of the topography and crystallographic structure in the damaged region revealed that the molten material re-solidified within a near surface, $\approx$330 $\mu$m thick, layer. This re-solidified region features a columnar grain structure with elongated grains oriented perpendicular to the surface, and the length of these grains typically matches the thickness of the re-solidified layer. 

Later, post the 2017, campaigns intervention continued to study RE beams created during disruptions, almost exclusively in limiter shapes.Patterns observed were similar to the previous campaigns, with no significant changes to the damage asymmetry map. Some limiters were badly affected, and some others completely unscathed, over the course of several tens of RE impacts.
Runaway heat loads were significantly reduced with the discovery of the so-called "benign termination" scenario in 2019~\cite{ReuxPRL_2021}, with heat loads barely measurable when the suppression method was used successfully. Some damage was still recorded on limiters and divertors following explorations of the validity domain of the ``benign termination'' scenario.

\subsection*{Current and future challenges}

During the final weeks of JET operation in November-December 2023, a series of dedicated experiments were conducted to examine the impact of runaway electrons on both the main chamber limiter tiles and the divertor. New damage evidence was found on the IWGL tiles. A high-resolution image survey conducted in the summer of 2024, following JET's final operational campaign, will generate an updated toroidal damage distribution map similar to those presented in Refs.~\cite{Reux_2015, Jepu_2024}. A single RE damage event, linked to a single purposely designed JET pulse, was identified in the last days of the operation, and its analysis will contribute to the toroidal assessment of inner limiter damage due to RE impacts. Additionally, for the first time in JET, clear evidence of RE impact on the inner divertor was observed. Divertor impacts are visually very different from limiter impacts in the sense that they produce many more ejecta from the impact point. The nature of the ejecta is not completely clear as some of the affected areas are covered with deposits. It is therefore possible that they contain a mixture of molten tungsten coatings, ablated deposits (mostly Be)~\cite{Brezinsek_2015} or C substrate.  A toroidal distribution assessment for the divertor damage is planned, with the preliminary results indicating similar localized behaviour and the absence of toroidal uniformity, consistent with the previously observed localized nature of REs~\cite{Reux_2015, Jepu_2024}. In addition, future studies will assess the W coating removal due to such events, its implication for the integrity of the PFC along with fuel retention and release mechanism behaviour.

\subsection*{Concluding remarks}

RE damage to PFCs in JET has evolved considerably between the \textit{JET-C} and \textit{JET-ILW} eras. In the carbon‐wall configuration, robust RE production during disruptions was observed, yet the carbon properties limited the severity of local damage. The resulting damage was less localized, with material losses distributed over larger areas, hence reducing the risk of major PFC failure. In contrast, the ILW configuration, comprising Be in the main chamber and W-based materials in the divertor, produced a different response to the RE production. While spontaneous runaway generation was drastically reduced, triggered events via massive gas injection was the way to produce high-current RE beams. These beams deposited energy in a highly localised manner, often resulting in intense, concentrated heat loads. This caused significant material displacement, and in some cases led to splashing of melt pools, with droplet ejection contributing to both PFC erosion and the overall dust inventory. The legacy of JET's comprehensive investigations in both configurations provides crucial information that will drive future design and operating methods for safely managing RE occurrences in next-generation fusion reactors. In particular, the so-called \textit{benign termination} scenario recently studied on JET~\cite{Reux_2021b} opens a hope for damage-free RE mitigation.

\clearpage 
\section{RE-induced damage of brittle in-vessel components}\label{:sec2}
\author{E. Hollmann$^1$, U. Sheikh$^2$, E. Tomesova$^3$, H. R. Koslowski$^4$ }
\address{
$^1$University of California - San Diego. San Diego, CA, 92093 USA\\
$^2$Swiss Plasma Center (SPC), Ecole Polytechnique Federale de Lausanne (EPFL), CH-1015 Lausanne, Switzerland\\
$^3$Institute of Plasma Physics of the CAS, CZ-18200 Praha 8, Czech Republic \\
$^4$Forschungszentrum J\"ulich GmbH, Institute of Fusion Energy and Nuclear Waste Management -- Plasma Physics, 52425 J\"ulich, Germany \\
}

\subsection*{Status}

\begin{figure*}
    \centering
    \includegraphics[width=0.56\textwidth]{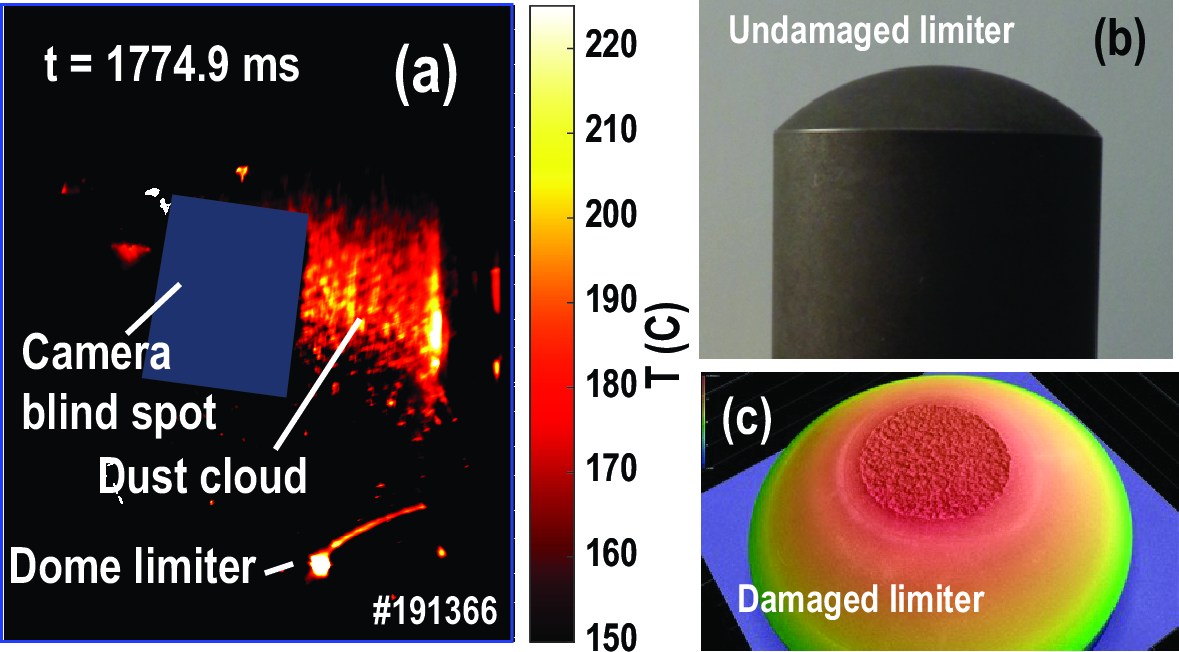}
    \caption{RE damage to sacrificial graphite limiter in DIII-D showing: IR image of resulting dust cloud (a), visible image of undamaged graphite limiter (b) , and confocal microscope image of damaged limiter (c) . Reprint with permission from Refs.~\cite{Hollmann_2025, Ratynskaia_2025}.
    }
    \label{fig:d3d_dimes}
\end{figure*}

\paragraph{DIII-D.}
Post-disruption runaway electron (RE) experiments in DIII-D typically study RE formation and dynamics~\cite{Hollmann_2011, Hollmann_2020}. During the experiments, inadvertent wall damage was occasionally observed; for example, in one experiment the outer midplane plunging probe was accidentally struck by a RE beam, leading to probe tip damage. More recently, experiments dedicated to studying RE-induced damage to brittle plasma facing components (PFCs) have been performed. These experiments have used domed ATJ graphite limiters with a peak height of 1 cm above the lower divertor floor, as shown in Fig.~\ref{fig:d3d_dimes}. These limiters have been intentionally struck with downward-moving high-current ($\approx$600 kA) post-disruption RE beams. Significant shot-shot variation in limiter damage has been observed; this variation has been attributed to shot-shot variation of the toroidal phase of the dominant RE loss to the floor. In all cases, the RE final loss event (instability) causes rapid ($<$ 1 ms) loss of REs into the limiter head. The REs hitting the limiter are expected to have a wide range of kinetic energies, but with an average of order 2 - 4 MeV \cite{Hollmann_2025}.  The initial in-plasma RE pitch angle (RE velocity vector relative to magnetic field vector) is thought to be fairly small, of order 0.1 - 0.2 on average; the experiments indicate that this may increase slightly (to perhaps 0.3 - 0.4) during the final loss process~\cite{Hollmann_2025}.

To-date RE graphite damage to the sacrificial limiters has been roughly grouped into three categories. At low RE fluences $q_0 < 100$ J/cm$^2$, no apparent change to the graphite is observed.  At intermediate fluence $q_0 > 100$ J/cm$^2$, an apparent surface morphology or phase change is observed on the sample surface, with the graphite appearing darker on visual inspection. At high RE fluences $q_0 > 1000$ J/cm$^2$, an explosive destruction of graphite and dust release is observed. The heat fluences are estimated from thermocouple and IR imaging data. The destruction depth is of order 1 mm, comparable to the RE surface penetration depth. The observed destroyed volume agrees with modeling of carbon bond failure in the graphite volume~\cite{Ratynskaia_2025}. Modeling is continuing to understand the observed dynamics of explosive dust release. For example, the dust appears to be released approximately 1 ms after RE impact at velocities in the range 100 to 150 m/s \cite{Hollmann_2025}. Example images of dust release and limiter damage are shown in Fig.~\ref{fig:d3d_dimes}.

\paragraph{TCV.}
The TCV tokamak, a carbon-walled machine capable of generating runaway electron beams exceeding 250 kA~\cite{Decker_2022, Sheikh_2024}, has been the focus of extensive research on RE generation mechanisms, transport, and mitigation strategies. Despite numerous experiments and high-current RE events, no direct damage to plasma-facing components has been observed~\cite{Decker_2004, Sheikh_2023, Sheikh_2024}. Explosive dust releases have been periodically observed on visible and infrared cameras, however, no systematic study has been conducted to link them to RE impact. Infrared measurements indicate that surface temperature increases from RE impacts on TCV typically range from 50 to 100 K, corresponding to heat fluxes on the order of 0.1–1 MW/m² with a total energy in the order of 10 kJ~\cite{Sheikh_2023, Sheikh_2024}.

\begin{figure*}
    \centering
    \includegraphics[width=0.5\textwidth]{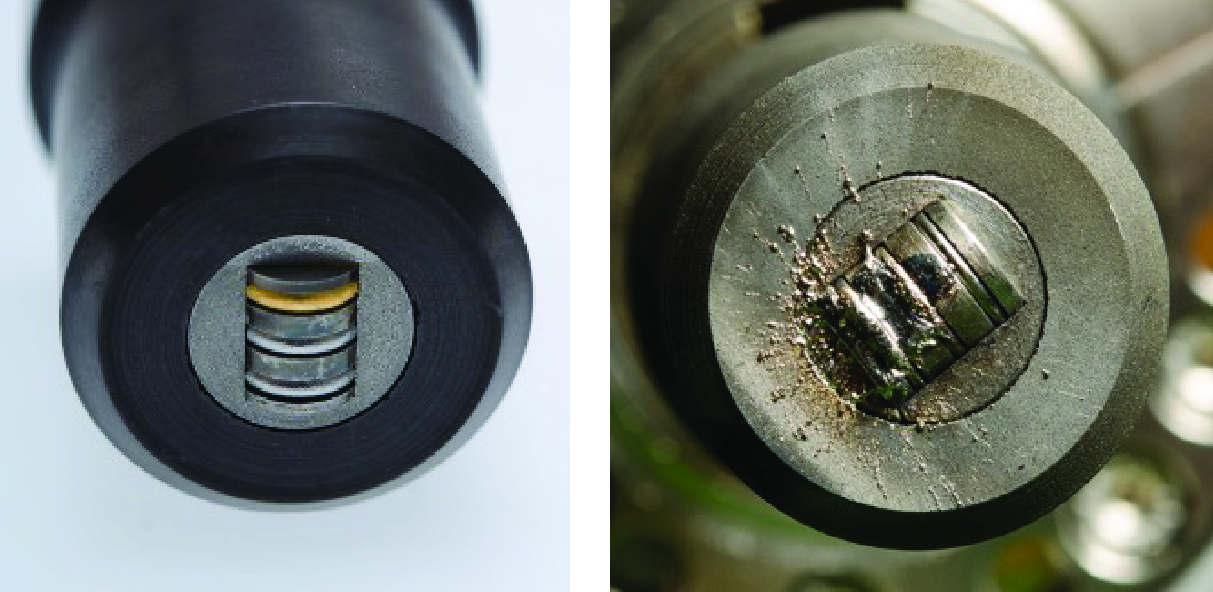}
    \caption{Cherenkov probe on TCV before (left) and after (right) RE impact during a VDE of a 250 kA current plasma with significant RE population. Melting observed on the molybdenum-based alloy spacer plates. 
    }
    \label{fig:tcv_probe}
\end{figure*}

The only documented case of RE-induced damage occurred not on a PFC but on a diagnostic component—the Cherenkov probe. It was associated with two consecutive pulses featuring a large RE seed population (expected to be over 50\%  of the pre-TQ current based on LUKE simulations~\cite{Decker_2004,Decker_2016}) and a thermal companion plasma with an electron temperature of approximately 1 keV~\cite{Sheikh_2020}. During the current ramp-up to the prescribed 250 kA flattop, plasma control was lost, resulting in a vertical displacement event (VDE). Consequently, the RE beam and plasma column shifted upwards, directly impacting the protruding Cherenkov probe, which extended approximately 1 cm beyond the front surface of the adjacent tiles. This exposure to the high-energy RE beam caused the localized melting visible in Fig.~\ref{fig:tcv_probe} (bottom), contrasting with the probe’s pre-installation condition shown in Fig.~\ref{fig:tcv_probe} (top). The outer diameter of the graphite head is 30mm and the inner diameter of the metal detector assembly is 14mm. The melting is predominantly observed on the spacer plates (distinguished by the front flat surfaces), which are made from TZM, a molybdenum-based alloy with 0.50\% titanium, 0.08\% zirconium and 0.03\% carbon. The detectors themselves, distinguished by the rounded front surfaces, are made from CVD diamond with a Ti/Pt/Au interlayer 1.3 $\mu\mathrm{m}$ thick, and 0, 25 or 53 $\mu\mathrm{m}$ of molybdenum as a variable energy filter.

\paragraph{COMPASS.}
COMPASS~\cite{Hron_2022} was a medium size tokamak with graphite and carbon PFCs and main experimental focus on RE studies. Following massive material injection triggered disruptions, RE currents reached up to 150\,kA. Average RE kinetic energies up to $ \approx 10\,\mathrm{MeV}$ were estimated, corresponding to a total kinetic energy of $\approx 5\,\mathrm{kJ}$ for $3 \times 10^{15}$ REs. Most RE-induced damage events occurred during dedicated RE campaigns, with typical RE impacts either on the calorimetry probe used as the protection LFS limiter~\cite{Caloud_2024} or on HFS midplane limiter tiles. Exceptions occurred when beam termination coincided with vertical displacement events or plasma positioning failures. 

\begin{figure*}
    \centering
    \includegraphics[width=0.475\textwidth]{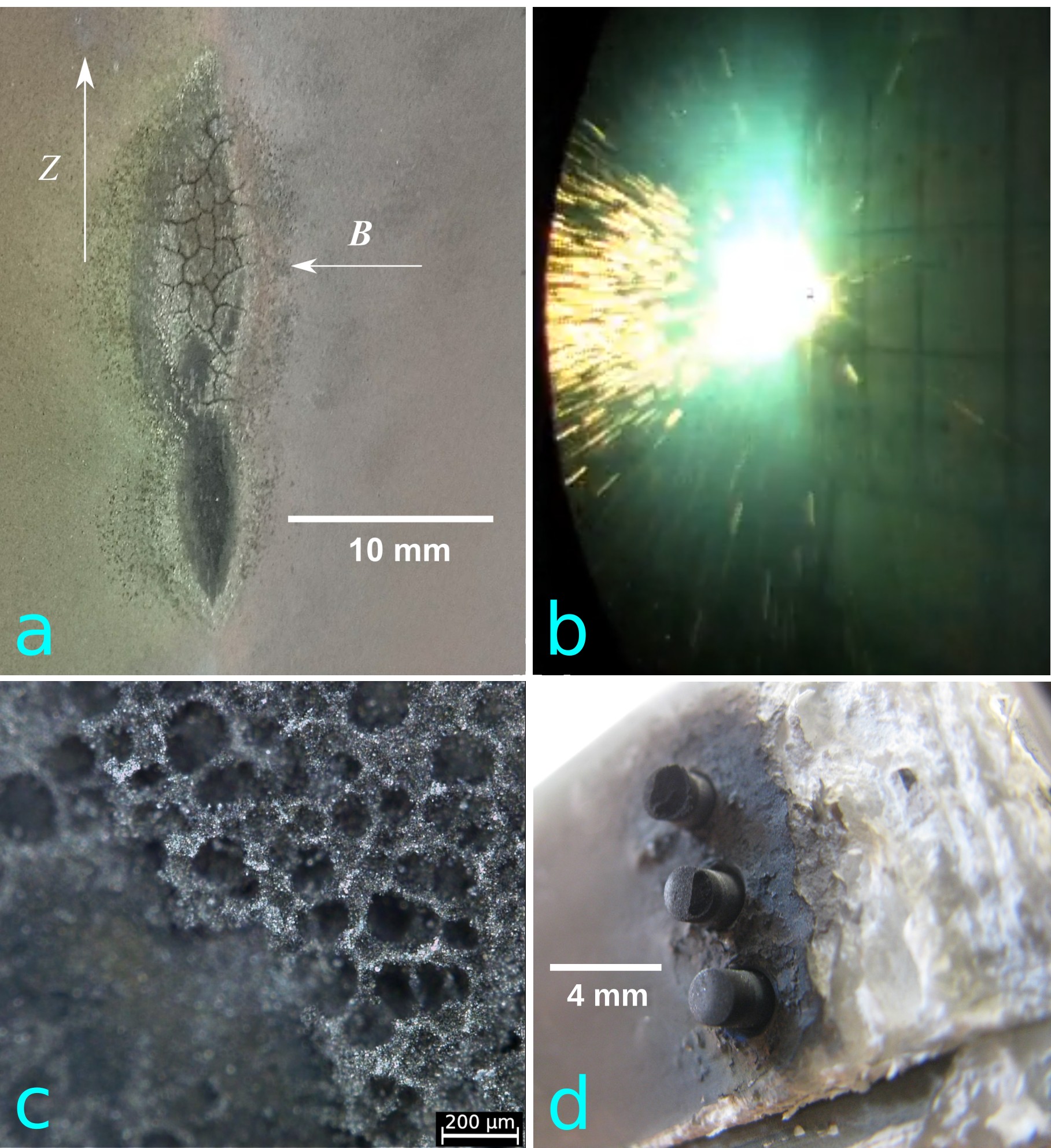}
    \caption{Examples of RE-induced damage of components from COMPASS. Damage to the recessed roof-shaped (RR) graphite limiter (a).  Reprint with permission from Ref.~\cite{Mlynar_2019}. Debris release during RE termination at the graphite RR tile in discharge 14494 (b). Craters on a carbon Langmuir probe (c).  Reprint with permission from Ref.~\cite{Caloud_2024}. Carbon U-probe,  mounted in a
 boron nitride support structure (seen as white), after an RE impact (d).
    }
    \label{fig:compass_probe}
\end{figure*}

Damage to the recessed roof-shaped (RR)~\cite{Horacek_2014} inner wall graphite limiter, which protruded into the vessel by 4\,mm (see Ref.~\cite{Mlynar_2019}) is shown in Fig~\ref{fig:compass_probe}(a). Two footprints corresponding to two distinct RE beam positions can be seen. The material loss exceeds 0.2\,cm$^{3}$. The upper footprint exhibits marks of significant thermal stress and intense ablation, with repeated heating cycles leading to a network of surface and bulk material cracks. The lower, darker footprint experienced shorter RE exposure. The surrounding region's texture suggests material re-deposition. Over 80 RE impacts to the high field side generated graphite dust release. An example is shown in Fig.~\ref{fig:compass_probe}(b): here, a 60\,kA RE beam with RE kinetic energy $\langle E \rangle \approx$ 12\,MeV struck the center post. Fast infrared cameras recorded heat fluxes reaching several GW/m$^2$, corresponding to several kJ of deposited RE beam kinetic energy.

Fig.~\ref{fig:compass_probe}(c) presents an optical microscope image of RE-induced damage on a carbon Langmuir probe located on the calorimetry head~\cite{Caloud_2024}. The probe, subjected to about 200 RE impacts, exhibited coalesced cracks resulting in material detachment over a large area, visible as a gray, slightly blurred region without craters. The remaining surface features craters approximately $100$--$200\,\mu\mathrm{m}$ in diameter and up to $100\,\mu\mathrm{m}$ deep. Energy dispersive x-ray spectroscopy (EDS) analysis confirmed silicon sublimation, indicating probe surface temperatures of at least 1200$^{\circ}$C.

U-probe~\cite{Kovarik_2017} carbon pin damage shown in Fig.~\ref{fig:compass_probe}(d) was probably inflicted in a single RE event. In contrast, the U-probe body, composed of boron nitride (white area), experienced multiple RE impacts. Debris ejection has been detected by visible cameras and boron nitride fragments (of various sizes, up to a few mm large) were later found throughout the lower section of the corresponding port and in the divertor region.

\begin{figure}
    \centering
    \includegraphics[width=0.8\linewidth]{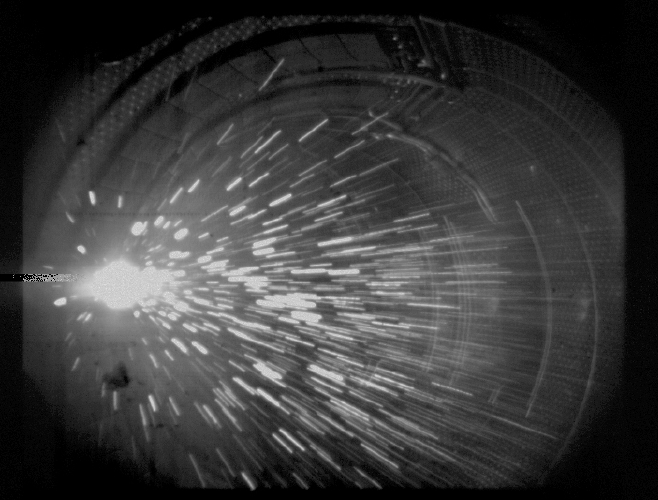}
    \caption{Runaway electron impact on Tore Supra carbon limiter. Courtesy of C.~Reux.}
    \label{fig:ToreSupra_impact}
\end{figure}

\paragraph{Tore Supra / WEST.}
Tore Supra was known for extensive RE production and seconds-long RE beams. The carbon-fiber composite wall absorbed most RE impacts, usually leading to dust release, as shown in Fig.~\ref{fig:ToreSupra_impact}. No quantitative analysis of the amount of ablated material was performed at the time. Visually, most of the affected tiles were left with black markings only and no other sign of major damage.
 
Tore Supra was later upgraded to WEST~\cite{Bucalossi_2022}, with tungsten plasma facing components. Between 2020 and 2024, the 4 central rows of the inner wall and the 5 central rows of the outer limiter were replaced  by boron nitride tiles to facilitate operation in the tungsten environment. These tiles were hit by up to $\sim$ 10 RE beams during dedicated experiments. As with the carbon impacts in Tore Supra, dust ejection was observed after the impacts. The tiles were damaged, with a visibly cracked surface similar to that of tree bark~\cite{Reux_2024}. A thickness of 1 mm of material was lost for the most badly damaged tile. Plasma operation was only moderately affected following impacts, with a hot spot probably resulting from the uneven surface of the tile when the plasma was put in contact with the damage. The hot spot disappeared after a few experimental sessions, most probably eroded by plasma thermal loads. 
The RE damage was found to be toroidally asymmetric, with one of the inner limiters badly damaged, three others moderately damaged and the last two nearly unscathed~\cite{Reux_2024}. This is similar to observations made on JET (see Section 1). The cause of this asymmetry remains unknown. Several hypotheses have been brought forward, including tile misalignments or the intrinsic 3D nature of the RE beam, but no conclusion has been reached yet.

\begin{figure}
    \centering
    \includegraphics[width=1\linewidth]{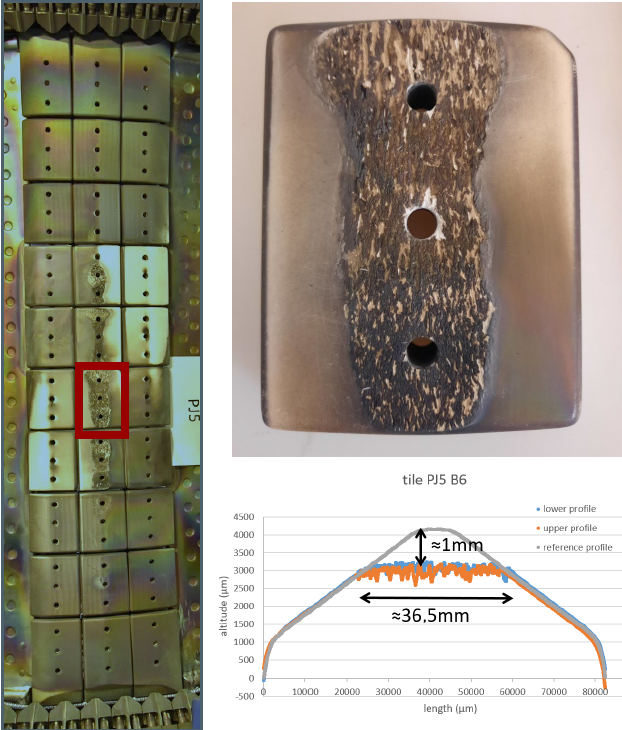}
    \caption{WEST inner limiter PJ5 damaged by multiple RE impacts, along with a zoom-in of the worse damaged tile B6 and its profilometry showing 1 mm deep material loss. Courtesy of M. Diez.}
    \label{fig:WEST_BN_damage}
\end{figure}

\paragraph{TEXTOR.}
The TEXTOR tokamak~\cite{Neubauer_2005} was a limiter machine with circular plasma cross section. The first wall was made of INCONEL and the ALT-II toroidal belt limiter~\cite{Kohlhass_1990} equipped with carbon tiles and three movable poloidal limiters (top, bottom and outboard) covered with EK98 graphite were utilized to control the plasma boundary and protect the first wall.
The first reported observations of RE damage to plasma facing components date back to the 1980s~\cite{Hoven_1989}. One graphite 'tooth' of the poloidal limiter at the outboard midplane was found to show erosion craters with a depth of approximately 1 mm. The surface of the limiter tooth showed a cracking pattern parallel to the surface. The damage was attributed to REs formed in low-density plasmas and during disruptions. 

RE damage was also observed to a scintillator probe in TEXTOR. The probe consisted of ten YSO scintillator crystals shielded by tungsten layers of various thicknesses. The entire probe was covered with 5 mm of graphite~\cite{Kudyakov_2008}. Inspection after about 30 RE measurements showed that the REs penetrated the graphite and deposited their energy on the inner part of the probe. Surprisingly, strong melting of stainless steel parts inside the probe was observed~\cite{Forster_2011}.

In addition, RE damage to a heat load probe was observed in TEXTOR. This probe had a bulk from epoxy resin with embedded copper particles which was contained in a graphite housing~\cite{Forster_2011}. The probe was designed to provide heat load information by post-mortem analysis of resin melt regions. However, the  heating was stronger than expected and the graphite housing cracked and the resin evaporated. Nevertheless, detailed analysis allowed an estimate of the mean RE kinetic energy in the range 8 MeV to 16 MeV to be made~\cite{Forster_2011}.

Finally, RE damage to a calorimeter probe was observed in TEXTOR. The calorimeter probe consisted of a thick piece of graphite along with a thinner piece of molybdenum forming the bulk, and a carbon fiber composite shield on the surface to protect the interior from the plasma. 
Both elements had thermocouples inserted to allow a temperature measurement~\cite{Forster_2012}. The probe showed quite severe damage when inspected after RE experiments. The CFC and graphite components were strongly eroded and had mass losses of up to 5 percent, while the molybdenum was partly molten but had lost only less than 1 percent of its mass.

\subsection*{Current and future challenges}

A variety of experimental challenges exist in the study of RE-induced damage of brittle in-vessel components. Foremost among these challenges is obtaining good data on incoming RE characteristics. Modeling of RE damage ideally begins with good data on the RE kinetic energy, pitch angle, and heat flux time history. These are hard to measure, however, as described in Section 4; and, in most instances, the characteristics of the impacting REs are not well-known. Nevertheless, recent analytical developments and advanced surface analysis techniques are beginning to offer promising opportunities to extract meaningful insights even from past, incompletely documented events of RE-induced cumulative damage. In dedicated experiments, progress is being made in diagnosing the characteristics of REs impacting instrumented in-vessel components. For example, calibrated fast thermocouples are being used to better quantify RE heat fluence, and embedded dosimeters are being implemented to help constrain incoming RE energy and pitch angle. 

\subsection*{Concluding remarks}

The study of brittle PFC RE impact is important for predicting possible RE damage to future brittle in-vessel components and for validation of physics models, which may be extended to metallic components. Across many machines, a clear hierarchy of damage is observed, ranging from no observable damage at low heat fluences, to surface morphology changes at medium heat fluences, to explosive dust release at high heat fluences. A wealth of information is available in present experiments, and future work will attempt to provide more quantitative measurements of brittle surface damage to provide better constraints on RE impact damage modeling. This work is expected to benefit RE damage modeling on W by first allowing model validation in the simpler no-melt-flow situation.

\clearpage 
\section{RE-induced damage of metal PFCs}\label{:sec3}
\author{M. De Angeli$^1$, R. Granetz$^2$, C. Reux$^3$, T. Tang$^4$ }
\address{
$^1$Institute for Plasma Science and Technology - CNR, Milan, Italy.\\
$^2$MIT PSFC, Cambridge, MA, USA\\
$^3$  IRFM, CEA Cadarache, F-13108 Saint-Paul-lez-Durance, France \\
$^4$Institute of Plasma Physics, Chinese Academy of Science, Hefei, China }

\subsection*{Status}

\begin{figure}[b]
  \centering
  \includegraphics[width=7cm]{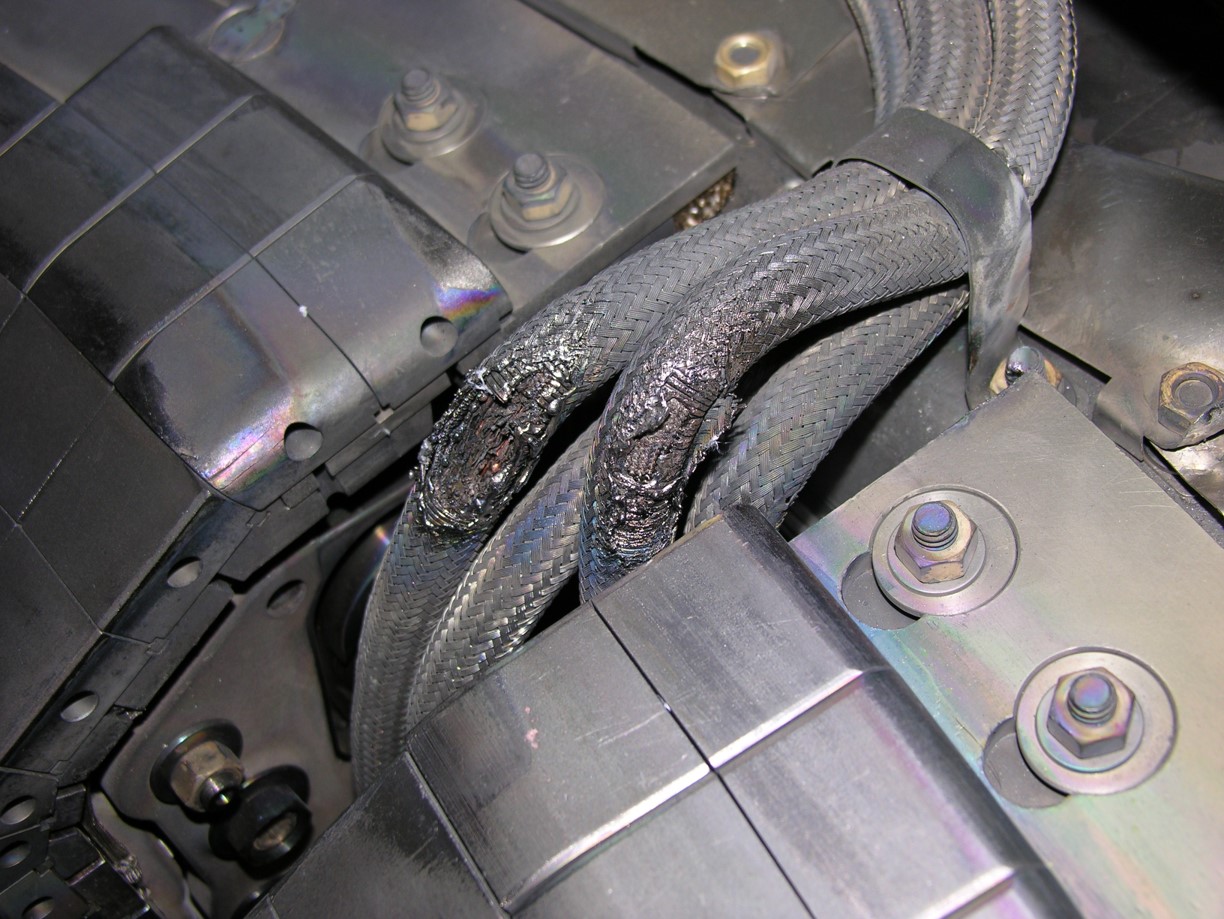}
  \caption{C-Mod SXR signal cables damaged by a RE strike}
  \label{C-Mod_damaged_SXR_cable}
\end{figure}

\paragraph{Alcator C-Mod.}
Alcator C-Mod's molybdenum tile first wall saw its share of melt
damage over its lifetime due to impacts by beams of runaway
electrons (REs).  C-Mod's first wall was not actively cooled, so there was no risk of coolant release, just melt damage, which had little or no effect on tokamak operations.
However, the worst RE event didn't involve the tiles on the
first wall, but rather the signal cables of C-Mod's soft x-ray
(SXR) tomography arrays. See figure~\ref{C-Mod_damaged_SXR_cable}.
These cables had a robust stainless steel
cladding, and were located in an area that was thought to be
protected from plasma contact.  An unusual plasma discharge occurred, which was
dominated by REs right from the beginning of the startup, and
it is likely that all 0.6 MA of plasma current was being carried
by runaway electrons.  The plasma disrupted without warning, and a
visible camera viewing the opposite side of the machine from the
SXR arrays recorded a huge spray of glowing hot molten droplets
emerging from behind the central column.  Half the channels in
one of the SXR arrays instantly went dead at that moment.  But
even worse, it took several days of discharge cleaning and
plasma attempts before normal plasma operation was
re-established.  During a subsequent planned shutdown and manned
access several months later, it was discovered that several
cubic centimeters of steel, copper, Teflon, and fiberglass had
been blasted out of the SXR signal cables.  Replacement signal
cables had to be fabricated and installed in order to return the
SXR tomography system back to full performance.

Regarding REs in Alcator C-Mod, a distinguishing feature was that despite having relatively high plasma current (up to 2.0 MA), disruptions \textit{never} generated relativistic RE beams or RE plateaus. REs occasionally occurred during plasma startup, sometimes continuing through ramp-up and into flattop. RE beams could also be generated during the discharge flattop by purposely reducing the density (so-called ``quiescent REs'').  This prescription allowed for controlled studies of synchrotron emission~\cite{Tinguely_2018}, and $E_{\rm crit}$~\cite{Granetz_2014} (the minimum electric field required for electrons to runaway).

\paragraph{FTU.}
FTU (Frascati Tokamak Upgrade) was a full metallic, cryogenic, high-magnetic-field tokamak with molybdenum-based alloy (TZM) external poloidal and internal toroidal limiters~\cite{Pizzuto_2004}. Generation of highly energetic RE beams was often observed during both start-up/steady state phases and disruptions. 
Typical values of RE energies and currents in FTU were 15-35 MeV and 150-230 kA respectively~\cite{Esposito_2017}, with incident angles (corresponding to the B field) of about 0-20 degrees depending on the toroidal profile of the tiles.

\begin{figure*}
\centering
\includegraphics[width=1.0\textwidth]{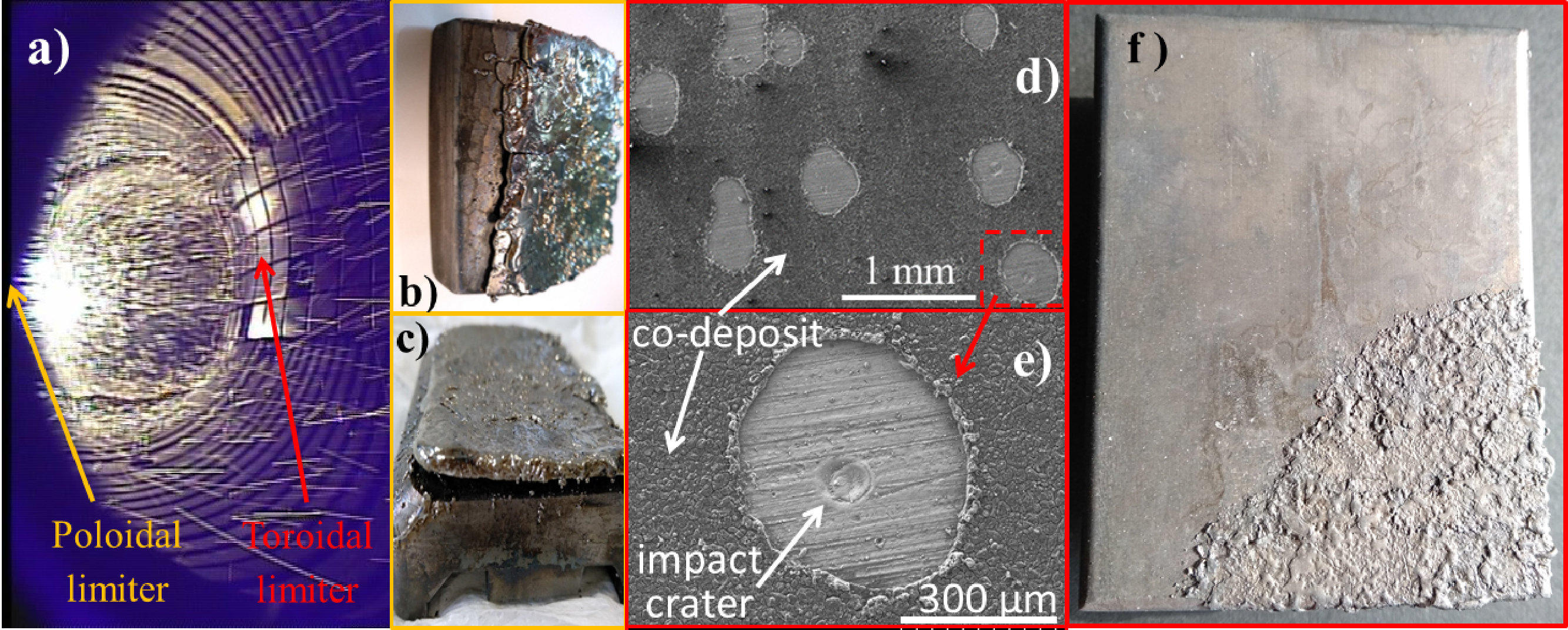}
\caption{Example of an RE-induced explosion event in FTU.  Debris trajectories recorded by VIS camera during shot 37761 (a). Damaged Mo-based tiles from the poloidal limiter showing several mm deep melting and layer de-attachment damage, (b)-(c). SEM images of craters, found on tiles from the TZM toroidal limiter, caused by fast debris impacts, (d)-(e). Reprint with permission from Ref.~\cite{DeAngeli_2023}.  Example of a  toroidal TZM tile damaged by RE beam terminating on the inward side showing $\sim$ 1 mm deep melting (f).
}\label{FTU_damage}
\end{figure*}

The most common terminal location for RE beams, generated during the start-up phase and accelerated in the stationary phase, was the midplane of the poloidal limiter, because of their outward orbit drift.
Such events cause the so-called \textit{primary localized damage} of PFCs by REs in FTU. 
The interaction between RE beams and the poloidal limiter is often, but not always, followed by an explosion-like event caused by the energetic RE beams dissipating their energy deeply inside the tiles bulk. 
In fact, the thermal shock caused by the RE energy dissipation drives material explosions leading to the ejection of fast solid dust~\cite{DeAngeli_2023, Ratynskaia_2025}.  The ejected dust, of $\sim 70$ $\mu$m of diameter and moving with velocities of $\sim 800$ $m s^{-1}$, hits the toroidal limiter tiles located across the vessel, almost in front of the poloidal limiter, resulting in the \textit{secondary non-localized damage}~\cite{DeAngeli_2023}.
Fig.~\ref{FTU_damage} shows an example of an explosive event, with release of fast debris (image (a)), along with damage induced to the poloidal and toroidal limiter TZM tiles, accumulated over several experimental campaigns.

The observed primary localized damage of poloidal tiles, exemplified in Figs.~\ref{FTU_damage}(b)-(c), include: (i) deep melting and material loss up to $\sim 6$ mm deep; (ii) de-attachments of several millimeters (up to 7 mm) thick layers; (iii) generation of  millimeters long intergranular cracks on the surface and in the bulk of tiles; and (iv) recrystallization and degradation of mechanical properties of the bulk material (i.e. hardness, measured by microhardness tests along cross-section profiles of some tiles). 
On the other hand, the secondary non-localized damage found on toroidal tiles, see Figs.~\ref{FTU_damage} (d)-(e), concern: (i) formation of craters, $\sim 100$ $\mu$m of diameter and $\sim 10$ $\mu$m in depth, due to fast dust impacts; (ii) cracks in the PFCs bulk; (iii) removal and migration of pre-existing co-deposited material present around the crater locations. 

In the case of RE beams generated during disruptions in FTU, the beams tend to terminate on the inward side of the vacuum vessel~\cite{Maddaluno_1999}, hitting the toroidal limiter tiles. Contrary to the outward RE beam trajectories, in this case the RE beams, having no preferential terminal point, could impact the toroidal limiter at any toroidal location. The resulting damage, about 1 mm deep and with no signs of layer de-attachment, shown in Fig.~\ref{FTU_damage} (f), is less destructive than that of poloidal tiles subjected to outward beams (compare with Figs ~\ref{FTU_damage}(b)-(c)).

FTU also had experimental campaigns with liquid tin contained in W capillary porous structures (CPS)~\cite{Vertkov_2017} wet by a layer between about 10 and 100 $\mu$m of liquid tin. These W meshes, extracted from the machine in 2019 after being exposed to hundreds of plasma discharges and to several RE beams, came out intact. Currently the work on identification of discharges that contained RE populations which could strike the meshes is ongoing though reaching an unambiguous conclusion on whether the RE beams fell on the liquid metal limiter does not appear feasible.
  
In general, it should be emphasized that the severe damages of the poloidal limiter tiles had almost no significant effect on tokamak operations and the damaged tiles could be replaced at the end of experimental campaigns or even during campaigns, since the poloidal limiter was removable without breaking the vacuum. Moreover, the limiter tiles in FTU were not actively cooled, thus excluding the risk of coolant accidents.

\paragraph{WEST.}
WEST is a long-pulse full tungsten machine~\cite{Bucalossi_2022}. It started out with W-coated CFC, but gradually moved to bulk tungsten components, including an ITER technology divertor. Aside from the divertor, WEST has 18 ribs of inner wall limiters (6 groups of 3 ribs in each sector) and one outer limiter, toroidally localized. WEST, especially in its early phase, tended to generate significant amounts of REs at startup. Many of these were not controlled in the first year of operation and terminated on the inner or outer limiters~\cite{Reux_2021} The W coatings hit by REs were quickly ablated, especially on the outer limiter, which is the sole point of interaction between REs and the outer wall. Post-mortem inspections of the outer limiter showed molten coatings sprayed and re-deposited around the interaction point, spread over 12 tiles, over a length of more than 50 cm in the poloidal direction ~\cite{Diez_2021}. The carbon substrate below the damaged tungsten coating was sometimes visible. Surprisingly, the inner limiter damage was less dramatic, with no large traces of extensive molten coating, but black markings. No systematic study of the impact features were conducted because most of the tiles accumulated dozens of impacts. The outer limiter itself suffered from about 64 impacts in that early phase, with one leading to the quench of a superconducting coil through gamma/neutron heating generated by the RE beam impact~\cite{Reux_2021, Torre_2019, Nicollet_2022}. Ejecta from the impact point were seen on almost every impact, with some particles large enough to bounce several times on the lower divertor or the baffle before cooling down. The damaged limiters did not prevent the operation of the tokamaks, and if they had a deleterious effect on operations, then it must have been a very slow and mild degradation because it went unnoticed. 

\begin{figure}
    \centering
    \includegraphics[width=0.8\linewidth]{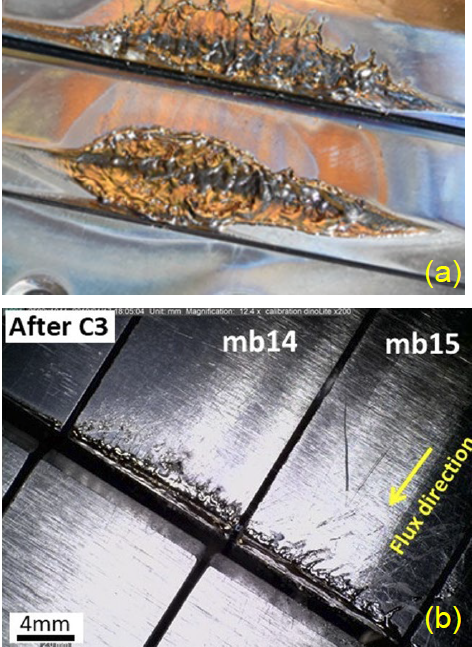}
    \caption{ RE-induced damage of W tiles of WEST outer limiter (a). Courtesy of M. Diez. The trailing edge of W monoblocks on the WEST divertor hit by REs, observations after the C3 campaign (b). Reprint with permission from Ref.~\cite{Diez_2021}.}
    \label{fig:WEST_tungsten_damage}
\end{figure}

A few more impacts were recorded on the outer limiter after its transition to bulk tungsten~\cite{Reux_2024}. Splashed molten pools on the limiter W tiles are depicted in Fig. \ref{fig:WEST_tungsten_damage} (a). 
The WEST divertor took much fewer visible hits since the beginning of WEST operations. A single impact was recorded in the first phase of WEST operations~\cite{Diez_2021} from 2017 to 2020. The tungsten tile, shown in Fig. \ref{fig:WEST_tungsten_damage} (b), reached melting temperatures on its trailing edge, as expected from the RE beam geometry. About five RE impacts were recorded on the divertor in the C7 campaign, again on trailing edges \cite{Houry_2024} and with strong toroidal asymmetries, similar to what is shown in Section 2 addressing damage to brittle PFCs on WEST. In all cases, clear traces of displaced liquid tungsten are visible in the post-mortem analysis over a radial length of about 3.5 cm~\cite{Reux_2024}, but did not cause significant operational problems. Most of these impacts were not noticed during operations.

\begin{figure}
    \centering
    \includegraphics[width = 7.4 cm]{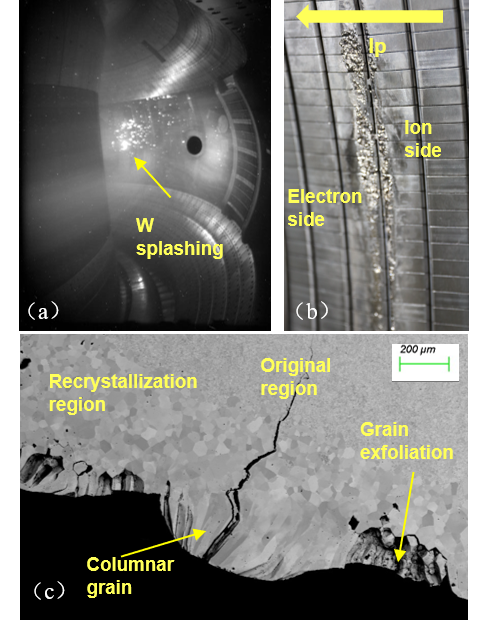}
    \caption{RE-induced damages in EAST main limiter. a) Hot spot and W splashing viewed by CCD camera; b) surface morphology of melting main limiter and c) cross section morphology of the melting region.
    }
    \label{EAST_damage}
\end{figure}

\paragraph{EAST.}
EAST is a fully superconducting tokamak with actively water-cooled tungsten-copper (W/Cu) PFCs for both the divertors and main limiter~\cite{Guo_2024}. As the closest PFC to the plasma at the low field side, the main limiter serves for absorbing and exhausting heat flux from the scrape-off layer to protect diagnostics and antennas during plasma operation. Consequently, interactions between plasma and main limiter occur frequently, predominately induced by REs during the beginning of each EAST campaign, particularly when REs are lost in disruptions. One of the examples is presented in Fig.~\ref{EAST_damage}, showing RE-induced tungsten splashing during disruption, as monitored by CCD camera, along with the resulting surface morphology of the molten tungsten limiter.

In EAST, RE currents have been typically measured in the range of 100-200 kA, with maximum energy reaching approximately 14 MeV. The sudden spike in HXR signal coincides with a sharp increase in temperature of the main limiter. Simultaneously, a hot spot appears, which eventually leads to W splashing, suggesting damages on the main limiter are predominately induced by runaway electrons. The splashed W material is ejected in all directions, reaching speeds of several tens m/s~\cite{Xuan_2025}.

Significant melting of the main limiter is found after 2023 EAST autumn experimental campaign, as seen in Fig.~\ref{EAST_damage} (b, c). Both damages were found on ion side and electron side, as noted in the direction of plasma current and in opposite direction in Fig.~\ref{EAST_damage}, however, more severe melting on electron side further confirmed that damages are predominately caused by REs. Cross-section morphology analysis reveals three distinct grain layers from the molten surface to deep region: columnar grain, recrystallization region and original grain respectively. The depth of columnar grain typically ranges from 100 to 300 $\mu m$, and the depth of the recrystallization region is of comparable depth. The grain distribution suggests a sharp temperature gradient, from the surface ($>$ 3695 K) to the bottom of recrystallization region ($>$ 1473 K), within several micrometer's region impacted by REs. The presence of columnar grain indicates near-surface melting of tungsten, as material tends to form columnar grain when transient heat fluxes are high enough to induce melting~\cite{Xuan_2025}.

\paragraph*{Other machines.}
Other noticeable machines equipped with metal PFCs are ASDEX Upgrade (AUG) and KSTAR. In over a decade of dedicated runaway electron beam experiments~\cite{Pautasso_2017, Heinrich_2024, Sheikh_2024}, and several cases of unintentional RE beams, so far the AUG team has not found any clear evidence of significant RE damages of PFCs. Nevertheless, a dedicated investigation on RE-material interaction is planned.

KSTAR replaced the carbon tiles in their lower divertor with tungsten tiles in 2023, and to date they have completed two plasma campaigns with the tungsten lower divertor.  No damage due to runaway electrons has been observed so far, and in fact, REs are rarely seen, even during disruptions.  This lack of disruption-generated REs is reminiscent of the experience on Alcator C-Mod, as mentioned previously.

\subsection*{Current and future challenges}

Although present-day metal machines are able to continue operating after one or more RE strikes, future machines such as ITER, ARC, or DEMO will have active cooling structures buried in their PFCs, which could potentially be compromised by an RE impacts causing deep energy deposition.  This could have major implications, such as forcing reactor operations to cease until repairs are made. This is a strong motivation to better understand the phenomena involved in the interaction of REs with metallic materials in order to characterize the impact geometry, i.e., in terms of the incident angle, toroidal symmetry, deposition area and especially energy deposition depth. This also exemplifies the need to investigate and develop methods to robustly avoid or mitigate runaway electrons, such as the passive RE mitigation coil that will be installed in SPARC~\cite{Rodriguez-Fernandez_2022}.

In dedicated investigations on the interaction of REs with PFCs, the main challenge is twofold. First, generation of controlled RE beams with known parameters is needed in order to provide input for the modeling of the energy deposition. Second, the beam should strike a precisely predicted point where an instrumented tile (a tile with embedded detectors) is located to enable measurements of various loading characteristics. 
In fact, the main concern in carrying out such activities in tokamaks is the safety issue related to the risk of damage from uncontrolled RE beams striking undesired locations.

Another issue, less harmful but still deserving further investigation is secondary non-localized damage in the surrounding PFCs due to the impacts of fast debris released upon explosive RE beam-PFCs interaction events, and the related dust generation problems.

Further RE experiments are planned on AUG and WEST tokamaks, mostly centered on RE mitigation through the benign termination scenario \cite{Reux_2021b} but also first dedicated RE-induced damaged exposure will take place in 2025. In particular, one of the WEST inner bumper tiles has been equipped with thermocouples to better diagnose the deposited energy. In AUG, the upcoming experiment will have an instrumented W sample which will be exposed by using the mid-plane manipulator system. 

\subsection*{Concluding remarks}

It is known that the generation of RE beams and their frequency depends on plasma configuration and size~\cite{Breizman_2019} and an infrequent generation of RE beams is common in present-day tokamaks. This trend is confirmed by the described metal machines. All metal tokamaks in this section have shown impacts reaching melting temperatures with splashing as seen from post-mortem analysis. High speed ejecta have also been observed on some of them, with secondary damage. Toroidally and poloidally localized impact damage is also a common feature of all observations.

Generally speaking, the positive lesson learned from metal machines is that the damage to metallic components by RE does not inhibit tokamak operation as long as no coolant is released. The situation may be different on larger machines with higher currents. RE beams could still cause major operation problems if they hit recessed components such as signal cables.

Overall, studies on RE-metal interaction are in their infancy and are ongoing on several machines. These investigations are needed to predict and prevent serious damage in future devices, which could stop their normal operation in case of severe damage.

\clearpage 
\section{RE diagnostics}\label{:sec4}
\author{O. Ficker$^1$, E. Hollmann$^2$, U. Sheikh$^3$, S.A. Silburn$^4$}
\address{
$^1$ Institute of Plasma Physics of the Czech Academy of Sciences, Prague, Czech Republic\\
$^2$ Center for Energy Research, University of California - San Diego, USA\\
$^3$ Swiss Plasma Center, École Polytechnique Fédérale de Lausanne, Switzerland\\
$^4$ UKAEA, Culham Campus, Abingdon, OX14 3DB, UK}

\subsection*{Status}
\begin{figure}[h!]
    \centering
    \includegraphics[width=0.8\linewidth]{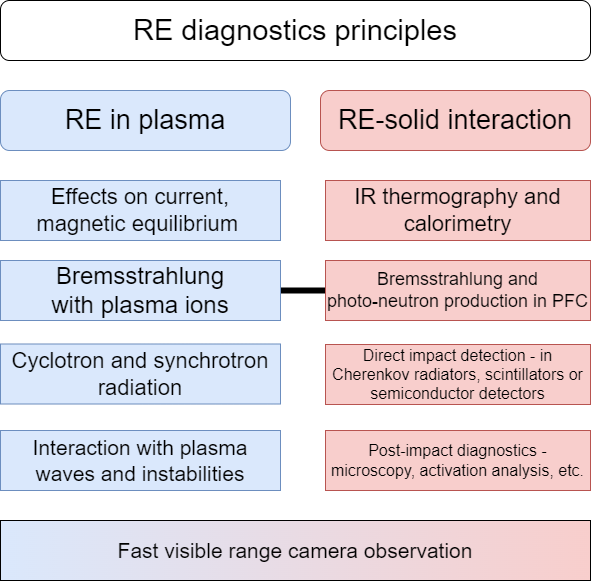}
    \caption{The RE diagnostics classification. Principles highlighted  in red are covered in this section.
    }
    \label{fig:diag_overview}
\end{figure}

The runaway electron (RE) diagnostics is a wide area of research at the intersection of high temperature plasma diagnostics and diagnostics of relativistic particle effects. Various methods are employed to diagnose the properties, mainly energy distribution, pitch angle and (current) density of REs in the tokamak. As the RE diagnostics in general were covered in recent overview publications~\cite{Breizman_2019,Ficker_2023} this section is focused on the means of RE impact diagnostics  - red boxes in Fig.~\ref{fig:diag_overview}. 

\subsubsection*{IR thermography. }
Infra-red (IR) imaging is a valuable diagnostic for measuring the spatial structure of heat loads on plasma-facing components (PFCs)~\cite{hill_1988,vondracek_2017,dunn_2020}, and is widely available on many tokamaks. For RE studies, IR imaging has also found use for study of in-plasma RE beam structure through measurement of synchrotron emission~\cite{Finken_1990,Tong_2016,Reux_2022,Zhang_2025}, and this application is being considered during the design process of IR diagnostics for near-future devices such as SPARC~\cite{Tinguely_2024}. IR imaging of PFC surface temperature excursions following RE-wall impacts was first reported from JET, where it showed that RE impact on the upper dump plate appeared to preferentially heat tile leading edges~\cite{Lehnen_2009}. 
Similar IR imaging of RE impacts have since been obtained in other tokamaks including TCV~\cite{Sheikh_2024}, ASDEX Upgrade~\cite{Sheikh_2024}, DIII-D~\cite{Beidler_2024} and COMPASS~\cite{Caloud_2024}. Fitting of interpretative models to the measured time-dependent PFC surface temperatures has been used to estimate the energy deposition densities in RE-wall impact events, allowing quantification and characterisation of the relative severity of impacts~\cite{Reux_2022, Hollmann_2017}. 
The ability to directly observe the spatial distribution and, more recently, also to quantify the heat deposition, has enabled the observation across several devices that RE beams accompanied by a low-Z (light impurity) background plasma tend to have a significantly (several times) larger poloidal wetted area than RE beams accompanied by high-Z (heavy impurity), with correspondingly lower heat deposition density~\cite{Reux_2021,Sheikh_2024,Hollmann_2025}. IR imaging provides not only accurate information on position of hot-spots due to RE-wall interaction but is also capable of detecting heated material ejected from the impact site with more sensitivity than visible cameras (at lower temperature). It thus serves as a useful diagnostics to relate findings from post-mortem analysis of PFCs and wall inspections with specific RE-wall impact events as well as  to study the large scale structure and symmetry of impacts. 
In principle, the measured surface temperature spatial and temporal distributions can be useful to benchmark RE-wall impact modelling, however uncertainties in the interpretation of the IR imaging, as described below, make detailed comparisons challenging.

\subsubsection*{Calorimetry. }
\label{subsec:calorimetry}
Calorimetry in tokamaks is a critical diagnostic technique used to measure the energy deposited on PFCs, where they can directly intercept the plasma or RE beams. Calorimeters are typically designed as thermal sensors embedded in tiles that measure temperature changes and infer energy deposition in the volume. The simplicity of these devices can allow the installation of numerous sensors in a tokamak, proving spatial information on the energy distribution.
The use of such systems has been demonstrated on TEXTOR, COMPASS, JET, DIII-D, AUG and TCV, where calorimetry has provided valuable insights into energy deposition and the spatial characteristics of RE beams. Experiments on TEXTOR, COMPASS and DIII-D developed protruding calorimetry probes to make direct measurements of impact energy deposition. In TEXTOR, a typical RE plateau with a current of 100kA resulted in a deposited energy of 30-35kJ, which represents a magnetic to kinetic energy conversion of 30\%~\cite{Forster_2012}. The deposited energy was found to vary linearly with RE current. Similarly, deposited energies of 2-12 kJ were measured in COMPASS at RE currents in the range 40-130 kA and were found to scale with the loop voltage and amount of injected impurities~\cite{Caloud_2024}. 

In DIII-D experiments, RE impact calorimetry has been performed with intentionally inserted domed graphite limiter heads. These heads were inserted during dedicated RE-forming disruptions using the DIII-D material exposure system (DiMES) which can insert samples into the
lower divertor floor of DIII-D~\cite{Wong_2007}. 
The graphite dome protruded 1 cm above lower divertor floor. A thermocouple is used to measure temperature inside the domed head (forming a calorimeter system), while a shunt resistor and current monitor lead are used to measured current flowing from the limiter head into vessel ground. Dosimeter chips are used to monitor the single-shot hard X-ray (HXR) fluence distribution inside the limiter. These heads were intentionally struck with high current (500 - 600 kA) low-Z RE plateaus and heat loads of order 1 - 10 kJ were measured. This large variation was attributed to the toroidal phase of the RE final loss MHD varying from shot to shot.


Work on AUG, JET and TCV is on-going within the RE benign termination database (ITPA MDC-23), where toroidally separated thermocouples embedded in tiles are being compared with infrared thermography measurements to infer heat flux and toroidal asymmetries~\cite{Sheikh_2023,Sheikh_2024}.

\subsubsection*{X-rays at RE impact.}
\begin{figure}
    \centering
    \includegraphics[width=0.7\linewidth]{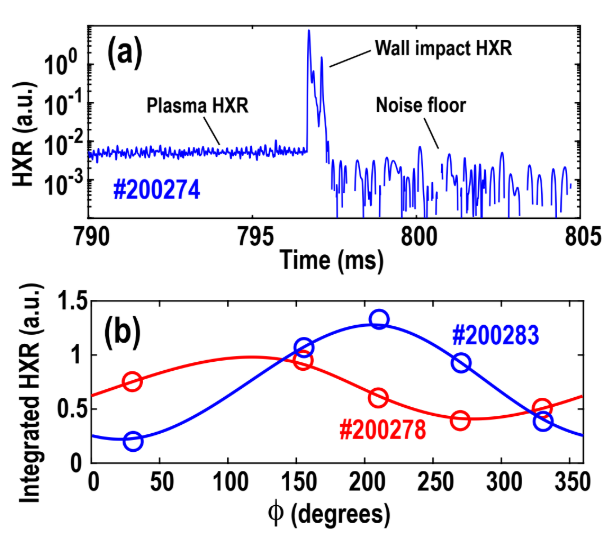}
    \caption{HXR time trace showing steady plasma emission initially and then a huge spike at the main RE impact event (a). Integral over the wall impact spike as a function of outer midplane detector toroidal position with approximate fit (b) . 
    }
    \label{fig:diag_HXR_asymmetry}
\end{figure}

As the impact of MeV-range electrons onto a solid surface is accompanied by very intensive bremsstrahlung radiation in the HXR and soft X-ray 
(SXR) spectral regions, various detectors of such radiation are instrumental in assessing the time evolution, duration or intensity of the impact and estimating the energy of the RE population. The radiation is also produced during interaction of RE with plasma ions, mainly impurities, however, unless very large impurity densities are reached, the HXR intensities from the RE-PFC interaction are dominating, especially within fast loss events. This radiation is typically measured by uncollimated or partially collimated HXR monitors and spectrometers based on suitable fast scintillators with a photo-multiplier. LaBr$_{3}$(Ce) or CeBr$_{3}$ are the most common scintillation crystals, significantly surpassing e.g., previously favourite NaI(Tl) in scintillation decay time. Such systems are in use at AUG~\cite{Dalmolin_2023}, TCV~\cite{Simons_2023,Simons_2025}, JET~\cite{Tardocchi_2008}, COMPASS~\cite{Cerovsky_2022}, DIII-D~\cite{Dalmolin_2021}, J-TEXT~\cite{Ma_2017} and many other machines. At JET~\cite{Rigamonti_2018}, DIII-D~\cite{Pace_2016} and WEST~\cite{Wongrach_2021}, collimated multi-LOS systems are used as well, however these are typically dedicated to measure RE-ion interaction bremsstrahlung and not RE-PFC interaction.
The comparison of steady state and termination relative HXR flux together with approximate measurement of toroidal asymmetry of HXR flux during the impact at DIII-D is shown in Fig.~\ref{fig:diag_HXR_asymmetry}.

\begin{figure}
    \centering
    \includegraphics[width=0.8\linewidth]{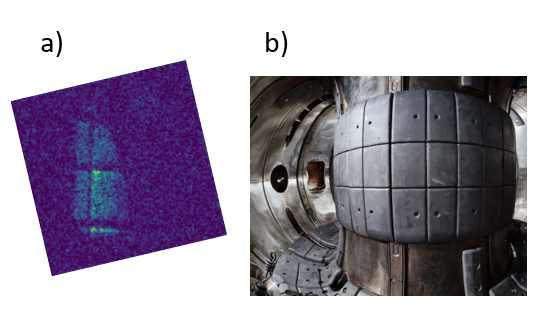}
    \caption{ SXR image of start-up REs impacting HFS carbon limiter tiles of COMPASS as recorded by TimPix3 with lead pinhole (a).  Interior of the COMPASS vacuum vessel for reference of the HFS tiles layout (b). Reprint with permission from Ref.~\cite{kulkov_2025}.
    }
    \label{fig:diag_tpix}
\end{figure}

The HXR spectrometers provide energy resolution via pulse-height analysis and the RE energy may be inferred from the HXR spectra using regularisation. This is however an inverse problem based on two subsequent processes, thus only limited information can be retrieved - general shape of the spectra and maximum or average energy with significant error bars. Furthermore, the potential of these methods is significantly suppressed by the high flux at termination leading to pile-ups.\\
At COMPASS and Golem~\cite{Grover_2016}, also high pixel resolution extended range SXR matrix chips derived from TimePix3~\cite{Poikela_2014} were tested~\cite{svihra_2019,kulkov_2022,kulkov_2025} for direct imaging of the impact locations, in an extended dynamic range or in a Compton Camera set-up.
It is possible to detect the impact of the start-up REs on the high field side tiles, as shown in Fig.~\ref{fig:diag_tpix}), however the small pixels absorb  only a minor part of the photon energy for $E_{HXR}>50\,\mathrm{keV}$. 

\subsubsection*{Photo-neutrons.}
The photo-neutrons as a product of bremsstrahlung-generated X-rays interacting with suitable nuclei can also be used for diagnostics~\cite{Jarvis_1988}. 
As photo-neutrons and PFC activation are covered  in Sec.\ref{:sec9}, we only mention the most important advantages for RE impact diagnostics;  (i) existence of energy threshold for given reaction provides energy discrimination for the RE population, (ii) due to the scattering of the neutrons before detection, the measurement of photo-neutrons can be considered less depended on the impact location than HXR detection, which is useful for a global comparison of impact events. 
The role of photo-neutron detection in comparison with other diagnostic means has been investigated, e.g., in Ref.~\cite{Ficker_2023}.

\subsubsection*{Cherenkov probes and other direct RE detection methods. }
The direct detection methods, i.e., those where the energy of REs is fully or partially deposited in the detector volume are very attractive, however their applications are limited. The advantages are precise spatial localization and no intermediate steps in the RE measurement process. The main disadvantage is the risk of fast deterioration of the detector due to exposure to REs. 
Cherenkov detectors are based on production of Cherenkov radiation in diamond or similar radiator covered by metallic layers to provide energy thresholds for impacting REs. Typically three channels with different energy thresholds in the range of energies 50-300 keV are used. Such detectors were applied at FTU~\cite{Zebrowski_2018}, COMPASS~\cite{Rabinski_2017}, TCV~\cite{Kwiatkowski_2021}, ISTTOK~\cite{Plyusnin_2008}, Tore Supra~\cite{Jakubowski_2010} and other machines. Due to the given range of energy thresholds in hundreds of keV, the Cherenkov detectors are best suited for measuring start-up REs or TQ RE seed losses. The detectors must be tuned to the suitable range of fluxes of RE and likely cannot capture the termination events, where fluxes increase by several orders of magnitude.
Similarly, scintillation detector directly exposed to RE flux or series of probes consisting of conversion layers producing bremsstrahlung and scintillators detecting it are utilised to inspect the properties of RE population directly. Such methods were applied at TEXTOR~\cite{Kudyakov_2008} and concepts developed, e.g., at Golem~\cite{Dhyani_2019}. Various silicon-based detectors can also be used for direct detection of lower energy RE population~\cite{Novotny_2020}. In principle, various electric probes, floating or biased can be also used to detect electrons of high energy. A planar electric probe was embedded in the COMPASS calorimeter~\cite{Caloud_2024}.
The pitch angle of the RE population can be estimated by the direct detection methods using suitable geometry of the detectors, e.g. adding various collimators or entrance slits that allow only RE with specific pitch angle to reach the detector volume.

\subsubsection*{Post-impact damage analysis. }
\begin{figure}
    \centering
    \includegraphics[width=1.0\linewidth]{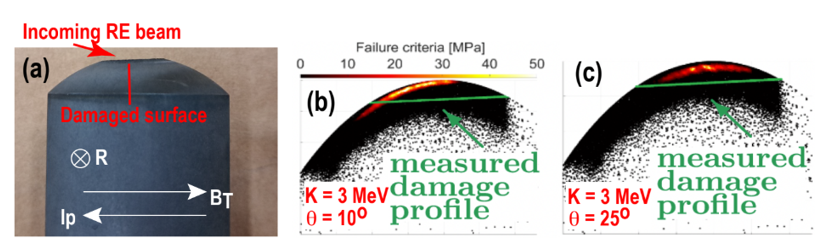}
    \caption{RE impact on graphite limiter as a pitch angle $\theta_{p}$ diagnostics: graphite limiter damage from RE impact (a)  and  modelling the graphite material failure, (b) and (c) . The model assumed REs with a single initial kinetic energy and pitch, the best matching parameters are $\mathrm{E}\approx 3 \mathrm{MeV}$ and $\theta_{p} \approx 25 ^{\circ}$. Reprint with permission from  Ref.~\cite{Ratynskaia_2025}.
    }
    \label{fig:diag-pitch-analysis}
\end{figure}

The impact locations on various PFCs or diagnostics are analysed by the means of optical microscopy~\cite{Caloud_2024, DeAngeli_2023} and scanning electron microscopy~\cite{DeAngeli_2023} in both impact surface view and cut through the sample. 
Various types of damage have been observed after the impact, including craters and re-deposition structures. The analysis of the chemical composition is performed using energy-dispersive X-ray spectroscopy (EDX), X-ray photoelectron spectroscopy (XPS) and other methods to investigate layer transitions in the PFC and oxidation of the impacted surface.
Furthermore, the impact pattern on special surface shape evaluated post-mortem can be used together with modelling to investigate the parameters of RE population, e.g., the pitch angle distribution, as shown in Fig.~\ref{fig:diag-pitch-analysis}.

\subsubsection*{Fast camera observation of the damage dynamics. }
Cameras with high pixel resolution and frame rate up to several 100 kHz (exceeding the performance of IR cameras) can provide imaging of melting, sputtering, ejection of debris and other processes accompanying the impact of REs on the PFCs. Relevant data has been recorded on multiple machines and the video processing, including mapping to magnetic field and CAD models, image analysis, particle tracking~\cite{Cowley_2020} etc.\ is crucial for the inference of physical quantities.

\subsection*{Current and future challenges}
\subsubsection*{IR thermography.} 
An overview of challenges involved in interpretation of IR images of RE-wall impacts was presented previously~\cite{Breizman_2019}. Well-known challenges for quantitative analysis of IR imaging even for steady-state plasmas include surface morphology~\cite{Delchambre_2009}, reflections~\cite{Aumeunier_2017} and volumetric emission of IR light by the plasma, all of which are exacerbated during RE experiments. Surface morphology role becomes particularly important for strong impact events where surface changes, e.g. by very rapid melting and resolidification of metal PFCs, might be significant.  This introduces large uncertainties in the wall surface profile and emissivity evolving on the same timescale as the desired measurement, potentially distorting the time history of the measured temperature. Application of multi-wavelength IR imaging~\cite{McLean_2012,Ushiki_2022,Zhang_2025} may be able to help address this challenge, however such systems are not yet common on existing devices and likely will still suffer uncertainties in such a complex measurement environment. 
Synchrotron emission, while enabling other diagnostic techniques as referenced above, is an additional very strong source of contaminating volumetric emission and associated reflections when considering wall temperature measurements. For example during recent RE experiments on JET, the radiance in the 3 - 3.5\,$\mu$m wavelength band when directly viewing the synchrotron emission was typically several times the radiance emitted by a Be surface at its melting point, making the surface temperature measurements barely possible. 
To some extent, this can be mitigated due to the imaging nature of the diagnostic, identifying suitable lines of sight in the image which can be used as a measure of the unwanted signal to be subtracted, and by comparing the time variation of different image regions to separate contributions to the IR signal from different effects, as was done in Ref.~\cite{Reux_2022,Hollmann_2017}. The significantly higher dynamic range requirements resulting from these constraints are met by the current IR cameras digitized at 14-bits.

\begin{figure}
    \centering
    \includegraphics[width=0.6\linewidth]{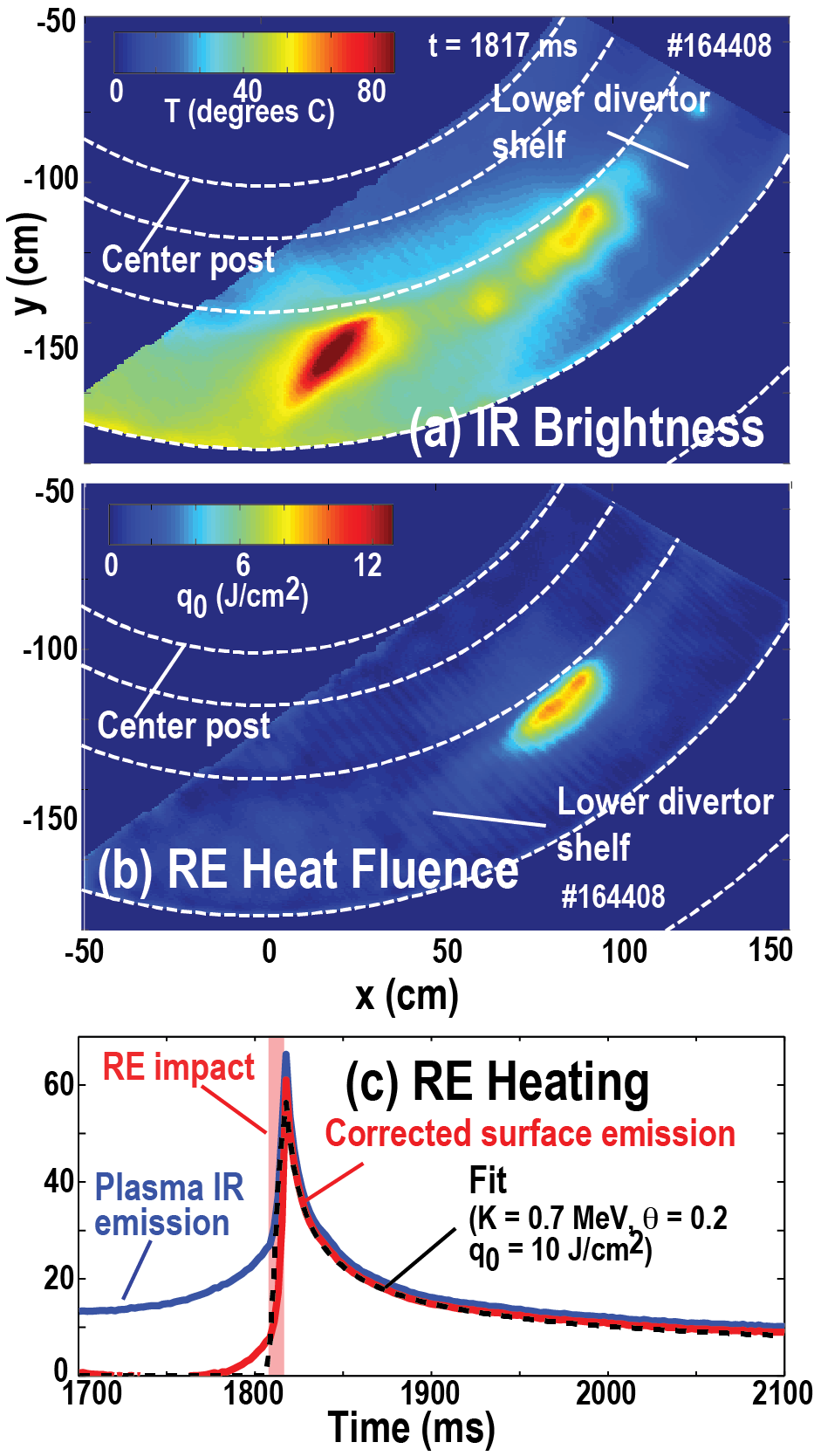}
    \caption{IR image of RE impact on lower divertor shelf of DIII-D (a), RE heat fluence extracted from IR image (b) and time decay analysis of single pixel to remove IR plasma emission and estimate RE impact heat fluence (c). Reprint with permission from Ref.~\cite{Hollmann_2017}
}
    \label{fig:diag_ir}
\end{figure}

Fig.~\ref{fig:diag_ir} illustrates the results of IR analysis of RE impact on the DIII-D divertor shelf where these challenges have been taken into account~\cite{Hollmann_2017}.  Fig.~\ref{fig:diag_ir}(a) shows an IR image of the lower divertor shelf of DIII-D during an RE-wall strike. Time decay analysis is used to isolate the RE heat fluence, giving a map of the RE heat fluence, Fig.~\ref{fig:diag_ir}(b), and removing the plasma line emission. The time history of a single point on the image is shown in Fig.~\ref{fig:diag_ir}(c). The initial plasma line emission (blue curve) is removed from the analysis. Fitting the resulting time decay (black dashed curve) can give the RE heat fluence. This time decay analysis method cannot accurately separate RE kinetic energy and pitch angle, and future techniques need to be developed to remove this degeneracy. The IR imaging discussed above was done in the ``mid-IR'' imaging range of $3 - 5 \mathrm{\mu m}$; it has not been investigated yet to what degree changing imaging wavelength range could benefit RE impact diagnosis. For example, moving to longer wavelengths could reduce sensitivity to plasma line emission but at the cost of reduced frame rate and spatial resolution.

The timescale for RE-wall impact is usually quite short ($\sim 1$ ms), which can be faster than the frame rate of the IR imaging, complicating the analysis due to the fast dynamics not being fully resolved. A further interpretation challenge is due to the volumetric nature of the heat deposition, which has to be reconstructed with the aid of modelling, since the diagnostic can only measure the surface temperature.

\subsubsection*{HXR, photo-neutron and direct detection.}
The diagnostics systems are subject to challenging design decisions, mainly in terms of collimation and scintillator size, balancing the count rate range and energy range/resolution. The flux of photons at the detector position may vary across more than 10 orders of magnitude depending on the scenario, thus multiple detectors with very different sensitivity are often preferred.

\subsection*{Concluding remarks}
Typically, different machines excel in different diagnostics systems. The progress in commercial detector performance (IR and visible range cameras, scintillators) further pushes the boundaries of RE-PFC interaction diagnostics 
The synthetic diagnostics approach applied to combine several different diagnostics methods is the most promising way to increase the relevance and accuracy of physical parameter inference and connection with the modelling.

\clearpage 
\section{Models for the characteristics of the RE beam prior to the impact}\label{:sec5}
\author{M. Hoppe$^1$, B. Breizman$^2$, V. Bandaru$^3$, G. Pokol$^4$}
\address{
$^1$Department of Electrical Engineering, KTH Royal Institute of Technology, SE-114 28, Stockholm, Sweden
$^2$ Institute for Fusion Studies, The University of Texas at Austin, USA
$^3$ Department of Mechanical Engineering, Indian Institute of Technology Guwahati, India
$^4$ Department of Nuclear Techniques, Faculty of Natural Sciences, Budapest Univ.\ of Technology and Economics, Hungary
}


Characterizing the wall damage caused by runaway electrons (RE) impact requires detailed knowledge of the RE beam. This involves the beam energy spectrum and the impact angles that depend on the pitch angle distribution before impact. This section summarizes the current status of simulations aimed at providing such a detailed characterization of the RE beam and some of the associated challenges.

\subsection*{Status}

The RE beam characteristics prior to impact are determined by the full history of events occurring from the point of initial RE generation up to the impact. Generation of REs requires the parallel electric field $E_\parallel$ to exceed a critical value, $E_{\rm c}$, and so REs are normally found in phases where the plasma current changes in time. Of particular concern to reactor-scale tokamaks are REs generated during startup/ramp-up and disruptions, with the latter receiving the most attention in literature. Although the plasma environment during startup and disruptions has some similarities, specialized models tailored to the specific conditions are usually employed. In the modeling of both phases, there is a need to account not only for the kinetics of the REs but also for the evolution of the background plasma and the way in which it is influenced by the REs. In many situations, REs carry a significant fraction of the plasma current and therefore strongly influence the electric field evolution and current-driven MHD instabilities. REs can also impact, or even dominate, the energy balance and ionization rate of the plasma, such that a reliable RE beam model must cover these effects.

Two complementary modeling approaches are used to gain predictive insights.
The kinetic approach deals with an equation for the energy and pitch angle distribution function $f_e$ of REs, namely,
\begin{eqnarray}
	\frac{\partial f_e}{\partial t} &=
		-eE_\parallel\frac{\partial f_e}{\partial p_\parallel} +
		\nabla_p\cdot\left(\boldsymbol{F}_{\rm rad}f_e\right) +\nonumber\\
		&+ C_{\rm FP}\left[f_e\right] +
		C_{\rm ava}\left[f_e\right] +
        Q_e\left[f_e\right].\label{eq:5:kinetic}
\end{eqnarray}
Here, $E_\parallel$ is the electric field component parallel to the magnetic field, and $\boldsymbol{F}_{\rm rad}$ denotes the reaction-force due to bremsstrahlung~\cite{Bakhtiari_2005,Fernandez-Gomez_2007,Embreus_2016} and synchrotron radiation~\cite{Andersson_2001,Decker_2016}. The Fokker-Planck collision operator $C_{\rm FP}[f_e]$ must be relativistic~\cite{Beliaev_1956,Braams_1987}, and a test particle approximation is commonly used for the electron-electron operator~\cite{Pike_2014}. In the electron-ion collision operator, partial ionization effects are essential for RE dynamics~\cite{Zhogolev_2014,Breizman_2017,Hesslow_2017,Hesslow_2018b}, especially in mitigated disruptions. In contrast to thermal electrons, REs are strongly influenced by large-angle collisions causing the avalanche-type exponential growth~\cite{Sokolov_1979}, as described by an avalanche source term~\cite{Rosenbluth_1997,Chiu_1998} in the collision operator denoted $C_{\rm ava}[f_e]$~\cite{Aleynikov_2014,Boozer_2015,Embreus_2018}. To describe the effect of wave-particle interactions, a quasi-linear operator $Q_e[f_e]$ is included~\cite{Parail_1978,Pokol_2008,Aleynikov_2015b}.

An analytical solution of Eq.~(\ref{eq:5:kinetic}) is feasible under certain conditions, most notably, to assess the RE growth at constant electric field. This includes both the primary Dreicer generation~\cite{Dreicer_1960,Connor_1975,Hesslow_2019b} and the secondary avalanche generation~\cite{Putvinski_1997,Hesslow_2019}. While analytical results provide useful estimates of the RE current and show parametric trends, there is also an early self-consistent disruption simulator GO~\cite{Smith_2006,Feher_2011,Papp_2013} that relaxes some limitations of the analytical estimates~\cite{Olasz_2021}.
Several codes have been developed for solving Eq.~(\ref{eq:5:kinetic}) directly. These include CODE~\cite{Landreman_2014,Stahl_2016}, CQL3D~\cite{Harvey_1992}, DREAM~\cite{Hoppe_2021b}, LAPS-RFP~\cite{Guo_2019}, LUKE~\cite{Decker_2004}, NORSE~\cite{Stahl_2017}, RAMc~\cite{McDevitt_2019}, and codes in Refs.~\cite{Aleynikov_2014,Aleynikov_2017,Daniel_2020}. Modules also exist for codes ASCOT~\cite{Hirvijoki_2014}, JOREK~\cite{Hoelzl_2021,Bergstrom_2025} and KORC~\cite{Carbajal_2017} to solve Eq.~(\ref{eq:5:kinetic}) using orbit-following techniques.


The fluid approach, in contrast, does not describe RE pitch angle and parallel momentum, but only the RE density ($n_\mathrm{re}$) via a rate equation of the form 
\begin{equation}
    \frac{\partial n_\mathrm{re}}{\partial t} + \nabla \cdot \left( \bm{v}_\mathrm{adv} n_\mathrm{re}\right) = S_\mathrm{seed} + S_\mathrm{avalanche},
\end{equation}
wherein advective transport and volumetric source terms dictate the RE evolution. Here, $\bm{v}_\mathrm{adv}$ involves the parallel advection and cross-field drifts. This evolution equation for $n_\mathrm{re}$ is electromagnetically coupled to the background plasma equations to form a closed set. While less accurate compared to the kinetic approach, the advantage of the fluid approach is in enabling also 3D simulations over long timescales (see Section~7) in realistic geometries. An RE fluid model was implemented in JOREK~\cite{Bandaru_2024_refluid}, M3D-C1~\cite{Liu_(Chang)_2021}, NIMROD~\cite{Sainterme_2024}, EXTREM~\cite{Matsuyama_2017}, GO~\cite{Smith_2006,Feher_2011,Papp_2013}, DREAM~\cite{Hoppe_2021b}, among others. These studies enable predictions not only of current driven instabilities with REs, but also address the effects of plasma vertical motion, impurity and deuterium injections, RE beam termination due to stochastization, etc.

The number of REs is typically much smaller than the total number of plasma electrons since most electrons cannot enter the runaway regime when the driving electric field is below approximately $21\%$ of the Dreicer field~\cite{Dreicer_1960}.  However, being much faster than the bulk electrons, REs can carry a significant part of the total plasma current in a tokamak. They can reach relativistic energies if the electric field exceeds the Connor-Hastie field~\cite{Connor_1975}, which is smaller than the Dreicer field by the factor $T/mc^2$. In the relativistic regime, the current density of REs is $en_\mathrm{re}c$, implying that a RE density of $n_\mathrm{re}=\SI{2e16}{m^{-3}}$ is sufficient to carry a current density of $\SI{1}{MA/m^2}$.
When the driving field exceeds the Connor-Hastie field, the RE energy gain is not limited by Coulomb collisions, but synchrotron losses and pitch-angle scattering.
The scattering rate increases as the pitch-angle distribution narrows during electron acceleration, and synchrotron losses increase accordingly for higher electron energies. 


REs are ubiquitous in startup plasmas during electrical breakdown, but their current usually remains small by the time the plasma heats up and becomes a good conductor~\cite{DeVries_2019,DeVries_2023,Breizman_2019}. They then gradually dissipate because the low loop voltage is unable to support them in the multi-keV range. A sufficiently long time of discharge shut-down in regular operational regimes can easily prevent reoccurrence of REs, but it becomes a clear danger in case of fast emergency shut-down. Copious production of REs can then be powered by the large energy of the poloidal magnetic field~\cite{Breizman_2019}. 

When $E_\parallel\gg E_{\rm c,eff}$, the characteristic time scale of the avalanche growth is typically much shorter than the lifetime of the plasma current~\cite{Martin-Solis_2010}. Thus, $E_\parallel$ relaxes quickly to its threshold level when REs start to dominate the current. However, REs cannot last long if $E_\parallel\ll E_{\rm c,eff}$ as Coulomb friction would slow them down. Consequently, $E_\parallel\sim E_{\rm c,eff}$ is required for a long-lasting RE population~\cite{Breizman_2014}.


RE generation is highly sensitive to the background plasma properties. Consequently, quantitative modeling of REs must cover this aspect~\cite{Putvinski_1997b,Smith_2006,Papp_2013,Linder_2021}. The Dreicer and hot-tail mechanisms of primary RE production depend significantly on the electric field, electron temperature and density~\cite{Dreicer_1960,Smith_2008}. The secondary avalanche mechanism is also sensitive to the ionization state of the plasma~\cite{Hesslow_2019}.

Models for the plasma heat content have similar forms for startup and disruption studies~\cite{Hollmann_2019,Kim_2020,Hoelzl_2021,Hoppe_2021b,Hoppe_2022}, with a rate equation for electrons
\begin{eqnarray}
	\frac{\partial W_e}{\partial t} &=
		\left(P_{\rm ohm} + P_{\rm re}\right) +
		\sum_i P_{ei,{\rm coll}} -\nonumber\\
		&-
		\left(P_{\rm line} + P_{\rm rec}\right)
		- P_{e,{\rm transp}},
\end{eqnarray}
and equations for each ion and neutral species $i$
\begin{equation}
	\frac{\partial W_i}{\partial t} =
		\sum_j \left(P_{ij,{\rm coll}} - P_{ij,\rm cx}\right) - P_{i,{\rm transp}}.
\end{equation}
Here, $P_{\rm ohm}$ denotes Ohmic heating, $P_{\rm re}$ collisional heating of the background plasma by REs, $P_{ij,{\rm coll}}$ collisional thermalization between species $i$ and $j$, $P_{ij,{\rm cx}}$ charge exchange processes, $P_{\rm line}$ line radiation losses, $P_{\rm rec}$ radiation losses due to recombination and bremsstrahlung, and $P_{e/i,{\rm transp}}$ the power loss due to spatial transport. The term $P_{\rm re}$ is often negligible during startup as well as in an early phase of disruptions due to the small RE content. In post-disruption RE termination scenarios, the term $P_{i,{\rm transp}}$ for neutrals is expected to play an important role~\cite{Sheikh_2024}.

The ion composition of the plasma is governed by a particle balance equation of the form
\begin{equation}
	\frac{\partial n_i^{(j)}}{\partial t} =
		S_{i,{\rm ioniz}}^{(j)} + S_{i,{\rm rec}}^{(j)} +
		S_{i,{\rm ioniz,re}}^{(j)} + \sum_j S_{ij,{\rm cx}}^{(j)},
\end{equation}
where $n_i^{(j)}$ is the density of charge state $j$ of species $i$, $S_{i,{\rm ioniz}}^{(j)}$ and $S_{i,{\rm rec}}^{(j)}$ are the particle production/destruction rate due to ionization and recombination, $S_{i,{\rm ioniz,re}}^{(j)}$ is the production rate due to RE impact ionization, and $S_{ij,{\rm cx}}^{(j)}$ is the production/destruction rate due to charge-exchange processes. In post-disruption RE beam termination scenarios, molecule formation can play an important role in particle and power balance, in which case the ionization and recombination terms should also be interpreted as molecular association/dissociation rates~\cite{Hollmann_2019,Hollmann_2020,Hollmann_2023}.

\subsection*{Current and future challenges}

Challenges remain in fully characterizing the spatial and momentum distributions of REs, requiring a collection of tools and techniques.

\subsubsection*{Multi-scale modeling.}
REs inherently span across several orders of magnitude in energy. Thermal electrons, with single-digit eV energies, can undergo large-angle collisions with REs and find themselves in the runaway range where they can be accelerated to hundreds of MeV. Simulations must cover electron dynamics across this entire energy range. This is further complicated by the sensitivity to critical energy value, above which electrons accelerate freely, which can vary significantly in a self-consistent simulation. One promising approach to addressing this issue is the fluid-kinetic models in which electrons are divided into different energy regions treated either as a fluid or kinetically~\cite{Aleynikov_2017, Hoppe_2021b}, 

Particle simulations of the RE dynamics, as conducted in, for example, Refs.~\cite{Bergstrom_2024,Bergstrom_2025} and~\cite{Beidler_2020,Beidler_2024}, are primarily limited by the different time scales involved. In particular, disruptions exhibit an evolution in which typically a small (but important) population of seed electrons are generated on the TQ time scale, multiplied by the avalanche mechanism on the longer CQ time scale, and finally relaxed in momentum space on an even longer time scale corresponding to the remainder of the RE beam life time. While time steps in particle codes are limited by either the gyro or guiding-center motion, continuum codes are limited by the time scales over which the various RE generation mechanisms play out. At the beginning of a disruption, this is the ionization time scale for the injected material, which can require sub-microsecond time steps, while later in the disruption, millisecond time steps may be admissible.

\subsubsection*{Energy balance and impurities.}
The fraction of the total current carried by REs depends critically on the bulk plasma resistivity and hence the bulk temperature. For nominal ITER parameters, a $\SI{1}{keV}$ electron temperature would allow the bulk to carry the total current at an inductive field lower than the Connor-Hastie field, i.e., below the runaway avalanche threshold. At lower temperatures, REs become a substantial part of the total current. 
It is worth noting that in a pure hydrogen plasma a sudden drop in electron temperature in the absence of subsequent heat losses would not allow the RE build-up because the stored magnetic energy of the toroidal plasma current allows the bulk electrons to reheat faster than the growth rate of the RE avalanche~\cite{Breizman_2019, deVries_2012}.  The spike in the inductive field would then be short and benign, as in the plasma startup regime. However, a significant presence of impurities may prevent this reheating. Rapid cooling of bulk electrons during TQ could also result from large electron heat flux to the wall along the stochastic magnetic field lines. The energy balance of the bulk plasma has to be modeled carefully because its ambiguity translates into a high uncertainty in the bulk plasma conductivity and inductive electric field.

\subsubsection*{Near-threshold dynamics.}
In situations where the electric field $E_\parallel$ is close to the threshold value $E_{\rm c,eff}$ required for net RE generation~\cite{Martin-Solis_2010,Aleynikov_2015a,Hesslow_2018a}, several effects can play an important role in shaping the RE distribution. Such a situation arises at the end of disruptions, in the RE plateau, where REs carry the entire current. Since all primary RE generation mechanisms are inefficient in this phase, large-angle (knock-on) collisions determine the growth and decay of the RE current, with a gradual decay for $E_\parallel< E_{\rm c,eff}$~\cite{Breizman_2014,Embreus_2018}. The time scale of the reduction of the electric field might be determined by the interaction with the wall current, as the wall times in large tokamaks approach or exceed the time scales relevant for RE beam formation. Pitch angle scattering and radiation also play an important role in shaping the distribution function in the near-threshold regime. Pitch angle scattering enhances energy losses due to synchrotron radiation~\cite{Hoppe_2021a}. Synchrotron losses and bremsstrahlung can produce bumps in the RE distribution function~\cite{Bakhtiari_2005,Aleynikov_2015a,Decker_2016} when $E>E_{\rm c,eff}$, which can trigger kinetic instabilities. In disruptions, as well as in certain startup scenarios, the electric field will eventually reduce from a peak value $E_\parallel>E_{\rm c,eff}$ to a sub-threshold value, causing REs to gradually lose energy via radiation. A quantitative description of this effect in reactor-scale tokamaks still remains to be developed.

\subsubsection*{RE-background interaction.}
The most significant contribution by REs to the tokamak plasma conditions is the electrical current that constitutes close to 100\% of the remnant plasma current in a post-disruption plateau. The REs can also affect plasma behavior in other ways. A significant RE population may act as a dominant heat source for the system~\cite{McDevitt_2023}.  Collisions of REs with (partially ionized) atoms can enhance the ionization state of the background plasma~\cite{Garland_2020,Garland_2022,Hoppe_2025}. Whereas models for these effects, suitable for fluid codes, have been derived, the validity of these models for post-disruption plasmas remains uncertain. In particular, the available models derived for high-energy weakly collisional REs miss the role of more collisional low-energy REs that could potentially enhance heating and ionization beyond the available predictions.

\subsubsection*{Access to benign termination.}
The benign termination scenario has emerged as a potential technique for safely terminating a RE beam~\cite{Reux_2021b,Paz-Soldan_2021}. However, as has been demonstrated experimentally~\cite{Sheikh_2024}, this scenario requires secondary injection of low-$Z$ material in order to bring the neutral pressure to within a specific range, wherein the plasma largely recombines and a final discharge-terminating instability can be triggered. While the bounds of this neutral pressure range have been characterized experimentally on specific tokamaks, significant uncertainty remains regarding the bounds in reactor-scale tokamaks such as ITER and SPARC. Further modelling is needed accounting for the atomic and molecular physics at play, as well as the complex interaction between REs, cold plasma with neutrals, and tokamak wall~\cite{Hollmann_2019,Hollmann_2023,Hoppe_2025}.

\subsubsection*{Control via wave-particle interactions}
There are conceivable ways to mitigate the RE population via self-excited or externally driven (injected) waves~\cite{Guo_2018,Martin-Solis_2004}. In the high-frequency range, REs can resonate with the electron plasma waves or whistler waves~\cite{Fulop_2006,Pokol_2008,Pokol_2014}. Both are difficult to drive from outside by antennas because they are under the cutoff frequencies of dense plasmas and subject to significant collisional damping at low electron temperatures. If attempted to affect the seed REs, such externally launched waves would heat the bulk electrons rather than impact the RE population via wave-particle resonance~\cite{Decker_2024}. A plausible alternative might be to employ  radiofrequency (RF) current drive~\cite{Reinke_2019} that may prevent RE build-up as the TQ and the resulting drop in Spitzer conductivity would not necessarily entail the high inductive electric field in the presence of RF-current drive. This motivates an effort to model the buildup of REs in the presence of the RF-current drive to assess the potential benefits of the technique for RE control~\cite{Breizman_2015}.

Among self-excited waves, whistlers receive extensive attention~\cite{Fulop_2009,Spong_2018,Heidbrink_2019}. They can enhance pitch-angle scattering of the REs, leading to the enhancement of synchrotron losses. REs excite whistlers via anomalous Doppler resonance, and the RE current must already be substantial to overcome the collisional damping of whistlers. More specifically, the minimum electron temperature for exciting whistlers by the nominal ITER current density of $\SI{1}{MA/m^2}$ is $\SI{25}{eV}$~\cite{Aleynikov_2015b}, well above the expected post-disruption temperature in ITER. In present-day RE flat-top experiments, whistlers are easier to excite because of higher electron temperatures~\cite{Spong_2018,Bin_2022} and can indeed enhance RE scattering~\cite{Liu_(Chang)_2018} keeping the RE current at a marginal stability level~\cite{Breizman_2023b,Breizman_2023a}.

\subsubsection*{Losses during VDE.}
Disruptions in large tokamaks can involve plasma motion in a vertical displacement event (VDE), during which REs can scrape against the wall~\cite{DINA-report:2016, Martin-Solis_2022}. Recent studies found that VDE motion and consequent RE scraping can lead to a substantial loss of REs on open outer flux surfaces reducing the RE current~\cite{Wang_2025,Bandaru_2025,Vallhagen_2025}. Although initial considerations show some promise, more work is needed to quantify the robustness of this avalanche-reducing mechanism and validate it experimentally. This RE current dependence on the competing processes of RE avalanche and vertical motion translates into sensitivity to impurity content, direction of vertical motion (up or down), rate of current decay, etc. However, scraping induces additional wall loads, potentially negating the benefits of RE mitigation. The scrape-off effect is expected to be beneficial only when it happens while the RE population is small enough not to cause wall damage. In typical ITER upward VDE situations, including scenarios with Neon injection or flush-out, predictions by DINA~\cite{DINA-report:2016} and JOREK~\cite{Bandaru_2025} do not show a clear advantage, as the cumulative RE energy dumped onto the PFCs does not change considerably. More work and more accurate models are needed to verify this observation. Work is also needed to quantify the momentum-space distribution of lost REs. Since scraping coincides with the most RE generation-intense phase of a disruption, the momentum distribution of lost REs may deviate significantly from that of confined REs at the end of a disruption.

\subsubsection*{Radial transport.}
Radial transport has long been considered attractive as a way to prevent detrimental avalanching. Different forms of radial transport were studied, including externally induced transport via resonant magnetic perturbation coils~\cite{Papp_2011a,Papp_2011b}, passive asymmetric coils~\cite{Boozer_2011,Smith_2013,Sweeney_2020}, MHD instabilities~\cite{Paz-Soldan_2019,Lvovskiy_2020}, and electron cyclotron waves~\cite{Martin-Solis_2004,Decker_2024}, among others. In support of such studies, work has also been done to relate the RE transport rate to the magnetic stochasticity~\cite{Hauff_2009,Saerkimaeki_2020,Svensson_2021}. What remains much less explored is how the naturally present radial transport arises and affects the REs.

Experimental measurements indicate that some radial RE transport is naturally present without external interference~\cite{Myra_1992,Kwon_1988,Esposito_1996}, and simulations show that even moderate radial losses can reduce the number of REs generated in a disruption~\cite{Martin-Solis_2021,Svensson_2021}. To predict the generation of REs accurately, theoretical models for their radial transport should be developed. These can either provide values for advective and convective transport coefficients~\cite{Saerkimaeki_2020} or be based on developments in chaos theory~\cite{Janosi_2024}. To accomplish this, a better understanding of the underlying physics of the radial transport is needed, specifically whether the transport in the TQ, CQ, and RE plateau, respectively, is driven by MHD or turbulence.



\subsection*{Concluding remarks}
State-of-the-art RE models have come a long way from early works aimed at estimating the RE number and momentum distribution in steady state. Present-day numerical tools incorporate key physics mechanisms affecting the RE behavior and background plasma self-consistently. Despite such progress, conclusive modeling of the TQ presents an outstanding challenge.

Other major open questions exist for quantitative predictions of RE properties in reactor-scale tokamaks. In particular, techniques for treating highly and moderately relativistic electrons simultaneously are required as the latter may play a crucial role in the near-threshold regime, typical of RE plateaus. Lower energy electrons couple stronger with the background plasma and can conceivably enhance ionization and heating by REs beyond present estimates.

Due to the potentially severe damage caused by even moderate amounts of REs, near-term model development should focus on potential RE avoidance and mitigation techniques, such as wave-particle interactions (including possible benefits from the RF current drive), Deuterium second injection, and RE losses during VDEs. The mechanisms driving radial RE transport in the different disruption phases are still poorly understood and remain crucial for creating robust quantitative RE distribution models.

\clearpage 
\section{Reduced RE beam dynamics and impact models}\label{:sec6}
\author{P. Aleynikov$^1$, M. Beidler$^2$, J.R. Martin-Solis$^3$ 
}
\address{
$^1$Max-Planck-Institut für Plasmaphysik, Wendelsteinstraße 1, 17491 Greifswald, Germany\\
$^2$Fusion Energy Division, Oak Ridge National Laboratory, Oak Ridge, TN 37831, USA\\
$^3$Universidad Carlos III de Madrid, Avenida de la Universidad 30, 28911-Madrid, Spain} 


\subsection*{Status}

The magnetic energy, $W_{mag}$, of MA runaway electron (RE) beams formed during the disruption current quench (CQ) is typically much larger than the kinetic energy of the REs, $W_{kin}$ (in ITER $W_{kin} \sim$ MJs, whereas $W_{mag} \sim$ hundreds MJs~\cite{Martin-Solis_2014, Martin-Solis_2015}) so that, when the RE current is lost during the termination phase of a disruption, conversion of magnetic energy into RE kinetic energy can occur, which could substantially increase the energy fluxes deposited by the REs in comparison with the kinetic energy gain during the CQ of the disruption. The importance of this issue was first pointed out in Ref.~\cite{Putvinski_1997}, and later in Ref.~\cite{Riemann_2012}, and experimental evidence has been reported in several devices~\cite{Hollmann_2013,Loarte_2011,Martin-Solis_2014,Martin-Solis_2015}, showing that a remarkable conversion of magnetic into RE kinetic energy can indeed take place. Here, a summary of lower dimensional models used to characterize the energy balance during a disruption is presented, focusing on the fraction of the magnetic energy of the runaway plasma, which might be transferred to REs. Particle-based kinetic models are also discussed as well as the essential parameters determining the RE impact on the plasma facing components (PFCs).

\subsubsection*{0D/1D fluid models.} 
Simplified 0D/1D fluid models for the termination phase of the disruption have been 
developed~\cite{Martin-Solis_2014,Martin-Solis_2017,Martin-Solis_2022} with simplistic RE kinetics, which allow to elucidate the mechanisms that govern the conversion of magnetic into RE kinetic energy, with a view to their evaluations for ITER. All of them include a coupling between models for the plasma current and the RE current evolution.

The 0D model presented in Ref.~\cite{Martin-Solis_2014} consists of two coupled equations for the plasma current, $I_{p}$, and the current induced in the vessel, $I_{v}$, together with an 
equation for the evolution of the RE current, $I_{r}$, accounting for the generation of REs by the avalanche mechanism, approximated
by~\cite{Rosenbluth_1997,Putvinski_1997,Jayakumar_1993},
\begin{equation}
 \left( \frac{dI_{r}}{dt} \right)_{avalanche} \approx
 \frac{e \, \left( E_{||} - E_{R} \right) \, I_{r}}{m_{e} c \, {\rm ln} \Lambda \, a_{z}} ,
\label{eq_Iravalanche}
\end{equation}
and the loss of REs during the termination phase, $- I_{r}/\tau_{d}$, with a characteristic loss time, $\tau_{d}$. In Eq.~(\ref{eq_Iravalanche}), $E_{R} = n_{e} e^{3} {\rm ln} \Lambda /4\pi \varepsilon_{0}^{2} m_{e} c^{2}$ is the critical field for RE generation~\cite{Connor_1975,Rosenbluth_1997}, $a_{z} \equiv \sqrt{3 \, (5 + Z)/\pi}$, and the parallel electric field, $E_{||}$, is determined by the resistive current in the plasma, $E_{||} = \eta \, (j_{p} - j_{r})$, where $j_{p.r} = I_{p,r} /\pi a^{2} \kappa$ are the average plasma and RE current densities, respectively, $a$ is the plasma minor radius, $\kappa$ the plasma elongation, and $\eta$ the plasma resistivity.

The energy transferred to the REs during the termination phase is estimated:
\begin{equation}
 \Delta W_{run} = 2 \, \pi \, R_{o} \, \int \, I_{r} \, \left( E_{||} - E_{R} \right) \, dt .
 \label{eq_Wrun}
\end{equation} 

Comparisons between models and several tokamak experiments (JET, DIII-D and FTU)~\cite{Martin-Solis_2014} suggest that the same physical model reasonably accounts for observations in present devices. The conversion efficiency is found to be largely determined by the ratio of the characteristic RE loss time, $\tau_{d}$, describing the radial RE transport, to the resistive time of the residual ohmic plasma, 
$\tau_{res}$, increasing with $\tau_{d} /\tau_{res}$. The avalanche generation of REs increases the conversion of magnetic energy into RE kinetic energy, particularly for large enough RE currents and long lasting termination phases. Furthermore, when the termination timescale is longer than the vacuum vessel resistive time, the penetration of the magnetic energy external to the vacuum vessel has to be considered for its conversion into RE kinetic energy.

1D models~\cite{Martin-Solis_2015,Martin-Solis_2017} taking into account the evolution of the plasma and RE current density profiles have also been applied, using a simple cylindrical plasma geometry, in which the evolution of the plasma current density ($j_{p} (t,r)$) is calculated solving the current diffusion equation,
\begin{equation}
\mu_{0} \, \frac{\partial j_{p}}{\partial t} =  \frac{1}{r}
\frac{\partial}{\partial r}\left[ r \frac{\partial E_{||}}
{\partial r}\right]  =  \frac{1}{r} \frac{\partial}{\partial r}\left[ r \frac{\partial
 \eta (j_{p} - j_{r})} {\partial r}\right] ,
\label{eq_jpmodel}
\end{equation}
together with an equation for the RE current density ($j_{r} (t,r)$) similar to the 0D model equation. The energy deposited on the REs,
\begin{equation}
\Delta W_{run} = \int_{termination} \, dt'  \int \, j_{r}  \, \left( E_{||} - E_{R} \right) \, dv ,
\label{eq_DeltaWrun1D}
\end{equation}
is found to increase in comparison with the 0D calculations as
the RE current density profile created during the CQ is expected to be
more peaked in the plasma center than the pre-disruption current density profile~\cite{Eriksson_2004,Martin-Solis_2017}, which will enhance significantly the growth of the RE population by avalanche in the plasma center during current termination and lead, particularly for the slowest terminations, to substantial energy conversion~\cite{Martin-Solis_2017}. 

In ITER, RE beams are expected to be vertically unstable and, hence, when the plasma hits the wall, the scraping-off of the RE plasma occurs and the current is terminated~\cite{Martin-Solis_2022}. A simple 0D model which mimics the plasma surrounded by the conducting structures~\cite{Kiramov_2017}, including self-consistently the vertical plasma motion and the generation of REs, has been used for an evaluation of the RE formation and termination during disruptions~\cite{Martin-Solis_2022}. The model approximates the plasma-wall system by a set of three parallel thin circular coaxial rings of radius $R_{0}$. The bottom and top conductors carry currents $I_{1}$, $I_{2}$, respectively, and  represent the current in the conducting wall, while the middle conductor represents the plasma current, $I_{p}$, which can move vertically. The model also includes a static external magnetic field created by two constant circular currents, $I_{e}$. Hence, the model consists of three circuit equations for the inductively coupled currents ($I_{1}$, $I_{2}$, $I_{p}$) plus one equation for the RE current which, once the plasma hits the wall,
in addition to the avalanche term, as given by Eq.~(\ref{eq_Iravalanche}), includes a loss term describing the scraping-off of the beam, $dI_{r} /dt \approx 2 \dot{a} \, I_{r}/a$~\cite{Lehnen_2018,Martin-Solis_2022}. For the vertical plasma motion, the force-free constraint~\cite{Kiramov_2017}, $\xi = (I_{1} - I_{2})/(I_{1} + I_{2} + 2 \, I_{e})$, is used
($\xi$ is the vertical plasma displacement normalized to $a_{w}$, where $2a_{w}$ is the distance between the two wall conductors).



During scraping-off, the plasma velocity and electric field can noticeably increase leading to the deposition of a remarkable amount of energy on the REs ($\sim$ hundreds of MJs), and it is found that larger temperatures of the residual ohmic plasma during scraping-off might be efficient in reducing the power fluxes onto the PFCs~\cite{Martin-Solis_2022}. Moreover, during scraping-off, the plasma reaches the $q_{a} = 2$ limit before the current is terminated.
Negligible additional conversion of magnetic into RE kinetic energy is predicted during the RE deconfinement following magnetic fluctuations after $q_{a} = 2$ is crossed for characteristic deconfinement times lower than 0.1 ms~\cite{Martin-Solis_2022}, leading to a benign termination of the disruption, whereas larger values of $\tau_{d}$ might result in a stronger energy conversion and a non-benign disruption termination.


\subsubsection*{2D fluid models.} 
Axisymmetric 2D plasma evolution models are a powerful and computationally efficient tool for studying vertical plasma motion in tokamaks. Compared to more complex 3D simulations, 2D models offer significantly lower computational costs and allow extensive parametric scans. At the same time, compared to 0D/1D models, 2D models are able to provide approximate predictions of the strike point locations~\cite{Pitts_2025}, taking into account the up/down asymmetry and the wall structure. Furthermore, they facilitate the characterization of the linear magnetohydrodynamic (MHD) stability of post-disruption plasma~\cite{Aleynikova_2016}, which is important for determining the likelihood of MHD events leading to prompt RE losses to the wall. Additionally, 2D models hold the potential to refine vertical disruption event (VDE) modelling through improved modelling of the halo currents, the currents surrounding the plasma during VDE.

To date, five codes capable of 2D plasma motion calculations have been equipped with RE models: DINA~\cite{Khayrutdinov_1993}, TSC~\cite{Jardin_1986, Bandyopadhyay_2012}, JOREK~\cite{Hoelzl_2021, Bandaru_2019, Bandaru_2024}, NIMROD~\cite{Sainterme_2024} and M3D-C1~\cite{Zhao_2021}. Of these, DINA and JOREK have been extensively used to model the impact of REs on PFCs in ITER~\cite{Konovalov_2016, Pitts_2025, Bandaru_2024, Bandaru_2025}. Early 2D simulations using DINA confirmed the expected high poloidal peaking factor of the RE strike wetted area, both with and without the presence of static 3D magnetic fields (error fields and field generated by RMP coils), highlighting the importance of 2D models in predicting localized heat loads on PFCs~\cite{Aleynikov_2010}.

For upward VDEs in ITER, similar to the 0D model, the 2D calculations show the formation of a RE plateau. The subsequent termination of the RE beam in the 2D simulations occurs mainly by scraping-off of the RE by the wall. Scraping results in RE loss and a simultaneous re-avalanche in the inner region of the plasma close to the edge, resulting in a skin current on the RE current profile~\cite{Putvinski_1997}. Note that the RE distribution function is expected to be different in the strongly re-avalanching skin layer, where the exponential distribution function is expected~\cite{Rosenbluth_1997}, and in the inner region, where the electric field is expected to remain close to the avalanche threshold and a non-monotonic distribution function can form~\cite{Aleynikov_2015a}. Different distribution functions should lead to different energy dissipation rates, but both DINA~\cite{Konovalov_2016} and JOREK~\cite{Bandaru_2024} currently estimate the un-dissipated RE energy using a version of Eq.~(\ref{eq_DeltaWrun1D}), which neglects this difference. Both codes predict that up to 100-150 MJ of poloidal magnetic energy is transferred to the REs and deposited on the wall during the scraping-off and the final collapse. This is in general in agreement with 1D models~\cite{Martin-Solis_2017}. Note that in the limit of an ideally conducting wall, the force-free constraint~\cite{Kiramov_2017} implies that the vertical position of the plasma depends solely on the total plasma current. As a result, ``acceleration'' of the CQ via high-Z impurity injection has a limited effect on reduction of magnetic energy transfer~\cite{Konovalov_2016}. 

Both DINA~\cite{Konovalov_2016} and JOREK~\cite{Bandaru_2025} calculations show that  downward VDEs in ITER are much faster than upward VDEs due to stronger magnetic field gradients below the X-point. Both models show that wall scraping-off limits avalanche growth, and no significant RE current is formed in low impurity content. For high impurity content, the avalanche is strong enough to form a multi-MA RE current during downward VDEs. 

A key advantage of axisymmetric 2D simulations is the ability to perform linear MHD stability analysis for the sequence of equilibria during the RE development and the RE beam termination. Such an analysis can help characterize the possible RE losses due to MHD activity and predict the final collapse without performing full 3D simulations~\cite{Aleynikova_2016}.

The parameters of the halo currents significantly impact the dynamics of the VDE in general and the RE beam termination in particular. Although advancements in halo current modeling have been made, such as in Refs.~\cite{Kiramov_2016, Artola_2021}, the physics fidelity and predictive capabilities of these models have not yet been systematically validated. Further verification is necessary to assess their reliability in disruption simulations.

\subsubsection*{Particle-based RE-kinetic models.}


Particle-based kinetic models operating with 2D axisymmetric electromagnetic fields emerge as a powerful tool. These models improve upon their counterparts discussed in the previous subsection by tracking individual RE trajectories in 6D phase space, capturing the local effects of electromagnetic fields, companion plasma, and impurities. This level of fidelity is indispensable for precisely evaluating wall damage, providing crucial parameters such as energy, pitch angle, and gyrophase, which are essential for characterizing RE impacts on PFCs. In particular, a detailed understanding of the RE beam's interaction with the leading edges of wall facets demands this level of precision. The volumetric energy deposition into PFCs is controlled by the total energy of the REs, as discussed in Sec.~\ref{:sec8}. Additionally, the deposition is also affected by the angle of incidence into the PFC $\Theta_{\rm PFC}$, which is controlled by the guiding center angle of incidence $\theta_{\rm GC}$ and the pitch angle relative to the magnetic field $\eta$ according to
\begin{equation}
    \sin\Theta_{\rm PFC}=-\cos\theta_{\rm GC}\sin\eta\sin\chi+\sin\theta_{\rm GC}\cos\eta,
\end{equation} 
for instantaneous gyroangle $\chi$. The resulting elliptical, gyro-averaged wetted area will scale as $A_{\rm w}=\pi r_{\rm L}^2/\sin\theta_{\rm GC}$, where the relativistic Larmor radius is $r_{\rm L}=p\sin\eta/eB$.


Electromagnetic fields serve as the primary drivers of the RE trajectories, dictating where these particles strike the wall. In axisymmetric configurations, losses typically arise from large drift-orbit effects at high energies or beam scrape-off due to failures in position control. However, the inherent non-uniformity of magnetic fields introduces changes in momentum space, as governed by fundamental conservation laws. Notably, REs with relatively low energies and perpendicular pitch angles can become trapped within these magnetic structures. Furthermore, the toroidal electric field accelerates REs, influencing their pitch angles. The interplay of plasma and impurity profiles further shapes the RE momentum distribution through collisional and radiative processes, including synchrotron and bremsstrahlung radiation. Collisional friction reduces the total momentum, while pitch-angle scattering broadens the pitch-angle distribution, potentially leading to trapping or detrapping events. Partially ionized impurities contribute to small-angle collisional friction, pitch-angle scattering, and avalanche source generation from large-angle collisions. 


While many particle-based methods exist \cite{Russo_1991,Heikkinen_1993,Guan_2010,Izzo_2011,Papp_2011a,Sommariva_2018,Liu_(Yueqiang)_2019}, most are primarily applied to non-axisymmetric field systems that will be discussed in Sec.~\ref{:sec7}. Here, we focus on the work done with the codes KORC~\cite{Carbajal_2017} and RAMc~\cite{McDevitt_2019}. 
KORC offers 6D full-orbit and 5D guiding-center simulations, leveraging dynamic experimental electromagnetic fields, plasma, and impurity profiles. Because KORC primarily uses experimental EM fields, self-consistent evolution of fields with the kinetic population is not enabled. Simulations of DIII-D discharges~\cite{Beidler_2020} have revealed RE beam deconfinement before substantial collisional friction energy loss, with pitch-angle scattering dominating during high-Z impurity injection, leading to rapid current evolution and eventual loss of position control. Conversely, simulations of JET pulses~\cite{Beidler_2021} demonstrate well-confined RE beams with significant collisional friction losses before position control failure. The loss of vertical stability during wall impact results in a broader energy deposition pattern. Subcritical electrons generated by the avalanche operator contribute to significant PFC surface heating, including counter-propagating REs~\cite{Beidler_2024}. During final loss events, an increase in the induced toroidal electric field accelerates avalanche growth, but many energetic electrons are generated in deconfined regions. The avalanche source preferentially generates a significant population of lower-energy electrons, counter-propagating to the initial beam, an important factor for PFC beveling design. 

RAMc employs 5D guiding-center RE trajectories in dynamic analytic field formulations, which allows for a self-consistent evolution with the kinetic population. It has been shown that toroidal geometry reduces the avalanche growth rate near the beam edge except when the electric field is significantly larger than the critical field~\cite{McDevitt_2019b}. The avalanche source generates trapped REs at the beam edge, which are not sufficiently accelerated by the electric field unless it is significantly larger than critical. High-Z impurities induce an inward Ware pinch~\cite{McDevitt_2019c} and oppose the local avalanche rate decrease near the RE beam edge~\cite{Arnaud_2024}. The increased fraction of trapped REs from pitch-angle scattering and avalanche generation are Ware pinched. Increased collisionality can also detrap REs in the beam edge, allowing them to be accelerated by the electric field. The increased trapping of REs by the avalanche operator in high-Z impurity RE plateaus, which are insensitive to 3D field transport, can provide a seed to reconstitute the RE beam after an MHD event~\cite{McDevitt_2023b}. 

The computational demand of particle-based models is substantial. Resolving the gyrofrequency for full-orbit particles or the trajectory of guiding-center particles, both occurring on rapid timescales due to relativistic energies, poses a significant challenge. Moreover, achieving statistically robust results for wall strikes necessitates tracking upwards of 10 million REs~\cite{Ratynskaia_2025}. Fortunately, discrete particle calculations are well-suited for GPU hardware, enabling nearly a 100-fold increase in the number of simulated REs compared to traditional CPU-based approaches. 

\subsection*{Current and future challenges}

Lower-dimensionality models, like those referenced above, are crucial for characterizing the global energy balance during disruptions. These models aim to predict the fraction of pre-quench magnetic energy that is channeled into REs and to describe the dissipation mechanisms through which this energy is lost to the wall. 


To obtain reliable predictions, there needs to be a stronger coupling between existing electromagnetic (EM) models for plasma current and position, such as 0D inductively coupled wire models~\cite{Martin-Solis_2022, Kiramov_2017} or 2D VDE models~\cite{Konovalov_2016, Bandaru_2025},  and the models for RE kinetics. 

Extrapolations to ITER are subject to large uncertainties mainly due to an incomplete understanding of the thermal plasma, the RE beam following the CQ or the instabilities leading to the RE loss. The energy, $W_{run}$, deposited by the REs on the PFCs has been estimated for the termination of ITER RE plateaus in the range of 2 - 10 MA, and characteristic RE loss times $\sim 0.1 - 10 \, {\rm ms}$ (covering the range observed in the experiments), showing that the magnetic energy converted into RE kinetic energy can be as much as a few hundreds MJ for large RE plateau currents and slow (large $\tau_{d}$) terminations~\cite{Martin-Solis_2014,Martin-Solis_2017,Martin-Solis_2022,Konovalov_2016}.

Note that these results are consistent with observations of benign RE terminations via low-Z injection~\cite{Paz-Soldan_2021,Reux_2021b,Reux_2022,Sheikh_2024}. 
Extremely strong MHD activity in such plasmas leads to a very fast deconfinement of the RE population, resulting in negligible conversion of magnetic to RE kinetic energy and a very large wetted area without significant heat loads on the PFCs. A sufficiently clean plasma is required to avoid RE regeneration and energy conversion~\cite{Reux_2021b}. More simulations are needed to evaluate RE regeneration in these events in ITER, where larger avalanche rates may change the situation.

Currently, reduced EM models rely on simplistic RE kinetic models, which are inadequate for a reliable characterization of RE energy loss channels.
On the other hand, existing detailed kinetic simulations are typically expensive, limiting their application. There is a need to develop plasma evolution models
that account for essential features of RE kinetics,
such as the significance of subcritical energetic electrons generated by RE avalanche
for surface heating~\cite{Beidler_2024}, or the survival of the trapped RE population
for RE current reconstitution after a MHD flush during benign termination of the disruption~\cite{McDevitt_2023b}.

Radial RE transport and the background plasma composition play key roles in determining the dominant RE energy loss channels. Regimes with slow radial transport correspond to near-threshold RE regimes, where the Ohmic current is negligible and the dominant RE energy loss is through collisions and radiation. In contrast, regimes with strong radial transport are characterized by high over-criticality ($E/E_{c}$) and significantly higher direct wall impact. Thus, refining transport and ambient plasma models are essential for reliable prediction of global energetics of the disruption and the RE beam termination. From this point of view, the development and integration of reduced self-consistent models for MHD-induced transport regarding the loss time scale and deposition width is of interest. Further validation of these reduced models with the experiment would be desirable.
Improvements are needed regarding plasma-wall interactions, including the reliable prediction of halo current formation during VDEs and plasma quasi-neutrality issues during large RE losses to the wall, as well as developing a better understanding of what determines the RE wetted area for scrape-off impacts. 


\subsection*{Concluding remarks}

Reduced models for the energy balance during the termination phase of the disruption have shown to be adequate for the identification of the main mechanisms leading to the conversion of magnetic into RE kinetic energy during current termination as well as to get simple estimates of the amount of energy that can be deposited on the PFCs due to the REs. Despite the uncertainties regarding the thermal residual plasma and the RE beam features following the CQ of the disruption, or the processes and instabilities leading to the loss of the RE plasma, it is expected that in ITER as much as a few hundreds of MJs might be transferred to the REs, particularly for slow current terminations. Nevertheless, for a confident assessment of the impact of the RE loads on the PFCs, it turns out to be essential that future reduced-dimensionality disruption models include a number of key physics elements at an appropriate fidelity level: elements of RE kinetics, including RE avalanching in strong fields induced during plasma scraping, adequate descriptions of background plasma and radial transport, or effects of plasma-wall interaction such as the halo currents.

\clearpage
\section{High fidelity RE beam dynamics and impact models}\label{:sec7}
\author{E. Nardon$^1$, M. Hoelzl$^2$, C. Liu$^3$, D. del-Castillo-Negrete$^4$}
\address{
$^1$IRFM, CEA Cadarache, F-13108 Saint-Paul-lez-Durance, France \\
$^2$Max Planck Institute for Plasma Physics, Boltzmannstr. 2, 85748 Garching b. M, Germany\\
$^3$State Key Laboratory of Nuclear Physics and Technology, School of Physics, Peking University, Beijing 100871, China\\
$^4$ Institute for Fusion Studies, University of Texas, Austin, Texas, USA}


\subsection*{Status}

There is clear experimental evidence that the 3D nature of the magnetic field plays an important role in a substantial fraction of runaway electron (RE) impacts. It might be a key ingredient to access benign terminations on one hand, but also the origin of highly localized depositions. One piece of evidence is the fact that RE impact patterns are typically non-axisymmetric to a degree not explicable by wall asymmetries as has been illustrated in Sections 1-3. Another one is that, e.g. in JET~\cite{Reux_2015}, the RE beam evolution when approaching termination is typically characterized by sudden drops in the RE current $I_{RE}$ accompanied by discrete bursts of magnetic fluctuations, hard X-rays (HXRs) and neutrons, with the final impact associated with a relatively low value (often approaching 2) of the safety factor at the edge of the beam $q_a$, suggesting a key role of MHD instabilities. Note that in these JET cases, as well as in comparable scenarios in DIII-D and ASDEX Upgrade (AUG), $q_a$ decreases over time down to low values. In other experiments, e.g., on AUG~\cite{Pautasso_2015}, $q_a$ is thought to remain relatively high, and in these cases a steady decay of $I_{RE}$ down to 0 is observed. Another aspect that points to an important role of the 3D nature of the magnetic field is the wetted area by the RE impact that exceeds the one explicable by a 2D scrape-off, in particular in benign termination scenarios.

Accurate simulation of REs in stochastic magnetic fields is also critical for the assessment of resonant magnetic perturbations in the control and confinement of REs~\cite{Cornille_2022}. In general, understanding the relationship between the stochasticity of the magnetic field and the stochasticity of the RE orbits can be a very nontrivial problem requiring full-orbit (FO) computations.  For example, as documented in Ref.~\cite{Carbajal_2020}, depending on the regime, guiding center (GC) simulations might significantly overestimate RE losses with respect to FO computations. In addition, when the RE Larmor radius is comparable to the island size, GC and FO can exhibit significant differences. 
A key question related to this is the role of magnetic stochasticity, which could explain such a broadening of the deposition. It is, on the other hand, important to note that a 3D magnetic field structure does not necessarily imply strong stochasticity or broad deposition: it may be that the RE beam gets distorted (i.e., kinks) while the flux surfaces remain topologically nearly intact. Such a distortion, combined with a motion into the wall due, e.g., to a vertical instability, is a hypothesis to explain why the impacts of the RE beam may be so localized in certain cases. On the other hand, significant -- and sometimes very strong -- magnetic stochasticity may appear, as suggested by numerical simulations discussed below, and could have a key influence on the RE beam impact. One may imagine a deleterious influence of stochasticity in a scenario where a large stochastic region is initially bounded by an annulus of good flux surfaces that gets destroyed over a short period of time, leading to an abrupt loss of all REs, potentially in a localized manner. Such a scenario could lead to rapid loss of confined REs in a narrow flux tube~\cite{Boozer_2016}, but has not been observed in 3D non-linear simulations so far, which instead suggest that stochasticity grows from the outside in, leading to a relatively broad deposition pattern. Such simulations have, however, focused on benign termination scenarios, and the inside-out formation of stochasticity is not necessarily excluded for other scenarios. A key concept to understand the impact of stochasticity is the magnetic footprint, that is, the region of the wall that is connected to the core of the beam by field lines~\cite{Abdullaev_2013}. This region is bounded by so-called ``manifolds''. Depending on how fast and how large the magnetic footprint grows, the wall surface where REs hit may vary significantly. A possibly important complication is that the orbits of highly energetic electrons may be less stochastic than the field lines themselves due to drift orbit shifts as well as a spatial averaging of the field over the gyro-orbits. Such a decoupling of transport from magnetic topology was described for instance in~\cite{deRover_1996,Hauff_2009,Papp_2012,Saerkimaeki_2020,Bergstrom_2025}.

Runaway electron dynamics in fusion plasmas exhibits a wide range of temporal scales, going from the fast gyro-period, $\sim10^{-11} {\rm s}$, to the observational time scales, 
$\sim10^{-3}\dots10^{-2} {\rm s}$.
Overcoming the computational challenges of this scale separation has motivated the development of approximate, perturbative orbit models like the extensively used GC description. Although these approximations have been useful for thermal plasmas, the simulation of the relativistic dynamics of high-energy REs in 3D magnetic fields requires the incorporation of FO effects, synchrotron radiation damping effects along with collisions and avalanche sources. Motivated by this need, numerical codes like the KORC (Kinetic Orbit Runaway electrons Code) have been developed to perform high fidelity simulations of REs using high-performance-computing platforms~\cite{Carbajal_2017, Beidler_2020}. Among other uses, full-orbit information is critical for the development of accurate synchrotron emission synthetic diagnostics~\cite{Carbajal_2017b,DelCastilloNegrete_2018}.


\begin{figure}
    \centering
    \includegraphics[width=23em]{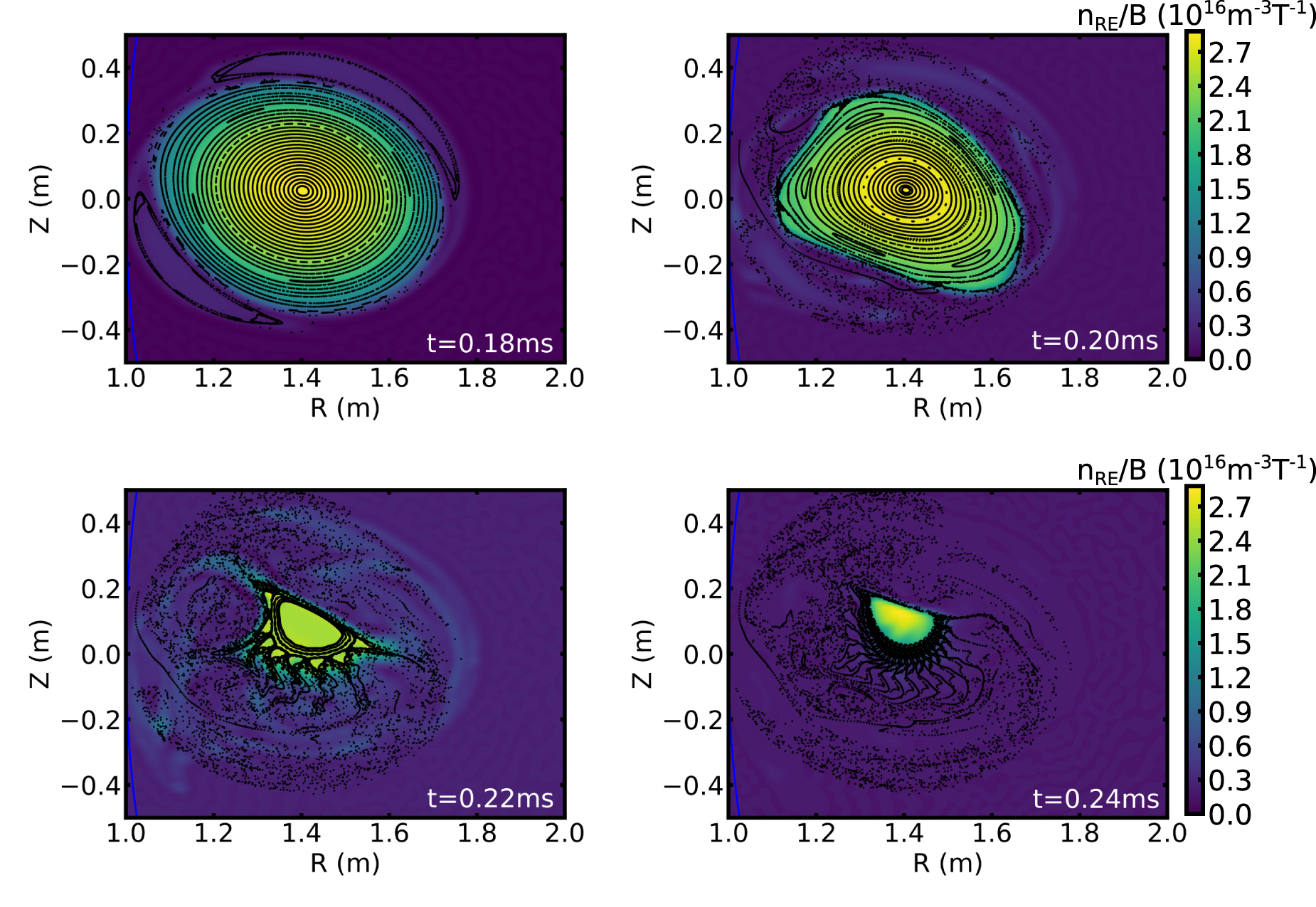}
    \caption{M3D-C1 simulation of a termination in the DIII-D tokamak. The time evolution of the RE density distribution is shown in color along with a Poincar{\'e} plot visualizing the dynamically changing magnetic field topology. Reprint with permission from Ref.~\cite{Liu_(Chang)_2021}.
    }
    \label{fig:diii-d-termination}
\end{figure}

The type of instability that develops during the evolution of the RE beam up to its impact on PFCs may vary from case to case. Since they are often related to $q_a \simeq 2$ \cite{Paz-Soldan_2021}, it appears likely that benign terminations in many DIII-D, JET, TCV and AUG experiments involve an $m=2$, $n=1$ resistive external kink mode ($m$ and $n$ being the poloidal and toroidal mode numbers, respectively). Also collapses at larger $q_a$ are observed, which may also be external kink modes but at $q$ values on axis substantially larger than unity. A decrease of $q_a$ with time is necessary to access these scenarios, which could be robust to achieve exploiting the scrape-off of outer flux surfaces during the motion into the wall, re-induction of Ohmic current~\cite{Artola_2020}, and conversion into RE current by avalanching. The physics of such instabilities, in particular for the case where the current is initially carried by REs, has been studied with M3D-C1~\cite{Jardin_2007,Ferraro_2016}, including a self-consistently coupled RE fluid model~\cite{Liu_(Chang)_2021}. Figure~\ref{fig:diii-d-termination} shows the evolution of the magnetic topology along with the RE density during the burst of MHD activity. Similar observations were made with the JOREK code~\cite{Huysmans_2007,Hoelzl_2021} for ITER~\cite{Bandaru_2024}. In some JET benign terminations, on the other hand, where synchrotron emission suggests a hollow current profile, the relevant mode appears to be an $m=4$, $n=1$ double tearing mode~\cite{Bandaru_2021} (the beam's synchrotron emission shows magnetic-island-like patterns before termination~\cite{Sommariva_2024}). 
In Ref.~\cite{DelCastilloNegrete_2018} it was observed that synchrotron emission in 3D stochastic magnetic fields, computed using NIMROD, is strongly influenced by magnetic islands. Also, in the presence of magnetic islands, trapped particles exhibit higher radiation damping than passing particles. In Ref.~\cite{Marini_2024} experimental measurements and KORC simulations of synchrotron emission in the presence of a (2/1) mode and magnetic stochasticity were used for current profile reconstructions in post-disruption runaway electron plateau plasmas in DIII-D.

The physics of tearing modes (TMs) in the presence of REs has been studied theoretically~\cite{Helander_2007} and numerically with the M3D-C1~\cite{Zhao_2021}, NIMROD~\cite{Sainterme_2024}, SCOPE3D~\cite{Singh_2023} and JOREK~\cite{Bergstrom_2025} codes. A key finding is that the growth rate of these modes is essentially determined by the resistivity $\eta$ and density $n_e$ of the background plasma. Furthermore, the growth rates follow the $\eta^{3/5}$ scaling up to much higher resistivity values than in a thermal plasma, effectively increasing the growth rates for very cold background plasmas~\cite{Helander_2007,Liu_(Chang)_2018,Bergstrom_2025,Liu_(ShiJie)_2025}. The same destabilizing effect at high $\eta$ of the companion plasma is also observed for other MHD modes like resistive internal kinks~\cite{Matsuyama_2017,Bandaru_2019} and is probably quite universal. The reason for the saturation of growth rates at very high values of $\eta$ is a consequence of the radial extent of the eigenmodes becoming comparable to the plasma minor radius. The presence of REs tends to reduce the radial extent of the mode structures such that this saturation is shifted to higher resistivity values. In addition, the saturated island width is enhanced by the presence of REs for otherwise identical plasma conditions. Additionally, simulations have shown that in highly resistive plasma (like a CQ after substantial high-$Z$ impurity injection), resistive hose instabilities may exhibit higher growth rates than tearing modes, leading to rapid RE transport and losses~\cite{Sainterme_2024}.

It appears likely that benign RE beam terminations are (at least partly) caused by the very fast growth of large magnetic footprints, the latter being due to a combination of the very high $\eta$ and low $n_e$ of the background plasma. The very high $\eta$ and low $n_e$ result from the massive injection of deuterium used to obtain the benign termination, which causes the plasma to cool down and recombine. The fast growth of large magnetic footprints at high $\eta$ and low $n_e$ is supported by JOREK~\cite{Bandaru_2024} and M3D-C1 simulations~\cite{Liu_(Chang)_2021}. In these simulations, the magnetic field is strongly stochastic with very short connection lengths to the walls almost throughout the beam volume, leading to a nearly full dumping of the REs onto the wall on the time scale of micro-seconds. Simulations have also been performed with the linear MHD code MARS-F, in which the amplitude of the MHD mode was prescribed based on experimental magnetic measurements for DIII-D cases~\cite{Liu_(Yueqiang)_2019}. During the RE dumping event, the current is transferred to Ohmic current carried by the background plasma with apparently weak RE re-avalanching, as discussed in Section~\ref{:sec6}, but 3D simulations including a fully self-consistent evolution of the background plasma (reheating, re-ionization, etc.) have not been performed yet. The fact that modes grow faster at high $\eta$ and low $n_e$ is in line with the above-mentioned finding from previous works that the modes growth rate is determined by the properties of the background plasma ($n_e$ influences the dynamics \textit{via} the Alfv\'en speed). The reason for the growth up to a larger amplitude is less clear. To some extent it may be a direct consequence of the presence of the REs, as shown for saturated TMs analytically~\cite{Helander_2007}. It might also be related to the mechanisms discussed in \cite{Nardon_2023}, although it should be noted that REs had not been included in that work. When REs carry the initial current prior to a termination, another mechanism that can come into play is that the radial RE transport associated to the increasing stochasticity may flatten the current density profile and thus reduce the drive for further mode growth. There seems to be evidence for this self-regulation from JOREK simulations conducted using different RE parallel transport models: when the RE parallel transport is mimicked by a relatively slow diffusive transport to save computational costs~\cite{Bandaru_2021}, modes reach a larger amplitude in comparison to other simulations where the RE parallel transport is realistically accounted for by advection at the speed of light~\cite{Bergstrom_2025,Singh_2025}.

One difference between benign and non-benign RE beam terminations is that the former are typically associated to a single burst of magnetic fluctuations, HXRs and neutrons, while the latter are associated to multiple (smaller) bursts. The origin of this behavior is not fully established, although stronger MHD activity in benign scenarios, causing a faster near-complete loss of the REs is likely to play a central role in many cases. The energy and pitch angle distribution of REs impacting the wall during DIII-D RE termination has been diagnosed and compared with KORC simulations \cite{Hollmann_2025}, revealing that MHD instabilities can lead not only to a broader wetted area, but also to increased RE pitch angles.

\begin{figure}
    \centering
    \includegraphics[width=23em]{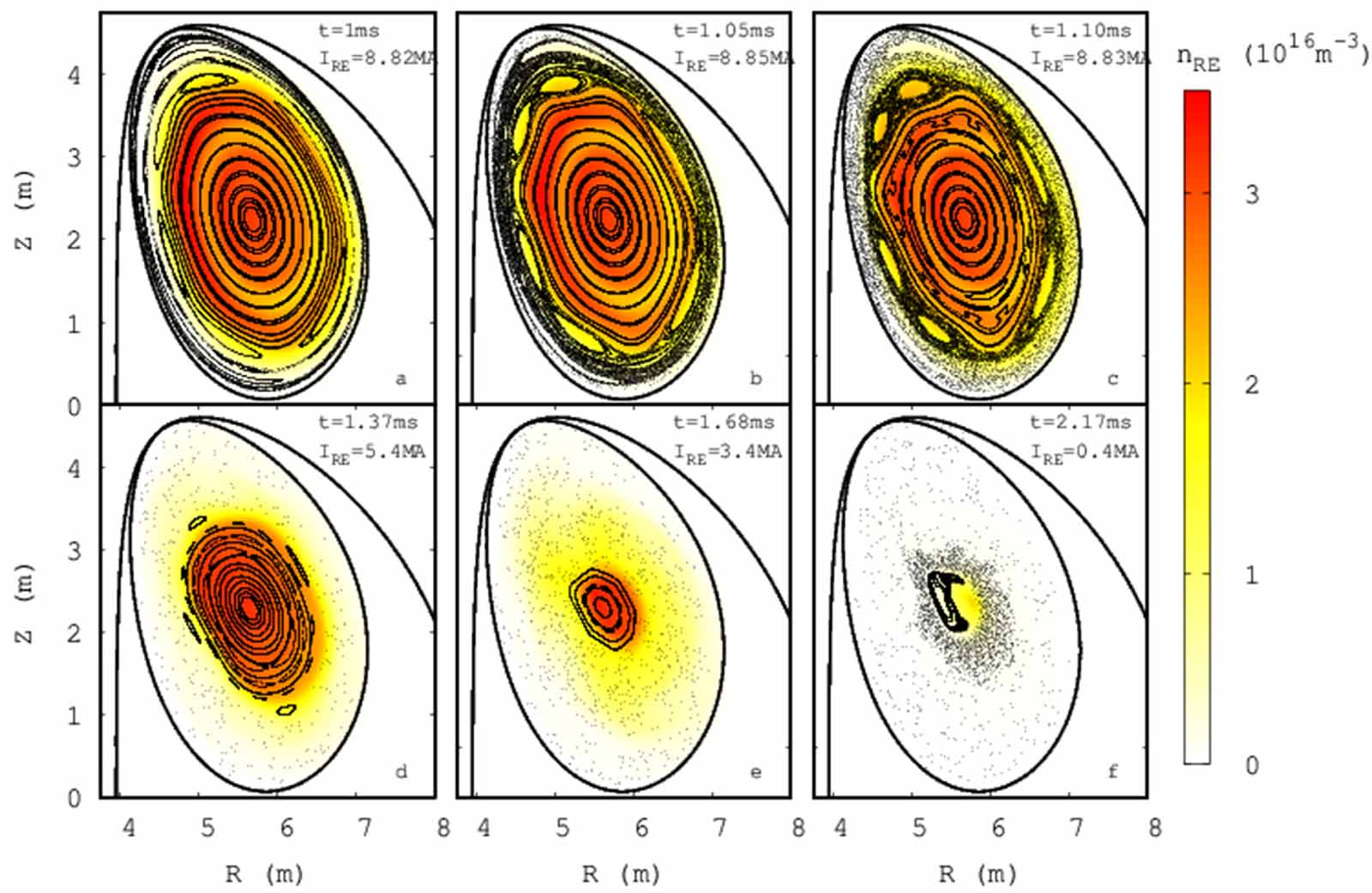}
    \includegraphics[width=23em]{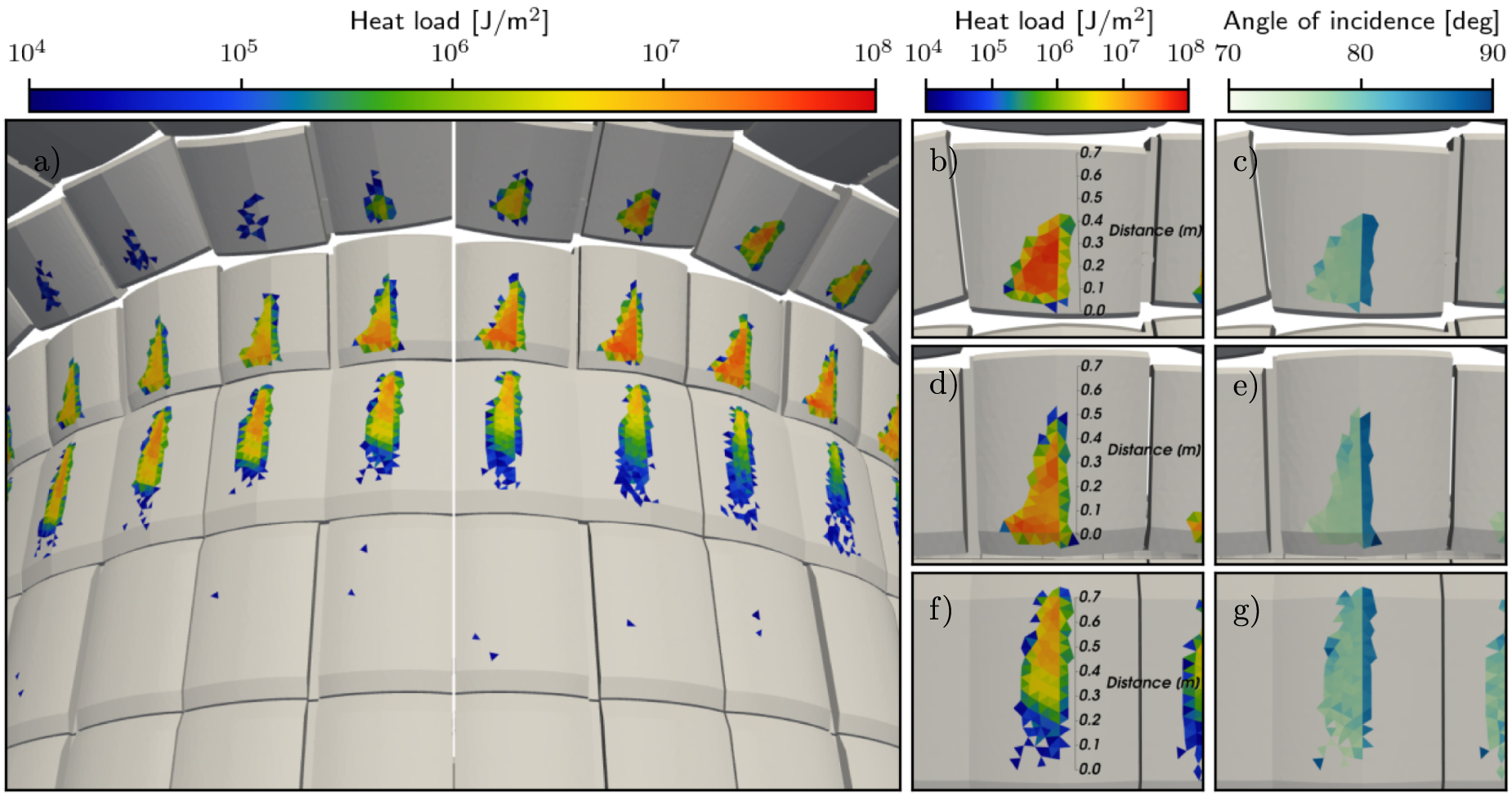}
    \caption{Top: JOREK simulation of a RE beam termination in ITER when the safety factor at the plasma edge has reached a value of $q_a=2.2$. Reprint with permission from Ref.~\cite{Bandaru_2024}.
    Bottom: RE induced wall loads in such a RE termination event in ITER. (a) Overview of tiles hit by REs at the top on the high field side. (b/d/f) Heat load for the tile with the strongest RE load in the top/middle/bottom row of (a). (c/e/g) Average angle of RE incidence corresponding to the tiles of panes (b/d/f). Reprint with permission from Ref.~\cite{Bergstrom_2024}.}
    \label{fig:iter-wall-load}
\end{figure}

Predictions on RE beam impacts have been made for ITER with JOREK~\cite{Bandaru_2024,Bergstrom_2024} and MARS-F~\cite{Liu_(Yueqiang)_2022} and for EU-DEMO with JOREK~\cite{Vannini_2025}. In the case of JOREK, the 3D simulations can be post-processed with kinetic test particles to obtain a load distribution to the realistic 3D wall structures. Figure~\ref{fig:iter-wall-load} shows both the loss of REs during a termination event in ITER and the resulting loads to 3D walls from such a calculation. The methodically comparable work on EU-DEMO~\cite{Vannini_2025} also assesses the efficiency of sacrificial limiters in protecting the wall from excessive RE loads and indicates that a stronger protrusion of the limiters from the wall might be needed than previously expected due to the massive 3D magnetic field perturbations arising.
Benign termination in DIII-D and JET was studied in Ref.~\cite{Paz-Soldan_2021} using MARS and KORC to elucidate the necessary levels of $\delta B/B$ to deconfine the full RE population and to estimate the wetted area over which the RE energy distributes during the loss events.
In addition to MHD instabilities, plasma waves associated with REs have been observed both during the flattop~\cite{Spong_2018,Bin_2022} and disruption phases~\cite{Lvovskiy_2018,Lvovskiy_2019}. Simulation studies have confirmed that these excited waves can induce additional pitch-angle scattering~\cite{Liu_(Chang)_2018} and spatial diffusion~\cite{liu_(Chang)_2023} of resonant REs, presenting a potential approach for RE mitigation through self-excited or externally launched RF waves. The full-wave code AORSA, in conjunction with KORC, has been used to investigate the phase-space diffusion of REs in the presence of these waves. However, whether this diffusion effect can effectively reduce the RE energy impact on PFCs remains an open question.

Particle-based plasma kinetic simulations of REs are time-consuming due to the multiscale dynamical processes involved and the need to follow large ensembles of particles to avoid statistical sampling errors. To overcome some of these computational challenges, Ref.~\cite{Yang_2024} proposed a generative artificial intelligence based Normalizing-Flow method for the efficient computation of hot-tail generation of REs. The proposed method learns the probability distribution function of the final state conditioned to the initial state, such that the model only needs to be trained once and then used to handle arbitrary initial conditions.

\subsection*{Current and future challenges}

Currently, the community is developing predictive modeling capabilities to support the design of future machines and/or mitigation strategies in such a way that RE beam impacts remain tolerable. The most effective way is probably to rely on 3D non-linear MHD codes including self-consistent RE models, in combination with realistic RE-wall interaction modeling (Section~\ref{:sec8}). In this area, a number of challenges need to be solved as discussed in the following paragraphs.

Termination simulations need to start from a (relatively) stable state instead of a highly unstable one, as has been done so far. To capture the explosive nature of termination, stabilizing effects may be needed beyond those already included in the models currently existing (similar to how the explosive onset of type-I ELM cycles could only be captured by incorporating stabilizing terms~\cite{Cathey_2020}).
  
The background plasma, which may be weakly ionized at the start of the termination event, needs to be self-consistently described, including its re-ionization and/or re-heating as the REs are lost. Kinetic models may be needed to capture the ballistic nature of neutral particle transport.

Realistic kinetic RE distributions at the start of impact need to be used for termination studies as well as for wall impact calculations, as they can affect the plasma dynamics (see the following two items) and drastically change the wall damage (as explained in Section~\ref{:sec8}). The companion plasma composition and the dynamics of MHD instabilities may affect the RE distribution function and thus directly affect the benign-ness of a termination event.

Since trapped REs could survive a stochastic phase and later act as seed for re-avalanching~\cite{McDevitt_2023b}, kinetic treatment is also needed in view of trapping and de-trapping like assessed in the context of 3D MHD dynamics in Ref.~\cite{Saerkimaeki_2022}.

The contribution of the REs to the major-radial force balance must be taken into account, as it can affect both the MHD stability and the wall contact point. The Shafranov-like shift of flux surfaces is neglected in the present RE fluid modeling but can be captured in kinetic modeling~\cite{Bergstrom_2025} or, under the assumption of mono-energetic RE beams, also in advanced fluid models~\cite{Bandaru_2023,Yuan_2023}.

The drift orbit shift of REs needs to be taken into account. Since this deviation of REs from flux surfaces depends on their energy and pitch angle, it may de-correlate the narrow current structures associated with the MHD modes and thus have a direct effect on the stability of the MHD modes as well as the RE deposition and losses.

The effect of the vapor and/or solid debris produced by the impact may need to be taken into account directly in the 3D simulations as it can have direct consequences on re-avalanching and RE distribution functions. This could require future self-consistent coupling of 3D MHD simulations to the wall damage models described in Section~\ref{:sec8}.

The interaction of REs with kinetic instabilities, which are not captured in the MHD framework, should be assessed as well, as they could play an additional role in transport and termination processes or open up alternative ways of mitigation.

Note that we focus on the interaction of REs with material structures here such that RE \textit{mitigation} or \textit{suppression} approaches like passive helical coils or active mitigation measures are not addressed, albeit they are often studied by the same codes and models.

\subsection*{Concluding remarks}
There are indications that, both, extremely localized RE depositions causing large damage as well as very broad depositions in case of benign termination scenarios are a result of 3D plasma dynamics. At the same time, REs can have direct consequences on the stability and non-linear saturation of plasma instabilities such that MHD and REs need to be studied self-consistently.
Major non-linear MHD codes are establishing hierarchies of models for these dynamics and interpretative as well as predictive simulations are substantially contributing to the advance of the field. Further effects not taken into account yet will require additional developments towards higher fidelity models likely describing REs, neutrals and impurities kinetically. The phase space distribution of REs may directly affect MHD dynamics and determines the volumetric energy deposition of REs in the wall structures, it is thus essential when aiming to assess the RE induced risk for material integrity. The 3D modeling profits from the strong advances regarding the complementary lower-dimensional models (Section~\ref{:sec5} and~\ref{:sec6}) and can provide accurate phase-space resolved deposition patterns for wall damage analysis (Section~\ref{:sec8}).

\clearpage 
\section{Modeling of thermo-mechanical PFC response under RE impacts}\label{:sec8}
\author{S. Ratynskaia$^1$, P. Tolias$^1$, V. Dimitriou$^2$}
\address{
$^1$ Space \& Plasma Physics, KTH Royal Institute of Technology, Stockholm, SE-100 44, Sweden \\
$^2$ Institute of Plasma Physics and Lasers - IPPL, University Research and Innovation Centre, Hellenic Mediterranean University, Rethymno, GR-74150 Greece}

\subsection*{Status}

As reviewed in Secs.~\ref{:sec1}--\ref{:sec3}, RE-PFC interaction is often an explosive event accompanied by the expulsion of melt or fast solid debris. PFC explosions are  triggered by the non-monotonic temperature response, with the beneath-surface maximum implying the build-up of internal stresses which leads to failure, fragmentation and fast material detachment. With the RE wetting characteristics provided, see Secs.~\ref{:sec5}--\ref{:sec7}, modeling of RE-induced PFC damage further requires: (i) Monte Carlo (MC) calculations of electron transport in the PFC to evaluate the energy deposition profile, (ii) continuum mechanics calculations of the thermo-mechanical PFC response to this volumetric heat source.

\subsubsection*{Energy deposition}

The transport of relativistic electrons inside condensed matter is accompanied by momentum losses owing to elastic scattering on the nuclei as well as by energy losses due to the ionization or the excitation of core or valence electrons (electronic stopping) and due to the emission of Bremsstrahlung photons in the field of the nuclei or core electrons (radiative stopping)\,\cite{ICRUReport_1984}. Therefore, a particle shower is induced that comprises fast atomic electrons (delta-rays), photons, positrons (via gamma conversion) and neutrons (via photo-nuclear reactions)\,\cite{Berger_1970}. The relevant electromagnetic and nuclear cross-sections strongly depend on the energy and the atomic number\,\cite{Bethe_1953,Carron_2007}. In W: radiative stopping overcomes electronic stopping for electron energies larger than $10\,$MeV\,\cite{Seltzer_1986}, gamma conversion overcomes Compton scattering, Rayleigh scattering and photo-absorption for photon energies larger than $6\,$MeV\,\cite{Hubbell_1980}, while photonuclear reactions have a $7\,$MeV photon threshold and $15\,$MeV maximum\,\cite{IAEAReport_2000}. In graphite: Bremsstrahlung losses dominate for electron energies above $100\,$MeV\,\cite{Seltzer_1986}, gamma conversion is the primary photon interaction for photon energies exceeding $30\,$MeV\,\cite{Hubbell_1980}, while photonuclear reactions have a $19\,$MeV threshold and $22.5\,$MeV maximum\,\cite{IAEAReport_2000}.

\begin{figure}[b]
  \centering
  \includegraphics[width=8.1cm]{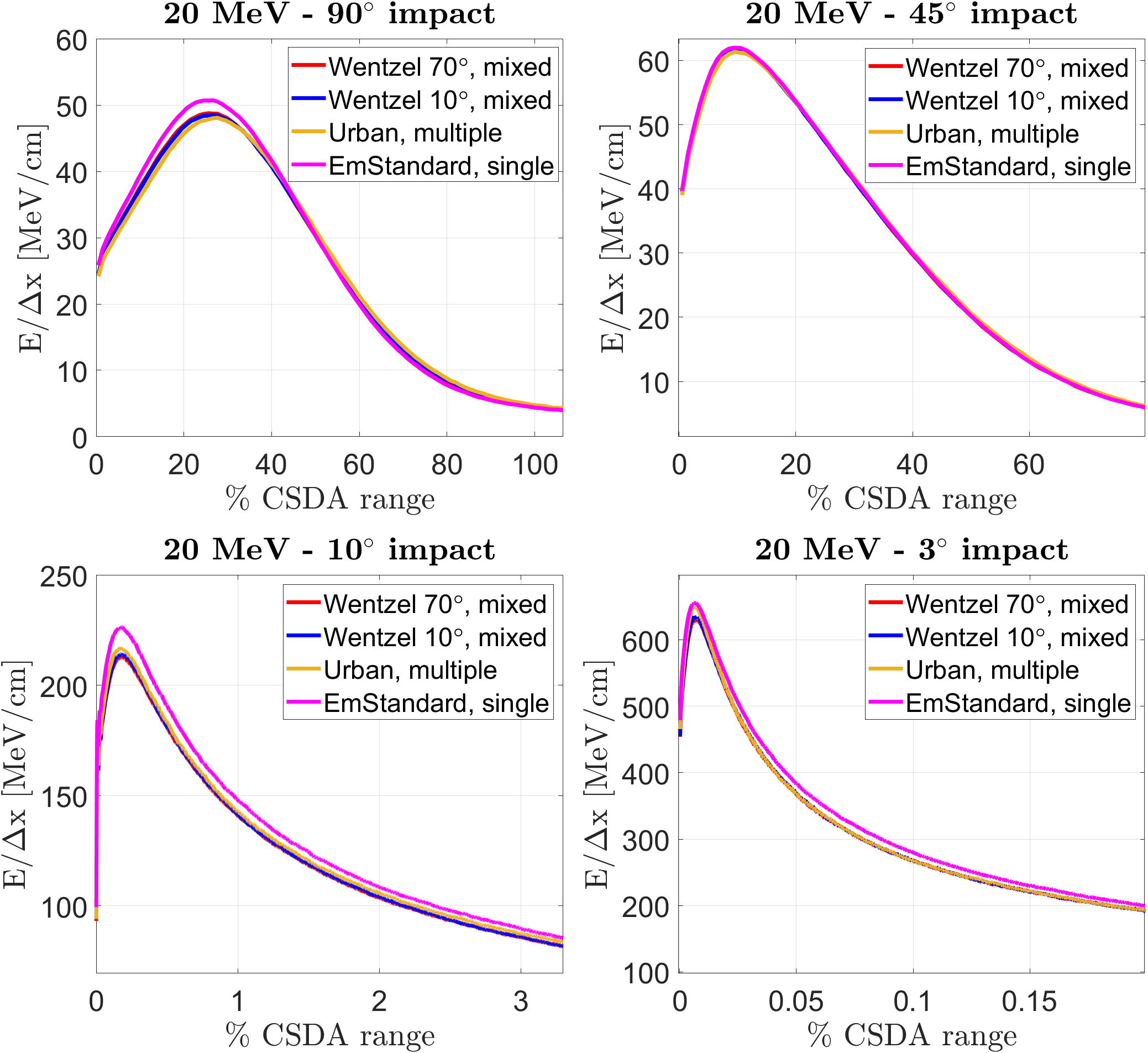}
  \caption{Geant4 results (with various scattering implementations) for the plane averaged energy density profiles of $20\,$MeV electrons impacting W at normal (90$^{\circ}$) to shallow (3$^{\circ}$) angles.
  }
  \label{profiles}
\end{figure}

Within the continuous-slowing down approximation (CSDA), electrons lose energy continuously along their path with a mean energy loss per unit path length given by the stopping power\,\cite{Fano_1963}. The CSDA range is the reciprocal of the stopping power and is defined as the average path length traveled by the incident electron until its complete thermalization\,\cite{Fano_1963}. It allows to appreciate the volumetric nature of the energy deposition. For typical RE energies from 1 to 50 MeV, the corresponding CSDA ranges vary within $0.04-0.78\,$cm for W and within $0.28-12.85\,$cm for C\,\cite{Berger_1992}. However, these values do not reflect the characteristic energy deposition depths in any scenario and especially in fusion scenarios due to the facts that: (i) the energy carried by delta electrons and Bremsstrahlung photons is non-locally deposited, (ii) the path length and depth generally differ due to scattering, (iii) the incident angles are generally shallow in tokamaks exacerbating the difference between path-length and depth (see Fig.~\ref{profiles}), (iv) the tabulated CSDA ranges do not consider the effect of the magnetic field on charged particle trajectories.

\begin{figure}[b]
    \centering
    \begin{subfigure}
        \centering
        \includegraphics[width=6.1cm]{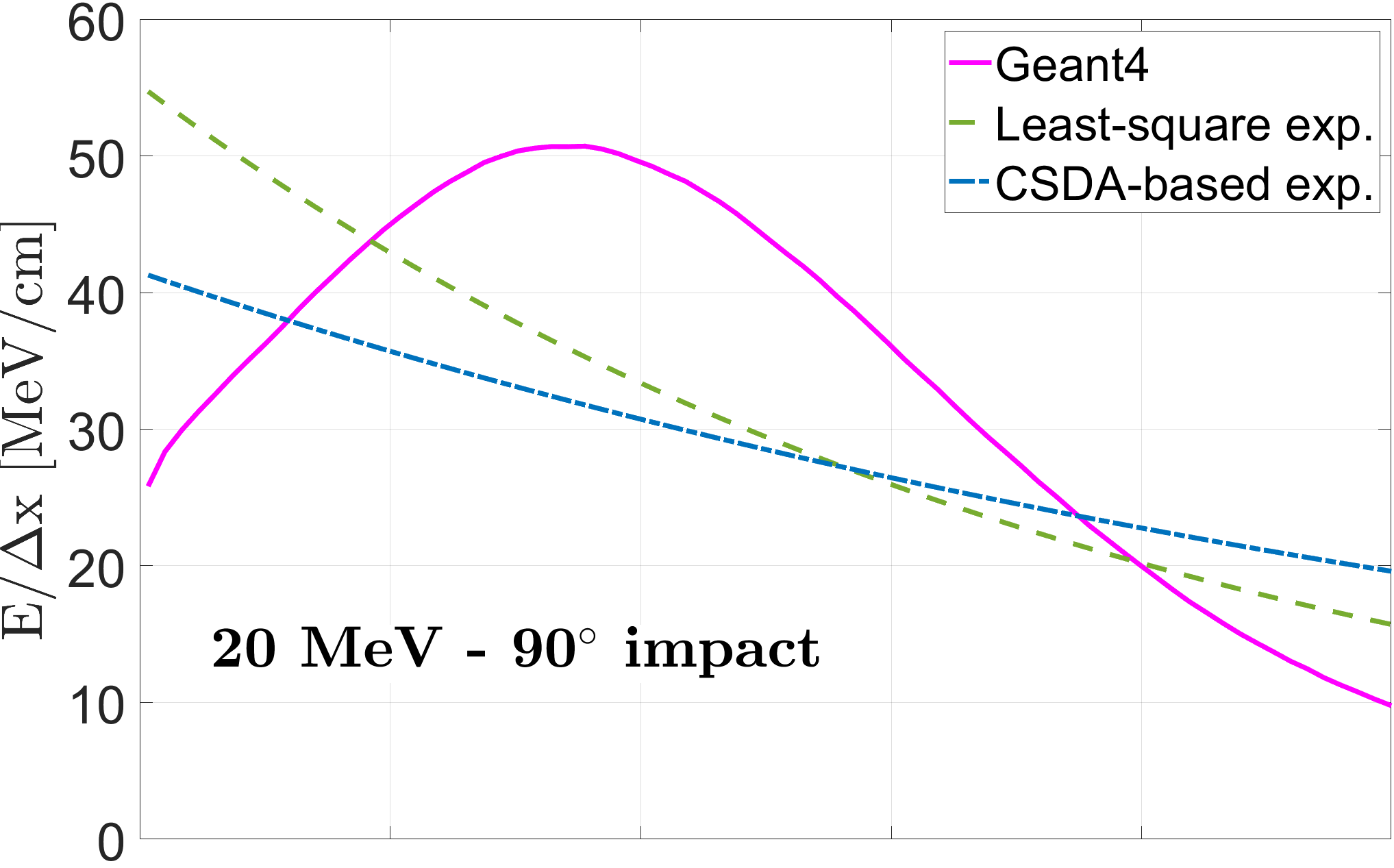}
    \end{subfigure}
    \vspace{0.1 mm}
    \begin{subfigure}
        \centering
        \includegraphics[width=6.1cm]{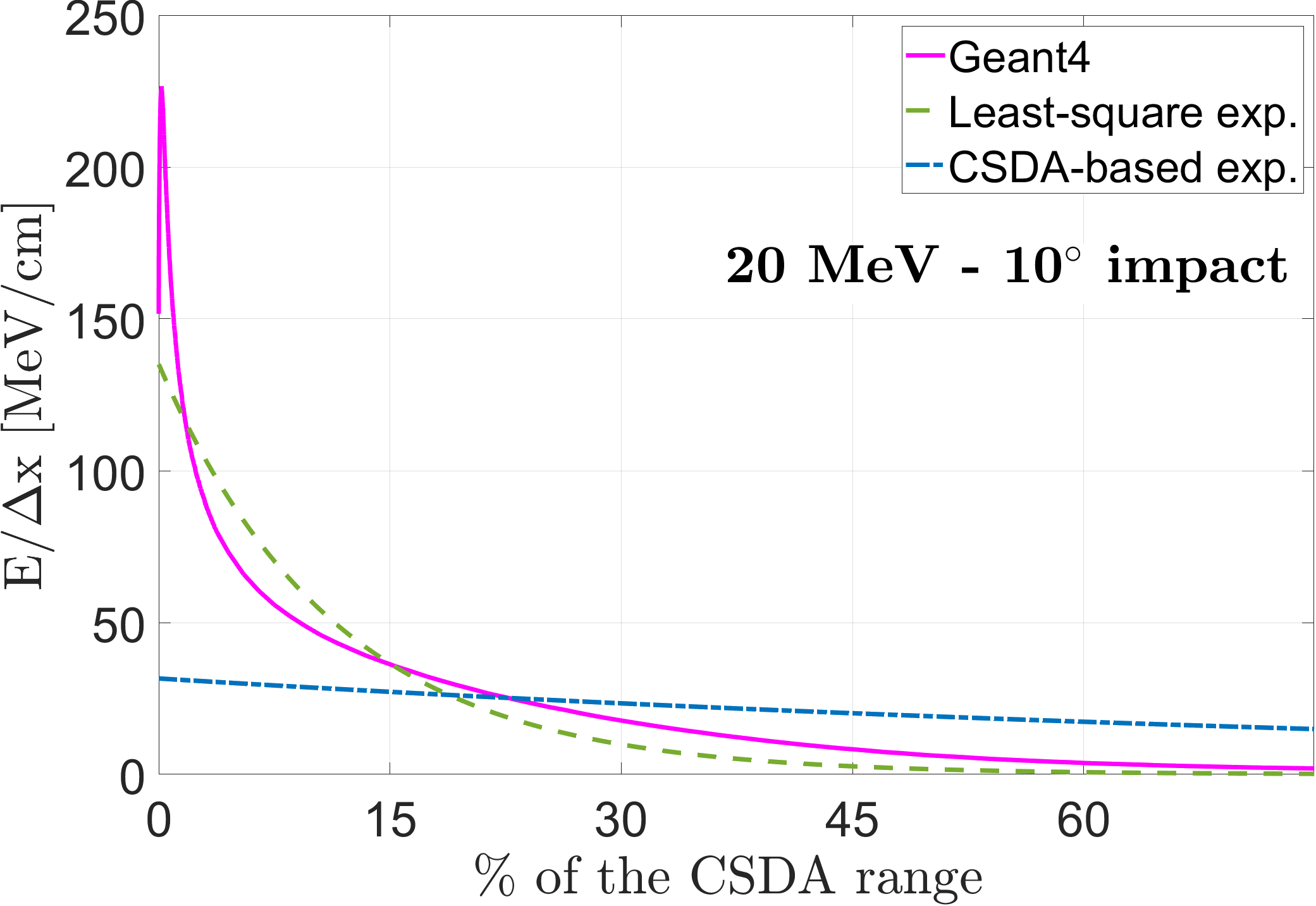}
    \end{subfigure}
    \caption{Comparison of Geant4 deposition profiles, for $20\,$MeV electrons impacting W at normal (90$^{\circ}$) and shallow (10$^{\circ}$) angles, with a least-square and a CSDA-based exponential decay.
    }
    \label{decay}
\end{figure}

Approximations of RE deposition as a surface load\,\cite{Bandaru_2024, Bergstrom_2024} allow use of heat flux factor thresholds to evaluate whether REs can cause PFC melting\,\cite{Yu_2015}. Such critical heat flux factors stem from the analytical solution for heat diffusion in a semi-infinite solid\,\cite{Carslaw_1959}. However, the validity of the surface loading assumption is very limited, e.g., to $<1$MeV incident energies for W. Approximations of RE deposition as an exponentially decaying volumetric source with an e-folding length equal to the CSDA range\,\cite{Lehnen_2009,Beidler_2024,Martín-Solís_2014} also exploit the availability of exact solutions of the 1D heat diffusion equation. Such analytical temperature profiles have been derived for an exponentially decaying source\,\cite{Carslaw_1959}, even including vaporization\,\cite{Dabby_1972} or convection\,\cite{Blackwell_1990} and have found practical use in laser-solid interactions and medical physics\,\cite{Behling_2025}. This exponentially decaying assumption also has strong limitations. Even under the gross simplification that energy is strictly deposited along the straight path of primary electrons (implying that the secondary particle depth ranges are all zero), the Bethe formula does not admit an exponentially decaying solution due to the $\propto\ln{E}/E$ basic scaling of the electronic stopping power (despite the $\propto{E}$ basic scaling of the radiative stopping power). In fact, it is known from calorimetry measurements\,\cite{Lockwood_1980} that the energy deposition profiles of MeV electrons in low- and high-Z materials have a well-pronounced maximum, see Fig.~\ref{profiles}. The maximum could reside near the surface, as for low energies and grazing angles, but still the CSDA based exponential decay remains a poor fit, see Fig.~\ref{decay}.

The first MC simulations of RE energy deposition that considered coupling between electron and photon transport can be traced back to the beginning of the 1990s and were performed with early versions of the ITS\,\cite{Gilligan_1990,Niemer_1991,Halbleib_1984}, GEANT\,\cite{Bolt_1991,Bartels_1994,Brun_1985,Brun_1987} and EGS\,\cite{Kunugi_1992,Kunugi_1993,Nelson_1988} codes. They were followed by MC investigations that are dated back to the 2000s and 2010s with the FLUKA\,\cite{Maddaluno_2003,Ferrari_2005} and EGSnrc\,\cite{Ward_2004,Krawrakow_2000} software packages as well as with the inaccessible ENDEP code\,\cite{Bazylev_2011b,Bazylev_2013} or with in-house modules\,\cite{Sizyuk_2009}. Nowadays, most MC studies are carried out with recent versions of Geant4\,\cite{Agostinelli_2003, Allison_2006, Allison_2016} or FLUKA\,\cite{Battistoni_2015, Ahdida_2022}; versatile toolkits that offer an extensive selection of differential cross-sections and scattering algorithms. The user's choice determines the accuracy and compute time that depend on the PFC material and RE wetting. 

The magnetic field effect on RE energy deposition was pointed out already in the 1990s\,\cite{Niemer_1991,Bartels_1994,Kunugi_1993}. The B-field presence outside the PFC implies that backscattered electrons can be redeposited during their gyration, thus more energy is absorbed\cite{Bartels_1994,Kunugi_1993}. This is the main effect unless the backscattering yield is negligible, which is the case for normal incidence and energies deep into the MeV range\,\cite{Dressel_1966,Tabata_1967}. The B-field presence inside the PFC implies that all charged particles are gyrating\,\cite{Bartels_1994,Kunugi_1993}. This is a secondary effect due to the collision-dominated transport. The B-field has been discussed in connection with the impact angles and phase randomization upon wetting\,\cite{Sizyuk_2009}. Backscattered electron re-deposition can lead to non-uniform loading even for uniform RE impacts\,\cite{Ratynskaia_2025a}.

\subsubsection*{Thermo-mechanical response}

Most investigations have focused on the thermal PFC response by solving the heat diffusion equation, which allows estimates of the instantaneous melt depth, erosion depth due to vaporization and coolant temperature\,\cite{Pitts_2025}. The latter constitutes a critical safety issue for fusion reactors, since elevated temperatures at the bond interface with the cooling substrate can lead to water leaks. As aforementioned, grazing angles are responsible for shifting the energy deposition maximum towards the plasma-facing surface. This results to very high energy density values in the vicinity of the free surface and, thus, to very high surface temperatures. Strong vaporization implies that the free surface deformation needs to be accounted for in the heat transfer simulations\,\cite{Dabby_1972, Ward_2004, Ratynskaia_2025a}.  At this point, we stress that there are two reasons behind the build-up of the non-monotonic temperature profiles; (i) the non-monotonic energy deposition profiles, which are naturally reflected in the temperature profiles courtesy of the relatively short loading times that limit the smoothing effect of diffusion\,\cite{DeAngeli_2023}, (ii) the intense surface cooling due to vaporization, which can efficiently reduce the surface temperature\,\cite{Dabby_1972}. 

The first investigations beyond the thermal PFC response were recently reported focusing on predictions for the onset of failure in brittle PFCs that do not possess a liquid phase (graphite and boron nitride)\,\cite{Ratynskaia_2025, Rizzi_2025a}. The governing partial differential equations were those of linear one-way coupled thermo-elasticity\,\cite{Hetnarski_2019}. In particular, within infinitesimal strain theory, the heat diffusion equation with a volumetric source was coupled to the Navier displacement equation via the thermal expansion term. The mechanical work carried out during thermal expansion was not included in the heat transfer equation\,\cite{Hetnarski_2019,Fung_2001}, being dominated by the RE volumetric heating. The equations were solved with the finite element method (FEM) and the brittle failure predictions were based on the Rankine criterion, which compares principal stresses with ultimate tensile and compressive strengths\,\cite{Budynas_1999}. A novel KORC - Geant4 - FEM workflow was validated against the first controlled RE-induced damage experiment carried out with a graphite sample in DIII-D\,\cite{Hollmann_2025} (see Sec.~3), which provided constraints on the loading time and total energy deposited. This PFC response model was also capable of describing aspects of RE-induced boron-nitride tile damage in WEST, but with less constraints due to the accidental nature of the event\,\cite{Rizzi_2025a}.

\subsection*{Current and future challenges}

\subsubsection*{Thermophysical and mechanical W properties}

Multiphysics modeling of the thermo-mechanical response of W PFCs must consider the hydrodynamic behavior, which requires a multiphase equation of state valid in a wide range of pressures and temperatures as well as the deviatoric behavior which requires a damage criterion and a viscoplastic flow rule that captures the effects of high strain rates, elevated temperatures and strain hardening\,\cite{Scapin_2012,Scapin_2014,Kaselouris_2017b,Kaselouris_2021,Scapin_2022}. Tabular equations of state are available for W\,\cite{Kerley_2003}. Empirical and semi-empirical constitutive models also exist, such as the Johnson-Cook\,\cite{Johnson_1983} and Zerilli-Armstrong model\,\cite{Zerilli_1987}, which feature various material constants whose values are available for W\,\cite{Lennon_2000}. The accurate determination of these constants from an ever increasing body of experimental data and the quantification of uncertainty propagation pose a great challenge.

\subsubsection*{Fracture and fragmentation}

It is beneficial to identify the volume patches of the continuum geometry and the exploded fragments. The former can be modeled with traditional FEM, while the latter require sophisticated techniques that overcome mesh limitations. 

Finite strain theory is essential for large displacements and strains ($\gtrsim2\%$)\cite{Belytschko_2014}. The Cauchy strain tensor is substituted with the Green-Lagrange strain tensor which contains a second-order term that describes large deformations and rotations but leads to nonlinear partial differential equations. Traditional FEM is a mesh-based approach developed on variational principles and/or weighted residual methods, capable of handling infinitesimal and finite strains\,\cite{Fung_2001}. Finite strain FEM, applied for high energy density problems involving explosions, requires iterative solvers like Newton-Raphson due to nonlinearity\,\cite{Fung_2001,Belytschko_2014}. Such simulations are limited only by the severe mesh distortions that introduce numerical instabilities. Therefore, mesh adaptivity and updating or remeshing of the discretized domain is essential to capture large deformations and strong gradients. Explicit FEM solvers can accommodate various damage criteria, since flags can be set to trigger failure of an integration point, erosion and element deletion\,\cite{Halquist_2006,Song_2008,Dimitriou_2013,Dimitriou_2015,Kaselouris_2016,Apostolova_2021}. 

Smoothed Particle Hydrodynamics (SPH) is a mesh-free numerical method suitable for highly dynamic problems involving extreme deformations, fragmentation as well as fluid-structure interaction\,\cite{Liu_2003}. SPH is based on kernel interpolation and a Lagrangian formulation, but instead of introducing a mesh, it represents the material as a collection of particles that move and interact. The kernel ensures greater contributions from nearest neighbor particles to the reference particle, achieved with a smoothing length. In particular, fluid-structure interaction allows for a better representation of the motion of high-speed debris. However, SPH is sensitive to tensile instability due to kernel approximation errors and requires special treatments like adaptive kernel corrections and artificial viscosity. 

The Arbitrary Lagrangian Eulerian (ALE) method is a hybrid of the Lagrangian and Eulerian approaches that combines their best features\,\cite{Messahel_2013,Richter_2017}. The mesh partially moves with the material but can also be adjusted to reduce distortion while still tracking material motion. ALE is excellent for fluid-structure interaction, blast \& explosion modeling, and high-speed impact problems\,\cite{Richter_2017,Kaselouris_2017}. However, ALE is computationally expensive due to the Lagrangian mesh motion and remapping and often introduces numerical diffusion during the remapping steps in the Eulerian mesh. 

The Phase Field Method (PFM) models fracture by diffusing a crack over a small length scale instead of treating it as a discontinuity\,\cite{Wu_2020}.  This is achieved with the phase field $\phi$ that continuously varies from 0 (pristine) to 1 (failed). Intermediate values correspond to a diffuse transition zone, thus eliminating the need to track an explicit crack surface\,\cite{Wu_2020}. The $(\boldsymbol{u},\phi)$ fields are found by minimizing a free energy functional that includes the elastic strain energy and fracture energy. PFM has found limited use in large deformations, since coupling with finite strain theory and damage criteria demands modifications that make it computationally costly. PFM requires fine meshes to resolve the diffuse zones, implicit solvers for the coupled $(\boldsymbol{u},\phi)$ system, time-stepping schemes for the crack evolution and pre-conditioning techniques to improve convergence.

Peridynamics Theory (PD) is a nonlocal formulation that replaces spatial derivatives with integrals to handle fracture and extreme deformations\,\cite{Silling_2010,Madenci_2014}. The material body is represented as a collection of discrete points, each interacting with other points within a horizon $\delta$ through bonds that can stretch or compress under deformation and break when a failure criterion is met\,\cite{Madenci_2014}. The horizon defines the nonlocal interaction range with larger $\delta$ increasing accuracy and the compute time. PD is mesh free and element connectivity is not required, avoiding mesh distortions. However, PD is difficult to implement and to efficiently parallelize.

\subsubsection*{Coupling between the RE deposition and PFC response}

MC simulations calculate the RE energy deposition into a pristine amorphous material at room temperature. As far as the pure thermal PFC response is concerned: (i) the direct effect of elevated temperatures in the stopping power is negligible, since the PFC temperature should be an order of magnitude lower than the electronic Fermi temperature ($\sim10\,$eV), (ii) the indirect effect of elevated temperatures in the stopping power through the mass density can locally reach $20\%$ (see the proportionality), which is also rather insignificant, (iii) in the case of strong vaporization, the vapor cloud could have a strong effect on energy deposition, provided that there is significant density depletion due to expansion into vacuum during RE wetting. As far as the thermo-mechanical PFC response is concerned, depending on the deformation scales and the wetting duration: the evolving free surface due to melt motion or fragmentation could affect energy deposition. Such problems require dynamic coupling between MC and FEM simulations, which is computationally costly.

\subsubsection*{Coupling between the RE transport and PFC response}

Transport of REs into PFCs is accompanied by the ejection of electrons, positrons, photons with a broad energy distribution. In case of strong vaporization, low energy impurity atoms are also emitted. In case of PFC explosions, debris with sizes less than the CSDA range are ejected in the vacuum vessel. It is conceivable that interactions with these products could affect the RE seeding, avalanche or termination phases. An iterative RE impact -- RE deposition -- PFC response workflow is unrealistic and toy models would need to be developed.

\subsubsection*{RE-induced stress}

In general, RE transport yields a volumetric heat source $Q$ in the heat transfer equation and an external stress term $\nabla\cdot\vec{\boldsymbol{\sigma}}$ in the displacement equation. MC simulations can calculate the cumulative internal stress field generated by the incident electrons and secondary particles. However, the post-processing is complicated and probably also prone to statistical \& grid errors. A simplified comparison between the strain contribution from particle momentum transfer and the strain contribution from thermal expansion should precede the development of post-processing routines.

\subsubsection*{Indirect RE-driven damage}

As mentioned in Sec.~3, high-velocity impacts of solid dust on the surrounding wall are a source of de-localized RE-induced damage. The high velocity regime of $0.2-4\,$km/s is distinguished by strong plastic deformation, projectile fragmentation, shallow target cratering and near-surface melting of both bodies\,\cite{Klinkov_2005,Hassani_2018,Tolias_2023,DeAngeli_2024}, while the hypervelocity regime of $>4\,$km/s is characterized by complete dust vaporization, deep crater formation and fast secondary solid ejecta production\,\cite{Eichhorn_1976,Burchell_1999,Ratynskaia_2008,Fraile_2022}. Such impacts are routinely realized in the laboratory with light gas guns, laser ablation systems and electrostatic accelerators\,\cite{Veysset_2021}. They share common physics with RE-driven explosions, as evidenced by their reliable modeling with PD\,\cite{Amani_2016,Ren_2024} or SPH\,\cite{Libersky_1997,Remington_2020}. Given the availability of experimental data, they can be employed to fine-tune and enhance the predictive power of tools that simulate RE-driven explosions.

\subsubsection*{Surrogate energy deposition profiles}

Even when considering only the thermal PFC response, the full workflow is too computationally heavy for scoping studies. It is important to discern whether REs lead to PFC melting without costly code-chains. Energy deposition maps (on a semi-infinite solid) for every tokamak relevant energy, impact angle, magnetic field combination can act as basis states that enable the approximate reconstruction of any realistic RE energy deposition map by superposition. Given the large number of combinations, interpolations are unavoidable. One-dimensional heat conduction solvers can then be coupled with codes that model the RE impact on the vessel wall (see e.g. recent JOREK code calculations\,\cite{Artola_2024}) allowing fast estimates and identifying cases of interest for more detailed coupled 3D MC and FEM simulations.

\subsection*{Concluding remarks}

Modeling of RE-driven damage in brittle materials with no liquid phase has largely progressed benefiting from controlled RE-impact experiments on graphite carried out in DIII-D. The thermo-elasticity equations supplemented with the maximum normal stress brittle failure criterion have provided an adequate description of the PFC response up to the onset of fragmentation. It is expected that the state-of-the-art of the modelling can be successfully extended to the explosive process by coupling a traditional FEM analysis of the continuum geometry to a more sophisticated analysis of the debris generation (SPH and ALE formulations). Modeling of RE-driven damage in W, which is ductile at elevated temperatures with a stable liquid phase, remains to be addressed. The W compressible hydrodynamic and viscoplastic behavior can be captured by a multiphase equation of state coupled with a multiphysics strength material model. Any progress has been impeded by the absence of controlled RE-impact tokamak experiments on W. The construction of an electron beam material testing facility that replicates RE impacts would speed up the evolution of high fidelity numerical models and the development of innovative wall designs.

The long term goal should be to translate set-ups that are currently incorporated in commercial general purpose MC and FEM software packages to a stand-alone open-source multi-physics code that is dedicated to predictions of RE-driven PFC damage.

\clearpage 
\section{Neutron production and activation}\label{:sec9}
\author{ M. Houry$^1$, Y. Peneliau$^1$, J. Eriksson$^2$, R. Villari$^3$}
\address{
$^1$ CEA, IRFM, F-13108, Saint-Paul-lez-Durance, France \\
$^2$ Department of Physics and Astronomy, Uppsala University, SE-75237 Uppsala, Sweden \\
$^3$ Nuclear Department, ENEA, Via E. Fermi 45, 00044 Frascati, Rome, Italy
}

\subsection*{Status}

\subsubsection*{Neutron production and PFC activation by RE impacts.}

Runaway electrons (REs) may activate the first wall and divertor components as observed in the tokamaks FT~\cite{Maddaluno_1987}, TFR~\cite{1980203}, PLT~\cite{Barnes_1981}. This might cause a non negligible dose rate in in-vessel components. The activation comes both directly from photonuclear reactions and from neutrons produced by the latter reactions~\cite{SUKEGAWA20181653}. The doses induced by REs are likely to be subdominant in experiments with significant fusion power. However, in experiments with limited fusion power such as the startup phase of ITER or present small- to medium-sized machines, the dose induced by REs could overcome the dose from fusion neutrons. Neutron production and PFC activation may also be exploited to constrain and validate RE impact models. This section will illustrate these aspects in particular by describing recent and ongoing work on WEST.

\subsubsection*{RE beams in WEST.}

\begin{figure}[h]
\centering
\includegraphics[scale=0.44]{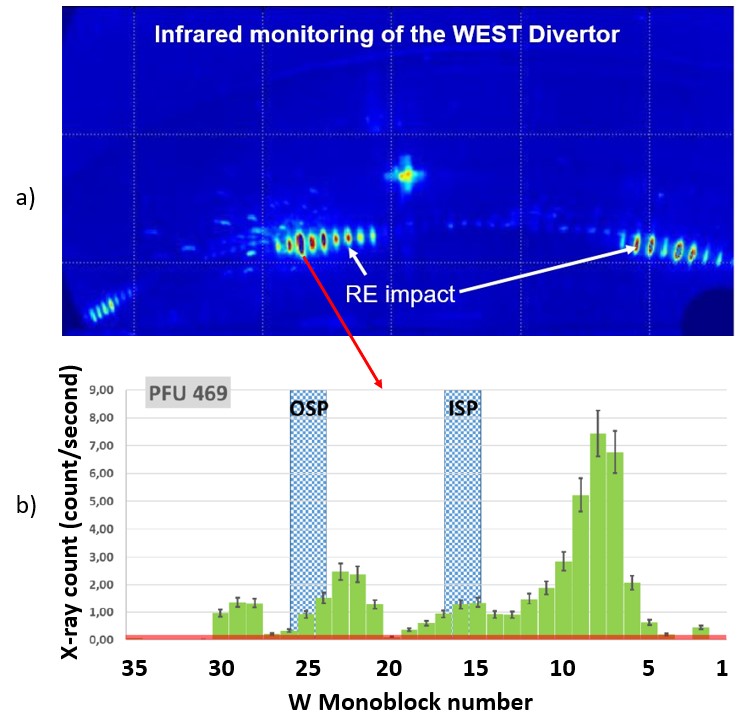}
\caption{ Infrared view of a RE beam impact in WEST for pulse \#58005 (a). Post mortem measurement of decay gammas in the monoblocks of the Plasma Facing Unit \#469  on the lower divertor of WEST possibly due to multiple impacts (b).}
\label{fig:Impact}  
\end{figure}

During WEST disruptions, RE beams carrying a current exceeding 100 kA and with an energy spectrum extending beyond 10 MeV can be generated.  WEST diagnostics allow observing its formation and evolution via e.g. infrared (IR) tangential imaging of synchrotron radiation and a direct IR view observing the RE crash onto the divertor, see FFig.~\ref{fig:Impact} (a).  Moreover, activation of W PFCs has been detected during post mortem analysis, as presented in Fig.~\ref{fig:Impact} (b).

\begin{figure}[h]
\centering
\includegraphics[scale=0.44]{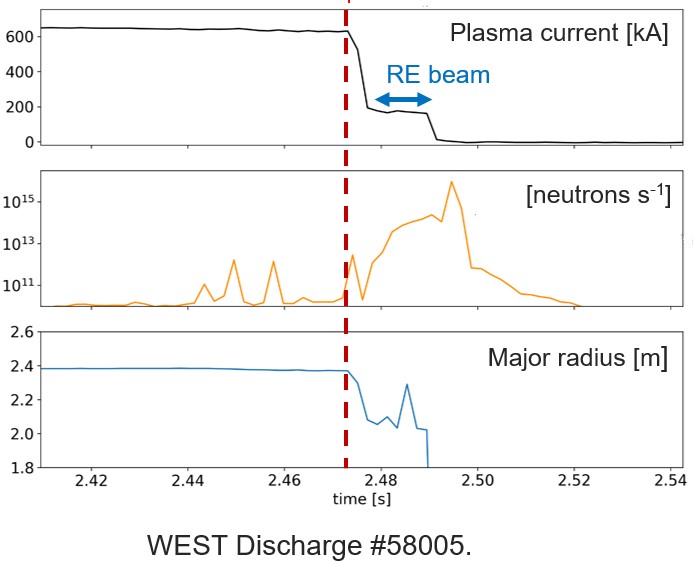}
\caption{Parameters during disruption of WEST for pulse \#58005. Current  in the plasma (in kA). Neutrons (D-D and RE) measured in the Vacuum Vessel thanks to Fission Chambers (in n/s). Localization of the electron current in the Vacuum Vessel (radius in m).}
\label{fig:Disruption}  
\end{figure}

Figure~\ref{fig:Disruption} shows key parameters recorded during pulse \#58005, where a RE event occurred. The plasma current features a plateau after the start of the disruption, indicating a RE beam duration of about 10 ms. Then the RE beam crashes onto the divertor, causing the current to drop to zero. The neutron rate measurement shows a sharp peak at this moment, indicating nuclear reactions with neutron production. Magnetic diagnostics provide the radial localization of the RE beam (the bottom plot). In this specific case, the measured position of the beam aligns with the impact location shown in Fig.~\ref{fig:Impact}. 

\begin{figure}[h]
\centering
\includegraphics[scale=0.37]{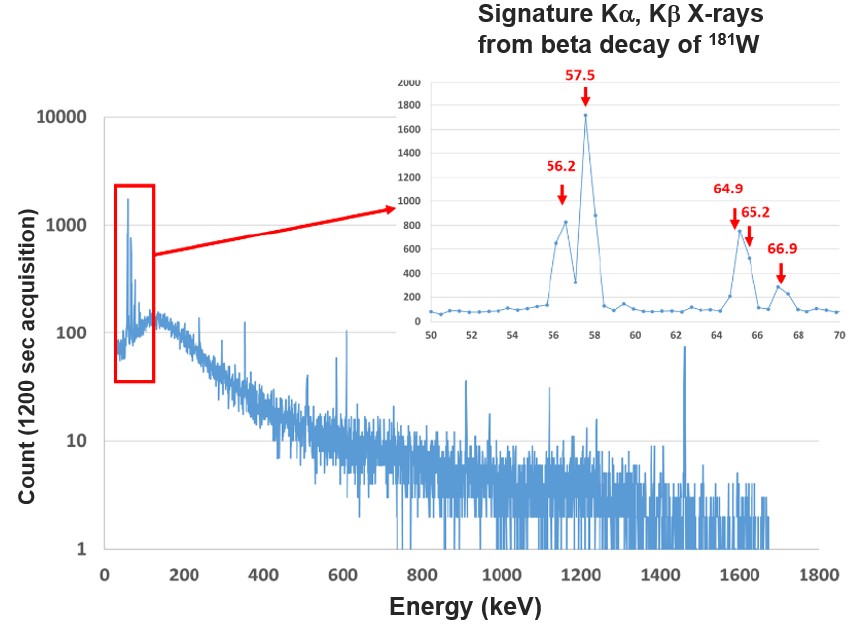}
\caption{Spectroscopy of the WEST PFCs}
\label{fig:Spectroscopy}  
\end{figure}

Gamma spectroscopy measurement on the W components of the WEST divertor reveals spectral lines that can be attributed to $^{181}$W . The spectrum shown in Figure~\ref{fig:Spectroscopy} clearly displays five peaks around 60 keV, identified as originating from $^{181}$W. The formation of this radionuclide, with a half-life of 121 days, can only result from nuclear reactions leading to the transmutation of $^{182}$W into $^{181}$W, either through photoneutron reactions or (n,2n) reactions. Monte Carlo simulations of the measurement configuration with the HPGe detector have been carried out in order to calculate the scaling factor between the W activity in the monoblock and the number of gamma ray counts. Conservatively assuming that the gamma source is homogeneous in the material leads to a possible 50 Bq/g specific activity for the most activated monoblock for which the gamma rays counts measurement was 10 cts/s, shown in Fig~\ref{fig:Impact} (b). Additional spectrometry measurements were conducted to analyze the spatial distribution of $^{181}$W within the component. Using shielding screens, gamma spectroscopy was performed in 7 mm depth increments. The results show a clean absence of radioisotopes beyond 7 mm, indicating that REs impact the W component at a shallow incidence angle. Thanks to the spectroscopy conducted on one divertor sector ($1\times0.5$ m$^2$), providing an activity map, it is observed that RE crashes lead to the formation of a narrow beam, approximately 5 cm in width, which toroidally propagates at a certain radial distance. In WEST, the corresponding activation zone is influenced by the ripple effect and thus appears with a periodic discontinuity.

The activation in W is a 2-step process. The RE impact can produce high energy bremsstrahlung photons due to slowing down of the REs. Then, high energy photons can have photonuclear reactions with the nuclei of the material: (X,n) reactions. The latter are threshold reactions happening above 8 MeV in W, in which photons can produce neutrons as well as W radioactive isotopes as the reaction removes one neutron or more from a stable nucleus. Regarding W isotopes, photonuclear reactions produce two radioisotopes, $^{181}$W and $^{185}$W, by removing neutrons to the initial stable isotopes. $^{181}$W is the most contributing isotope to dose rate because its X-ray decay intensities per disintegration are of the order of dozens of \%, whereas X-ray decay intensities of $^{185}$W are of the order of 0.01\%. However, regarding neutron production, as the photonuclear cross sections and the relative fraction of parent isotopes are similar, they contribute almost equally.

\subsubsection*{Monte Carlo simulation of the electromagnetic shower.}
Simulations of PFC activation by RE impacts consist first in simulating electron transport and the electromagnetic shower in the material. This is usually performed thanks to specific Monte Carlo codes like MCNP-6 \cite{TechReport_2023_LANL_LA-UR-22-33103Rev.1_RisingArmstrongEtAl}, TRIPOLI-4 \cite{HUGOT2024} and Geant4\,\cite{Agostinelli_2003, Allison_2006, Allison_2016}. These codes simulate the production and transport of bremsstrahlung photons. They can implement the photonuclear reactions in the same simulation, producing neutrons that can themselves be transported and produce activation on one hand, or new secondary photons due to neutron reactions like radiative capture on the other hand. Photonuclear reactions are usually inelastic reactions ($\gamma$,n') (also noted as (X,n) in this article) in the energy range of interest. Fission reactions ($\gamma$,f) or (at high enough photon energy) neutron multiplicity reactions ($\gamma$,2n), ($\gamma$,3n) are also possible.

\subsubsection*{Photonuclear reactions in W.}
\begin{figure}[h]
\centering
\includegraphics[scale=0.35]{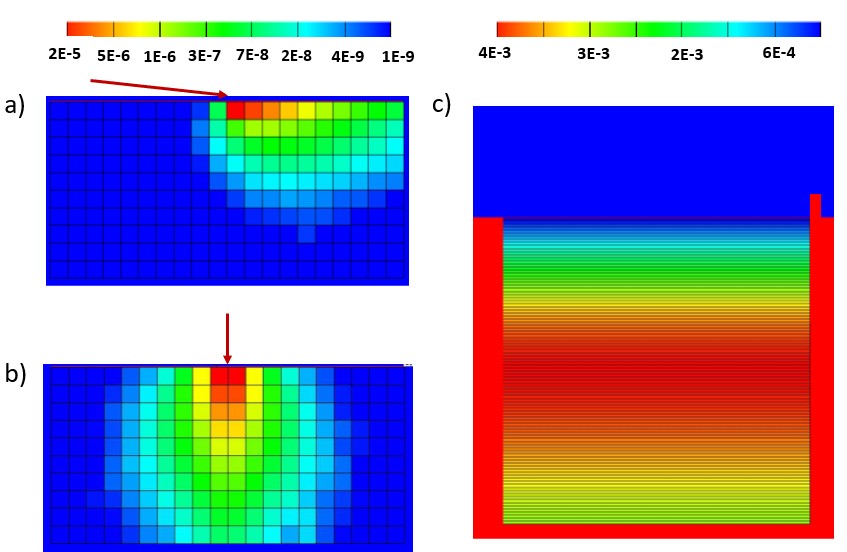}
\includegraphics[scale=0.20]{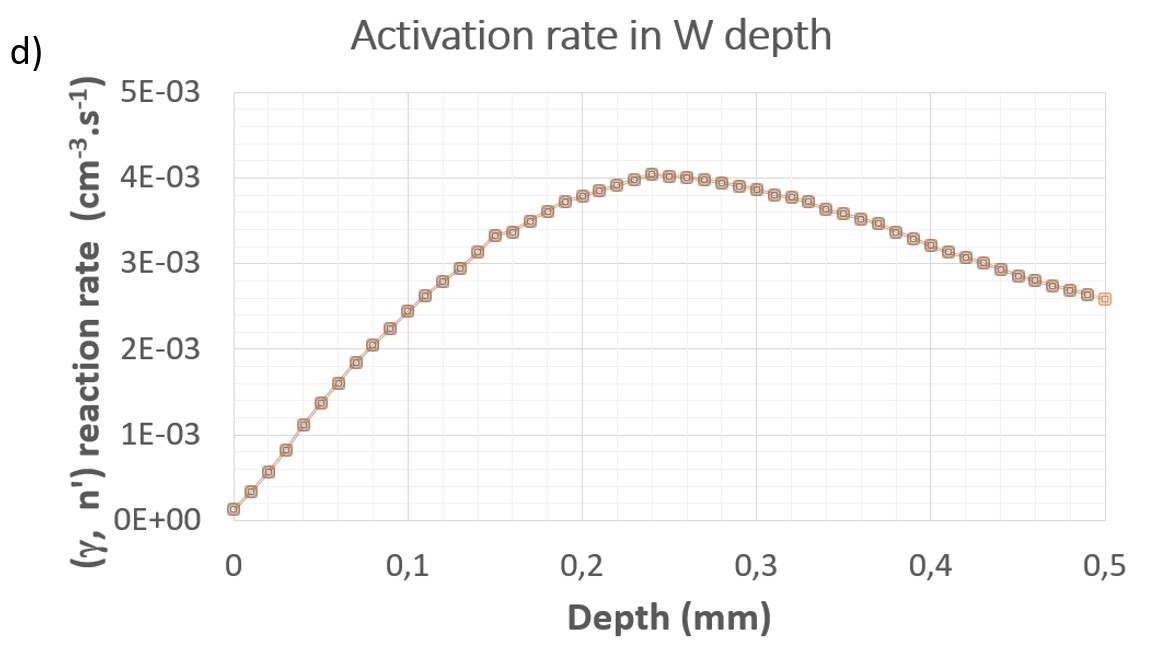}
\caption{ Activation rate map in W for 12 MeV electrons impinging tangentially (a) or  normally  (b) for a 5 mm voxel size. Activation rate map in W for 30 MeV electrons for a 0.05 mm voxel size in Z axis and 5 mm in other axes (c) . Reaction rate profile corresponding to the former activation rate map  (d).}
\label{fig:ActivationRate_Section9}  
\end{figure}

Parametric studies on the incidence energy and angle of the REs help to elucidate the phenomenon of activation. Fig.\ref{fig:ActivationRate_Section9}  shows the activation rate (i.e. the number of reactions per second and cm$^3$) due to 12 MeV electrons on a 5 mm mesh for a tangential incidence of the electron (a) and  for a normal incidence (b) in bulk W. The activation rate is calculated for a source of 1 e$^{-}$/s and the activation rate ranges from $10^{-5}$ (red) to $10^{-9}$ cm$^{-3}$.s$^{-1}$ (blue) on the figure (logarithmic scale). Both configurations are of interest in a tokamak as REs can crash tangentially on the divertor but also normally on an antenna protection limiter for instance. Other angles should also be considered as the RE pitch angle and shape of the W components in a tokamak can vary. A finer mesh analysis with 0.05 mm large voxels shows that a pile-up phenomenon occurs. The latter is illustrated for 30 MeV electrons and normal incidence in Fig.\ref{fig:ActivationRate_Section9} (c), that presents a refinement of the distribution in the most activated voxel of the previous coarse mesh. The electrons produce bremsstrahlung photons in the first layers. The photons contribute to the activation rate in the following layers as well, as long as their energy is high enough to exceed the reaction threshold. When the electron energy loss is significant, the electron cannot produce photons with energy above the threshold, and from layer to layer, the number of photons able to contribute to photonuclear reactions decreases. It should be noted that the refinement of the mesh is important as the W linear attenuation coefficient for 60 keV gammas is 71 cm$^{-1}$ and the gamma flux drops drastically with penetration. 

\subsubsection*{Neutron production.}
An analytical formula for the (photo)neutron yield was proposed by Jarvis \textit{et al.}~\cite{Loarte_2011}. In early JET studies with a carbon wall and inconel vacuum vessel, the neutron yield for RE energies of $E_0\gg15$ MeV was given as $Y_n = \frac{E_0}{60 \times 2.9 \times 10^{-3}}$ n/e$^{-}$ (with $E_0$ in MeV). For a W wall, recent systematic radiation transport simulations were conducted with GEANT~\cite{Ataeiseresht2023}. The energy spectra of photons and neutrons produced in electron collisions were calculated. Photoneutron production in W occurs at photon energies above 8 MeV. Neutron production increases with the electron energy, occurring in the 10-100 MeV range, with a 100 MeV beam generating fast neutrons (average energy ~1.3 MeV). High-energy neutrons from these interactions escape the target and may impact tokamak equipment. For electron energies above 20 MeV, many neutrons exit the W wall. If neutron production occurs in the initial target layers, slowing down and thermalization may cause neutron activation in PFCs.

\subsubsection*{Activation calculation schemes.}
Like for standard dose rate calculations based on D1S~\cite{VALENZA2001411} or R2S~\cite{CHEN2002107} schemes, appropriate methodologies can be implemented to calculate activation dose rates. From the electron simulation, it is possible to calculate a photon ($\gamma$,n') reaction rate map as it was done in the investigations for the WEST reactor and the activation of $^{181}$W. It is also possible to calculate neutron production maps and neutron production spectra to define photoneutron sources for further neutron simulations. From the activation map, a secondary decay photon source is defined for all the radionuclides (like for $^{181}$W here below) from the photon (X,n') activation rate at the end of the beam: 

\begin{equation}
S_{\gamma} (\vec{r}, E_{\gamma}) = \lambda^{^{181}W} \times \tau^{^{182}W (X, n)^{181}W} (\vec{r}) \times I_{e^{-}} \times f(E_{\gamma})
\label{eq:1}
\end{equation}

where $\lambda^{^{181}W}$ is the decay constant of the radioactive isotope $^{181}$W in s$^{-1}$, $\tau^{^{182}W (X, n)^{181}W}$ is the activation dose rate of $^{181}$W production in $(e^-)^{-1} \times$ cm$^{-3}$, $I_{e^{-}}$ is the total number of electrons impacting the PFC, and $f(E_{\gamma})$ is the X and $\gamma$ energy distribution in MeV$^{-1}$.

\subsection*{Current and future challenges}

\subsubsection*{Extrapolation of W activation by REs to ITER.}
ITER DD first plasmas during start of research operations (SRO) are designed to induce a low dose rate in the vacuum vessel in order to allow hands-on maintenance during the phase of first wall replacement between SRO and DT-1. Thus, the activation of the first wall due to RE beams in the ITER vacuum vessel requires attention. The extrapolation from WEST to ITER is not straightforward as the RE beam characteristics differ. The expected RE current will be one to two orders of magnitude larger, and the number of photons produced above the W photonuclear reaction threshold could be significantly higher depending on the RE energy, substantially affecting the contact dose rate. On the contrary, the RE wetted surface is expected to be larger due to the machine size and the mitigation strategy~\cite{Lehnen_2015, Polevoi_2021}. Thus, the effect on contact dose rate could be one order of magnitude lower. It is also expected that more bremsstrahlung photons will have an energy higher than 25 MeV, where the ($\gamma$,n) cross sections significantly drop.

\subsubsection*{Diagnostics.}
The neutron rate monitors employed to measure DD and DT neutron emission from tokamaks can also often pick up the signal from photoneutrons, and can thus provide direct evidence of the formation of REs from the plasma (see e.g. Figure~\ref{fig:Disruption} and~\cite{Plyusnin_2006, Causa_2015}). However, even more detailed information about the interaction between REs and PFCs could potentially be available if the energy spectrum of the photo-neutrons could be measured too. This is a challenging task that would require a neutron detector with spectroscopic capabilities, capable of measuring a short burst of photo-neutrons in a significant background of X-rays and gamma-rays. Furthermore, in order to minimize the amount of scattering the photo-neutrons undergo before reaching the detector, the detector would need to view the machine through a suitably positioned diagnostic port and care must be taken to ensure that the field of view of the detector covers the part of the PFCs where the RE beam is expected to hit. Hence, the first step in order to install a photo-neutron spectrometer on a given machine would be a more detailed modelling/assessment of the expected characteristics (time duration, location, energy range) of the photo-neutron emission, in order to determine more specific design requirements of the spectrometer. Depending on the spatial constraints of the machine, both compact spectrometers - such as scintillators and semiconductor detectors - and more specialized instruments - such as time-of-flight and proton recoil systems - could be considered and evaluated; see e.g.~\cite{Ericsson_2019} for an overview of different neutron spectroscopy techniques. In experiments where photo-neutrons are the only source of neutron emission (i.e. if the plasma does not contain D or T), the energy spectrum could potentially also be reconstructed from activation measurements~\cite{Stancar_2021}.

\subsubsection*{Photonuclear data.}
Available photonuclear data are ENDF/B-VII~\cite{CHADWICK20112887} or ENDF/B-VIII~\cite{BROWN20181} more recently, JENDL-5~\cite{Iwamoto02012023}, IAEA-PD-2019~\cite{KAWANO2020109} and TENDL-2023~\cite{KONING20191} data. Some issues arise with these various sources. The first issue is the energy range covered by the evaluation process. ENDF/B-VII or ENDF/B-VIII provide inelastic cross sections in the range 8 MeV - 30 MeV only. Other libraries cover the energy range until 200 MeV, but IAEA and JENDL provide production data for the sum of all inelastic reactions, without distinction of ($\gamma$,n'), ($\gamma$,2n), ... In ITER or larger tokamaks, higher electron energies are expected and thus modelling may lack crucial information for the simulations with ENDF/B series. Moreover, a significant validation effort~\cite{Tuyet01022024} should undertaken for the isotopes of interest for fusion applications, and mainly for W isotopes due to anticipated use of this material as first wall in reactors.

\subsubsection*{Accounting for the magnetic field.}
As discussed in Sec.~\ref{:sec8}, presence of the magnetic field leads to redeposition of the backscattered electrons due to gyration. It is of utmost importance to take into account this effect and a validation campaign should also be carried out with specific benchmarks.

\subsubsection*{Specific D1S methodology development.}
As it was done for neutron activation for the plasma D-T source, similar activation schemes should be developed. The D1S methodology~\cite{SAUVAN2020111399, VILLARI20142083} could be applied to photons interacting with the material. Each time a photon has a photonuclear reaction and produces a radioisotope, the code simulates the decay photon production and assesses the subsequent dose rate. Specific libraries are needed in which the production of prompt neutrons due to photonuclear interactions is replaced by decay photon production. The same time factors as the ones used for plasma neutrons are compliant for X or gamma simulations for the specific radioisotopes.

\subsubsection*{RE dose rate in ITER and further fusion reactors.}
As stated previously, the assessment of the possible impact of RE activation in ITER is relevant with regard to the operations planned between the SRO phase and the DT-1 phase. The components to be analyzed are mainly the divertor and the first wall, both of which have been made of W since the re-baselining of ITER. Subsequently, depending on the assembly schedule for the more voluminous machines, this work will also be carried out on DEMO or other facilities.

\subsubsection*{WEST dedicated experiment.}
An dedicated experimen was conducted at WEST in April 2025 to test the control~\cite{Carnevale_2019} of the RE beam position on the facility. It opens the opportunity to plan a specific activation and neutron production experiment in a sacrificial component in order to have experimental results for the validation of the complex simulation process of the phenomenon. The measurements will consist in activation foils dosimetry to characterize the neutron flux and potentially to derive the neutron spectrum at different locations, but also in activated components spectrometry and dose rate measurement. The final aim of the experiment is the validation of the Monte Carlo- based computational methodology.

\subsection*{Concluding remarks}

REs and the subsequent photoneutron production not only generate damages to the components but also contribute to their activation. Recent spectrometry analyses on WEST divertor components impacted by RE beams have shown W radionuclide activity. The measurements performed on these components (decay X ray counts) at different locations can serve as experimental data source for obtaining information on the characteristics of the RE beam. The activity and subsequent dose rate in a larger scale fusion reactor are expected to be much lower than the ones due to the direct plasma neutron interactions in the full power D-T pulses. Nevertheless, the phenomenon deserves being further studied as it can be significant enough in the early stages of the reactor operations, before D-T nuclear phase.

\clearpage 
\section{ITER perspective}\label{:sec10}
\author{F.J. Artola$^1$, R. A. Pitts$^1$} 
\address{
$^1$ ITER Organization, Route de Vinon-sur-Verdon, CS 90 046, 13067 St. Paul Lez Durance Cedex, France}



\subsection*{Status}
\subsubsection*{Runaway impact at ITER.}
As discussed in Section~\ref{:sec6}, runaway electron beams in ITER could deposit kinetic energies on the order of tens of MJs and magnetic energies of several hundred MJs, with typical average RE energies around 20 MeV. Simulations using a workflow based on the codes DINA-SMITER–GEANT4–MEMENTO have shown that energy densities of 150–200 kJ per tungsten First Wall Panel (FWP) can result in melt depths of 0.5–1.5 mm~\cite{Pitts_2025, Ratynskaia_2025a}. Assuming a 9 MA RE beam carries  $\sim$24 MJ of kinetic energy ~\cite{Bandaru_2024}, and neglecting additional magnetic energy conversion, this would correspond to a RE current of $\sim$75kA if all energy were deposited on a single FWP. Alternatively, uniform deposition across the 36 FWPs in one toroidal row would correspond to a RE current of $\sim$2.7MA.

\begin{figure*}[t]
    \centering
    \includegraphics[width=0.9\textwidth]{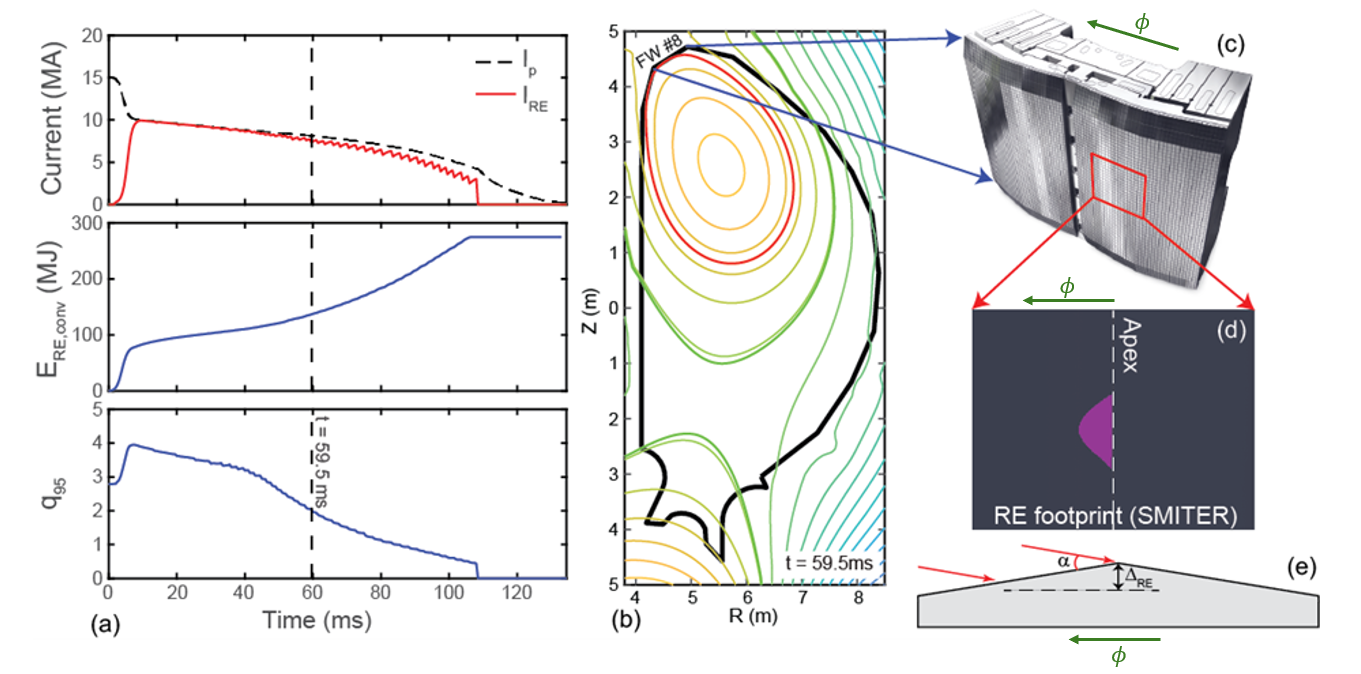}
    \caption{Composite describing the key input parameters for the new RE impact simulations performed with the DINA-SMITER-GEANT4-MEMENTO code workflow: (a) DINA simulation (IMAS disruption database shot 100097, run 1) including RE conversion during an upward going CQ at 15 MA, 5.3 T showing the RE current and magnetic energy conversion with $q_{95} = 2$ reached at $t = 59.5$ ms; (b) magnetic equilibrium at $q_{95} = 2$ assumed as the instant at which the RE beam is destabilized with impact on FWP \#8; (c) Blanket Module CAD model illustrating the double winged apex structure of the FWP; (d) SMITER field line tracing calculation of the RE beam wetted area for $\Delta_{RE} = 4$ mm; (e) schematic representation of the FWP apex modelled as a simple rooftop with given fixed RE beam impact angle. Green arrows in (c-d) represent the toroidal direction. Reprint with permission from Ref.~\cite{Pitts_2025}.}
    \label{fig:ITER-DINA-SMITER}
\end{figure*}

Fig.~\ref{fig:ITER-DINA-SMITER} illustrates the scenario used to provide the RE inputs necessary for the study. The DINA simulation results in a $\sim$9 MA RE beam formed during an upward current quench (CQ), following the conversion of the available magnetic energy within the ITER vacuum vessel into RE kinetic energy during a slow termination event (Fig.~\ref{fig:ITER-DINA-SMITER}a). The beam impacts FWP \#8 near the top of the vessel, with geometry and field line tracing (via SMITER) used to define the impact location and area (Fig.~\ref{fig:ITER-DINA-SMITER}d). The impact area was taken at the time instant corresponding to $q_{95}=2$, where the beam is expected to terminate after crossing the MHD stability limit. The RE energy distribution is taken as exponential with $E_0=15$ MeV (range 1–50 MeV), and the beam width $\Delta_{RE}$ (or deposition depth) is fixed at 4 mm, consistent with the Larmor radius at this value of $E_0$. In addition, all electrons were assumed to strike at 5$^{\circ}$ grazing angle based on the magnetic field geometry provided by DINA (Fig.~\ref{fig:ITER-DINA-SMITER} e). The choice of such inputs was encouraged by a successful study at JET~\cite{Chen_2021}, in which the same workflow was applied to pulse \#86801 and the results were qualitatively consistent with the post-mortem RE damage analysis. The assumption on the beam width is consistent with the RE Larmor radius at ITER as minimal possible width. In this respect, it is expected that for RE pitch angle distributions occurring when $E\gg E_ c$ \cite{Hesslow_2018a} and $E\sim E_ c$ \cite{Aleynikov_2015a}, the average Larmor radius of 20 MeV electrons for full ITER toroidal field operation exceeds 2 mm.

It is also important to note that MHD termination events may lead to considerably larger energy deposition areas (or RE beam widths), as also in  the case of benign RE termination. To assess this, the study summarized in Figure \ref{fig:ITER-DINA-SMITER} was repeated replacing the DINA-SMITER components by JOREK MHD calculations and RE orbit tracing on the resulting time-varying 3D fields \cite{Bergstrom_2024, Bandaru_2024}. In this case, the RE deposition area is notably much broader as shown in Figure~\ref{fig:iter-wall-load} and discussed in Section~\ref{:sec7}. When this input was considered, RE beams with 24 MJ kinetic energy deposited over 1 ms, and a maximum of 1.2 MJ per FWP, yield melt and vaporization depths of 1.1 mm and 28 $\mu$m respectively~\cite{Ratynskaia_2025a}. For that particular case, a mono-energetic distribution considering 26 MeV REs and a single pitch of $v_{\parallel}/v=0.99$ were used for the orbit tracing with JOREK, but RE distributions expected during the avalanche phase were also analyzed \cite{Bergstrom_2024,Ratynskaia_2025a}. In common with the calculations for the scenario shown in Figure \ref{fig:ITER-DINA-SMITER}, these studies also indicate significant heating at the PFC cooling interface layer (CuCrZr) for an 8 mm W armour thickness (which is the nominal thickness chosen for the original Be first wall armour). Too high a temperature at this bond interface constitutes a risk for the initiation of water leaks, which would entail severe cost and operational consequences.  Given that ITER RE beams may in fact deposit substantially more energy than assumed in the scenarios modelled thus far, even a handful of such events may not be tolerable.

Most RE impact studies at ITER will take place during the Start of Research Operation phase (SRO), where the  Disruption Mitigation System (DMS) will be commissioned and optimized up to 15 MA before the beginning of fusion power operation. During this phase, an inertially cooled Temporary First Wall (TFW) will facilitate RE impact studies without the risk of water leaks~\cite{Pitts_2025}. However, particular caution is required to avoid RE impact on the divertor, which is actively cooled and will be in place from the beginning of the SRO phase. For the final, actively cooled W FW, the code workflow studies described above have been used to recommend an increased thickness of W armour in specific regions of the main chamber on which the most probable RE impact is expected. There is, however, always a trade-off between the conflicting requirements of stationary and transient power handling and cost/design complexity. Increased armour thickness can be very effective at preserving lifetime if the deposited RE energies are within a certain range.  Once these energies are surpassed, the worst affected areas of the FWPs will still never be capable of sustaining many impacts before panel replacement is required. Effective RE avoidance and mitigation strategies will thus always be mandatory on ITER and achieving this is a major component of the re-baselined ITER Research Plan~\cite{Loarte_2025}.

\medskip

\subsubsection*{Runaway avoidance at ITER.}
The first line of defence to minimize RE impact on the ITER PFCs is RE avoidance. Its primary challenge in ITER arises from the avalanche generation mechanism, which produces seed amplification factors many orders of magnitude higher than those in present tokamaks. The avalanche multiplication scales exponentially with plasma current, following $10^{\alpha_{av} I_p}$ \cite{Putvinski_1997}, where $\alpha_{av}$ is estimated to be on the order of 1 $\textrm{MA}^{-1}$  \cite{Martin-Solis_2017, Vallhagen_2024}. This implies that each additional MA of $I_p$-decay can increase the RE current by roughly an order of magnitude.  Simulations with JOREK incorporating vertical displacements during the CQ suggest a reduction of the amplification from $\sim 10^{16}$ down to $\sim 10^{10}$ \cite{Wang_2025}. Although this reduction eases RE avoidance strategies, such an amplification factor would still amplify tritium and Compton-induced RE seeds (of the order of $10$mA \cite{Martin-Solis_2017, Vallhagen_2024}) to multi-MA beams. 

Achieving electron densities exceeding $3 \times 10^{22}  \textrm{m}^{-3}$, required for avalanche suppression, is expected to be unfeasible. Assuming perfect material assimilation, this would demand the injection of approximately 16 out of the 27  full-size pellets available in the ITER DMS. However, assimilation is limited by pre-disruption thermal energy and by drift effects such as plasmoid motion and the rocket effect~\cite{Jachmich_2024, Vallhagen_2024}. Even if sufficient material were assimilated, strong plasma recombination could lead to excessively short CQ durations, leading to intolerable eddy current loads on the ITER blanket modules. Furthermore, the unfavorable ratio between total (free + bound) and free electrons is expected to enhance, rather than suppress avalanche growth  \cite{Vallhagen_2020}.

\begin{figure*}[t]
    \centering
    \includegraphics[scale=0.3]{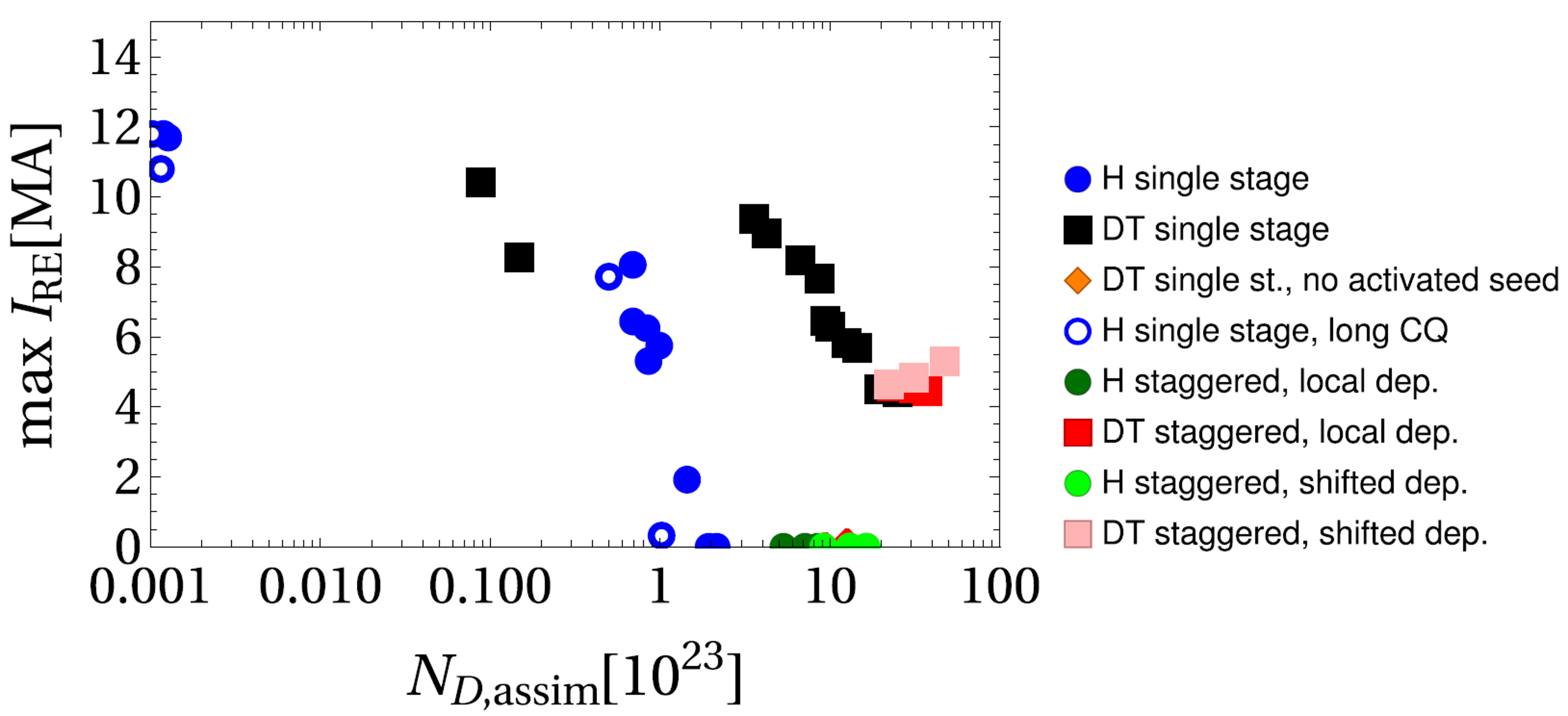}
    \caption{Representative RE current (linear) vs. assimilated material (logarithmic). The cases marked with "H" (hydrogen) correspond to scenarios without activated RE sources (Compton and tritium), while the DT scenarios contain all source (except the orange diamonds). Note that the ITER DMS will inject hydrogen instead of deuterium, as used in these studies. However, isotopic effects are not expected to affect RE generation. Reprint with permission from Ref.~\cite{Vallhagen_2024}. }
    \label{fig:ITER-DREAM-suppression}
\end{figure*}

Given the challenges discussed above, alternative mechanisms for RE loss and deceleration are necessary. As outlined in Section~\ref{:sec5}, whistler waves could potentially deconfine REs; however, their excitation requires post-TQ electron temperatures of at least 25 eV, which may not be achievable in mitigated ITER plasmas. With regard to externally applied magnetic perturbations, the ITER design is now frozen and does not include passive 3D coils for RE de-confinement that have been proposed for other devices ~\cite{Boozer_2011,Smith_2013,Sweeney_2020}. The potential use of the existing resonant magnetic perturbation (RMP) coils for this purpose was studied in Ref.~\cite{Papp_2011a}, which showed only partial deconfinement. Improved performance may, though, be achievable in later stages of the CQ, when $I_p$ is lower and with optimized RMP configurations. Intrinsic MHD activity during this phase may also lead to significant RE deconfinement. Simulations with JOREK using realistic CQ parameters have shown encouraging results for RE suppression~\cite{Artola_2022, sarkimaki_2022}, though the post-TQ profiles used in these studies were ad hoc and are currently being revisited. Finally, an alternative mitigation strategy involving the injection of a train of solid fragments to deplete the RE seeds during the CQ has been proposed~\cite{Nardon_2022} and is under re-evaluation for application in ITER.

In view of the current uncertainties in controlling RE losses, the ITER strategy for RE avoidance primarily focuses on suppressing RE seed generation through H injection. Simulations using the DREAM code for various ITER scenarios mitigated by SPI indicate that Dreicer and hot-tail seeds can be sufficiently reduced to achieve full avoidance of RE avalanches if sufficient plasma densification is achieved as shown in Figure \ref{fig:ITER-DREAM-suppression} for an L-mode Hydrogen scenario (blue circles). This suppression is enabled by a two-stage plasma cooling process~\cite{Nardon_2020,Vallhagen_2022,Vallhagen_2024}, consisting of an initial dilution phase at nearly constant thermal energy, followed by a rapid radiative collapse. However, the triggering of the TQ in DREAM remains based on an ad hoc criterion, and the associated uncertainties need to be carefully assessed. In addition, the effects of plasmoid and fragment (rocket) drifts must be incorporated more realistically to refine predictions of seed suppression efficacy. While tritium seeds may be suppressed by sufficient plasma densification and the increase of the critical energy beyond 18.6 keV, Compton generated electrons can have energies in the range of MeV \cite{Martin-Solis_2017}, and RE avoidance does not therefore seem possible in the absence of CQ transport. As shown in Figure \ref{fig:ITER-DREAM-suppression}, even in the best assimilation conditions for nominal DT H-mode plasmas, multi-MA beams are formed.

\subsubsection*{Runaway  Mitigation and Termination in ITER}
In the event of multi-MA RE beam formation in ITER, vertical control of the plasma is expected to be lost unless more than two-thirds of the pre-disruptive $I_p$ is converted into RE current~\cite{lukash2013study}. For lower RE current fractions, vertical instabilities will result in RE beam impact on the PFCs irrespective of the RE current decay time (see Section~\ref{:sec6}). Given the inherent difficulty of RE position control and ramp-down, the primary strategy to mitigate RE impact is H injection into the RE beam to induce plasma recombination and impurity flush-out. This process facilitates a benign termination through an MHD-triggered collapse, as discussed in Section~\ref{:sec7}. Preliminary studies suggest that a single H ITER-sized pellet could be sufficient to access this benign termination regime for RE impact~\cite{Hollmann_2023}, but its physics basis is not yet fully established. However, achieving compatibility between RE avoidance and RE mitigation remains challenging, particularly due to plasma re-ionization above an upper neutral pressure limit, which may compromise the benignness of the termination~\cite{Sheikh_2024b}.

\subsubsection*{Runaway relevant diagnostics at ITER}
The assessment of RE impact on PFCs at ITER relies on a combination of diagnostics. Infra-red (IR) cameras and thermography systems, covering approximately 70\% of the divertor and FWP surfaces~\cite{Aumeunier_2024}, will provide surface temperature measurements with a time resolution of the order of 1 ms to identify RE deposition patterns. The runaway energy deposition could be diagnosed with thermocouples installed in the TFW, if present at the impact location. In the event of significant RE impact, post-shot inspections will be performed using the In-Vessel Viewing System (IVVS), which provides full coverage of the divertor and FWPs with spatial resolution below 3 mm. During the first fusion power operation phase (DT-1), the in-vessel lighting system will also support damage evaluation. Post-mortem inspection opportunities will be available during the decommissioning of the TFW following the SRO phase.

During SRO, the primary diagnostic for RE detection will be the Hard X-Ray Monitor (HXRM), which is expected to detect RE currents as low as 10 kA in the 0.1–20 MeV energy range and provide key insights into the RE distribution function up to 100 MeV~\cite{Pandya_2018,Patel_2023}. Additional RE detection will be attempted using the Radial X-ray Camera, as well as ECE and visible/IR cameras, which can capture synchrotron radiation to further characterize the RE distribution (see Section~\ref{:sec4}). Following the SRO phase, the HXRM is unlikely to be operational, and its function will be taken over by Gamma-Ray Spectrometers~\cite{Nocente_2017}. Runaway electron losses will also be monitored using radial and vertical neutron cameras~\cite{Esposito_2022}. 

\subsection*{Current and future challenges}
Given the high risk of RE generation and its potentially severe consequences for ITER, it is essential to quantify the damage caused by RE impact on PFCs. Accurate damage assessments will be critical for defining operational limits and setting RE avoidance and mitigation targets to ensure PFC integrity. This is necessary to prevent unexpected delays and to support the timely achievement of ITER’s research goals. During the SRO phase, dedicated experiments with controlled RE beams at low plasma current will be conducted to validate predictions of RE impact and mitigation. A key challenge is the development of a low-$I_p$ RE scenario triggered by neon SPI, since such beams have not yet been successfully generated in existing tokamaks using SPI. Instead, they have only been achieved with MGI, a system that will not be available at ITER. Moreover, it remains unclear how results from low-current experiments will extrapolate to full-current operation, where intentional RE impacts would likely not be permissible.

The general challenges of modelling RE impacts have already been outlined in previous sections, including RE characteristics prior and during the impact (sections~\ref{:sec5}, \ref{:sec6} and~\ref{:sec7}) and the complex thermo-mechanical response of the PFCs (section~\ref{:sec8}). The multi-physics nature of RE-MHD coupling, especially when impurities, kinetic effects and losses are included, is particularly difficult to model.  At ITER, the long CQ duration ($\sim$100 ms), driven by machine size, further complicates the problem by pushing current solvers to prohibitive computational limits; even simplified fluid CQ MHD simulations with today's solvers require millions of core-hours. Development of more efficient and scalable sparse-matrix solvers would be highly beneficial in this respect.

On the materials side, the high RE energy deposition expected at ITER may require the implementation of advanced PFC response workflows capable of capturing phenomena such as fracturing and explosive material failure approaches not yet applied to ITER.  If such fragmentation occurs, the reintroduction of PFC debris into the plasma could significantly alter both RE dynamics and MHD behavior, creating feedback loops that complicate impact modelling and prediction. Such RE beam-wall interaction, will probably be very localized, posing even higher complexity for modelling. Finally, evaluating cumulative damage from multiple RE terminations also presents challenges, since evolving surface geometry may strongly influence subsequent energy deposition and failure thresholds.

Due to current uncertainties on RE avoidance and impact at ITER, a stepwise increase in plasma current and magnetic field during SRO will allow safe development of RE control strategies, with controlled RE impacts directed to the inertially cooled TFW using the in-vessel vertical control coils to reduce risk of water leaks on the actively cooled divertor. However, it cannot be ensured that all disruptions will be successfully guided upward ~\cite{Loarte_2025}. Further analysis is therefore needed to assess RE impact on divertor structures, particularly the outer baffle, which is the most likely impact location.

Dedicated experiments during DT-1 will progressively assess the impact of tritium and Compton-induced RE seeds. Approximately one month of operation has been allocated early in DT-1 to demonstrate that the ITER DMS can effectively mitigate disruptions up to 15 MA in hydrogen-tritium (HT) plasmas, retiring the risk from tritium $\beta$-decay seeds before fusion power operation \cite{Loarte_2025}. The influence of Compton seeds will be evaluated through targeted DMS pulses during the staged fusion power ramp-up in the development of DT H-mode plasmas. If efficient thermal and electromagnetic load mitigation proves incompatible with RE avoidance at ITER, compromises may be necessary (e.g., such as reducing the injected neon quantity and accepting lower radiated energy fractions) to support RE suppression by limiting the avalanche-enhancing effect of bound electrons from neon. Furthermore, the RE benign termination scheme may not compatible with the quantity of injected material for RE avoidance due to the upper neutral pressure limit for recombination.

\subsection*{Concluding remarks}
Due to the significant challenges in achieving reliable RE avoidance at ITER, assessing RE impact remains essential for validating models, defining operational limits, and managing disruption budget consumption.  Although challenging, continued efforts to develop predictive workflows for RE impact will be critical to define tolerable RE events and refine mitigation strategies. These activities are necessary to prevent excessive component damage, avoid operational delays, and ensure ITER meets its research goals within acceptable RE exposure limits.

\clearpage

\newcommand{\Ip}{I_p}
\renewcommand{\ne}{n_e}
\newcommand{\Te}{T_e}
\newcommand{\Ro}{R_0}
\newcommand{\Bo}{B_0}

\section{DEMO and other future machines perspective}\label{:sec11}
\author{R. Ding$^1$, G. Pautasso$^2$,  R.A. Tinguely$^3$}
\address{
$^1$ Institute of Plasma Physics, HFIPS, Chinese Academy of Sciences, Hefei 230031, China
$^2$ Max Planck Institute for Plasma Physics, Boltzmannstrasse 2, 85748 Garching, Germany
$^3$ Plasma Science and Fusion Center, Massachusetts Institute of Technology, Cambridge, MA, USA }

In this section, we focus on runaway electrons (REs) and their impact in future tokamaks, planned or under construction, around the globe, while ITER is discussed separately in Section 10. Fast electron populations in stellarators with non-zero toroidal currents~\cite{Helander_2011} or levitated dipoles~\cite{Garnier_2006} are outside the scope of this article.

\subsection*{Status}\label{sec:future-status}

\subsubsection*{In Asia.}\label{sec:future-status-asia}

\begin{figure}
    \centering
    \includegraphics[width=0.85\columnwidth]{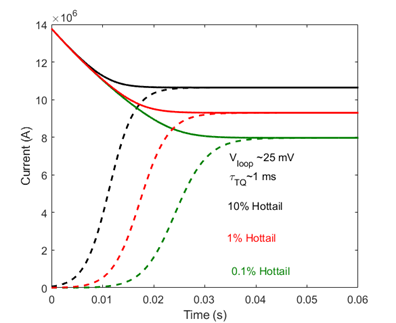}
    \caption{Evolution of plasma current and RE current in cases with $I_p = 13.78$ MA for CFETR. A simple 0-D model is applied here including Dreicer generation, hot-tail generation, avalanching\cite{Tang_2021}.}
    \label{fig:RE_CFETR}
\end{figure}

Recent simulation results for a typical plasma discharge in the China Fusion Engineering Test Reactor (CFETR) reveal that a majority of plasma current, $\Ip$, would be converted into RE current during a disruption. The assumed pre-disruption plasma parameters are based on a reference hybrid mode 1~GW scenario, with $\Ip = 13.78$ MA and electron density $\ne \sim 9 \times 10^{19}$~m$^{-3}$. The pre-disruption loop voltage is 25~mV, the characteristic thermal quench (TQ) duration is estimated to be 1~ms, the post-disruption electron temperature is $\Te \sim$~10~eV, and the electron density is 10 times the pre-disruption density. $Z_{\text{eff}} \sim 4$ is assumed during disruptions, and $l_i$ is the same as in the ITER 15~MA scenario. 
The majority of the RE seed is provided by the hot-tail mechanism.
RE losses are considered as an overall reduction of the hot-tail seeds to a fraction of 0.1\%, 1\%, and 10\%. 
RE calculations based on these assumptions are shown in Fig.~\ref{fig:RE_CFETR}. Almost 75\% of the plasma current is converted into RE current during the disruption, with RE seeds of tens of kA leading to ${\sim}$10.6~MA of RE current when 90\% RE seed loss is considered. 
However, even with 99.9\% loss of the RE seed, the RE current is estimated to be $\sim$8~MA due to the high avalanche gain.

\subsubsection*{In Europe.}\label{sec:future-status-europe}

The European Demonstration Fusion Power Plant (EU-DEMO) is a project aimed at designing a tokamak
that can produce a few hundred MWs of electricity, demonstrate reliability and availability, and work with
a closed fuel cycle.
The EU-DEMO 2018 design~\cite{Siccinio_2022} foresees a plasma with 
a major radius $\Ro = 9.07$~m, aspect ratio $A=3.1$,
and $\Ip = 17.75$~MA. Because of the large current and associated high avalanche gain, the formation of a substantial RE beam is expected in every disruption.
The different types of disruptions expected in DEMO are described in Ref.~\cite{Maviglia_2022} and follow the guidelines of the ITER Physics Basis~\cite{Hender_2007}.

The study of RE generation in DEMO~\cite{Martin-Solis_2017b,Martin-Solis_2018} was adapted from the simulations done for ITER~\cite{Martin-Solis_2017}.
For the shortest current quenches (CQs) following pure Ar injection for disruption mitigation, a substantial RE current of $13{-}16$~MA forms.
For the longest CQs, RE currents up to $15{-}17$~MA are estimated after Ar and Ne injection, due to tritium decay, Compton scattering, and hot-tail mechanisms. The 0D simulations, carried out for deuterium injection up to $\sim$15~kPa m$^3$, 
suggest that the formation of the RE current due to the tritium and hot-tail
RE sources could be very efficiently controlled by injection of a sufficient amount of deuterium ($\ge$10~kPa m$^3$). In contrast, the control of the RE current due to the Compton scattering seed might require a substantially larger amount of deuterium ($\ge$20~kPa m$^3$ corresponding to the so-called ``critical density'').

A disruption mitigation system has not yet been developed for the EU-DEMO. 
A preliminary study~\cite{Pautasso_2019} pointed out that RE suppression requires 
a volume-averaged density increase of the order of $4 \times 10^4$ (critical density), which should take place in a time interval of the order of 10~ms. In addition, the density increase must occur in the plasma center where the current density peaks and the toroidal electric field is maximum. It is unlikely that this is
physically and technically doable.
A RE beam would not be a danger if it were position-controlled to avoid interaction with plasma facing components (PFCs) and
if its current were ramped down by the Ohmic system or by massive impurity injection. 
It is the interaction of the RE beam with structures and the slowing-down 
of the RE in the material that cause damage.

The temporal and spatial evolution of the RE interaction with the PFC surface must be known in detail to  evaluate the damage.
The EU-DEMO first wall protection relies on limiters
placed on the wall at specific toroidal and poloidal positions, where a
direct contact with the boundary of a disrupting plasma is
expected. The limiters
should protrude  $3{-}7$~cm from the first wall to protect it from transient 
heat and particle fluxes. 
The size, number and thickness of the limiters will have to be adjusted on the basis of numerical simulations to prevent first wall damage. On the other hand, the total area covered by the limiters should be kept as small as possible to maximize the tritium breeding capacity. 
Individual limiters shall be located at  maintenance ports to allow replacement~\cite{You_2022}. The engineering design and the integration of the upper limiter has been carried out with some detail~\cite{Richiusa_2025}. 
The rationale behind the design choices of the limiter sub-system is supported by an
extensive assessment workflow, based on neutronics and thermal-
hydraulics evaluations, electromagnetic studies, and structural assessment, but does not yet take into account RE impacts.


Several numerical codes and relative physics models are required to estimate the RE thermal load on the PFCs and the related damage. The different codes form a workflow, which delivers information needed to design the PFCs. The present workflow consists of (1)~preliminary geometry (and material, later on) of the PFCs;
(2)~codes able to evolve the magnetic equilibrium; (3)~a field-line tracing code, if not yet part of the codes above; (4)~a RE kinetic model to evolve the RE energy distribution; (5)~a RE-PFC interaction model; and (6)~a code to simulate the PFC material's temperature evolution and eventual damage.

The codes CarMa0NL~\cite{Ramogida_2015} and MAXFEA~\cite{Lombroni_2021} typically provide the 2D evolution of the plasma equilibrium, 
which identifies the plasma-PFC contact areas and
allows to position the limiters and to modify the magnetic equilibrium when needed. 
These codes have simple halo current recipes to be replaced in the future by more physics-based models. 
Nevertheless, they do not account for RE generation. 
The 3D field-line tracing codes PFCflux~\cite{Firdaouss_2013} and SMARDDA~\cite{Wayne_2015} calculate the thermal loads due to charged 
particles. They use as inputs magnetic equilibria, assumptions 
on power density crossing the separatrix, and the e-folding length for different 
power channels of the near and far scrape-off layer. 

Since the whole equilibrium evolution determines the evolution of the plasma-PFC 
contact points -- i.e., the RE deposition map --
time-dependent equilibrium codes must take into account RE generation.
The nonlinear MHD JOREK code can evolve the equilibrium with a halo region, while also modeling RE generation and losses.
A recent simulation of an EU-DEMO plasma CQ is presented in Ref.~\cite{Vannini_2025}: 
Here, a 17~MA RE beam forms after the TQ and undergoes vertical instability. 
During the upward vertical movement, the beam suffers
an MHD instability, and most of the REs are lost along the stochastic field lines, which differs substantially from 2D scrape-off (see Section~7 for more details on 3D impact modelling).
Particle tracing has been used to evolve the initialized
RE markers during the RE beam termination phase and to deposit their energy onto the upper limiter and first wall.

A model for the RE-structure interaction is then required to simulate the RE slowing down and 
evaluate their deposition in the structure. For this purpose, simpler (stopping-power) 
or complex and high-fidelity models (e.g. FLUKA) exist.   
FLUKA was used in Ref.~\cite{Maddaluno_2019} for studies of RE damage in EU-DEMO. 
Recent modeling of RE-PFC interaction with FLUKA is summarized in Ref.~\cite{Singh_2021}:
A sample geometry of the tungsten EU-DEMO first wall is irradiated
with 20~MeV REs with an incidence angle of 1 and
10~degrees on a flat surface; the RE energy is found to be deposited within a few mm of the surface and is around
6.4~MeV and 10.5~MeV for 1$^\circ$ and 10$^\circ$, respectively. 
Since true RE beams are not mono-energetic (see Section~5) and their energy distribution plays a key role in determining the 
heat deposition profile in the structure (see Section~8), the code 
DREAM~\cite{Hoppe_2021b} has been used to evaluate the RE distribution function. Preliminary results~\cite{Pokol_2024} are now available for new FLUKA calculations.
The use of the MEMENTO melt dynamics code~\cite{Ratynskaia_2025} 
can then be used to study eventual RE-induced damage of the EU-DEMO PFCs. Further details on models for the RE-structure interaction are described in Section~8.

\subsubsection*{In North America.}\label{sec:future-status-america}

In 2021, a toroidal field ``model'' coil, wound with high temperature superconducting (HTS) tape, was successfully built and tested in a collaboration between the Massachusetts Institute of Technology (MIT) and Commonwealth Fusion Systems (CFS) \cite{Hartwig_2024}. Today, the first HTS coils are being completed for the SPARC tokamak.
This HTS technology enables SPARC's high magnetic field strength, $\Bo = 12.2~\mathrm{T}$; high plasma current, $\Ip = 8.7~\mathrm{MA}$; and compact size with major and minor radii, $\Ro = 1.85~\mathrm{m}$ and $a = 0.57~\mathrm{m}$, respectively~\cite{Creely_2020}. The planned highest-performing plasma, the ``Primary Reference Discharge'' (PRD), expects electron temperatures ${>}20~\mathrm{keV}$ and DT fusion power ${>}100~\mathrm{MW}$ \cite{Rodriguez-Fernandez_2022}. For disruptions, the ITPA database scaling~\cite{Eidietis_2015} predicts a fastest CQ time of ${\sim}3.2~\mathrm{ms}$; more on SPARC disruptions, halo currents, thermal and electromagnetic loads, and mitigation systems can be found in Ref.~\cite{Sweeney_2020}.

Runaway electron (RE) seeds from the hot electron tail, tritium beta decay, and Compton scattering are expected in SPARC, as well as strong avalanching from the relatively high current. The first DREAM~\cite{Hoppe_2021b} simulations of RE generation~\cite{Tinguely_2021} in a SPARC PRD disruption, mitigated with 
Ne MGI, indicated a final electron temperature ${<}10~\mathrm{eV}$, RE plateau current ${\sim}5.5~\mathrm{MA}$, and average RE energy ${\sim}8~\mathrm{MeV}$. 
However, a sufficiently high deuterium MGI density and moderate Ne MGI density could bring the RE current below $1~\mathrm{MA}$ \cite{Ekmark_2025}, and the novel Runaway Electron Mitigation Coil (REMC) may prevent the beam altogether, depending on the reformation of magnetic flux surfaces during the CQ and RE confinement therein~\cite{Tinguely_2023}. However, it is important to note that both MGI and REMC are not machine protection systems for SPARC, which has been designed to be robust to plasma disruptions and REs; instead, they are part of the broader experimental mission to validate and optimize such systems for future tokamak power plants, with upgrades possible.

Yet REs may still cause some damage to the tungsten-based PFCs in SPARC. 
Since the threshold energy density for W damage is ${\sim}3{-}5~\mathrm{MJ/m}^2$ \cite{Lehnen_2009}, RE beams with currents ${>}2~\mathrm{MA}$ would likely cause melting to PFCs for wetted areas ${<}1~\mathrm{m}^2$ \cite{Feyrer_2024}. We note that because SPARC discharges will be ${<}10~\mathrm{s}$ in duration and operate at low duty cycle, its PFCs are inertially cooled, i.e. having no cooling channels near the PFCs. This precludes any RE-induced loss-of-coolant accidents. The Heat flux Engineering Analysis Toolkit (HEAT) toolkit~\cite{Looby_2022} has recently added REs to study their losses and impact on CAD PFCs. 
From realistic vertical displacement event (VDE) simulations from M3D-C1, REs were found to ``scrape off'' on PFCs during the vertical motion, and a RE-impact footprint with poloidal extent ${\sim}0.5~\mathrm{m}$ was estimated in HEAT, resulting in a total wetted area ${>}1~\mathrm{m}^2$, assuming toroidal symmetry \cite{Feyrer_2024}.
Further HEAT development is ongoing, including RE diffusion and drifts, as well as the coupling of open-source heat transfer software. Planned diagnostics for SPARC early campaigns~\cite{Reinke_2024} -- such as magnetics, visible and infrared imaging and spectroscopy systems, neutron and hard x-ray detectors -- will provide insight into RE beam currents, energies, and impacts.

\begin{figure}[t]
        \centering
        \includegraphics[width=\linewidth]{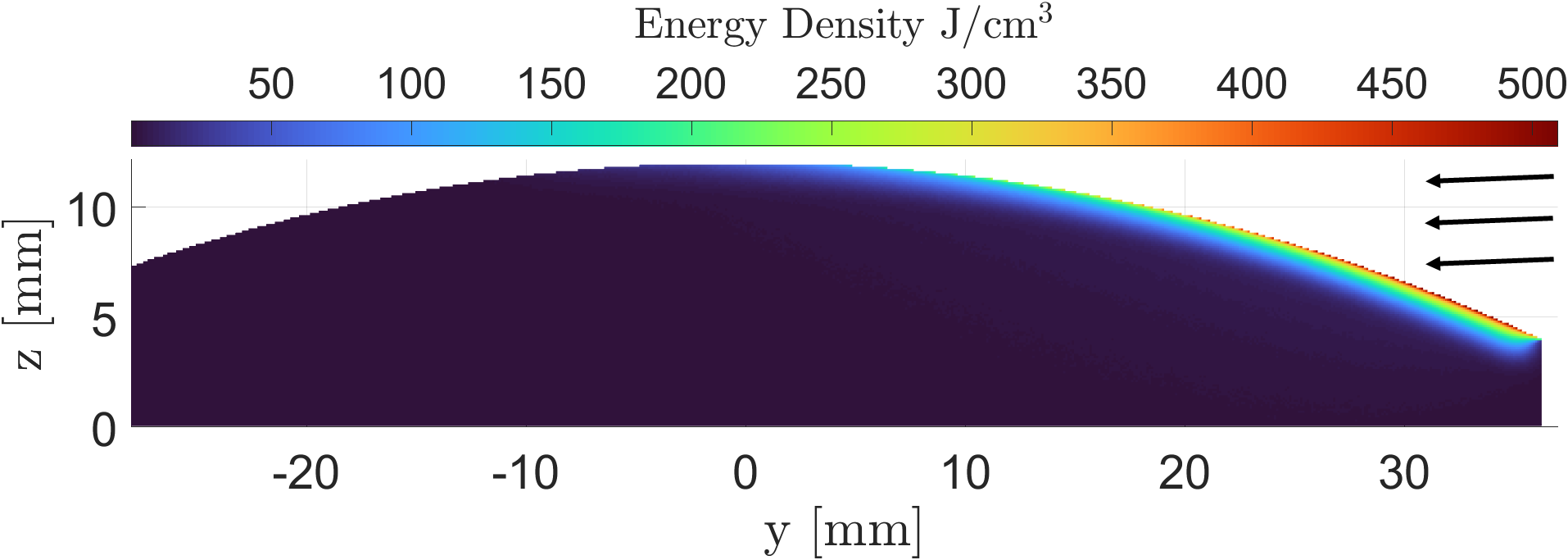}
        \caption{Monte Carlo simulations of RE deposited energy density (normalized to 1~kJ total energy) within a cross-sectional cut of a realistic SPARC plasma facing component~\cite{KTH_2025}. The incoming REs (10~MeV, zero pitch angle) are indicated by black arrows. Magnetic field inclination angle increases from 1~degree at the tile surface's topmost point to 25~degrees at the edge, with $\vert B \vert = 13.6$~T. Courtesy of S.~Ratynskaia.
        }
        \label{fig:sparc-memento}
    \end{figure}

In addition to the total RE beam energy, individual electron energies and impact angles are important parameters defining resulting damage. An initial scoping study of energy deposition and the resulting thermal response was carried out by the GEANT4-MEMENTO workflow~\cite{Ratynskaia_2025a}. A scan in individual RE energy and pitch angle, total RE beam energy, and impact duration was performed for a realistic SPARC VDE (same as above) and a realistic tile geometry, with a cross-section shown in Fig.~\ref{fig:sparc-memento}. The curved tile geometry implies (i)~steep impact angles and (ii)~that even with a uniform RE source, the loading is nonuniform, with REs depositing more energy in the edge of the tile as clearly visible from Fig.~\ref{fig:sparc-memento}. For the case shown (10~MeV, zero pitch angle REs), 100~kJ of total energy deposited over $1{-}10$~ms leads to ${\sim}1$~mm-deep melting at the tile's edge, but negligible vaporization.

\subsection*{Current and future challenges}\label{sec:future-challenges}

\subsubsection*{In Asia.}\label{sec:future-challenges-asia}

In a power plant like CFETR, disruptions are expected to occur infrequently. This is based on the assumption that well-qualified plasma scenarios and operational procedures will be established and adhered to following experimental campaigns at ITER and other fusion facilities. Yet REs generated during plasma disruptions can cause substantial damage to the wall, not to mention the resulting electromagnetic forces. Thus, proper disruption detection, prevention, and mitigation methods are still needed. 

Detection involves real-time monitoring of various plasma parameters to identify early signs of instability. Advanced diagnostics and sensors are expected to be used to detect changes in current, density, temperature, and magnetic fields. 
When a disruption is unavoidable, mitigation techniques are employed to minimize its impact. Shattered pellet injection (SPI), like in ITER, is being considered for CFETR. 
Other disruption mitigation systems like massive gas injection or electromagnetic particle injector are  promising.
Disruption detection, prevention and mitigation methods in CFETR have not been decided yet. It is believed that tokamaks like ITER will provide more insight about the disruption physics and test the reliability of these methods. Additionally, the current modeling is restricted to the thermal PFC response. The lack of understanding regarding tungsten damage by REs presents challenges in evaluating component damaged and impact on plasma operation. More efforts should be directed toward designing and conducting dedicated experiments to gain deeper insights into this phenomenon.

\subsubsection*{In Europe.}\label{sec:future-challenges-europe}

Besides EU-DEMO, other fully superconducting tokamaks are being built or are planned to be built in Europe. These are the Divertor Tokamak Test (DTT, $\Ip \leq 5.5$~MA, major radius $\Ro=2.19$~m, and minor radius $a =0.70$~m) in Italy and the Spherical Tokamak for Energy Production (STEP, $\Ip =20$~MA, $\Ro=3.6$~m and $a=2$~m) in the United Kingdom. 
The CQ phase of DTT and the likelihood for RE generation and multiplication in early plasma current (up to 2~MA) plasmas were studied using the non-linear MHD code JOREK~\cite{Emanuelli_2025}. The results indicate that in this initial low-current scenario, the RE generation is negligible as long as the impurities injected by the disruption mitigation system are limited.
Unmitigated RE generation in STEP was modeled using the code DREAM~\cite{Fil_2024}, with hot-tail generation found to be the dominant primary generation mechanism and avalanche multiplication of REs found to be extremely high. To mitigate potential RE beams, the STEP concept design already includes multiple SPI systems (with MGI as another option) dedicated to RE  mitigation. This includes redundancy and assumes repetitive injections during the RE plateau to keep conditions prone to benign termination of the RE beam.

\subsubsection*{In North America.}\label{sec:future-challenges-america}

In 2023, the US Department of Energy announced award recipients for its milestone-based public-private partnership program~\cite{DOE_Milestone_2023} to accelerate efforts toward a Fusion Pilot Plant (FPP) as part of a ``Bold Decadal Vision for Fusion Energy'' \cite{DOE_Vision_2024}. As one of the awardees, CFS has further developed its electricity-producing ARC tokamak design, which is now planned to be sited in Virginia, USA~\cite{CFS-ARC}. 
From Ref.~\cite{Hillesheim_2024}, ARC will have tokamak parameters
$R_0\sim4~\mathrm{m}$, $a\sim1~\mathrm{m}$, $B_0\sim11.5~\mathrm{T}$, and $I_p\sim10~\mathrm{MA}$,
producing ${\sim}0.5{-}1~\mathrm{GW}$ of DT fusion power. Due to its increased size, ARC CQ durations are expected to be ${>}3$ times longer than SPARC; while stored thermal and magnetic energies are also larger, the resulting disruption fluxes and stresses are -- in some ways -- balanced by the increased surface area. More details on ARC can be found in upcoming publications of the ARC Physics Basis~\cite{ARC-Physics-Basis}.

REs will have similar generation mechanisms in ARC compared to SPARC, although one new challenge could be the increased level of activation, due to higher fusion power and neutron fluence, and its impact on Compton scattering. Simulations are underway to assess this. In addition, the several MA increase in plasma current from SPARC to ARC could lead to higher avalanching, even by a factor ${>}100$. PFCs will also be actively cooled in ARC; while the final design is not yet set, current design iterations are considering wall thicknesses of order ${\sim}15~\mathrm{mm}$. It needs to be assessed whether REs could cause melt (or vaporize) a hole into a cooling channel. Furthermore, RE energy deposition into the coolant or into the blanket material, planned to be the liquid molten salt FLiBe, behind PFCs must be assessed. Secondary particles, such as high energy gamma-rays from bremsstrahlung, may penetrate the blanket and shielding to deposit energy in the HTS coils. A Low Temperature Superconducting (LTS) magnet was quenched before in the Tore-Supra/WEST tokamak after a beam impact~\cite{Torre_2019}; a worst case scenario for SPARC, analyzed with GEANT4, estimated a manageable temperature rise in HTS with a wider operational space than LTS. However, this still must be evaluated for ARC.

Yet the mission of SPARC is to map learnings to ARC, influencing aspects of the design and operation, e.g. PFC geometry or sacrificial limiters. Model validation of SPARC REs will provide more accurate predictions for ARC REs: generation, evolution dynamics, mitigation strategies, PFC impacts and damage. Even sooner, upcoming experiments at an ${\sim}$MeV electron accelerator will provide empirical data for benchmarking with codes like MEMENTO and HEAT. 
While MGI is a baseline disruption mitigation system for ARC~\cite{ARC-Physics-Basis}, other options could be explored as SPARC upgrades to then install on ARC, e.g. SPI. 
Furthermore, the REMC physics experiment on SPARC will inform the need for an ARC REMC. However, we note that an important goal of any tokamak-based FPP is a stable, reproducible plasma discharge with ideally fewer disruptions and REs inherently. ARC's planned ability to replace the entire vacuum vessel also allows some flexibility in design iteration as well as its overall robustness to disruption loads and RE impacts~\cite{Hillesheim_2024}. 

\subsection*{Concluding remarks}\label{sec:future-conclusion}

Several complex and reliable physical models are required to evaluate the potential damage of REs interacting with PFCs. In spite of the level of detail reached by most of the numerical codes involved, large uncertainties affect the evaluation of the damage because the magnitude of the RE flux is unknown. Limiters and first wall tiles melted by RE beams exist,  but much work remains to understand the phenomena  responsible for the losses.
Therefore, it is presently impossible to describe the plasma conditions leading to RE damages and to specify the largest RE flux expected in a given device. Well diagnosed experiments and dedicated simulations are mandatory in the near future to gather the needed missing information.


\newpage 
\clearpage
\section*{Acknowledgments}
Some of the work has been performed within the framework of the EUROfusion Consortium, funded by the European Union via the Euratom Research and Training Programme (Grant Agreement No\,101052200 - EUROfusion). The views and opinions expressed are however those of the authors only and do not necessarily reflect those of the European Union, the European Commission or the ITER Organization. The European Union, European Commission or ITER Organization cannot be held responsible for them. 

R.A.T. acknowledges support by Commonwealth Fusion Systems. O.F. acknowledges support for COMPASS related work by CR-MEYS projects LM2023045 and 9D22001. J.R.M-S. acknowledges support of work related to the Universidad Carlos III de Madrid in section 6 by Ministerio de Ciencia e Innovacion (Spain), Project PID2022-137869OB-I00. For work contributed by E.M.H., this material is based upon work supported by the U.S. Department of Energy, Office of Science, Office of Fusion Energy Sciences, using the DIII-D National Fusion Facility, a DOE Office of Science user facility, under Awards DE-FC02-04ER54698 and DE-FG02-07ER54917. B.B and D.d.C.N acknowledge support from the U.S. Department of Energy under Contract Nos. DE-FG02-04ER-54742 and DESC001683.

S.R, M.H and E. N. thank Dr.\ F.\ Leuterer (Max Planck Institute for Plasma Physics, Garching, Germany) for bringing to their attention the evidence of RE-induced damage to antennas in 1983 ASDEX experiments.

Dr. \ M. \ Diez and Dr. \ Y.\ Corre (CEA, Saint-Paul-lez-Durance, France) are gratefully acknowledged for providing damage evidence from WEST presented in Secs. \ 2 and 3. 

\textbf{Disclaimer:} This report was prepared as an account of work partially sponsored by an agency of the United States Government. Neither the United States Government nor any agency thereof, nor any of their employees, makes any warranty, express or implied, or assumes any legal liability or responsibility for the accuracy, completeness, or usefulness of any information, apparatus, product, or process disclosed, or represents that its use would not infringe privately owned rights. Reference herein to any specific commercial product, process, or service by trade name, trademark, manufacturer, or otherwise does not necessarily constitute or imply its endorsement, recommendation, or favoring by the United States Government or any agency thereof. The views and opinions of authors expressed herein do not necessarily state or reflect those of the United States Government or any agency thereof.



\clearpage
\section*{Appendix: List of acronyms}
    \small
    \begin{tabular}{|l|l|}
    \hline
    Acronym & Explanation \\
    \hline
    0D-6D  & Zero- to Six-dimensional  \\
    ALE & Arbitrary Lagrangian Eulerian \\
    ATJ & Private company, acronym unknown \\
    AUG & ASDEX Upgrade \\
    CFC & Carbon Fibre Composites \\
    CFS & Commonwealth Fusion Systems \\
    CFETR & China Fusion Engineering Test Reactor \\
    CFS & Commonwealth Fusion Systems \\
    CSDA & Continuous-slowing down approximation \\
    DMS & Disruption mitigation system \\
    DMV & Disruption mitigation valve \\
    EDS & Electron dispersive x-ray spectroscopy \\
    EDX & Energy-dispersive X-ray spectroscopy \\
    ELM & Edge localized mode \\
    EM & Electromagnetic \\
    FEM & Finite element method \\
    FPP & Fusion power plant \\
    FWP & First wall panel \\
    GC & Gyro-center \\
    HFS & High field side \\
    HXR & Hard x-ray \\
    HTS & High temperature superconductor \\
    ILW & so-called ITER like wall (JET) \\
    IR & Infrared \\
    IVIS & In-vessel visual inspection \\
    \hline
    \end{tabular}
    \begin{tabular}{|l|l|}
    \hline
    Acronym & Explanation \\
    \hline
    IVVS & In-Vessel Viewing System  (in ITER)\\
    IWGL & Inner wall guard limiter \\
    JET & Joint European Torus \\
    LFS & Low field side \\
    LTS & Low temperature superconductor \\
    MC & Monte Carlo \\
    MGI & Massive gas injection \\
    MHD & Magneto-hydrodynamics \\
    MIT & Massachusetts Institute of Technology \\
    MMI & Massive material injection \\
    PD & Peridynamic theory \\
    PFC & Plasma facing component \\
    PFM & Phase field method \\
    PRD & Primary reference discharge (SPARC) \\
    SPH & Smoothed particle hydrodynamics \\
    RE & Runaway electron \\
    RMP & Resonant magnetic perturbation \\
    SPI & Shattered pellet injection \\
    SRO & Start of research operation (in ITER) \\
    SXR & Soft x-ray \\
    TFW & Temporary first wall (in ITER) \\
    TM & Tearing mode \\
    UDP & Upper dump plate \\
    VDE & Vertical displacement event \\
    XPS & X-ray photoelectron spectroscopy \\
    \hline
    \end{tabular}

\clearpage
\bibliographystyle{iop-art}
{\small
\bibliography{bibliography}

\providecommand{\noopsort}[1]{}\providecommand{\singleletter}[1]{#1}%
\providecommand{\newblock}{}
\begin{thebibliography}{100}
\expandafter\ifx\csname url\endcsname\relax
  \def\url#1{{\tt #1}}\fi
\expandafter\ifx\csname urlprefix\endcsname\relax\def\urlprefix{URL }\fi
\providecommand{\eprint}[2][]{\url{#2}}

\bibitem{Federici_2001}
G.~Federici, C.~Skinner, J.~Brooks, J.~Coad et~al. 2001 {\em Nuclear Fusion\/}
  {\bf 41} 1967 \urlprefix\url{https://dx.doi.org/10.1088/0029-5515/41/12/218}

\bibitem{Bandaru_2024}
V.~Bandaru, M.~Hoelzl, H.~Bergström, F.~Artola et~al. 2024 {\em Nuclear
  Fusion\/} {\bf 64} 076053
  \urlprefix\url{https://dx.doi.org/10.1088/1741-4326/ad50ea}

\bibitem{Bergstrom_2024}
H.~Bergström, K.~Särkimäki, V.~Bandaru, M.~M. Skyllas et~al. 2024 {\em
  Plasma Physics and Controlled Fusion\/} {\bf 66}(9) 095001
  \urlprefix\url{https://doi.org/10.1088/1361-6587/ad5fb5}

\bibitem{Ratynskaia_2025}
S.~Ratynskaia, P.~Tolias, T.~Rizzi, K.~Paschalidis et~al. 2025 {\em Nuclear
  Fusion\/} {\bf 65} 024002
  \urlprefix\url{https://dx.doi.org/10.1088/1741-4326/adab05}

\bibitem{ICF_2024}
{The Indirect Drive ICF Collaboration} 2024 {\em Phys. Rev. Lett.\/} {\bf
  132}(6) 065102
  \urlprefix\url{https://link.aps.org/doi/10.1103/PhysRevLett.132.065102}

\bibitem{Maslov_2023}
M.~Maslov, E.~Lerche, F.~Auriemma, E.~Belli et~al. 2023 {\em Nuclear Fusion\/}
  {\bf 63} 112002 \urlprefix\url{https://dx.doi.org/10.1088/1741-4326/ace2d8}

\bibitem{Dinklage_2025}
A.~Dinklage, R.~Buttery, K.~Crombé, A.~B. del Cerro~Gordo et~al. 2025 {\em
  Plasma Physics Control. Fusion\/} {\bf 67} 063701
  \urlprefix\url{https://iopscience.iop.org/article/10.1088/1361-6587/add621}

\bibitem{Pitts_2017}
R.~Pitts, S.~Bardin, B.~Bazylev, M.~{van den Berg} et~al. 2017 {\em Nuclear
  Materials and Energy\/} {\bf 12} 60--74
  \urlprefix\url{https://www.sciencedirect.com/science/article/pii/S2352179116302885}

\bibitem{Pitts_2019}
R.~Pitts, X.~Bonnin, F.~Escourbiac, H.~Frerichs et~al. 2019 {\em Nuclear
  Materials and Energy\/} {\bf 20} 100696
  \urlprefix\url{https://www.sciencedirect.com/science/article/pii/S2352179119300237}

\bibitem{Pitts_2025}
R.~Pitts, A.~Loarte, T.~Wauters, M.~Dubrov et~al. 2025 {\em Nuclear Materials
  and Energy\/} {\bf 42} 101854
  \urlprefix\url{https://www.sciencedirect.com/science/article/pii/S2352179124002771}

\bibitem{Krieger_2018}
K.~Krieger, M.~Balden, J.~Coenen, F.~Laggner et~al. 2018 {\em Nuclear Fusion\/}
  {\bf 58} 026024 \urlprefix\url{https://dx.doi.org/10.1088/1741-4326/aa9a05}

\bibitem{Ratynskaia_2022_b}
S.~Ratynskaia, L.~Vignitchouk and P.~Tolias 2022 {\em Plasma Physics and
  Controlled Fusion\/} {\bf 64} 044004
  \urlprefix\url{https://dx.doi.org/10.1088/1361-6587/ac4b94}

\bibitem{Ratynskaia_2022_c}
S.~Ratynskaia, A.~Bortolon and S.~Krasheninnikov 2022 {\em Reviews of Modern
  Plasma Physics\/} {\bf 6} 20
  \urlprefix\url{https://doi.org/10.1007/s41614-022-00081-5}

\bibitem{Beckers_2023}
J.~Beckers, J.~Berndt, D.~Block, M.~Bonitz et~al. 2023 {\em Physics of
  Plasmas\/} {\bf 30} 120601 \urlprefix\url{https://doi.org/10.1063/5.0168088}

\bibitem{Krieger_2025}
K.~Krieger, S.~Brezinsek, J.~Coenen, H.~Frerichs et~al. 2025 {\em Nuclear
  Fusion\/} {\bf 65} 043001
  \urlprefix\url{https://dx.doi.org/10.1088/1741-4326/adaf42}

\bibitem{Berger_1992}
{Berger, M. J.} 1992 {ESTAR, PSTAR and ASTAR: computer programs for calculating
  stopping-power and range tables for electrons, protons and helium ions} Tech.
  rep. {Technical Report NISTIR 4999, National Institute of Standards and
  Technology, Gaithersburg, MD}

\bibitem{Coenen_2013}
J.~Coenen, K.~Krieger, B.~Lipschultz, R.~Dux et~al. 2013 {\em Journal of
  Nuclear Materials\/} {\bf 438} S27--S33
  \urlprefix\url{https://www.sciencedirect.com/science/article/pii/S0022311513000135}

\bibitem{Coenen_2017}
J.~W. Coenen, G.~F. Matthews, K.~Krieger, D.~Iglesias et~al. 2017 {\em Physica
  Scripta\/} {\bf 2017} 014013
  \urlprefix\url{https://dx.doi.org/10.1088/1402-4896/aa8789}

\bibitem{Krieger_2017}
K.~Krieger, B.~Sieglin, M.~Balden, J.~W. Coenen et~al. 2017 {\em Physica
  Scripta\/} {\bf 2017} 014030
  \urlprefix\url{https://dx.doi.org/10.1088/1402-4896/aa8be8}

\bibitem{Jepu_2019}
I.~Jepu, G.~Matthews, A.~Widdowson, M.~Rubel et~al. 2019 {\em Nuclear Fusion\/}
  {\bf 59}(8) 086009
  \urlprefix\url{https://iopscience.iop.org/article/10.1088/1741-4326/ab2076}

\bibitem{Corre_2021}
Y.~Corre, A.~Grosjean, J.~P. Gunn, K.~Krieger et~al. 2021 {\em Physica
  Scripta\/} {\bf 96} 124057
  \urlprefix\url{https://dx.doi.org/10.1088/1402-4896/ac326a}

\bibitem{Corre_2023}
Y.~Corre, M.-H. Aumeunier, A.~Durif, J.~Gaspar et~al. 2023 {\em Nuclear
  Materials and Energy\/} {\bf 37} 101546
  \urlprefix\url{https://www.sciencedirect.com/science/article/pii/S2352179123001850}

\bibitem{Ratynskaia_2020}
S.~Ratynskaia, E.~Thorén, P.~Tolias, R.~A. Pitts et~al. 2020 {\em Nuclear
  Fusion\/} {\bf 60} 104001
  \urlprefix\url{https://dx.doi.org/10.1088/1741-4326/abadac}

\bibitem{Ratynskaia_2021}
S.~Ratynskaia, E.~Thorén, P.~Tolias, R.~A. Pitts et~al. 2021 {\em Physica
  Scripta\/} {\bf 96} 124009
  \urlprefix\url{https://dx.doi.org/10.1088/1402-4896/ac1cf4}

\bibitem{Thoren_2018}
E.~Thorén, S.~Ratynskaia, P.~Tolias, R.~Pitts et~al. 2018 {\em Nuclear
  Materials and Energy\/} {\bf 17} 194--199
  \urlprefix\url{https://www.sciencedirect.com/science/article/pii/S235217911830098X}

\bibitem{Thoren_2021}
E.~Thorén, S.~Ratynskaia, P.~Tolias, R.~A. Pitts et~al. 2021 {\em Plasma
  Physics and Controlled Fusion\/} {\bf 63} 035021
  \urlprefix\url{https://dx.doi.org/10.1088/1361-6587/abd838}

\bibitem{Ratynskaia_2022_a}
S.~Ratynskaia, K.~Paschalidis, P.~Tolias, K.~Krieger et~al. 2022 {\em Nuclear
  Materials and Energy\/} {\bf 33} 101303
  \urlprefix\url{https://www.sciencedirect.com/science/article/pii/S2352179122001843}

\bibitem{Ratynskaia_2024}
S.~Ratynskaia, K.~Paschalidis, K.~Krieger, L.~Vignitchouk et~al. 2024 {\em
  Nuclear Fusion\/} {\bf 64} 036012
  \urlprefix\url{https://dx.doi.org/10.1088/1741-4326/ad219b}

\bibitem{Paschalidis_2023}
K.~Paschalidis, S.~Ratynskaia, F.~{Lucco Castello} and P.~Tolias 2023 {\em
  Nuclear Materials and Energy\/} {\bf 37} 101545
  \urlprefix\url{https://www.sciencedirect.com/science/article/pii/S2352179123001849}

\bibitem{Paschalidis_2024}
K.~Paschalidis, F.~{Lucco Castello}, S.~Ratynskaia, P.~Tolias et~al. 2024 {\em
  Fusion Engineering and Design\/} {\bf 206} 114603
  \urlprefix\url{https://www.sciencedirect.com/science/article/pii/S092037962400454X}

\bibitem{Coburn_2020}
J.~Coburn, E.~Thoren, R.~A. Pitts, H.~Anand et~al. 2020 {\em Physica Scripta\/}
  {\bf 2020} 014076 \urlprefix\url{https://dx.doi.org/10.1088/1402-4896/ab4c6b}

\bibitem{Coburn_2021}
J.~Coburn, M.~Lehnen, R.~Pitts, E.~Thorén et~al. 2021 {\em Nuclear Materials
  and Energy\/} {\bf 28} 101016
  \urlprefix\url{https://www.sciencedirect.com/science/article/pii/S2352179121000922}

\bibitem{Coburn_2022}
J.~Coburn, M.~Lehnen, R.~Pitts, G.~Simic et~al. 2021 {\em Nuclear Fusion\/}
  {\bf 62} 016001 \urlprefix\url{https://dx.doi.org/10.1088/1741-4326/ac38c7}

\bibitem{Paschalidis_2024b}
K.~Paschalidis, S.~Ratynskaia, P.~Tolias and R.~Pitts 2024 {\em Nuclear
  Fusion\/} {\bf 64} 126022
  \urlprefix\url{https://dx.doi.org/10.1088/1741-4326/ad7f6b}

\bibitem{Lehnen_2015}
M.~Lehnen, K.~Aleynikova, P.~Aleynikov, D.~Campbell et~al. 2015 {\em Journal of
  Nuclear Materials\/} {\bf 463} 39--48
  \urlprefix\url{https://www.sciencedirect.com/science/article/pii/S0022311514007594}

\bibitem{Boozer_2015}
A.~H. Boozer 2015 {\em Physics of Plasmas\/} {\bf 22}(3) None
  \urlprefix\url{https://doi.org/10.1063/1.4913582}

\bibitem{Boozer_2017}
A.~H. Boozer 2017 {\em Nuclear Fusion\/} {\bf 57} 056018
  \urlprefix\url{https://dx.doi.org/10.1088/1741-4326/aa6355}

\bibitem{Breizman_2019}
B.~N. Breizman, P.~Aleynikov, E.~M. Hollmann and M.~Lehnen 2019 {\em Nuclear
  Fusion\/} {\bf 59} 083001
  \urlprefix\url{https://dx.doi.org/10.1088/1741-4326/ab1822}

\bibitem{Reux_2021b}
C.~Reux, C.~Paz-Soldan, P.~Aleynikov, V.~Bandaru et~al. 2021 {\em Physical
  Review Letters\/} {\bf 126} 175001
  \urlprefix\url{https://doi.org/10.1103/PhysRevLett.126.175001}

\bibitem{Paz-Soldan_2021}
C.~Paz-Soldan, C.~Reux, K.~Aleynikova, P.~Aleynikov et~al. 2021 {\em Nuclear
  Fusion\/} {\bf 61} 116058
  \urlprefix\url{https://dx.doi.org/10.1088/1741-4326/ac2a69}

\bibitem{Loarter_2024}
A.~Loarte, R.~Pitts, T.~Wauters, I.~Nunes et~al. 2024 {Initial evaluations in
  support of the new ITER baseline and Research Plan}

\bibitem{Hollmann_2025}
E.~M. Hollmann, C.~Marini, D.~L. Rudakov, E.~Martinez-Loran et~al. 2025 {\em
  Plasma Physics and Controlled Fusion\/} {\bf 67} 035020
  \urlprefix\url{https://dx.doi.org/10.1088/1361-6587/adb5b6}

\bibitem{Hoppe_2021b}
M.~Hoppe, O.~Embreus and T.~Fülöp 2021 {\em Computer Physics
  Communications\/} {\bf 268} 108098
  \urlprefix\url{https://www.sciencedirect.com/science/article/pii/S0010465521002101}

\bibitem{Martin-Solis_2017}
J.~R. Martin-Solis, A.~Loarte and M.~Lehnen 2017 {\em Nuclear Fusion\/} {\bf
  57} 066025 \urlprefix\url{https://doi.org/10.1088/1741-4326/aa6939}

\bibitem{Liu_(Chang)_2018}
C.~Liu, E.~Hirvijoki, G.-Y. Fu, D.~P. Brennan et~al. 2018 {\em Phys. Rev.
  Lett.\/} {\bf 120} 265001
  \urlprefix\url{https://link.aps.org/doi/10.1103/PhysRevLett.120.265001}

\bibitem{Bandaru_2019}
V.~Bandaru, M.~Hoelzl, F.~J. Artola, G.~Papp et~al. 2019 {\em Phys. Rev. E\/}
  {\bf 99}(6) 063317
  \urlprefix\url{https://link.aps.org/doi/10.1103/PhysRevE.99.063317}

\bibitem{Breizman_2017}
B.~Breizman and P.~Aleynikov 2017 {\em Nuclear Fusion\/} {\bf 57}(12) 125002
  \urlprefix\url{https://doi.org/10.1088/1741-4326/aa8c3f}

\bibitem{Vallhagen_2024}
O.~Vallhagen, L.~Hanebring, F.~Artola, M.~Lehnen et~al. 2024 {\em Nuclear
  Fusion\/} {\bf 64} 086033
  \urlprefix\url{https://dx.doi.org/10.1088/1741-4326/ad54d7}

\bibitem{TFR_1975}
E.~TFR 1975 {\em Control. Fusion and Plasma Physics\/} {\bf 1} 2

\bibitem{Barber_1959}
W.~C. Barber and W.~D. George 1959 {\em Phys. Rev.\/} {\bf 116}(6) 1551--1559
  \urlprefix\url{https://link.aps.org/doi/10.1103/PhysRev.116.1551}

\bibitem{Strachan_1977}
J.~Strachan, E.~Meservey, W.~Stodiek, R.~Naumann et~al. 1977 {\em Nuclear
  Fusion\/} {\bf 17} 140
  \urlprefix\url{https://dx.doi.org/10.1088/0029-5515/17/1/015}

\bibitem{Maddaluno_1987}
G.~Maddaluno and A.~Vannucci 1987 {\em Journal of Nuclear Materials\/} {\bf
  145-147} 697--699

\bibitem{ASDEX_1983}
K.~Pinkau and {et al} 1983 Annual report 1983 Tech. rep. Max-Planck-Institut
  für Plasmaphysik Garching bei München

\bibitem{Leuterer_1985}
F.~Leuterer, F.~Soldner, D.~Eckhartt, A.~Eberhagen et~al. 1985 {\em Plasma
  Physics and Controlled Fusion\/} {\bf 27} 1399
  \urlprefix\url{https://dx.doi.org/10.1088/0741-3335/27/12A/007}

\bibitem{Nygren_1997}
R.~Nygren, T.~Lutz, D.~Walsh, G.~Martin et~al. 1997 {\em Journal of Nuclear
  Materials\/} {\bf 241-243} 522--527
  \urlprefix\url{https://www.sciencedirect.com/science/article/pii/S002231159780092X}

\bibitem{Torre_2019}
A.~Torre, D.~Ciazynski, S.~Girard, B.~Lacroix et~al. 2019 {\em IEEE
  Transactions on Applied Superconductivity\/} {\bf 29} 1--5

\bibitem{Chen_2021}
L.~Chen, R.~Pitts and M.~Lehnen 2021 {\em 5th Asia-Pacific Conference on Plasma
  Physics, MF2-II4\/}

\bibitem{Keilhacker_1999}
M.~Keilhacker, M.~L. Watkins and {JET Team} 1999 {\em Journal of Nuclear
  Materials\/} {\bf 266-269} 1--13
  \urlprefix\url{https://doi.org/10.1016/S0022-3115(98)00811-3}

\bibitem{Litaudon_2017}
X.~Litaudon and {et al} 2017 {\em Nuclear Fusion 57 102001\/}
  \urlprefix\url{https://doi.org/10.1088/1741-4326/aa5e28}

\bibitem{Loarte_2007}
A.~Loarte, B.~Lipschultz, A.~Kukushkin, G.~Matthews et~al. 2007 {\em Nuclear
  Fusion\/} {\bf 47} S203
  \urlprefix\url{https://iopscience.iop.org/article/10.1088/0029-5515/47/6/S04/pdf}

\bibitem{Federici_2016}
G.~Federici, C.~Bachmann, W.~Biel, L.~Boccaccini et~al. 2016 {\em Fusion
  Engineering and Design\/} {\bf 109-111} 1464--1474
  \urlprefix\url{https://doi.org/10.1016/j.fusengdes.2015.11.050}

\bibitem{Coad_1997}
J.~P. Coad, M.~Rubel and C.~H. Wu 1997 {\em Journal of Nuclear Materials\/}
  {\bf 241-243} 408--413
  \urlprefix\url{https://doi.org/10.1016/S0022-3115(97)80073-6}

\bibitem{Matthews_2007}
G.~F. Matthews, P.~Edwards, T.~Hirai, M.~Kear et~al. 2007 {\em Physica
  Scripta\/} {\bf T128} 137--143
  \urlprefix\url{https://doi.org/10.1088/0031-8949/2007/T128/027}

\bibitem{Matthews_2011}
G.~F. Matthews, M.~Beurskens, S.~Brezinsek, M.~Grothand et~al. 2011 {\em
  Physica Scripta\/} {\bf 2011}(T145)
  \urlprefix\url{https://doi.org/10.1088/0031-8949/2011/T145/014001}

\bibitem{Matthews_2013}
G.~F. Matthews, {JET EFDA Contributors} and {ASDEX-Upgrade Team} 2013 {\em
  Journal of Nuclear Materials\/} {\bf 438} S2--S10
  \urlprefix\url{https://doi.org/10.1016/j.jnucmat.2013.01.282}

\bibitem{Gill_2002}
R.~D. Gill, B.~Alper, M.~de~Baar, T.~C. Hender et~al. 2002 {\em Nuclear
  Fusion\/} {\bf 42}(8) 1039--1044

\bibitem{Helander_2002}
P.~Helander, L.-G. Eriksson and F.~Andersson 2002 {\em Plasma Physics and
  Controlled Fusion\/} {\bf 44}(12B)
  \urlprefix\url{https://iopscience.iop.org/article/10.1088/0741-3335/44/12B/318}

\bibitem{Riccardo_2010}
V.~Riccardo, G.~Arnoux, P.~Cahyna, T.~C. Hender et~al. 2010 {\em Plasma Physics
  and Controlled Fusion\/} {\bf 52}(12) 124018
  \urlprefix\url{https://iopscience.iop.org/article/10.1088/0741-3335/52/12/124018}

\bibitem{Plyusnin_2012}
V.~V. Plyusnin, V.~Kiptily, B.~Bazylev, A.~Shevelev et~al. 2012 {\em IAEA
  FEC\/} {\bf EX/P8–05}
  \urlprefix\url{https://www-pub.iaea.org/MTCD/Meetings/PDFplus/2012/cn197/cn197_Programme.pdf}

\bibitem{Gill_2000}
R.~D. Gill, B.~Alper, A.~Edwards, M.~J. L.C.~Ingesson et~al. 2000 {\em Nuclear
  Fusion\/} {\bf 40}(2) 163
  \urlprefix\url{https://iopscience.iop.org/article/10.1088/0029-5515/40/2/302}

\bibitem{Hender_2007}
T.~Hender, J.~Wesley, J.~Bialek, A.~Bondeson et~al. 2007 {\em Nuclear Fusion\/}
  {\bf 47} S128 \urlprefix\url{https://dx.doi.org/10.1088/0029-5515/47/6/S03}

\bibitem{Lehnen_2009}
M.~Lehnen, S.~Abdullaev, G.~Arnoux, S.~Bozhenkov et~al. 2009 {\em J. Nucl.
  Mater.\/} {\bf 390-391} 740--746
  \urlprefix\url{https://doi.org/10.1016/j.jnucmat.2009.01.200}

\bibitem{Arnoux_2011}
G.~Arnoux, B.~Bazylev, M.~Lehnen, A.~Loarte et~al. 2011 {\em Journal of Nuclear
  Materials\/} {\bf 415} S817–S820
  \urlprefix\url{https://doi.org/10.1016/j.jnucmat.2010.11.042}

\bibitem{Plyusnin_2006}
V.~Plyusnin, V.~Riccardo, R.~Jaspers, B.~Alper et~al. 2006 {\em Nuclear
  Fusion\/} {\bf 46} 277
  \urlprefix\url{https://dx.doi.org/10.1088/0029-5515/46/2/011}

\bibitem{Bazylev_2011}
B.~Bazylev, G.~Arnoux, W.~Fundamenski, Y.~Igitkhanov et~al. 2011 {\em Journal
  of Nuclear Materials 415 (2011)\/} {\bf 415} S841–S844
  \urlprefix\url{https://www.sciencedirect.com/science/article/pii/S0022311510007944}

\bibitem{Reux_2015}
C.~Reux, V.~Plyusnin, B.~Alper, D.~Alves et~al. 2015 {\em Nuclear Fusion\/}
  {\bf 55} 093013
  \urlprefix\url{https://dx.doi.org/10.1088/0029-5515/55/9/093013}

\bibitem{Papp_2013}
G.~Papp, T.~Fülöp, T.~Fehér, P.~de~Vries et~al. 2013 {\em Nuclear Fusion\/}
  {\bf 53}(12) 123017
  \urlprefix\url{https://doi.org/10.1088/0029-5515/53/12/123017}

\bibitem{Lehnen_2013}
M.~Lehnen, G.~Arnoux, S.~Brezinsek, J.~Flanagan et~al. 2013 {\em Nuclear
  Fusion\/} {\bf 53}(9) 093007
  \urlprefix\url{https://iopscience.iop.org/article/10.1088/0029-5515/53/9/093007}

\bibitem{ReuxJNM_2015}
C.~Reux, V.~Plyusnin, B.~Alper, D.~Alves et~al. 2015 {\em Journal of Nuclear
  Materials\/} {\bf 463} 143--149
  \urlprefix\url{https://doi.org/10.1016/j.jnucmat.2014.10.002}

\bibitem{deVries_2012}
P.~C. de~Vries, G.~Arnoux, A.~Huber, J.~Flanagan et~al. 2012 {\em Plasma
  Physics and Controlled Fusion\/} {\bf 54} 124032
  \urlprefix\url{https://iopscience.iop.org/article/10.1088/0741-3335/54/12/124032}

\bibitem{Jepu_2024}
I.~Jepu, A.~Widdowson, G.~Matthews, J.~Coad et~al. 2024 {\em Nuclear Fusion\/}
  {\bf 64}(10) 106047
  \urlprefix\url{https://iopscience.iop.org/article/10.1088/1741-4326/ad6614}

\bibitem{Moon_2019}
S.~Moon, P.~Petersson, M.~Rubel, E.~Fortuna-Zalesna et~al. 2019 {\em Nuclear
  Materials and Energy\/} {\bf 19} 59--66
  \urlprefix\url{https://www.sciencedirect.com/science/article/pii/S2352179118300814?via%3Dihub}

\bibitem{Rubel_2018}
M.~Rubel, A.~Widdowson, J.~Grzonka, E.~Fortuna-Zalesna et~al. 2018 {\em Fusion
  Engineering and Design\/} {\bf 136}(Part A) 579--586
  \urlprefix\url{https://www.sciencedirect.com/science/article/pii/S0920379618302370?via%3Dihub}

\bibitem{Fortuna_2017}
E.~Fortuna-Zaleśna, S.~M. J.~Grzonka, M.~Rubel, P.~Petersson et~al. 2017 {\em
  Physica Scripta\/} {\bf T170}(9pp) 014038
  \urlprefix\url{https://iopscience.iop.org/article/10.1088/1402-4896/aa8ddf}

\bibitem{ReuxPRL_2021}
C.~Reux, C.~Paz-Soldan, P.~Aleynikov, V.~Bandaru et~al. 2021 {\em Physical
  Review Letters\/} {\bf 126} 175001
  \urlprefix\url{https://journals.aps.org/prl/abstract/10.1103/PhysRevLett.126.175001}

\bibitem{Brezinsek_2015}
S.~Brezinsek, A.~Widdowson, M.~Mayer, V.~Philipps et~al. 2015 {\em Nuclear
  Fusion\/} {\bf 55}(6) 063021
  \urlprefix\url{https://iopscience.iop.org/article/10.1088/0029-5515/55/6/063021}

\bibitem{Hollmann_2011}
E.~Hollmann, P.~Parks and H.~D. et~al 2011 {\em Nuclear Fusion\/} {\bf 53}
  103026

\bibitem{Hollmann_2020}
E.~M. Hollmann, I.~Bykov, N.~W. Eidietis, J.~L. Herfindal et~al. 2020 {\em
  Physics of Plasmas\/} {\bf 27}(4) None
  \urlprefix\url{https://doi.org/10.1063/5.0003299}

\bibitem{Decker_2022}
J.~Decker and et~al 2022 {\em Nuclear Fusion\/} {\bf 62} 076038

\bibitem{Sheikh_2024}
U.~Sheikh, J.~Decker, M.~Hoppe, M.~Pedrini et~al. 2024 {\em Plasma Physics and
  Controlled Fusion\/} {\bf 66}(3) 035003
  \urlprefix\url{https://doi.org/10.1088/1361-6587/ad1e31}

\bibitem{Decker_2004}
J.~Decker and Y.~Peysson 2004 Dke: a fast numerical solver for the 3d drift
  kinetic equation Tech. rep. no. EUR-CEA-FC-1736
  \urlprefix\url{https://spcsrv18.epfl.ch/luke/LaTeX/Project_DKE/Doc/NoticeDKE.pdf}

\bibitem{Sheikh_2023}
U.~A. Sheikh and et~al 2023 {\em Theory and simulation of disruptions
  workshop\/}

\bibitem{Decker_2016}
J.~Decker, E.~Hirvijoki, O.~Embreus, Y.~Peysson et~al. 2016 {\em Plasma Physics
  and Controlled Fusion\/} {\bf 58}(2) 025016
  \urlprefix\url{https://doi.org/10.1088/0741-3335/58/2/025016}

\bibitem{Sheikh_2020}
U.~A. Sheikh and et~al 2020 {\em IAEA technical meeting on plasma disruptions
  and their mitigation\/}

\bibitem{Hron_2022}
M.~Hron, J.~Adámek, J.~Cavalier, R.~Dejarnac et~al. 2022 {\em Nuclear
  Fusion\/} {\bf 62} 042021
  \urlprefix\url{https://dx.doi.org/10.1088/1741-4326/ac301f}

\bibitem{Caloud_2024}
J.~Caloud, E.~Tomesova, O.~Ficker, J.~Cerovsky et~al. 2024 {\em Review of
  Scientific Instruments\/} {\bf 95} 113512
  \urlprefix\url{https://doi.org/10.1063/5.0222211}

\bibitem{Mlynar_2019}
J.~Mlynar, O.~Ficker, E.~Macusova, T.~Markovic et~al. 2018 {\em Plasma Physics
  and Controlled Fusion\/} {\bf 61} 014010
  \urlprefix\url{https://dx.doi.org/10.1088/1361-6587/aae04a}

\bibitem{Horacek_2014}
J.~Horacek, P.~Vondracek, R.~Panek, R.~Dejarnac et~al. 2015 {\em Journal of
  Nuclear Materials\/} {\bf 463} 385--388
  \urlprefix\url{https://www.sciencedirect.com/science/article/pii/S0022311514009398}

\bibitem{Kovarik_2017}
K.~Kovarik, I.~Duran, J.~Stockel, J.~Seidl et~al. 2017 {\em Review of
  Scientific Instruments\/} {\bf 88} 035106
  \urlprefix\url{https://doi.org/10.1063/1.4977591}

\bibitem{Bucalossi_2022}
J.~Bucalossi, J.~Achard, O.~Agullo, T.~Alarcon et~al. 2022 {\em Nuclear
  Fusion\/} {\bf 62} 042007
  \urlprefix\url{https://dx.doi.org/10.1088/1741-4326/ac2525}

\bibitem{Reux_2024}
C.~Reux, M.~Diez, J.~Gerardin and Y.~Corre 2024 {\em EPS 2024 Satellite meeting
  on runaway electron impacts\/}

\bibitem{Neubauer_2005}
O.~Neunauer, G.~Czymek, B.~Giesen, P.~W. Hüttemann et~al. 2005 {\em Fusion
  Sci. Technol.\/} {\bf 47} 76--86

\bibitem{Kohlhass_1990}
K.~Kohlhass and et~al 1990 {\em Fusion Engineering and Design\/} {\bf 13} 261

\bibitem{Hoven_1989}
H.~Hoven and et~al 1989 {\em Journal of Nuclear Materials\/} {\bf 162} 970

\bibitem{Kudyakov_2008}
T.~Kudyakov and et~al 2008 {\em Review of Scientific Instruments\/} {\bf 79}
  10F126

\bibitem{Forster_2011}
M.~Forster and et~al 2011 {\em Nuclear Fusion\/} {\bf 51} 043003

\bibitem{Forster_2012}
M.~Forster and et~al 2012 {\em Physics of Plasmas\/} {\bf 19} 052506

\bibitem{Tinguely_2018}
R.~Tinguely, R.~Granetz, M.~Hoppe and O.~Embréus 2018 {\em Nuclear Fusion\/}
  {\bf 58} 076019 \urlprefix\url{https://dx.doi.org/10.1088/1741-4326/aac444}

\bibitem{Granetz_2014}
R.~S. Granetz, B.~Esposito, J.~H. Kim, R.~Koslowski et~al. 2014 {\em Physics of
  Plasmas\/} {\bf 21} 072506 \urlprefix\url{https://doi.org/10.1063/1.4886802}

\bibitem{Pizzuto_2004}
A.~Pizzuto, C.~Annino, M.~Baldarelli, L.~Bettinali et~al. 2004 {\em Fusion
  Science and Technology\/} {\bf 45} 422--436
  \urlprefix\url{https://doi.org/10.13182/FST04-A523}

\bibitem{Esposito_2017}
B.~Esposito, L.~Boncagni, P.~Buratti, D.~Carnevale et~al. 2016 {\em Plasma
  Physics and Controlled Fusion\/} {\bf 59} 014044
  \urlprefix\url{https://dx.doi.org/10.1088/0741-3335/59/1/014044}

\bibitem{DeAngeli_2023}
M.~De~Angeli, P.~Tolias, S.~Ratynskaia, D.~Ripamonti et~al. 2022 {\em Nuclear
  Fusion\/} {\bf 63} 014001
  \urlprefix\url{https://dx.doi.org/10.1088/1741-4326/ac8a04}

\bibitem{Maddaluno_1999}
G.~Maddaluno and B.~Esposito 1999 {\em Journal of Nuclear Materials\/} {\bf
  266-269} 593--597
  \urlprefix\url{https://www.sciencedirect.com/science/article/pii/S0022311598005911}

\bibitem{Vertkov_2017}
A.~Vertkov, I.~Lyublinski, M.~Zharkov, G.~Mazzitelli et~al. 2017 {\em Fusion
  Engineering and Design\/} {\bf 117} 130--134
  \urlprefix\url{https://www.sciencedirect.com/science/article/pii/S0920379617300522}

\bibitem{Reux_2021}
C.~Reux, E.~PETIT, A.~TORRE, S.~NICOLLET et~al. 2021 {\em Virtual Event 28th
  IAEA Fusion Energy Conf.\/} pp 286--1055

\bibitem{Diez_2021}
M.~Diez, Y.~Corre, E.~Delmas, N.~Fedorczak et~al. 2021 {\em Nuclear Fusion\/}
  {\bf 61} 106011 \urlprefix\url{https://dx.doi.org/10.1088/1741-4326/ac1dc6}

\bibitem{Nicollet_2022}
S.~Nicollet, A.~Torre, B.~Lacroix, A.~Louzguiti et~al. 2022 {\em Cryogenics\/}
  {\bf 125} 103493
  \urlprefix\url{https://www.sciencedirect.com/science/article/pii/S0011227522000753}

\bibitem{Houry_2024}
M.~Houry, P.~Malard, F.-P. Pellissier and Y.~Peneliau 2024 {\em Proceedings of
  the 50th EPS Conference on Plasma Physics, Salamanca\/}

\bibitem{Guo_2024}
Z.~Guo, D.~Zhu, R.~Yan, C.~Xuan et~al. 2024 {\em Nuclear Fusion\/} {\bf 64}
  076026

\bibitem{Xuan_2025}
C.~Xuan, D.~Zhu, Y.~Wang, B.~Gao et~al. 2025 {\em Nuclear Fusion\/} {\bf 65}
  046027

\bibitem{Pautasso_2017}
G.~Pautasso, M.~Bernert, M.~Dibon, B.~Duval et~al. 2016 {\em Plasma Physics and
  Controlled Fusion\/} {\bf 59} 014046
  \urlprefix\url{https://dx.doi.org/10.1088/0741-3335/59/1/014046}

\bibitem{Heinrich_2024}
P.~Heinrich, G.~Papp, P.~Lauber, G.~Pautasso et~al. 2024 {\em Nuclear Fusion\/}
  {\bf 64} 076044 \urlprefix\url{https://dx.doi.org/10.1088/1741-4326/ad502b}

\bibitem{Rodriguez-Fernandez_2022}
P.~Rodriguez-Fernandez, A.~Creely, M.~Greenwald, D.~Brunner et~al. 2022 {\em
  Nuclear Fusion\/} {\bf 62} 042003
  \urlprefix\url{https://dx.doi.org/10.1088/1741-4326/ac1654}

\bibitem{Ficker_2023}
O.~Ficker, U.~Sheikh, C.~Reux, J.~Cerovsky et~al. 2023 {\em 29th IAEA Fusion
  Energy Conference (FEC 2023)\/} (London, UK)

\bibitem{hill_1988}
D.~N. Hill, R.~Ellis, W.~Ferguson, D.~E. Perkins et~al. 1988 {\em Review of
  Scientific Instruments\/} {\bf 59} 1878--1880

\bibitem{vondracek_2017}
P.~Vondracek, E.~Gauthier, O.~Ficker, M.~Hron et~al. 2017 {\em Fusion
  Engineering and Design\/} {\bf 123} 764--767

\bibitem{dunn_2020}
M.~J. Dunn, T.~W. Morgan, J.~W. Genuit, T.~Loewenhoff et~al. 2020 {\em Nuclear
  Materials and Energy\/} {\bf 25} 100832

\bibitem{Finken_1990}
K.~Finken, J.~Watkins, D.~Rusbüldt, W.~Corbett et~al. 1990 {\em Nuclear
  Fusion\/} {\bf 30} 859
  \urlprefix\url{https://dx.doi.org/10.1088/0029-5515/30/5/005}

\bibitem{Tong_2016}
R.~H. Tong, Z.~Y. Chen, M.~Zhang, D.~W. Huang et~al. 2016 {\em Review of
  Scientific Instruments\/} {\bf 87} 11E113
  \urlprefix\url{https://doi.org/10.1063/1.4960311}

\bibitem{Reux_2022}
C.~Reux, C.~Paz-Soldan, N.~Eidietis, M.~Lehnen et~al. 2022 {\em Nuclear
  Fusion\/} {\bf 64} 034002
  \urlprefix\url{https://doi.org/10.1088/1361-6587/ac48bc}

\bibitem{Zhang_2025}
Y.~Zhang, R.~Zhou, L.~Hu, S.~Lin et~al. 2025 {\em Fusion Engineering and
  Design\/} {\bf 211} 114738
  \urlprefix\url{https://www.sciencedirect.com/science/article/pii/S0920379624005891}

\bibitem{Tinguely_2024}
R.~A. Tinguely, A.~M. Rosenthal, M.~Silva~Sa, M.~Jean et~al. 2024 {\em Review
  of Scientific Instruments\/} {\bf 95} 113503
  \urlprefix\url{https://doi.org/10.1063/5.0219477}

\bibitem{Beidler_2024}
M.~Beidler, D.~del Castillo-Negrete, D.~Shiraki, L.~Baylor et~al. 2024 {\em
  Nuclear Fusion\/} {\bf 64}(7) 076038
  \urlprefix\url{https://doi.org/10.1088/1741-4326/ad4c77}

\bibitem{Hollmann_2017}
E.~M. Hollmann, N.~Commaux, N.~W. Eidietis, C.~J. Lasnier et~al. 2017 {\em
  Physics of Plasmas\/} {\bf 24} 062505
  \urlprefix\url{https://doi.org/10.1063/1.4985086}

\bibitem{Wong_2007}
C.~P.~C. Wong, D.~L. Rudakov, J.~P. Allain, R.~J. Bastasz et~al. 2007 {\em
  Journal of Nuclear Materials\/} {\bf 363--365} 276--281

\bibitem{Dalmolin_2023}
A.~Dal~Molin, M.~Nocente, M.~Dalla~Rosa, E.~Panontin et~al. 2023 {\em
  Measurement Science and Technology\/} {\bf 34} 085501

\bibitem{Simons_2023}
L.~Simons, U.~Sheikh, J.~Decker, B.~P. Duval et~al. 2023 {\em 5th European
  Conference on Plasma Diagnostics\/} (Rethymno)
  \urlprefix\url{https://ecpd2023.eventsadmin.com/Home/Welcome}

\bibitem{Simons_2025}
L.~Simons, J.~Cerovsk{\'{y}}, J.~Decker, B.~P. Duval et~al. 2025 {\em submitted
  to Review of Scientific Instruments\/}

\bibitem{Tardocchi_2008}
M.~Tardocchi, L.~I. Proverbio, G.~Gorini, G.~Grosso et~al. 2008 {\em Review of
  Scientific Instruments\/} {\bf 79} 10E524

\bibitem{Cerovsky_2022}
J.~Cerovsky, O.~Ficker, V.~Svoboda, E.~Macusova et~al. 2022 {\em Journal of
  Instrumentation\/} {\bf 17} C01033

\bibitem{Dalmolin_2021}
A.~Dal~Molin, L.~Fumagalli, M.~Nocente, D.~Rigamonti et~al. 2021 {\em Review of
  Scientific Instruments\/} {\bf 92} 043517

\bibitem{Ma_2017}
T.~K. Ma, Z.~Y. Chen, D.~W. Huang, R.~H. Tong et~al. 2017 {\em Nuclear
  Instruments and Methods in Physics Research Section A: Accelerators,
  Spectrometers, Detectors and Associated Equipment\/} {\bf 856} 81--85

\bibitem{Rigamonti_2018}
D.~Rigamonti, A.~Broslawski, A.~Fernandes, J.~Figueiredo et~al. 2018 {\em
  Review of Scientific Instruments\/} {\bf 89} 10I116

\bibitem{Pace_2016}
D.~C. Pace, C.~M. Cooper, D.~Taussig, N.~W. Eidietis et~al. 2016 {\em Review of
  Scientific Instruments\/} {\bf 87} 043507

\bibitem{Wongrach_2021}
K.~Wongrach, D.~Mazon, J.~Morales, L.~Fleury et~al. 2021 {\em AIP Advances\/}
  {\bf 11} 085313

\bibitem{Grover_2016}
O.~Grover, J.~Kocman, M.~Odstrcil, T.~Odstrcil et~al. 2016 {\em Fusion
  Engineering and Design\/} {\bf 112} 1038--1044

\bibitem{Poikela_2014}
T.~Poikela, J.~Plosila, T.~Westerlund, M.~Campbell et~al. 2014 {\em Journal of
  Instrumentation\/} {\bf 9} C05013

\bibitem{svihra_2019}
P.~Svihra, D.~Bren, A.~Casolari, J.~Cerovsky et~al. 2019 {\em Fusion
  Engineering and Design\/} {\bf 146} 316--319

\bibitem{kulkov_2022}
S.~Kulkov, M.~Marcisovsky, P.~Svihra, M.~Tunkl et~al. 2022 {\em Journal of
  Instrumentation\/} {\bf 17} P02030

\bibitem{kulkov_2025}
S.~Kulkov 2025 {\em Semiconductor Pixel Detectors for Nuclear Physics and
  Quantum Astrometry\/} Ph.D. thesis Czech Technical University Prague, Czech
  Republic

\bibitem{Jarvis_1988}
O.~N. Jarvis, G.~Sadler and J.~L. Thompson 1988 {\em Nuclear Fusion\/} {\bf 28}
  1981

\bibitem{Zebrowski_2018}
J.~Zebrowski, L.~Jakubowski, M.~Rabinski, M.~J. Sadowski et~al. 2018 {\em
  Journal of Physics: Conference Series\/} {\bf 959} 012002

\bibitem{Rabinski_2017}
M.~Rabinski, L.~Jakubowski, K.~Malinowski, M.~Sadowski et~al. 2017 {\em Journal
  of Instrumentation\/} {\bf 12} C10014

\bibitem{Kwiatkowski_2021}
R.~Kwiatkowski, M.~Rabinski, M.~J. Sadowski, J.~Zebrowski et~al. 2021 {\em The
  European Physical Journal Plus\/} {\bf 136} 1070

\bibitem{Plyusnin_2008}
V.~V. Plyusnin, L.~Jakubowski, J.~Zebrowski, H.~Fernandes et~al. 2008 {\em
  Review of Scientific Instruments\/} {\bf 79} 10F505

\bibitem{Jakubowski_2010}
L.~Jakubowski, M.~J. Sadowski, J.~Zebrowski, M.~Rabinski et~al. 2010 {\em
  Review of Scientific Instruments\/} {\bf 81} 013504

\bibitem{Dhyani_2019}
P.~Dhyani, V.~Svoboda, V.~Istokskaia, J.~Mlyn{\'a}{\v r} et~al. 2019 {\em
  Journal of Instrumentation\/} {\bf 14}

\bibitem{Novotny_2020}
L.~Novotny, J.~Cerovsky, P.~Dhyani, O.~Ficker et~al. 2020 {\em Journal of
  Instrumentation\/} {\bf 15} C07015--C07015

\bibitem{Cowley_2020}
C.~Cowley, P.~Fuller, Y.~Andrew, L.~James et~al. 2020 {\em Physical Review E\/}
  {\bf 102} 043311

\bibitem{Delchambre_2009}
E.~Delchambre, G.~Counsell and A.~Kirk 2009 {\em Plasma Physics and Controlled
  Fusion\/} {\bf 51} 055012

\bibitem{Aumeunier_2017}
M.~H. Aumeunier, M.~Ko{\v c}an, R.~Reichle and E.~Gauthier 2017 {\em Nuclear
  Materials and Energy\/} {\bf 12} 1265--1269

\bibitem{McLean_2012}
A.~G. McLean, J.-W. Ahn, R.~Maingi, T.~K. Gray et~al. 2012 {\em Review of
  Scientific Instruments\/} {\bf 83} 053706
  \urlprefix\url{https://doi.org/10.1063/1.4717672}

\bibitem{Ushiki_2022}
T.~Ushiki, R.~Imazawa, H.~Murakami, K.~Shimizu et~al. 2022 {\em Review of
  Scientific Instruments\/} {\bf 93} 084905
  \urlprefix\url{https://doi.org/10.1063/5.0089269}

\bibitem{Bakhtiari_2005}
M.~Bakhtiari, G.~J. Kramer, M.~Takechi, H.~Tamai et~al. 2005 {\em Physical
  Review Letters\/} {\bf 94}(21) 215003
  \urlprefix\url{https://doi.org/10.1103/physrevlett.94.215003}

\bibitem{Fernandez-Gomez_2007}
I.~Fernández-Gómez, J.~R. Martín-Solís and R.~Sánchez 2007 {\em Physics of
  Plasmas\/} {\bf 14}(7) None \urlprefix\url{https://doi.org/10.1063/1.2746219}

\bibitem{Embreus_2016}
O.~Embréus, A.~Stahl and T.~Fülöp 2016 {\em New Journal of Physics\/} {\bf
  18} 093023 \urlprefix\url{https://doi.org/10.1088/1367-2630/18/9/093023}

\bibitem{Andersson_2001}
F.~Andersson, P.~Helander and L.~Eriksson 2001 {\em Physics of Plasmas\/} {\bf
  8}(12) 5221--5229 \urlprefix\url{https://doi.org/10.1063/1.1418242}

\bibitem{Beliaev_1956}
S.~T. Beliaev and G.~I. Budker 1956 {\em Soviet Physics Doklady\/} {\bf 1} 218

\bibitem{Braams_1987}
B.~J. Braams and C.~F.~F. Karney 1987 {\em Physical Review Letters\/} {\bf
  59}(16) 1817--1820
  \urlprefix\url{https://doi.org/10.1103/physrevlett.59.1817}

\bibitem{Pike_2014}
O.~J. Pike and S.~J. Rose 2014 {\em Physical Review E\/} {\bf 89} 053107
  \urlprefix\url{https://doi.org/10.1103/PhysRevE.89.053107}

\bibitem{Zhogolev_2014}
V.~Zhogolev and S.~Konovalov 2014 {\em Problems of Atomic Science and
  Technology, Ser. Thermonuclear Fusion\/} {\bf 37}(3) 71--88
  \urlprefix\url{https://doi.org/10.21517/0202-3822-2014-37-3-71-88}

\bibitem{Hesslow_2017}
L.~Hesslow, O.~Embr\'eus, A.~Stahl, T.~C. DuBois et~al. 2017 {\em Phys. Rev.
  Lett.\/} {\bf 118}(25) 255001
  \urlprefix\url{https://doi.org/10.1103/PhysRevLett.118.255001}

\bibitem{Hesslow_2018b}
L.~Hesslow, O.~Embréus, M.~Hoppe, T.~DuBois et~al. 2018 {\em Journal of Plasma
  Physics\/} {\bf 84} 905840605

\bibitem{Sokolov_1979}
Y.~A. Sokolov 1979 {\em JETP Letters\/} {\bf 29}(4) 244
  \urlprefix\url{http://jetpletters.ru/ps/0/article_22066.shtml}

\bibitem{Rosenbluth_1997}
M.~N. Rosenbluth and S.~V. Putvinski 1997 {\em Nuclear Fusion\/} {\bf 37} 1355
  \urlprefix\url{https://doi.org/10.1088/0029-5515/37/10/I03}

\bibitem{Chiu_1998}
S.~C. Chiu, M.~N. Rosenbluth, R.~W. Harvey and V.~S. Chan 1998 {\em Nuclear
  Fusion\/} {\bf 38} 1711

\bibitem{Aleynikov_2014}
P.~Aleynikov, K.~Aleynikova, B.~Breizman, G.~Huijsmans et~al. 2014 {\em
  International Atomic Energy Agency 25th Fusion Energy Conference\/} pp 13--18
  (\textit{Preprint} \eprint{https://nucleus.iaea.org/sites/fusionportal/Shared
  Documents/FEC 2014/fec2014-preprints/319_THP338.pdf})

\bibitem{Embreus_2018}
O.~Embréus, A.~Stahl and T.~Fülöp 2018 {\em Journal of Plasma Physics\/}
  {\bf 84} 905840102

\bibitem{Parail_1978}
V.~Parail and O.~Pogutse 1978 {\em Nuclear Fusion\/} {\bf 18}(3) 303--314
  \urlprefix\url{https://doi.org/10.1088/0029-5515/18/3/001}

\bibitem{Pokol_2008}
G.~Pokol, T.~Fülöp and M.~Lisak 2008 {\em Plasma Physics and Controlled
  Fusion\/} {\bf 50}(4) 045003
  \urlprefix\url{https://doi.org/10.1088/0741-3335/50/4/045003}

\bibitem{Aleynikov_2015b}
P.~Aleynikov and B.~Breizman 2015 {\em Nuclear Fusion\/} {\bf 55} 043014

\bibitem{Dreicer_1960}
H.~Dreicer 1960 {\em Physical Review\/} {\bf 117} 329
  \urlprefix\url{https://doi.org/10.1103/PhysRev.117.329}

\bibitem{Connor_1975}
J.~Connor and R.~Hastie 1975 {\em Nuclear Fusion\/} {\bf 15} 415
  \urlprefix\url{https://doi.org/10.1088/0029-5515/15/3/007}

\bibitem{Hesslow_2019b}
L.~Hesslow, L.~Unnerfelt, O.~Vallhagen, O.~Embreus et~al. 2019 {\em Journal of
  Plasma Physics\/} {\bf 85}(6) 475850601
  \urlprefix\url{https://doi.org/10.1017/s0022377819000874}

\bibitem{Putvinski_1997}
S.~Putvinski, P.~Barabaschi, N.~Fujisawa, N.~Putvinskaya et~al. 1997 {\em
  Plasma Physics and Controlled Fusion\/} {\bf 39} B157
  \urlprefix\url{https://doi.org/10.1088/0741-3335/39/12B/013}

\bibitem{Hesslow_2019}
L.~Hesslow, O.~Embréus, O.~Vallhagen and T.~Fülöp 2019 {\em Nuclear
  Fusion\/} {\bf 59}(8) 084004 \urlprefix\url{http://arxiv.org/abs/1904.00602}

\bibitem{Smith_2006}
H.~Smith, P.~Helander, L.~Eriksson, D.~Anderson et~al. 2006 {\em Physics of
  Plasmas\/} {\bf 13}(10) None
  \urlprefix\url{https://doi.org/10.1063/1.2358110}

\bibitem{Feher_2011}
T.~Fehér, H.~M. Smith, T.~Fülöp and K.~Gál 2011 {\em Plasma Physics and
  Controlled Fusion\/} {\bf 53}(3) 035014

\bibitem{Olasz_2021}
S.~Olasz, O.~Embreus, M.~Hoppe, M.~Aradi et~al. 2021 {\em Nuclear Fusion\/}
  {\bf 61} 066010 \urlprefix\url{https://dx.doi.org/10.1088/1741-4326/abf0de}

\bibitem{Landreman_2014}
M.~Landreman, A.~Stahl and T.~Fülöp 2014 {\em Computer Physics
  Communications\/} {\bf 185}(3) 847--855
  \urlprefix\url{https://doi.org/10.1016/j.cpc.2013.12.004}

\bibitem{Stahl_2016}
A.~Stahl, O.~Embréus, G.~Papp, M.~Landreman et~al. 2016 {\em Nuclear Fusion\/}
  {\bf 56}(11) 112009
  \urlprefix\url{https://doi.org/10.1088/0029-5515/56/11/112009}

\bibitem{Harvey_1992}
 1992 {\em The {CQL3D} code\/}
  \urlprefix\url{http://www.compxco.com/cql3d_manual.pdf}

\bibitem{Guo_2019}
Z.~Guo, C.~Mcdevitt and X.~Tang 2019 {\em Physics of Plasmas\/} {\bf 26}(8)
  None \urlprefix\url{https://doi.org/10.1063/1.5055874}

\bibitem{Stahl_2017}
A.~Stahl, M.~Landreman, O.~Embréus and T.~Fülöp 2017 {\em Computer Physics
  Communications\/} {\bf 212}(3) 269--279
  \urlprefix\url{https://doi.org/10.1016/j.cpc.2016.10.024}

\bibitem{McDevitt_2019}
C.~J. McDevitt, Z.~Guo and X.~Tang 2019 {\em Plasma Physics and Controlled
  Fusion\/} {\bf 61}(5) 054008
  \urlprefix\url{https://doi.org/10.1088/1361-6587/ab0d6d}

\bibitem{Aleynikov_2017}
P.~Aleynikov and B.~N. Breizman 2017 {\em Nuclear Fusion\/} {\bf 57}(4) 046009
  \urlprefix\url{https://doi.org/10.1088/1741-4326/aa5895}

\bibitem{Daniel_2020}
D.~Daniel, W.~T. Taitano and L.~Chacón 2020 {\em Computer Physics
  Communications\/} {\bf 254} 107361
  \urlprefix\url{https://doi.org/10.1016/j.cpc.2020.107361}

\bibitem{Hirvijoki_2014}
E.~Hirvijoki, O.~Asunta, T.~Koskela, T.~Kurki-Suonio et~al. 2014 {\em Computer
  Physics Communications\/} {\bf 185}(4) 1310--1321
  \urlprefix\url{https://doi.org/10.1016/j.cpc.2014.01.014}

\bibitem{Hoelzl_2021}
M.~Hoelzl, G.~Huijsmans, S.~Pamela, M.~Bécoulet et~al. 2021 {\em Nuclear
  Fusion\/} {\bf 61}(6) 065001
  \urlprefix\url{https://doi.org/10.1088/1741-4326/abf99f}

\bibitem{Bergstrom_2025}
H.~Bergström, S.~Liu, V.~Bandaru, M.~Hoelzl et~al. 2025 {\em Plasma Physics
  and Controlled Fusion\/} {\bf 67}(3) 035004
  \urlprefix\url{https://doi.org/10.1088/1361-6587/adaee7}

\bibitem{Carbajal_2017}
L.~Carbajal, D.~del Castillo-Negrete, D.~Spong, S.~Seal et~al. 2017 {\em
  Physics of Plasmas\/} {\bf 24}(4) None
  \urlprefix\url{https://doi.org/10.1063/1.4981209}

\bibitem{Bandaru_2024_refluid}
V.~Bandaru, M.~Hoelzl, F.~J. Artola, O.~Vallhagen et~al. 2024 {\em Physics of
  Plasmas\/} {\bf 31} 082503 \urlprefix\url{https://doi.org/10.1063/5.0213962}

\bibitem{Liu_(Chang)_2021}
C.~Liu, C.~Zhao, S.~C. Jardin, N.~M. Ferraro et~al. 2021 {\em Plasma Physics
  and Controlled Fusion\/} {\bf 63} 125031
  \urlprefix\url{https://dx.doi.org/10.1088/1361-6587/ac2af8}

\bibitem{Sainterme_2024}
A.~P. Sainterme and C.~R. Sovinec 2024 {\em Physics of Plasmas\/} {\bf 31}
  010701 \urlprefix\url{https://doi.org/10.1063/5.0183530}

\bibitem{Matsuyama_2017}
A.~Matsuyama, N.~Aiba and M.~Yagi 2017 {\em Nuclear Fusion\/} {\bf 57} 066038
  \urlprefix\url{https://dx.doi.org/10.1088/1741-4326/aa6867}

\bibitem{DeVries_2019}
P.~de~Vries and Y.~Gribov 2019 {\em Nuclear Fusion\/} {\bf 59}(9) 096043
  \urlprefix\url{https://doi.org/10.1088/1741-4326/ab2ef4}

\bibitem{DeVries_2023}
P.~de~Vries, Y.~Lee, Y.~Gribov, A.~Mineev et~al. 2023 {\em Nuclear Fusion\/}
  {\bf 63}(8) 086016 \urlprefix\url{https://doi.org/10.1088/1741-4326/acdd11}

\bibitem{Martin-Solis_2010}
J.~R. Martín-Solís, R.~Sánchez and B.~Esposito 2010 {\em Physical Review
  Letters\/} {\bf 105}(18) 185002
  \urlprefix\url{https://doi.org/10.1103/physrevlett.105.185002}

\bibitem{Breizman_2014}
B.~N. Breizman 2014 {\em Nuclear Fusion\/} {\bf 54}(7) 072002
  \urlprefix\url{https://doi.org/10.1088/0029-5515/54/7/072002}

\bibitem{Putvinski_1997b}
S.~Putvinski, N.~Fujisawa, D.~Post, N.~Putvinskaya et~al. 1997 {\em Journal of
  Nuclear Materials\/} {\bf 241-243} 316--321
  \urlprefix\url{https://doi.org/10.1016/s0022-3115(97)80056-6}

\bibitem{Linder_2021}
O.~Linder, G.~Papp, E.~Fable, F.~Jenko et~al. 2021 {\em Journal of Plasma
  Physics\/} {\bf 87}(3) 905870301
  \urlprefix\url{https://doi.org/10.1017/s0022377821000416}

\bibitem{Smith_2008}
H.~M. Smith and E.~Verwichte 2008 {\em Physics of Plasmas\/} {\bf 15}(7) None
  \urlprefix\url{https://doi.org/10.1063/1.2949692}

\bibitem{Hollmann_2019}
E.~Hollmann, N.~Eidietis, J.~Herfindal, P.~Parks et~al. 2019 {\em Nuclear
  Fusion\/} {\bf 59}(10) 106014
  \urlprefix\url{https://doi.org/10.1088/1741-4326/ab32b2}

\bibitem{Kim_2020}
H.~Kim, A.~Mineev, D.~Ricci, J.~Lee et~al. 2020 {\em Nuclear Fusion\/} {\bf
  60}(12) 126049 \urlprefix\url{https://doi.org/10.1088/1741-4326/abb95c}

\bibitem{Hoppe_2022}
M.~Hoppe, I.~Ekmark, E.~Berger and T.~Fülöp 2022 {\em Journal of Plasma
  Physics\/} {\bf 88}(3) 905880317
  \urlprefix\url{https://arxiv.org/abs/2203.09900}

\bibitem{Hollmann_2023}
E.~Hollmann, L.~Baylor, A.~Boboc, P.~Carvalho et~al. 2023 {\em Nuclear
  Fusion\/} {\bf 63}(3) 036011
  \urlprefix\url{https://doi.org/10.1088/1741-4326/acb4aa}

\bibitem{Beidler_2020}
M.~T. Beidler, D.~del Castillo-Negrete, L.~R. Baylor, D.~Shiraki et~al. 2020
  {\em Physics of Plasmas\/} {\bf 27}(11) 112507
  \urlprefix\url{https://doi.org/10.1063/5.0022072}

\bibitem{Aleynikov_2015a}
P.~Aleynikov and B.~N. Breizman 2015 {\em Physical Review Letters\/} {\bf
  114}(15) 155001
  \urlprefix\url{https://doi.org/10.1103/physrevlett.114.155001}

\bibitem{Hesslow_2018a}
L.~Hesslow, O.~Embréus, G.~J. Wilkie, G.~Papp et~al. 2018 {\em Plasma Physics
  and Controlled Fusion\/} {\bf 60}(7) 074010
  \urlprefix\url{https://doi.org/10.1088/1361-6587/aac33e}

\bibitem{Hoppe_2021a}
M.~Hoppe, L.~Hesslow, O.~Embreus, L.~Unnerfelt et~al. 2021 {\em Journal of
  Plasma Physics\/} {\bf 87}(1) 855870102
  \urlprefix\url{https://doi.org/10.1017/s002237782000152x}

\bibitem{McDevitt_2023}
C.~J. McDevitt, X.~Tang, C.~J. Fontes, P.~Sharma et~al. 2023 {\em Nuclear
  Fusion\/} {\bf 63}(2) 024001
  \urlprefix\url{https://doi.org/10.1088/1741-4326/acae38}

\bibitem{Garland_2020}
N.~A. Garland, H.~Chung, C.~J. Fontes, M.~C. Zammit et~al. 2020 {\em Physics of
  Plasmas\/} {\bf 27}(4) None \urlprefix\url{https://doi.org/10.1063/5.0003638}

\bibitem{Garland_2022}
N.~A. Garland, H.~Chung, M.~C. Zammit, C.~J. McDevitt et~al. 2022 {\em Physics
  of Plasmas\/} {\bf 29}(1) None
  \urlprefix\url{https://doi.org/10.1063/5.0071996}

\bibitem{Hoppe_2025}
M.~Hoppe, J.~Decker, U.~Sheikh, S.~Coda et~al. 2025 {\em Plasma Physics and
  Controlled Fusion\/} \urlprefix\url{https://doi.org/10.1088/1361-6587/adbcd5}

\bibitem{Guo_2018}
Z.~Guo, C.~J. McDevitt and X.~Tang 2018 {\em Physics of Plasmas\/} {\bf 25}(3)
  None \urlprefix\url{https://doi.org/10.1063/1.5019381}

\bibitem{Martin-Solis_2004}
J.~Martín-Solís, B.~Esposito, R.~Sánchez and G.~Granucci 2004 {\em Nuclear
  Fusion\/} {\bf 44}(9) 974--981
  \urlprefix\url{https://doi.org/10.1088/0029-5515/44/9/005}

\bibitem{Fulop_2006}
T.~Fülöp, G.~Pokol, P.~Helander and M.~Lisak 2006 {\em Phys. Plasmas\/} {\bf
  13} 062506
  \urlprefix\url{http://scitation.aip.org/content/aip/journal/pop/13/6/10.1063/1.2208327}

\bibitem{Pokol_2014}
G.~Pokol, A.~Kómár, A.~Budai, A.~Stahl et~al. 2014 {\em Physics of Plasmas\/}
  {\bf 21}(10) 102503 \urlprefix\url{https://doi.org/10.1063/1.4895513}

\bibitem{Decker_2024}
J.~Decker, M.~Hoppe, U.~Sheikh, B.~Duval et~al. 2024 {\em Nuclear Fusion\/}
  {\bf 64}(10) 106027 \urlprefix\url{https://doi.org/10.1088/1741-4326/ad6c61}

\bibitem{Reinke_2019}
M.~Reinke, S.~Scott, R.~Granetz, J.~Hughes et~al. 2019 {\em Nuclear Fusion\/}
  {\bf 59} 066003

\bibitem{Breizman_2015}
B.~Breizman 2015 Runaway electrons (whitepaper) Tech. rep. USBPO group
  \urlprefix\url{https://burningplasma.org/resources/ref/Workshops2015/IS/A_breizman_b.pdf}

\bibitem{Fulop_2009}
T.~Fülöp, H.~M. Smith and G.~Pokol 2009 {\em Physics of Plasmas\/} {\bf
  16}(2) None \urlprefix\url{https://doi.org/10.1063/1.3072980}

\bibitem{Spong_2018}
D.~Spong, W.~Heidbrink, C.~Paz-Soldan, X.~Du et~al. 2018 {\em Phys. Rev.
  Lett.\/} {\bf 120} 155002
  \urlprefix\url{https://link.aps.org/doi/10.1103/PhysRevLett.120.155002}

\bibitem{Heidbrink_2019}
W.~W. Heidbrink, C.~Paz-Soldan, D.~A. Spong, X.~D. Du et~al. 2019 {\em Plasma
  Physics and Controlled Fusion\/} {\bf 61}(1) 014007
  \urlprefix\url{https://doi.org/10.1088/1361-6587/aae2da}

\bibitem{Bin_2022}
W.~Bin, C.~Castaldo, F.~Napoli, P.~Buratti et~al. 2022 {\em Phys. Rev. Lett.\/}
  {\bf 129} 045002
  \urlprefix\url{https://link.aps.org/doi/10.1103/PhysRevLett.129.045002}

\bibitem{Breizman_2023b}
B.~Breizman and D.~I. Kiramov 2023 {\em Physics of Plasmas\/} {\bf 30}

\bibitem{Breizman_2023a}
B.~Breizman and D.~Kiramov 2023 {\em Physics of Plasmas\/} {\bf 30}

\bibitem{DINA-report:2016}
V.~Lukash and R.~Khayrutdinov 2016 Final report {IO/15/CT/4300001189} Tech.
  rep. {ITER}

\bibitem{Martin-Solis_2022}
J.~R. Martin-Solis, J.~A. Mier, M.~Lehnen and A.~Loarte 2022 {\em Nuclear
  Fusion\/} {\bf 62} 076013
  \urlprefix\url{https://doi.org/10.1088/1741-4326/ac637b}

\bibitem{Wang_2025}
C.~Wang, E.~Nardon, F.~Artola, V.~Bandaru et~al. 2025 {\em Nuclear Fusion\/}
  {\bf 65} 016012 \urlprefix\url{https://doi.org/10.1088/1741-4326/ad8d66}

\bibitem{Bandaru_2025}
V.~Bandaru, M.~Hoelzl, F.~Artola, M.~Lehnen et~al. 2025 {\em Journal of Plasma
  Physics\/} {\bf 91}(1) E27
  \urlprefix\url{https://doi.org/10.1017/s0022377824001661}

\bibitem{Vallhagen_2025}
O.~Vallhagen, L.~Hanebring, T.~Fülöp, M.~Hoppe et~al. 2025 {\em Journal of
  Plasma Physics\/} {\bf 91} E78

\bibitem{Papp_2011a}
G.~Papp, M.~Drevlak, T.~Fülöp, P.~Helander et~al. 2011 {\em Plasma Physics
  and Controlled Fusion\/} {\bf 53}(9) 095004
  \urlprefix\url{https://doi.org/10.1088/0741-3335/53/9/095004}

\bibitem{Papp_2011b}
G.~Papp, M.~Drevlak, T.~Fülöp and P.~Helander 2011 {\em Nuclear Fusion\/}
  {\bf 51}(4) 043004
  \urlprefix\url{https://doi.org/10.1088/0029-5515/51/4/043004}

\bibitem{Boozer_2011}
A.~H. Boozer 2011 {\em Plasma Physics and Controlled Fusion\/} {\bf 53}(8)
  084002 \urlprefix\url{https://doi.org/10.1088/0741-3335/53/8/084002}

\bibitem{Smith_2013}
H.~M. Smith, A.~H. Boozer and P.~Helander 2013 {\em Physics of Plasmas\/} {\bf
  20}(7) None \urlprefix\url{https://doi.org/10.1063/1.4813255}

\bibitem{Sweeney_2020}
R.~Sweeney, A.~J. Creely, J.~Doody, T.~Fülöp et~al. 2020 {\em Journal of
  Plasma Physics\/} {\bf 86} 865860507

\bibitem{Paz-Soldan_2019}
C.~Paz-Soldan, N.~W. Eidietis, Y.~Q. Liu, D.~Shiraki et~al. 2019 {\em Plasma
  Physics and Controlled Fusion\/} {\bf 61}(5) 054001
  \urlprefix\url{https://doi.org/10.1088/1361-6587/aafd15}

\bibitem{Lvovskiy_2020}
A.~Lvovskiy, C.~Paz-Soldan, N.~Eidietis, P.~Aleynikov et~al. 2020 {\em Nuclear
  Fusion\/} {\bf 60}(5) 056008
  \urlprefix\url{https://doi.org/10.1088/1741-4326/ab78c7}

\bibitem{Hauff_2009}
T.~Hauff and F.~Jenko 2009 {\em Physics of Plasmas\/} {\bf 16} 102308
  \urlprefix\url{https://doi.org/10.1063/1.3243494}

\bibitem{Saerkimaeki_2020}
K.~Särkimäki, O.~Embreus, E.~Nardon, T.~Fülöp et~al. 2020 {\em Nuclear
  Fusion\/} {\bf 60} 126050
  \urlprefix\url{https://dx.doi.org/10.1088/1741-4326/abb9e9}

\bibitem{Svensson_2021}
P.~Svensson, O.~Embreus, S.~L. Newton, K.~Särkimäki et~al. 2021 {\em Journal
  of Plasma Physics\/} {\bf 87}(2) 905870207
  \urlprefix\url{https://arxiv.org/abs/2010.07156}

\bibitem{Myra_1992}
J.~R. Myra, P.~J. Catto, A.~J. Wootton, R.~D. Bengtson et~al. 1992 {\em Physics
  of Fluids B: Plasma Physics\/} {\bf 4}(7) 2092--2097
  \urlprefix\url{https://doi.org/10.1063/1.860016}

\bibitem{Kwon_1988}
O.~Kwon, P.~Diamond, F.~Wagner, G.~Fussmann et~al. 1988 {\em Nuclear Fusion\/}
  {\bf 28}(11) 1931--1943
  \urlprefix\url{https://doi.org/10.1088/0029-5515/28/11/002}

\bibitem{Esposito_1996}
B.~Esposito, R.~M. Solis, P.~van Belle, O.~N. Jarvis et~al. 1996 {\em Plasma
  Physics and Controlled Fusion\/} {\bf 38} 2035
  \urlprefix\url{https://dx.doi.org/10.1088/0741-3335/38/12/001}

\bibitem{Martin-Solis_2021}
J.~R. Martín-Solís 2021 {\em Physics of Plasmas\/} {\bf 28}(3) None
  \urlprefix\url{https://doi.org/10.1063/5.0032283}

\bibitem{Janosi_2024}
D.~Jánosi and G.~Károlyi 2024 {\em Chaos: An Interdisciplinary Journal of
  Nonlinear Science\/} {\bf 34} 081104
  \urlprefix\url{https://doi.org/10.1063/5.0216731}

\bibitem{Martin-Solis_2014}
J.~R. Martin-Solis, A.~Loarte, E.~M. Hollmann, B.~Esposito et~al. 2014 {\em
  Nuclear Fusion\/} {\bf 54} 083027
  \urlprefix\url{https://doi.org/10.1088/0029-5515/54/8/083027}

\bibitem{Martin-Solis_2015}
J.~R. Martín-Solís, A.~Loarte and M.~Lehnen 2015 {\em Physics of Plasmas\/}
  {\bf 22} 082503 \urlprefix\url{https://doi.org/10.1063/1.4927773}

\bibitem{Riemann_2012}
J.~Riemann, H.~M. Smith and P.~Helander 2012 {\em Physics of Plasmas\/} {\bf
  19} 012057 \urlprefix\url{https://doi.org/10.1063/1.3671974}

\bibitem{Hollmann_2013}
E.~Hollmann, M.~Austin, J.~Boedo, N.~Brooks et~al. 2013 {\em Nuclear Fusion\/}
  {\bf 53} 083004 \urlprefix\url{https://doi.org/10.1088/0029-5515/53/8/083004}

\bibitem{Loarte_2011}
A.~Loarte, V.~Riccardo, J.~R. Martin-Solís, J.~Paley et~al. 2011 {\em Nuclear
  Fusion\/} {\bf 51} 073004
  \urlprefix\url{https://doi.org/10.1088/0029-5515/51/7/073004}

\bibitem{Jayakumar_1993}
R.~Jayakumar, H.~Fleischmann and S.~Zweben 1993 {\em Physics Letters A\/} {\bf
  172} 447 \urlprefix\url{https://doi.org/10.1016/0375-9601(93)90237-T}

\bibitem{Eriksson_2004}
L.-G. Eriksson, P.~Helander, F.~Andersson, D.~Anderson et~al. 2004 {\em
  Physical Review Letters\/} {\bf 92} 205004
  \urlprefix\url{https://doi.org/10.1103/PhysRevLett.92.205004}

\bibitem{Kiramov_2017}
D.~I. Kiramov and B.~N. Breizman 2017 {\em Physics of Plasmas\/} {\bf 24}
  100702 \urlprefix\url{https://doi.org/10.1063/1.4993071}

\bibitem{Lehnen_2018}
M.~Lehnen 2018 {\em Private Communication\/}

\bibitem{Aleynikova_2016}
K.~Aleynikova, G.~T. Huijsmans and P.~Aleynikov 2016 {\em Plasma Physics
  Reports\/} {\bf 42} 486--494

\bibitem{Khayrutdinov_1993}
R.~Khayrutdinov and V.~Lukash 1993 {\em Journal of Computational Physics\/}
  {\bf 109} 193--201
  \urlprefix\url{https://www.sciencedirect.com/science/article/pii/S0021999183712118}

\bibitem{Jardin_1986}
S.~Jardin, N.~Pomphrey and J.~Delucia 1986 {\em Journal of Computational
  Physics\/} {\bf 66} 481--507
  \urlprefix\url{https://www.sciencedirect.com/science/article/pii/002199918690077X}

\bibitem{Bandyopadhyay_2012}
I.~Bandyopadhyay, A.~Singh, M.~Sugihara and S.~Jardin 2012 {\em International
  Atomic Energy Agency 24th Fusion Energy Conference\/}

\bibitem{Zhao_2021}
C.~Zhao, C.~Liu, S.~C. Jardin and N.~M. Ferraro 2020 {\em Nuclear Fusion\/}
  {\bf 60} 126017 \urlprefix\url{https://dx.doi.org/10.1088/1741-4326/ab96f4}

\bibitem{Konovalov_2016}
S.~Konovalov, P.~Aleynikov, K.~Aleynikova, Y.~Gribov et~al. 2016 {\em
  International Atomic Energy Agency 26th Fusion Energy Conference\/}
  \urlprefix\url{https://conferences.iaea.org/indico/event/98/session/23/contribution/321/material/slides/0.pdf}

\bibitem{Aleynikov_2010}
P.~B. Aleynikov, A.~A. Ivanov, R.~R. Khayrutdinov, S.~V. Konovalov et~al. 2010
  {\em 37th EPS Conference on Plasma Physics 2010\/} vol~1 pp 281--284

\bibitem{Kiramov_2016}
D.~Kiramov, M.~Lehnen, R.~Khayrutdinov and V.~Lukash 2016 {\em Proc. 43rd EPS
  Conf. Plasma Physics\/} p~P4

\bibitem{Artola_2021}
F.~J. Artola, A.~Loarte, E.~Matveeva, J.~Havlicek et~al. 2021 {\em Plasma
  Physics and Controlled Fusion\/} {\bf 63} 064004

\bibitem{Russo_1991}
A.~Russo 1991 {\em Nuclear Fusion\/} {\bf 31} 117
  \urlprefix\url{https://dx.doi.org/10.1088/0029-5515/31/1/011}

\bibitem{Heikkinen_1993}
J.~Heikkinen, S.~Sipilä and T.~Pättikangas 1993 {\em Computer Physics
  Communications\/} {\bf 76} 215--230
  \urlprefix\url{https://www.sciencedirect.com/science/article/pii/001046559390133W}

\bibitem{Guan_2010}
X.~Guan, H.~Qin and N.~J. Fisch 2010 {\em Physics of Plasmas\/} {\bf 17} 092502
  \urlprefix\url{https://doi.org/10.1063/1.3476268}

\bibitem{Izzo_2011}
V.~Izzo, E.~Hollmann, A.~James, J.~Yu et~al. 2011 {\em Nuclear Fusion\/} {\bf
  51} 063032 \urlprefix\url{https://dx.doi.org/10.1088/0029-5515/51/6/063032}

\bibitem{Sommariva_2018}
C.~Sommariva, E.~Nardon, P.~Beyer, M.~Hoelzl et~al. 2017 {\em Nuclear Fusion\/}
  {\bf 58} 016043 \urlprefix\url{https://dx.doi.org/10.1088/1741-4326/aa95cd}

\bibitem{Liu_(Yueqiang)_2019}
Y.~Liu, P.~Parks, C.~Paz-Soldan, C.~Kim et~al. 2019 {\em Nuclear Fusion\/} {\bf
  59} 126021 \urlprefix\url{https://dx.doi.org/10.1088/1741-4326/ab3f87}

\bibitem{Beidler_2021}
M.~T. Beidler, D.~del Castillo-Negrete, L.~R. Baylor, J.~L. Herfindal et~al.
  2021 {\em Preprint: 2020 IAEA Fusion Energy Conference (2021) TH/P1-9\/}

\bibitem{McDevitt_2019b}
C.~J. McDevitt and X.-Z. Tang 2019 {\em {EPL} (Europhysics Letters)\/} {\bf
  127} 45001 \urlprefix\url{https://doi.org/10.1209/0295-5075/127/45001}

\bibitem{McDevitt_2019c}
C.~J. McDevitt, Z.~Guo and X.-Z. Tang 2019 {\em Plasma Physics and Controlled
  Fusion\/} {\bf 61} 024004
  \urlprefix\url{https://doi.org/10.1088%2F1361-6587%2Faaf4d1}

\bibitem{Arnaud_2024}
J.~S. Arnaud and C.~J. McDevitt 2024 {\em Physics of Plasmas\/} {\bf 31} 062509
  \urlprefix\url{https://doi.org/10.1063/5.0198338}

\bibitem{McDevitt_2023b}
C.~J. McDevitt and X.~Tang 2023 {\em Physical Review E\/} {\bf 108}(4) L043201
  \urlprefix\url{https://doi.org/10.1103/PhysRevE.108.L043201}

\bibitem{Pautasso_2015}
G.~Pautasso et~al. 2015 {\em Proceedings of the 42nd EPS Conference on Plasma
  Physics, Lisbon\/}

\bibitem{Cornille_2022}
B.~S. Cornille, M.~T. Beidler, S.~Munaretto, B.~E. Chapman et~al. 2022 {\em
  Physics of Plasmas\/} {\bf 29} 052510
  \urlprefix\url{https://doi.org/10.1063/5.0087314}

\bibitem{Carbajal_2020}
L.~Carbajal, D.~del Castillo-Negrete and J.~J. Martinell 2020 {\em Physics of
  Plasmas\/} {\bf 27} 032502 \urlprefix\url{https://doi.org/10.1063/1.5135588}

\bibitem{Boozer_2016}
A.~H. Boozer and A.~Punjabi 2016 {\em Physics of Plasmas\/} {\bf 23} 102513
  \urlprefix\url{https://doi.org/10.1063/1.4966046}

\bibitem{Abdullaev_2013}
S.~Abdullaev 2013 {\em Magnetic Stochasticity in Magnetically Confined Fusion
  Plasmas\/} (Springer Cham)

\bibitem{deRover_1996}
M.~de~Rover, N.~J. Lopes~Cardozo and A.~Montvai 1996 {\em Physics of Plasmas\/}
  {\bf 3} 4478--4488 \urlprefix\url{https://doi.org/10.1063/1.871582}

\bibitem{Papp_2012}
G.~Papp, M.~Drevlak, T.~Fülöp and G.~I. Pokol 2012 {\em Plasma Physics and
  Controlled Fusion\/} {\bf 54} 125008
  \urlprefix\url{https://dx.doi.org/10.1088/0741-3335/54/12/125008}

\bibitem{Carbajal_2017b}
L.~Carbajal and D.~del Castillo-Negrete 2017 {\em Plasma Physics and Controlled
  Fusion\/} {\bf 59} 124001
  \urlprefix\url{https://dx.doi.org/10.1088/1361-6587/aa883e}

\bibitem{DelCastilloNegrete_2018}
D.~del Castillo-Negrete, L.~Carbajal, D.~Spong and V.~Izzo 2018 {\em Physics of
  Plasmas\/} {\bf 25} 056104 \urlprefix\url{https://doi.org/10.1063/1.5018747}

\bibitem{Artola_2020}
F.~J. Artola, K.~Lackner, G.~T.~A. Huijsmans, M.~Hoelzl et~al. 2020 {\em
  Physics of Plasmas\/} {\bf 27} 032501
  \urlprefix\url{https://doi.org/10.1063/1.5140230}

\bibitem{Jardin_2007}
S.~Jardin, J.~Breslau and N.~Ferraro 2007 {\em Journal of Computational
  Physics\/} {\bf 226} 2146--2174
  \urlprefix\url{https://www.sciencedirect.com/science/article/pii/S0021999107003038}

\bibitem{Ferraro_2016}
N.~M. Ferraro, S.~C. Jardin, L.~L. Lao, M.~S. Shephard et~al. 2016 {\em Physics
  of Plasmas\/} {\bf 23} 056114
  \urlprefix\url{https://doi.org/10.1063/1.4948722}

\bibitem{Huysmans_2007}
G.~Huysmans and O.~Czarny 2007 {\em Nuclear Fusion\/} {\bf 47} 659
  \urlprefix\url{https://dx.doi.org/10.1088/0029-5515/47/7/016}

\bibitem{Bandaru_2021}
V.~Bandaru, M.~Hoelzl, C.~Reux, O.~Ficker et~al. 2021 {\em Plasma Physics and
  Controlled Fusion\/} {\bf 63} 035024
  \urlprefix\url{https://dx.doi.org/10.1088/1361-6587/abdbcf}

\bibitem{Sommariva_2024}
C.~Sommariva, A.~Pau, S.~Silburn, C.~Reux et~al. 2024 {\em Nuclear Fusion\/}
  {\bf 64} 106050 \urlprefix\url{https://dx.doi.org/10.1088/1741-4326/ad6e03}

\bibitem{Marini_2024}
C.~Marini, E.~Hollmann, S.~Tang, J.~Herfindal et~al. 2024 {\em Nuclear
  Fusion\/} {\bf 64} 076039
  \urlprefix\url{https://dx.doi.org/10.1088/1741-4326/ad4db6}

\bibitem{Helander_2007}
P.~Helander, D.~Grasso, R.~J. Hastie and A.~Perona 2007 {\em Physics of
  Plasmas\/} {\bf 14} 122102 \urlprefix\url{https://doi.org/10.1063/1.2817016}

\bibitem{Singh_2023}
L.~Singh, D.~Borgogno, F.~Subba and D.~Grasso 2023 {\em Physics of Plasmas\/}
  {\bf 30} 122114 \urlprefix\url{https://doi.org/10.1063/5.0174167}

\bibitem{Liu_(ShiJie)_2025}
L.~S-J and {et al} 2025 {\em Journal of Plasma Physics\/} {\bf in preparation}

\bibitem{Nardon_2023}
E.~Nardon, V.~Bandaru, M.~Hoelzl, F.~J. Artola et~al. 2023 {\em Physics of
  Plasmas\/} {\bf 30} 092502 \urlprefix\url{https://doi.org/10.1063/5.0162608}

\bibitem{Singh_2025}
L.~Singh Impact of runaway electrons generated during disruptions on the first
  wall of the tokamak reactors, phd thesis https://arxiv.org/abs/2410.03512

\bibitem{Liu_(Yueqiang)_2022}
Y.~Liu, K.~Aleynikova, C.~Paz-Soldan, P.~Aleynikov et~al. 2022 {\em Nuclear
  Fusion\/} {\bf 62} 066026
  \urlprefix\url{https://dx.doi.org/10.1088/1741-4326/ac5d62}

\bibitem{Vannini_2025}
F.~Vannini, V.~K. Bandaru, H.~Bergström, N.~Schwarz et~al. 2025 {\em Nuclear
  Fusion\/}
  \urlprefix\url{http://iopscience.iop.org/article/10.1088/1741-4326/adac77}

\bibitem{Lvovskiy_2018}
A.~Lvovskiy, C.~Paz-Soldan, N.~W. Eidietis, A.~D. Molin et~al. 2018 {\em Plasma
  Phys. Control. Fusion\/} {\bf 60} 124003
  \urlprefix\url{http://stacks.iop.org/0741-3335/60/i=12/a=124003}

\bibitem{Lvovskiy_2019}
A.~Lvovskiy, W.~W. Heidbrink, C.~Paz-Soldan, D.~A. Spong et~al. 2019 {\em Nucl.
  Fusion\/} {\bf 59} 124004
  \urlprefix\url{https://doi.org/10.1088%2F1741-4326%2Fab4405}

\bibitem{liu_(Chang)_2023}
C.~Liu, A.~Lvovskiy, C.~Paz-Soldan, S.~C. Jardin et~al. 2023 {\em Phys. Rev.
  Lett.\/} {\bf 131} 085102
  \urlprefix\url{https://link.aps.org/doi/10.1103/PhysRevLett.131.085102}

\bibitem{Yang_2024}
M.~Yang, P.~Wang, D.~del Castillo-Negrete, Y.~Cao et~al. 2024 {\em SIAM Journal
  on Scientific Computing\/} {\bf 46} C508--C533
  \urlprefix\url{https://epubs.siam.org/doi/abs/10.1137/23M1585635}

\bibitem{Cathey_2020}
A.~Cathey, M.~Hoelzl, K.~Lackner, G.~Huijsmans et~al. 2020 {\em Nuclear
  Fusion\/} {\bf 60} 124007
  \urlprefix\url{https://dx.doi.org/10.1088/1741-4326/abbc87}

\bibitem{Saerkimaeki_2022}
K.~Särkimäki, J.~Artola, M.~Hoelzl and the JOREK~Team 2022 {\em Nuclear
  Fusion\/} {\bf 62} 086033
  \urlprefix\url{https://dx.doi.org/10.1088/1741-4326/ac75fd}

\bibitem{Bandaru_2023}
V.~Bandaru and M.~Hoelzl 2023 {\em Physics of Plasmas\/} {\bf 30} 092508
  \urlprefix\url{https://doi.org/10.1063/5.0165240}

\bibitem{Yuan_2023}
L.~Yuan and D.~Hu 2023 {\em Chinese Phys. B\/} {\bf 32} 075208
  \urlprefix\url{https://dx.doi.org/10.1088/1674-1056/acc1d7}

\bibitem{ICRUReport_1984}
ICRU-Report-37 1984 Stopping powers for electrons and positrons

\bibitem{Berger_1970}
M.~J. Berger and S.~M. Seltzer 1970 {\em Phys. Rev. C\/} {\bf 2} 621--631

\bibitem{Bethe_1953}
H.~A. Bethe and J.~Ashkin 1953 {\em Experimental Nuclear Physics Volume I\/} ed
  E.~Segre (New York: John Wiley \& Sons)

\bibitem{Carron_2007}
N.~J. Carron 2007 {\em An introduction to the passage of energetic particles
  through matter\/} ({Boca Raton}: {Taylor \& Francis})

\bibitem{Seltzer_1986}
S.~M. Seltzer and M.~J. Berger 1986 {\em Atomic data and nuclear data tables\/}
  {\bf 35} 345--418

\bibitem{Hubbell_1980}
J.~H. Hubbell, H.~A. Gimm and I.~Overbo 1980 {\em J. Phys. Chem. Ref. Data\/}
  {\bf 9} 1023

\bibitem{IAEAReport_2000}
IAEA-TECDOC-1178 2000 Handbook on photonuclear data for applications:
  cross-sections and spectra

\bibitem{Fano_1963}
U.~Fano 1963 {\em Annu. Rev. Nucl. Part. Sci.\/} {\bf 13} 1--66

\bibitem{Yu_2015}
J.~H. Yu, G.~{De Temmerman}, R.~P. Doerner, R.~A. Pitts et~al. 2015 {\em Nucl.
  Fusion\/} {\bf 55} 093027

\bibitem{Carslaw_1959}
H.~S. Carslaw and J.~C. Jaeger 1959 {\em Conduction Of Heat In Solids\/}
  (Oxford University Press)

\bibitem{Dabby_1972}
F.~Dabby and U.-C. Paek 1972 {\em IEEE J. Quantum Electron.\/} {\bf 8} 106--111

\bibitem{Blackwell_1990}
B.~F. Blackwell 1990 {\em J. Heat Trans.\/} {\bf 112} 567--571

\bibitem{Behling_2025}
R.~Behling, C.~Hulme, P.~Tolias and M.~Danielsson 2025 {\em Med. Phys.\/} {\bf
  52} 814--825

\bibitem{Lockwood_1980}
G.~J. Lockwood, L.~E. Ruggles, G.~H. Miller and J.~A. Halbleib 1980
  {SAND79-0414 Report: Calorimetric measurement of electron energy deposition
  in extended media. Theory vs experiment}

\bibitem{Gilligan_1990}
J.~Gilligan, K.~Niemer, M.~Bourham, C.~Croessmann et~al. 1990 {\em J. Nucl.
  Mater.\/} {\bf 176-177} 779--785

\bibitem{Niemer_1991}
K.~Niemer, J.~Gilligan and C.~Croessmann 1991 {\em [Proceedings] The 14th
  IEEE/NPSS Symposium Fusion Engineering\/} pp 372--376 vol.1

\bibitem{Halbleib_1984}
J.~A. {Halbleib} and T.~A. {Mehlhorn} 1984 {ITS: The Integrated TIGER Series of
  coupled electron/photon Monte Carlo transport codes}

\bibitem{Bolt_1991}
H.~Bolt and H.~Calén 1991 {\em J. Nucl. Mater.\/} {\bf 179-181} 360--363

\bibitem{Bartels_1994}
H.~W. Bartels 1994 {\em Fus. Eng. Des.\/} {\bf 23} 323

\bibitem{Brun_1985}
R.~Brun, L.~Urb{\`a}n, G.~Patrick, M.~Caillat et~al. 1985 {The GEANT3
  electromagnetic shower program and a comparison with the EGS3 code}

\bibitem{Brun_1987}
R.~Brun, F.~Bruyant, M.~Maire, A.~C. McPherson et~al. 1987 {\em {GEANT 3:
  user's guide Geant 3.10, Geant 3.11; rev. version}\/} (Geneva: CERN)

\bibitem{Kunugi_1992}
T.~Kunugi, M.~Akiba, M.~Ogawa, O.~Sato et~al. 1992 {\em Fusion Technol.\/} {\bf
  21} 1868--1872

\bibitem{Kunugi_1993}
T.~Kunugi 1994 {\em Fus. Eng. Des.\/} {\bf 23} 329--339

\bibitem{Nelson_1988}
W.~R. Nelson and D.~W. Rogers 1988 {\em Monte Carlo transport of electrons and
  photons\/} (Springer) pp 287--305

\bibitem{Maddaluno_2003}
G.~Maddaluno, G.~Maruccia, M.~Merola and S.~Rollet 2003 {\em J. Nucl. Mater.\/}
  {\bf 313-316} 651

\bibitem{Ferrari_2005}
A.~Ferrari, J.~Ranft, P.~R. Sala and A.~Fass{\`o} 2005 Fluka: A multi-particle
  transport code (program version 2005)

\bibitem{Ward_2004}
R.~C. Ward and D.~Steiner 2004 {\em Fusion Sci. Technol.\/} {\bf 45} 529--548

\bibitem{Krawrakow_2000}
I.~Krawrakow and D.~W.~O. Rogers 2000 {{The EGSnrc Code System: Monte Carlo
  Simulation of Electron and Photon Transport}}

\bibitem{Bazylev_2011b}
B.~Bazylev, Y.~Igitkhanov, I.~Landman, S.~Pestchanyi et~al. 2011 {\em J. Nucl.
  Mater.\/} {\bf 417} 655--658

\bibitem{Bazylev_2013}
B.~Bazylev, G.~Arnoux, S.~Brezinsek, Y.~Igitkhanov et~al. 2013 {\em J. Nucl.
  Mater.\/} {\bf 438} S237

\bibitem{Sizyuk_2009}
V.~Sizyuk and A.~Hassanein 2009 {\em Nucl. Fusion\/} {\bf 49} 095003

\bibitem{Agostinelli_2003}
S.~Agostinelli, J.~Allison, K.~Amako, J.~Apostolakis et~al. 2003 {\em Nucl.
  Instrum. Meth. Phys. Res. A\/} {\bf 506} 250--303

\bibitem{Allison_2006}
J.~Allison, K.~Amako, J.~Apostolakis, H.~Araujo et~al. 2006 {\em IEEE Trans.
  Nucl. Sci.\/} {\bf 53}

\bibitem{Allison_2016}
J.~Allison, K.~Amako, J.~Apostolakis, P.~Arce et~al. 2016 {\em Nucl. Instrum.
  Meth. Phys. Res. A\/} {\bf 835}

\bibitem{Battistoni_2015}
G.~Battistoni, T.~Boehlen, F.~Cerutti, P.~W. Chin et~al. 2015 {\em Ann. Nucl.
  Energy\/} {\bf 82}

\bibitem{Ahdida_2022}
C.~Ahdida, D.~Bozzato, D.~Calzolari, F.~Cerutti et~al. 2022 {\em Front.
  Phys.\/} {\bf 9}

\bibitem{Dressel_1966}
R.~W. Dressel 1966 {\em Phys. Rev.\/} {\bf 144} 332

\bibitem{Tabata_1967}
T.~Tabata 1967 {\em Phys. Rev.\/} {\bf 162} 336

\bibitem{Ratynskaia_2025a}
S.~Ratynskaia, P.~Tolias, K.~Paschalidis, T.~Rizzi et~al. to be submitted {\em
  Plasma Phys. Control. Fusion\/}

\bibitem{Rizzi_2025a}
T.~Rizzi, S.~Ratynskaia, P.~Tolias et~al. to be submitted {\em Plasma Phys.
  Control. Fusion\/}

\bibitem{Hetnarski_2019}
R.~B. Hetnarski and M.~{Reza Eslami} 2019 {\em {Thermal Stresses - Advanced
  Theory and Applications}\/} (Switzerland: Springer Nature)

\bibitem{Fung_2001}
W.~C. Fung and P.~Tong 2001 {\em {Classical and computational solid
  mechanics}\/} (Singapore: World Scientific Publishing)

\bibitem{Budynas_1999}
R.~G. Budynas 1999 {\em Advanced Strength and Applied Stress Analysis\/} ({New
  York}: {McGraw-Hill})

\bibitem{Scapin_2012}
M.~Scapin, L.~Peroni and A.~Dallocchio 2012 {\em J. Nucl. Mater.\/} {\bf 420}
  463--472

\bibitem{Scapin_2014}
M.~Scapin, L.~Peroni, V.~Boccone and F.~Cerutti 2014 {\em Comput. Struct.\/}
  {\bf 141} 74--83

\bibitem{Kaselouris_2017b}
E.~Kaselouris, V.~Dimitriou, I.~Fitilis, A.~Skoulakis et~al. 2017 {\em Nat.
  Commun.\/} {\bf 8} 1713

\bibitem{Kaselouris_2021}
E.~Kaselouris, G.~Tamiolakis, I.~Fitilis, A.~Skoulakis et~al. 2021 {\em Plasma
  Phys. Control. Fusion\/} {\bf 63} 085010

\bibitem{Scapin_2022}
M.~Scapin and L.~Peroni 2022 {\em Metals\/} {\bf 12} 670

\bibitem{Kerley_2003}
G.~I. Kerley 2003 {Equations of State for Be, Ni, W, and Au}

\bibitem{Johnson_1983}
G.~R. Johnson and W.~H. Cook 1983 A constitutive model and data for metals
  subjected to large strains, high strain rates and high temperatures

\bibitem{Zerilli_1987}
F.~J. Zerilli and R.~W. Armstrong 1987 {\em J. Appl. Phys.\/} {\bf 61}
  1816--1825

\bibitem{Lennon_2000}
A.~M. Lennon and K.~T. Ramesh 2000 {\em Mater. Sci. Eng.\/} {\bf A276} 9--21

\bibitem{Belytschko_2014}
T.~Belytschko, W.~K. Liu, B.~Moran and K.~Elkhodary 2014 {\em Nonlinear Finite
  Elements for Continua and Structures\/} ({New York}: {John Wiley \& Sons})

\bibitem{Halquist_2006}
J.~O. Hallquist 2006 {LS-DYNA Theory Manual}

\bibitem{Song_2008}
J.-H. Song, H.~Wang and T.~Belytschko 2008 {\em Comput. Mech.\/} {\bf 42}
  239--250

\bibitem{Dimitriou_2013}
V.~Dimitriou, E.~Kaselouris, Y.~Orphanos, M.~Bakarezos et~al. 2013 {\em Appl.
  Phys. Lett.\/} {\bf 103} 114104

\bibitem{Dimitriou_2015}
V.~Dimitriou, E.~Kaselouris, Y.~Orphanos, M.~Bakarezos et~al. 2015 {\em Appl.
  Phys. A\/} {\bf 118} 739--748

\bibitem{Kaselouris_2016}
E.~Kaselouris, I.~K. Nikolos, Y.~Orphanos, M.~Bakarezos et~al. 2016 {\em Int.
  J. Damage Mech.\/} {\bf 25} 42--55

\bibitem{Apostolova_2021}
T.~Apostolova, J.~Kohanoff, N.~Medvedev, E.~Oliva et~al. 2021 {\em Tools for
  investigating electronic excitation: experiment and multi-scale modelling\/}
  ({Universidad Politécnica de Madrid}: {Instituto de Fusión Nuclear
  "Guillermo Velarde"})

\bibitem{Liu_2003}
G.~Liu and M.~Liu 2003 {\em Smoothed Particle Hydrodynamics: a Meshfree
  Particle Method\/} (World Scientific, Singapore)

\bibitem{Messahel_2013}
R.~Messahel and M.~Souli 2013 {\em Comput. Model. Eng. Sci.\/} {\bf 96}(6)
  435--455

\bibitem{Richter_2017}
T.~Richter 2017 {\em {Lecture Notes in Computational Science and Engineering:
  Fluid-structure Interactions}\/} ({Cham, Switzerland}: {Springer
  International Publishing})

\bibitem{Kaselouris_2017}
E.~Kaselouris, T.~Papadoulis, E.~Variantza, A.~Baroutsos et~al. 2017 {\em Solid
  State Phenom.\/} {\bf 261} 339

\bibitem{Wu_2020}
J.-Y. Wu, V.~P. Nguyen, C.~T. Nguyen, D.~Sutula et~al. 2020 {\em Adv. Appl.
  Mech.\/} {\bf 53} 1--183

\bibitem{Silling_2010}
S.~A. Silling and R.~B. Lehoucq 2010 {\em Adv. Appl. Mech.\/} {\bf 44} 73--168

\bibitem{Madenci_2014}
E.~Madenci and E.~Oterkus 2014 {\em Peridynamic Theory and Its Applications\/}
  (Springer, New York)

\bibitem{Klinkov_2005}
S.~Klinkov, V.~Kosarev and M.~Rein 2005 {\em Aerospace Sci. Technol.\/} {\bf 9}
  582--591

\bibitem{Hassani_2018}
M.~Hassani-Gangaraj, D.~Veysset, K.~A. Nelson and C.~A. Schuh 2018 {\em Scr.
  Mater.\/} {\bf 145} 9--13

\bibitem{Tolias_2023}
P.~Tolias, M.~{De Angeli}, D.~Ripamonti, S.~Ratynskaia et~al. 2023 {\em Fus.
  Eng. Des.\/} {\bf 195} 113938

\bibitem{DeAngeli_2024}
M.~{De Angeli}, P.~Tolias, F.~Suzuki-Vidal, D.~Ripamonti et~al. 2024 {\em Nucl.
  Mater. Energy\/} {\bf 41} 101735

\bibitem{Eichhorn_1976}
G.~Eichhorn 1976 {\em Planet. Space Sci.\/} {\bf 24} 771--781

\bibitem{Burchell_1999}
M.~J. Burchell, M.~J. Cole, J.~A.~M. McDonnell and J.~C. Zarnecki 1999 {\em
  Meas. Sci. Technol.\/} {\bf 10} 41

\bibitem{Ratynskaia_2008}
S.~Ratynskaia, C.~Castaldo, K.~Rypdal, G.~Morfill et~al. 2008 {\em Nucl.
  Fusion\/} {\bf 48} 015006

\bibitem{Fraile_2022}
A.~Fraile, P.~Dwivedi, G.~Bonny and T.~Polcar 2022 {\em Nucl. Fusion\/} {\bf
  62} 026034

\bibitem{Veysset_2021}
D.~Veysset, J.-H. Lee, M.~Hassani, S.~E. Kooi et~al. 2021 {\em Appl. Phys.
  Rev.\/} {\bf 8} 011319

\bibitem{Amani_2016}
J.~Amani, E.~Oterkus, P.~Areias, G.~Zi et~al. 2016 {\em Int. J. Impact Eng.\/}
  {\bf 87} 83

\bibitem{Ren_2024}
B.~Ren and J.~Song 2024 {\em Surf. Coat. Technol.\/} {\bf 493} 131257

\bibitem{Libersky_1997}
L.~D. Libersky, P.~W. Randles, T.~C. Carney and D.~L. Dickinson 1997 {\em Int.
  J. Impact Eng.\/} {\bf 20} 525--532

\bibitem{Remington_2020}
T.~P. Remington, J.~M. Owen, A.~M. Nakamura, P.~L. Miller et~al. 2020 {\em
  Earth Space Sci.\/} {\bf 7} e2018EA000474

\bibitem{Artola_2024}
F.~Artola et~al. 2024 Thermal loads in unmitigated disruptions 44th ITPA MDC
  meeting, Gandhinagar, India

\bibitem{1980203}
T.~T. Group 1980 {\em Journal of Nuclear Materials\/} {\bf 93-94} 203--209
  \urlprefix\url{https://www.sciencedirect.com/science/article/pii/0022311580903232}

\bibitem{Barnes_1981}
C.~Barnes, J.~Stavely and J.~Strachan 1981 {\em Nuclear Fusion\/} {\bf 21} 1469
  \urlprefix\url{https://dx.doi.org/10.1088/0029-5515/21/11/011}

\bibitem{SUKEGAWA20181653}
A.~M. Sukegawa, K.~Okuno and S.~Tanaka 2018 {\em Fusion Engineering and
  Design\/} {\bf 136} 1653--1657
  \urlprefix\url{https://www.sciencedirect.com/science/article/pii/S0920379618305635}

\bibitem{TechReport_2023_LANL_LA-UR-22-33103Rev.1_RisingArmstrongEtAl}
M.~E. Rising, J.~C. Armstrong, S.~R. Bolding, F.~B. Brown et~al. 2023
  {MCNP\textsuperscript{\textregistered} Code Version 6.3.0 Release Notes}
  Tech. Rep. LA-UR-22-33103, Rev.~1 Los Alamos National Laboratory Los Alamos,
  NM, USA \urlprefix\url{https://www.osti.gov/biblio/1909545}

\bibitem{HUGOT2024}
F.-X. Hugot, A.~Jinaphanh, C.~Jouanne, C.~Larmier et~al. 2024 {\em EPJ -
  Nuclear Sciences \& Technologies\/} {\bf 10}
  \urlprefix\url{https://www.sciencedirect.com/science/article/pii/S2491929224000165}

\bibitem{Ataeiseresht2023}
L.~Ataeisereht, M.~R. Abdi, B.~Pourshahab and C.~Rasouli 2023 {\em Nature
  Scientific Reports\/} {\bf 13}
  \urlprefix\url{https://doi.org/10.1038/s41598-023-48672-7}

\bibitem{VALENZA2001411}
D.~Valenza, H.~Iida, R.~Plenteda and R.~T. Santoro 2001 {\em Fusion Engineering
  and Design\/} {\bf 55} 411--418
  \urlprefix\url{https://www.sciencedirect.com/science/article/pii/S0920379601001880}

\bibitem{CHEN2002107}
Y.~Chen and U.~Fischer 2002 {\em Fusion Engineering and Design\/} {\bf 63-64}
  107--114
  \urlprefix\url{https://www.sciencedirect.com/science/article/pii/S0920379602001448}

\bibitem{Polevoi_2021}
A.~Polevoi and {et al} 2021 {\em 28th IAEA Fusion Energy Conference -
  IAEA-CN-TH/P2-8\/}

\bibitem{Causa_2015}
F.~Causa, P.~Buratti, B.~Esposito, G.~Pucella et~al. 2015 {\em Nuclear
  Fusion\/} {\bf 55} 123021
  \urlprefix\url{https://dx.doi.org/10.1088/0029-5515/55/12/123021}

\bibitem{Ericsson_2019}
G.~Ericsson 2019 {\em Journal of Fusion Energy\/} {\bf 38} 330--355

\bibitem{Stancar_2021}
Z.~Štancar, Z.~Ghani, J.~Eriksson, A.~Žohar et~al. 2021 {\em Nuclear
  Fusion\/} {\bf 61} 126030
  \urlprefix\url{https://dx.doi.org/10.1088/1741-4326/ac3021}

\bibitem{CHADWICK20112887}
M.~Chadwick, M.~Herman, P.~Obložinský, M.~Dunn et~al. 2011 {\em Nuclear Data
  Sheets\/} {\bf 112} 2887--2996
  \urlprefix\url{https://www.sciencedirect.com/science/article/pii/S009037521100113X}

\bibitem{BROWN20181}
D.~Brown, M.~Chadwick, R.~Capote, A.~Kahler et~al. 2018 {\em Nuclear Data
  Sheets\/} {\bf 148} 1--142
  \urlprefix\url{https://www.sciencedirect.com/science/article/pii/S0090375218300206}

\bibitem{Iwamoto02012023}
O.~Iwamoto, N.~Iwamoto, S.~Kunieda, F.~Minato et~al. 2023 {\em Journal of
  Nuclear Science and Technology\/} {\bf 60} 1--60
  \urlprefix\url{https://doi.org/10.1080/00223131.2022.2141903}

\bibitem{KAWANO2020109}
T.~Kawano, Y.~Cho, P.~Dimitriou, D.~Filipescu et~al. 2020 {\em Nuclear Data
  Sheets\/} {\bf 163} 109--162
  \urlprefix\url{https://www.sciencedirect.com/science/article/pii/S0090375219300699}

\bibitem{KONING20191}
A.~Koning, D.~Rochman, J.-C. Sublet, N.~Dzysiuk et~al. 2019 {\em Nuclear Data
  Sheets\/} {\bf 155} 1--55
  \urlprefix\url{https://www.sciencedirect.com/science/article/pii/S009037521930002X}

\bibitem{Tuyet01022024}
T.~K. Tuyet, A.~Jinaphanh, C.~Jouanne, F.~Gérardin et~al. 2024 {\em Nuclear
  Science and Engineering\/} {\bf 198} 319--335
  \urlprefix\url{https://doi.org/10.1080/00295639.2023.2195925}

\bibitem{SAUVAN2020111399}
P.~Sauvan, R.~Juárez, G.~Pedroche, J.~Alguacil et~al. 2020 {\em Fusion
  Engineering and Design\/} {\bf 151} 111399
  \urlprefix\url{https://www.sciencedirect.com/science/article/pii/S0920379619308956}

\bibitem{VILLARI20142083}
R.~Villari, U.~Fischer, F.~Moro, P.~Pereslavtsev et~al. 2014 {\em Fusion
  Engineering and Design\/} {\bf 89} 2083--2087
  \urlprefix\url{https://www.sciencedirect.com/science/article/pii/S0920379614000726}

\bibitem{Carnevale_2019}
D.~Carnevale, M.~Ariola, G.~Artaserse, F.~Bagnato et~al. 2018 {\em Plasma
  Physics and Controlled Fusion\/} {\bf 61} 014036
  \urlprefix\url{https://dx.doi.org/10.1088/1361-6587/aaef53}

\bibitem{Loarte_2025}
A.~Loarte submitted {\em Plasma Physics and Controlled Fusion\/}

\bibitem{Jachmich_2024}
S.~Jachmich 2024 {\em Proceedings of the 50th EPS Conference on Plasma Physics,
  8-12 July, Salamanca\/}

\bibitem{Vallhagen_2020}
O.~Vallhagen, O.~Embreus, I.~Pusztai, L.~Hesslow et~al. 2020 {\em Journal of
  Plasma Physics\/} {\bf 86} 475860401

\bibitem{Artola_2022}
F.~Artola, A.~Loarte, M.~Hoelzl, M.~Lehnen et~al. 2022 {\em Nuclear Fusion\/}
  {\bf 62} 056023

\bibitem{sarkimaki_2022}
K.~S{\"a}rkim{\"a}ki, J.~Artola, M.~Hoelzl, J.~Team et~al. 2022 {\em Nuclear
  Fusion\/} {\bf 62} 086033

\bibitem{Nardon_2022}
E.~Nardon, A.~Matsuyama, D.~Hu and F.~Wieschollek 2021 {\em Nuclear Fusion\/}
  {\bf 62} 026003 \urlprefix\url{https://dx.doi.org/10.1088/1741-4326/ac3ac6}

\bibitem{Nardon_2020}
E.~Nardon, D.~Hu, M.~Hoelzl, D.~Bonfiglio et~al. 2020 {\em Nuclear Fusion\/}
  {\bf 60} 126040 \urlprefix\url{https://dx.doi.org/10.1088/1741-4326/abb749}

\bibitem{Vallhagen_2022}
O.~Vallhagen, I.~Pusztai, M.~Hoppe, S.~Newton et~al. 2022 {\em Nuclear
  Fusion\/} {\bf 62} 112004
  \urlprefix\url{https://dx.doi.org/10.1088/1741-4326/ac667e}

\bibitem{lukash2013study}
V.~Lukash, A.~Kavin, Y.~Gribov, R.~Khayrutdinov et~al. 2013 {\em 40th EPS Conf.
  on Plasma Physics\/} pp P5--167

\bibitem{Sheikh_2024b}
U.~Sheikh, S.~Jachmich, G.~Bodner, J.~Decker et~al. 2024 {\em Third IAEA
  Technical Meeting on Plasma Disruptions and their Mitigation, 3-6
  September\/}
  \urlprefix\url{https://inis.iaea.org/records/5a0r0-4bh93/files/55090923.pdf}

\bibitem{Aumeunier_2024}
M.-H. Aumeunier, A.~Juven, J.~Gerardin, C.-M.~B. Cisse et~al. 2024 {\em Nuclear
  Fusion\/} {\bf 64} 086044
  \urlprefix\url{https://dx.doi.org/10.1088/1741-4326/ad5a1f}

\bibitem{Pandya_2018}
S.~P. Pandya, L.~Core, R.~Barnsley, J.~Rosato et~al. 2018 {\em Physica
  Scripta\/} {\bf 93} 115601
  \urlprefix\url{https://dx.doi.org/10.1088/1402-4896/aaded0}

\bibitem{Patel_2023}
A.~Patel, S.~P. Pandya, A.~E. Shevelev, E.~M. Khilkevitch et~al. 2023 {\em
  Physica Scripta\/} {\bf 98} 085604
  \urlprefix\url{https://dx.doi.org/10.1088/1402-4896/ace135}

\bibitem{Nocente_2017}
M.~Nocente, M.~Tardocchi, R.~Barnsley, L.~Bertalot et~al. 2017 {\em Nuclear
  Fusion\/} {\bf 57} 076016
  \urlprefix\url{https://dx.doi.org/10.1088/1741-4326/aa6f7d}

\bibitem{Esposito_2022}
B.~Esposito, D.~Marocco, G.~Gandolfo, F.~Belli et~al. 2022 {\em Journal of
  Fusion Energy\/} {\bf 41} 22

\bibitem{Helander_2011}
P.~Helander, J.~Geiger and H.~Maaßberg 2011 {\em Physics of Plasmas\/} {\bf
  18} 092505 \urlprefix\url{https://doi.org/10.1063/1.3633940}

\bibitem{Garnier_2006}
D.~Garnier, A.~Hansen, J.~Kesner, M.~Mauel et~al. 2006 {\em Fusion Engineering
  and Design\/} {\bf 81} 2371--2380
  \urlprefix\url{https://www.sciencedirect.com/science/article/pii/S0920379606001876}

\bibitem{Tang_2021}
T.~Tang, L.~Zeng, D.~Chen, Y.~Sun et~al. 2021 {\em Nuclear Fusion\/} {\bf 61}
  076003 \urlprefix\url{https://dx.doi.org/10.1088/1741-4326/abf62f}

\bibitem{Siccinio_2022}
M.~Siccinio, J.~Graves, R.~Kembleton, H.~Lux et~al. 2022 {\em Fusion
  Engineering and Design\/} {\bf 176} 113047
  \urlprefix\url{https://www.sciencedirect.com/science/article/pii/S0920379622000473}

\bibitem{Maviglia_2022}
F.~Maviglia, C.~Bachmann, G.~Federici, T.~Franke et~al. 2022 {\em Fusion
  Engineering and Design\/} {\bf 178} 113125
  \urlprefix\url{https://www.sciencedirect.com/science/article/pii/S0920379622001259}

\bibitem{Martin-Solis_2017b}
J.~R. Martín-Solís 2017 {\em Report\/}
  \urlprefix\url{https://idm.euro-fusion.org/?uid=2N49CW}

\bibitem{Martin-Solis_2018}
J.~R. Martín-Solís 2018 {\em Report\/}
  \urlprefix\url{https://idm.euro-fusion.org/?uid=2NFCXH}

\bibitem{Pautasso_2019}
G.~Pautasso, E.~Fble, F.~Koechl, P.~Lang et~al. 2019 {\em 46th EPS Conference
  on Plasma Physics, Poster P4.1045\/}

\bibitem{You_2022}
J.~You, C.~Bachmann, V.~Belardi, M.~Binder et~al. 2022 {\em Fusion Engineering
  and Design\/} {\bf 174} 112988
  \urlprefix\url{https://www.sciencedirect.com/science/article/pii/S0920379621007638}

\bibitem{Richiusa_2025}
M.~Richiusa, J.~Lyytinen, A.~Sinha and G.~Spagnuolo 2025 {\em Fusion
  Engineering and Design\/} {\bf 214} 114890
  \urlprefix\url{https://www.sciencedirect.com/science/article/pii/S0920379625000924}

\bibitem{Ramogida_2015}
G.~Ramogida, G.~Maddaluno, F.~Villone, R.~Albanese et~al. 2015 {\em Fusion
  Engineering and Design\/} {\bf 96-97} 348--352
  \urlprefix\url{https://www.sciencedirect.com/science/article/pii/S0920379615301836}

\bibitem{Lombroni_2021}
R.~Lombroni, F.~Giorgetti, G.~Calabrò, P.~Fanelli et~al. 2021 {\em Fusion
  Engineering and Design\/} {\bf 170} 112697
  \urlprefix\url{https://www.sciencedirect.com/science/article/pii/S0920379621004737}

\bibitem{Firdaouss_2013}
M.~Firdaouss, V.~Riccardo, V.~Martin, G.~Arnoux et~al. 2013 {\em Journal of
  Nuclear Materials\/} {\bf 438} S536--S539
  \urlprefix\url{https://www.sciencedirect.com/science/article/pii/S0022311513001190}

\bibitem{Wayne_2015}
W.~Arter, E.~Surrey and D.~B. King 2015 {\em IEEE Transactions on Plasma
  Science\/} {\bf 43} 3323--3331

\bibitem{Maddaluno_2019}
G.~Maddaluno 2019 {\em Report\/}
  \urlprefix\url{https://idm.euro-fusion.org/?uid=2M9Y9Y}

\bibitem{Singh_2021}
L.~Singh October 2021 {\em Master of Science in Energy and Nuclear
  Engineering\/}

\bibitem{Pokol_2024}
G.~Pokol 2024 {\em Report\/}

\bibitem{Hartwig_2024}
Z.~S. Hartwig, R.~F. Vieira, D.~Dunn, T.~Golfinopoulos et~al. 2024 {\em IEEE
  Transactions on Applied Superconductivity\/} {\bf 34} 1--16

\bibitem{Creely_2020}
A.~J. Creely, M.~J. Greenwald, S.~B. Ballinger, D.~Brunner et~al. 2020 {\em
  Journal of Plasma Physics\/} {\bf 86} 865860502

\bibitem{Eidietis_2015}
N.~Eidietis, S.~Gerhardt, R.~Granetz, Y.~Kawano et~al. 2015 {\em Nuclear
  Fusion\/} {\bf 55} 063030
  \urlprefix\url{https://dx.doi.org/10.1088/0029-5515/55/6/063030}

\bibitem{Tinguely_2021}
R.~Tinguely, V.~Izzo, D.~Garnier, A.~Sundström et~al. 2021 {\em Nuclear
  Fusion\/} {\bf 61} 124003
  \urlprefix\url{https://dx.doi.org/10.1088/1741-4326/ac31d7}

\bibitem{Ekmark_2025}
I.~Ekmark, M.~Hoppe, R.~A. Tinguely, R.~Sweeney et~al. 2025 Runaway electron
  generation in disruptions mitigated by deuterium and noble gas injection in
  {SPARC} (\textit{Preprint} \eprint{2502.19891})
  \urlprefix\url{https://arxiv.org/abs/2502.19891}

\bibitem{Tinguely_2023}
R.~A. Tinguely, I.~Pusztai, V.~A. Izzo, K.~Särkimäki et~al. 2023 {\em Plasma
  Physics and Controlled Fusion\/} {\bf 65} 034002
  \urlprefix\url{https://dx.doi.org/10.1088/1361-6587/acb083}

\bibitem{Feyrer_2024}
A.~Feyrer, T.~Looby, R.~Sweeney and R.~Tinguely 2024 {\em Bulletin of the
  American Physical Society\/}

\bibitem{Looby_2022}
T.~Looby, M.~Reinke, A.~Wingen, J.~Menard et~al. 2022 {\em Fusion Science and
  Technology\/} {\bf 78} 10--27
  \urlprefix\url{https://doi.org/10.1080/15361055.2021.1951532}

\bibitem{Reinke_2024}
M.~L. Reinke, I.~Abramovic, A.~Albert, K.~Asai et~al. 2024 {\em Review of
  Scientific Instruments\/} {\bf 95} 103518
  \urlprefix\url{https://doi.org/10.1063/5.0218254}

\bibitem{KTH_2025}
{KTH Team} et~al. 2025 Modeling of runaway electron impacts on tungsten plasma
  facing components in {SPARC}

\bibitem{Emanuelli_2025}
E.~Emanuelli, F.~Vannini, M.~Hoelzl, N.~Schwarz et~al. submitted {\em Fusion
  Engineering and Design\/}

\bibitem{Fil_2024}
A.~Fil, L.~Henden, S.~Newton, M.~Hoppe et~al. 2024 {\em Nuclear Fusion\/} {\bf
  64} 106049 \urlprefix\url{https://dx.doi.org/10.1088/1741-4326/ad73e9}

\bibitem{DOE_Milestone_2023}
{Fusion Industry Association} 2023 Department of energy announces milestone
  public-private partnership awards
  \url{https://www.fusionindustryassociation.org/department-of-energy-announces-milestone-public-private-partnership-awards/}

\bibitem{DOE_Vision_2024}
{US Department of Energy} 2024 Doe announces new decadal fusion energy strategy
  \url{https://www.energy.gov/articles/doe-announces-new-decadal-fusion-energy-strategy}

\bibitem{CFS-ARC}
{Commonwealth Fusion Systems} 2024 Commonwealth fusion systems to build
  world’s first commercial fusion power plant in virginia
  \url{https://cfs.energy/news-and-media/commonwealth-fusion-systems-to-build-worlds-first-commercial-fusion-power-plant-in-virginia}

\bibitem{Hillesheim_2024}
J.~Hillesheim, A.~Creely, D.~Battaglia, T.~Body et~al. 2024 {\em Bulletin of
  the American Physical Society\/}

\bibitem{ARC-Physics-Basis}
J.~Hillesheim et~al. 2025 {ARC Physics Basis}

\end{thebibliography}
}

\end{document}